\documentclass[SIG,biber]{nowfnt}   

\usepackage[utf8]{inputenc}
\usepackage{booktabs}
\usepackage{url}
\usepackage{amstext}
\usepackage{amssymb}
\usepackage{graphicx}
\usepackage{tikz}
\usepackage{quantikz}

\theoremstyle{definition}

\title{An Introduction to Quantum Machine Learning for Engineers}

\maintitleauthorlist{
}

\issuesetup
{%
 copyrightowner={...},
 volume        = xx,
 issue         = xx,
 pubyear       = 2022,
 isbn          = xxx-x-xxxxx-xxx-x,
 eisbn         = xxx-x-xxxxx-xxx-x,
 doi           = 10.1561/XXXXXXXXX,
 firstpage     = 1, 
 lastpage      = 18
 }

\usepackage{mwe}

\author[1]{Osvaldo Simeone}

\affil[1]{King's College London}

\articledatabox{\nowfntstandardcitation}

\begin{document}

\makeabstracttitle

\begin{abstract}
In the current noisy intermediate-scale quantum (NISQ) era, quantum machine learning is emerging as a dominant paradigm to program gate-based quantum computers. In quantum machine learning, the gates of a quantum circuit are parametrized, and the parameters are tuned via classical optimization based on data and on measurements of the outputs of the circuit. Parametrized quantum circuits (PQCs) can efficiently address combinatorial optimization problems, implement probabilistic generative models, and carry out inference (classification and regression). This monograph provides a self-contained introduction to quantum machine learning for an audience of engineers with a background in probability and linear algebra. It first describes the necessary background, concepts, and tools necessary to describe quantum operations and measurements. Then, it covers parametrized quantum circuits, the variational quantum eigensolver, as well as unsupervised and supervised quantum machine learning formulations.

\end{abstract}

\chapter*{Preface}

\section*{Motivation}
As many engineers, I have developed an early fascination for quantum theory -- for its history, its counterintuitive predictions, its central role in the development of many existing technologies (semiconductors, lasers, MRI, atomic clocks) and, perhaps above all, its promise to unlock future, revolutionary, paradigms in materials, chemical, industrial, computer, and  communication engineering. 

At first, the topic is inviting for an engineer with my background on electrical and information engineering: The mathematical formalism is familiar, based as it is on linear algebra and probability; and concepts with wide-ranging and intriguing implications, such as superposition and entanglement, can be easily described on paper. Spend more time with it, however, and the field reveals its complexity, becoming for many, the former me included,  too abstruse to invite further study. Particularly  unfamiliar are ideas and architectures underlying key quantum algorithms, such as Shor's factorization method. As if that was not enough, the impressions that most algorithmic breakthroughs are by now textbook material, and that all the ``action'' is currently focused on scaling hardware implementations, have kept me from engaging with the state of the art on quantum computing.

This monograph is motivated by a number of recent developments that appear to define a possible new role for researchers with an engineering profile similar to mine. First, there are now several software libraries -- such as IBM's Qiskit, Google's Cirq, and Xanadu's PennyLane -- that make programming quantum algorithms more accessible, while also providing cloud-based access to actual quantum computers. Second, a new framework is emerging for programming quantum algorithms to be run on current quantum hardware: quantum machine learning.

\section*{Quantum Machine Learning}

Quantum computing algorithms have been traditionally designed by hand
assuming the availability of \textbf{fault-tolerant quantum processors} that can reliably support a large number of \textbf{qubits} and quantum operations, also known as \textbf{quantum gates}. A qubit is the basic unit of quantum information and computing, playing the role of a bit in classical computers.  In practice, current  quantum computers implement a few tens of qubits, with quantum gates that are inherently imperfect and noisy. \textbf{Quantum machine learning} refers to an emerging, alternative design paradigm that is tailored for  current \textbf{noisy intermediate-scale quantum (NISQ)} computers. The approach follows a two-step methodology akin to classical machine learning. In it, one first fixes a priori a, possibly generic, parametrized architecture for the quantum gates defining a quantum algorithm, and then uses classical optimization to tune the parameters of the gates. 

\begin{figure}
	\begin{center}
		\includegraphics[clip,scale=0.30]{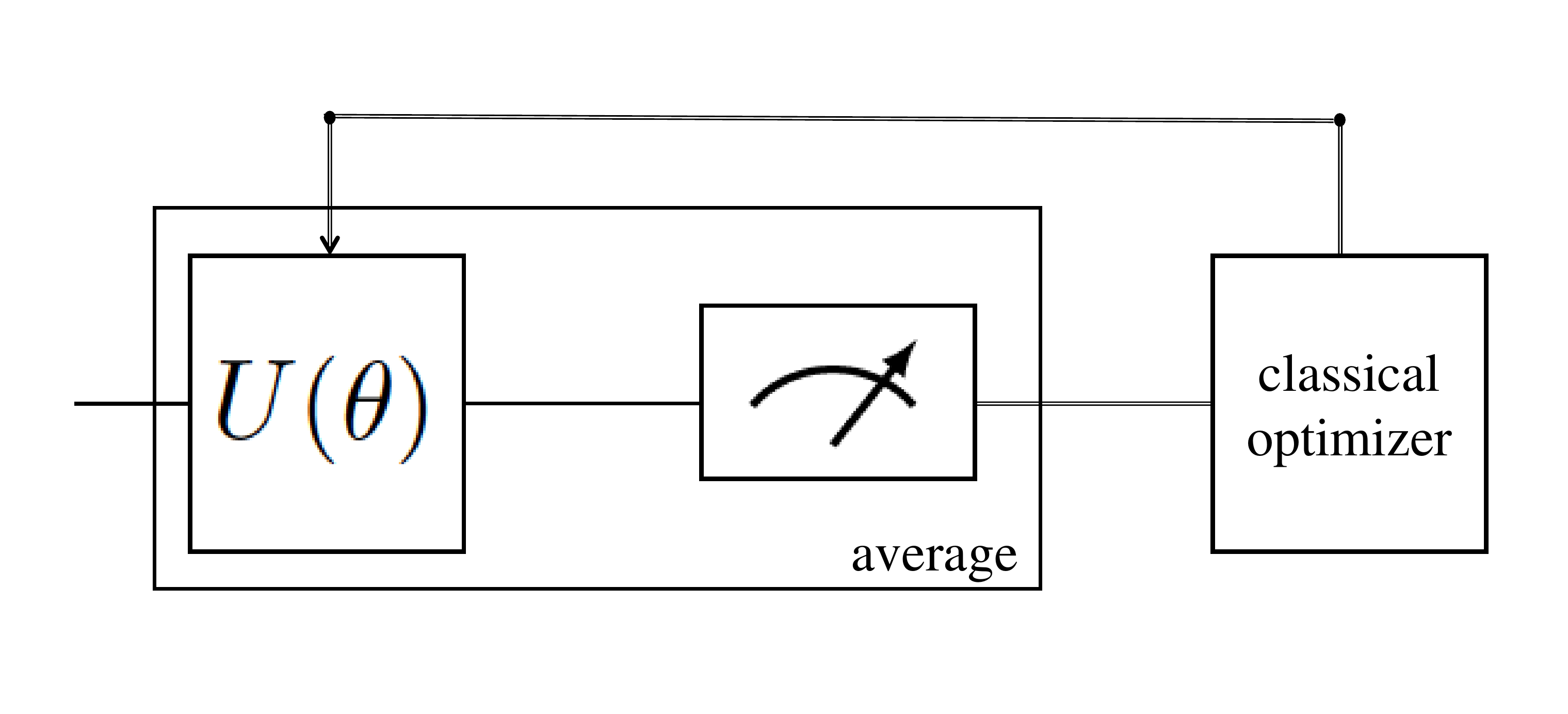}
		\par\end{center}\vspace{-0.4cm}\caption{Illustration of the quantum machine learning design methodology: A parametrized quantum circuit with a pre-specified architecture is optimized via its vector of parameters, $\theta$, by a classical optimizer based on data and measurements of its outputs. As we will see in this monograph, the operation of a parametrized quantum circuit is defined by a unitary matrix $U(\theta)$ dependent on vector $\theta$. The block marked with a gauge sign represents quantum measurements, which convert quantum information produced by the quantum circuit into classical information. This conversion is inherently random, and measurement outputs are typically averaged before being fed to the classical optimizer. \label{fig:ch1:qmlintro}}
\end{figure}

In more detail, as sketched in Fig. \ref{fig:ch1:qmlintro}, in quantum machine learning,  the quantum algorithm is defined by a \textbf{quantum circuit} -- denoted as $U(\theta)$ in the figure -- whose constituent quantum gates implement operations that depend on a vector $\theta$ of free parameters. Measurements of the quantum state produced by the quantum circuit produce classical information that is fed to a classical processor, along with data. The classical optimizer produces updates to the vector $\theta$ with the goal of minimizing some designer-specified cost function. 

The quantum machine learning architecture of Fig. \ref{fig:ch1:qmlintro} has a number of potential advantages over the traditional approach of handcrafting quantum algorithms assuming fault-tolerant quantum computers: \begin{itemize} 
	\item By keeping the quantum computer in the loop, the classical optimizer 
	can directly account for the non-idealities and limitations of quantum operations via measurements of the output of the quantum computer.
	\item If the parametrized quantum algorithm is sufficiently flexible and the classical optimizer sufficiently effective, the approach may automatically design well-performing quantum algorithms that would have been hard to optimize by hand via traditional formal methods.
\end{itemize}

Quantum machine learning, intended as the study of applications of parametrized quantum circuits, is distinct from the related topic of \emph{quantum-aided} classical machine learning. The aim of this older line of work is to speed up classical machine learning methods  by leveraging traditional quantum computing subroutines. This monograph will focus solely on quantum machine learning as illustrated in Fig. \ref{fig:ch1:qmlintro}.

Important open research questions in the field of quantum machine learning are discussed at the end of this text.  It is my hope that researchers who may not have otherwise contributed to these research directions would be motivated to do so upon reading these pages.

\section*{Goal and Organization}

The main goal of this monograph is to present a self-contained introduction to quantum information processing and quantum machine learning for a readership of engineers with a background in linear algebra and probability. My ambition in presenting this text is to offer a resource that may allow more researchers with no prior exposure to quantum theory to contribute to the field of quantum machine learning with new ideas and methods.

The monograph is written as a textbook, with no references except for end-of-chapter sections. References are kept to a minimum, and are mostly limited to books that the reader may peruse for additional information on different topics introduced in these pages. I have also included problems at the end of each chapter with the main aims of reviewing some key ideas described in the text and of inviting the reader to explore topics beyond this monograph.

It may be worth emphasizing that the text is meant to be read sequentially, as I have attempted to introduce notations and concepts progressively from the first page to the last page. 

The monograph does not include discussions about specific applications and use cases. There are several reasons for this. First, many applications are domain specific, pertaining fields like quantum chemistry, and are deemed to be outside the scope of this text, which focuses on concepts and tools. Second, many existing generic tasks and data sets currently used in the quantum machine learning literature are quite simplistic, and they arguably yield little insight into the potential of the technology. The reader is referred to research papers, appearing on a daily basis on repositories like arXiv, for up-to-date results, including new benchmarks and experiments.

The rest of the text is organized in the following chapters.

\textbf{Chapter 1. Classical bit (cbit) and quantum bit (qubit)}: This chapter introduces the concept of qubit through an algebraic generalization of random classical bits (cbits). A qubit can evolve in quantum systems via reversible linear (unitary) transformations -- also known as quantum gates -- or via measurements. The mathematical formalism underlying the description of both quantum gates and measurements is also covered in the chapter. Finally, the chapter illustrates a key difference in the behavior of random cbits and qubits, namely the phenomenon of interference. 

\textbf{Chapter 2. Classical bits (cbits) and quantum bits (qubits)}: This chapter extends the concepts introduced in the previous chapter, including quantum gates and measurements, to systems comprising multiple qubits. The new phenomenon of entanglement -- a form of correlation between quantum systems with no classical counterpart -- is introduced, and superdense coding is presented as an application of entanglement.

\textbf{Chapter 3. Generalizing quantum measurements (Part I)}: The third chapter presents two important generalizations of quantum measurements, namely measurements in an arbitrary basis and non-selective measurements. Decoherence, density matrices, and partial trace are also presented as concepts arising naturally from the introduction non-selective measurements.

\textbf{Chapter 4. Quantum computing}: Chapter 4 presents a brief introduction to the traditional approach for the design of quantum algorithms in gate-based quantum computers. This presentation culminates in the description of  Deutsch's algorithm, the first example of a quantum solution that can provably improve over classical algorithms. The chapter also describes the no cloning theorem, which sets important constraints on the design of quantum computing algorithms. 

\textbf{Chapter 5. Generalizing quantum measurements (Part II)}: This chapter presents two further extensions of quantum measurements: projective measurements and positive operator-valued measurements (POVMs). POVMs represent the most general form of quantum measurement. As an example of the application of projective measurements, the problem of quantum error correction is briefly introduced; while unambiguous state detection is presented as technique enabled by POVMs.  Observables are covered, and the chapter ends with a description of quantum channels as non-selective quantum measurements.

\textbf{Chapter 6. Quantum machine learning}: The final chapter provides an introduction to quantum machine learning that builds on the material covered in the previous chapters. After a description of the taxonomy of quantum machine learning methods, the concepts of parametrized quantum circuits and ansatz are introduced, along with the definition of cost functions used in quantum machine learning. These are leveraged to describe the variational quantum eigensolver (VQE), as well as unsupervised and supervised learning strategies for settings in which data are classical and processing is quantum. An outlook is also provided pointing to more advanced techniques and directions for research.

\chapter{Classical Bit (Cbit) and Quantum Bit (Qubit)}

\section{Introduction}

	This chapter introduces the qubit as the basic unit of quantum information and computing. To this end, we start by reviewing classical bits (cbits) and random cbits. Then, we present the qubit as an algebraic ``extension'' of the two-dimensional probability
	distribution of a random cbit to a complex, normalized, two-dimensional vector. The chapter will also present the two main ways in which a qubit can evolve over time: unitary operations, also known as quantum gates, and measurements. While quantum gates describe the evolution of a qubit in a closed system, measurements convert quantum information to classical information by coupling the qubit with a measurement instrument.

\section{Random Classical Bit}

We start by introducing some basic notation and key ideas used in quantum theory by considering the reference case of a classical bit.

\subsection{A Classical Bit as a One-Hot Vector}\label{eq:ch1:cbitoh}
A \textbf{classical bit (cbit)} is a system
		that can be in two \emph{unambiguously distinguishable, or orthogonal}, levels.  Examples include on-off switches and up-down magnets. Mathematically, the \textbf{state} of a cbit is represented by a logical binary digit taking value $0$ or $1$. 
		
		Alternatively, the state of a cbit can be described by  a two-dimensional \textbf{``one-hot'' amplitude vector}.  As illustrated in Table \ref{table:ch1:first}, the one-hot amplitude vector contains a single ``1'' digit, whose position indicates whether the cbit takes value $0$ or $1$. Specifically, a cbit taking value $0$ is encoded by a one-hot vector with a ``1'' digit in the first position, while a cbit taking value $1$ is encoded by one-hot vector with a ``1'' digit in the second position.
		  
	\begin{table}
\begin{center}
	{\small{}}%
	\begin{tabular}{cc}
		\toprule 
		{\small{}cbit} & {\small{}amplitude vector}\tabularnewline
		\midrule
		\midrule 
		{\small{}0} & {\small{}$|0\rangle=\left[\begin{array}{c}
				1\\
				0
			\end{array}\right]$}\tabularnewline
		\midrule 
		{\small{}1} & \textrm{\small{}$|1\rangle=\left[\begin{array}{c}
				0\\
				1
			\end{array}\right]$}\tabularnewline
		\bottomrule
	\end{tabular}{\small\par}
	\par\end{center}\vspace{-0.4cm}\caption{A cbit can be represented as a binary digit or as a two-dimensional one-hot vector, with the latter being described using Dirac's ket notation.\label{table:ch1:first}}
\end{table}

\subsection{Dirac's Ket and Bra Notations}\label{eq:ch1:Dirac}

Dirac's \textbf{ket} notation is conventionally used in quantum theory to identify \emph{column} vectors. 	Accordingly, a column vector is represented as $|a\rangle$, where $a$ is an identifier for the vector. For a single cbit, as illustrated in Table \ref{table:ch1:first}, the ket vector representing the cbit value $0$, which serves as identifier, is the one-hot amplitude vector\begin{equation}
	|0\rangle=\left[\begin{array}{c}
		1\\
		0
	\end{array}\right],
\end{equation} while the ket vector representing the cbit value $1$ is\begin{equation}
		|1\rangle=\left[\begin{array}{c}
			0\\
			1
		\end{array}\right].
	\end{equation}

Dirac's \textbf{bra} notation is used to identify \emph{row} vectors. Given a ket $|a\rangle$, the bra $\langle a|$ is defined as  the Hermitian transpose of the ket vector $|a\rangle$, i.e.,
		\begin{equation}\label{eq:ch1:bra}
		\langle a|=|a\rangle^{\dagger},
		\end{equation}
		where $\dagger$ represents the Hermitian transpose operation. The Hermitian transpose is given by the cascade of a transposition operation, denoted as $(\cdot)^T$, and of an element-wise complex conjugation, denoted as $(\cdot)^*$, i.e., $(\cdot)^\dagger=((\cdot)^T)^*$. For example, the bra vector corresponding to the bit value 1 is
		\begin{equation}
		\langle1|=[\begin{array}{cc}
			0 & 1\end{array}].
		\end{equation} We will see later in this chapter that quantum states involve complex numbers, making it important to use a Hermitian transpose, rather than a standard transpose, operation to define the bra in (\ref{eq:ch1:bra}).

The \textbf{inner product} between two kets $|a\rangle$
		and $|b\rangle$ is defined as
		\begin{equation}
		|a\rangle^{\dagger}|b\rangle=\langle a||b\rangle=\langle a|b\rangle,
		\end{equation}
		where the last expression is known as Dirac's \textbf{bra-ket} notation for the inner product. (The pun is intended.) Note that we have the equality \begin{equation}
			\langle b | a \rangle = \langle a | b \rangle^{*},
		\end{equation} where we recall that $(\cdot)^{*}$ represents the complex conjugate operation. 
	
	The \textbf{squared $\ell_2$ norm} of a ket  $|a\rangle$ is accordingly defined as  \begin{equation}|||a\rangle||_{2}^{2}=\langle a|a\rangle.
	\end{equation} Note that the subscript in  $||\cdot||_{2}$ identifies the type of norm. In this monograph, we will only use the $\ell_2$ norm, and hence we will use the term norm for the operation $||\cdot||_{2}$.
		
			The kets $|0\rangle$ and $|1\rangle$ 
		define an \textbf{orthonormal basis} for the linear space of two-dimensional vectors. In fact, the two vectors are orthogonal, i.e., \begin{equation}
			\langle 0 | 1 \rangle = \langle 1 | 0 \rangle =0,
			\end{equation} and they have unitary norm, i.e., \begin{equation}
			\langle 0 | 0 \rangle = \langle 1 | 1 \rangle =1.
		\end{equation} 
		
	Throughout this monograph, for reasons that will be made clear later in this chapter, we will take all bra and ket vectors (not only $|0\rangle$ and $|1\rangle$) to have \textbf{unitary norm}. The assumption of unitary norm vectors amounts to the condition
		\begin{equation}
		|||a\rangle||_{2}^{2}=||\langle a|||_{2}^{2}=\langle a|a\rangle=1
		\end{equation} for all kets $|a\rangle$.

\subsection{A Random Cbit as a Probability Vector}\label{sec:ch1:rcbitprob}

	The amplitude, one-hot, vector representation of a cbit is clearly
		less efficient than the direct specification in terms of a single logical bit. In fact, the one-hot vector requires two binary digits to describe a single cbit. Despite this shortcoming, amplitude vectors are routinely used in machine
		learning when dealing with discrete random
		variables. 
		
		The \textbf{probability distribution, or probability mass function},
		of a random cbit is given by the two-dimensional vector 
		\begin{equation}\label{eq:ch1:probvecsq}
		p=\left[\begin{array}{c}
			p_{0}\\
			p_{1}
		\end{array}\right],
		\end{equation}
		where $p_{x}\geq0$ represents the probability of the random cbit taking value $x\in\{0,1\}$, and we have the condition  \begin{equation}\label{eq:ch1:normp}p_{0}+p_{1}=1.
			\end{equation} Note that the norm of a probability vector is generally different from 1. Hence, following the convention described in the previous subsection, the ket notation is not used for vector $p$. 
		
		By the definition (\ref{eq:ch1:probvecsq}), the probability of observing value $x\in\{0,1\}$
		can be computed via the inner product
		\begin{equation}\label{eq:ch1:innprod}
		|x\rangle^{\dagger}p=\langle x|p=p_{x},
		\end{equation} where we recall that $|x\rangle$ is the one-hot amplitude vector representing cbit value $x\in\{0,1\}$ (see Table \ref{table:ch1:first}). Geometrically, the inner product (\ref{eq:ch1:innprod}) can be interpreted as the projection of the probability vector $p$ into the direction defined by vector $|x\rangle$.

Being two dimensional, the probability vector (\ref{eq:ch1:probvecsq}) for a single random cbit can be written as a linear combination of the orthonormal basis vectors $\{|0\rangle,|1\rangle\}$ as
		\begin{equation}\label{eq:ch1:rcbit}
		p=\left[\begin{array}{c}
			p_{0}\\
			p_{1}
		\end{array}\right]=p_{0}|0\rangle+p_{1}|1\rangle.
		\end{equation}
		In words, the  probability vector (\ref{eq:ch1:probvecsq}) can be viewed as the \textbf{``superposition''} of the two orthogonal
		vectors $|0\rangle$ and $|1\rangle$, each representing
		one of the two possible states of the system. The weights of this superposition are given by the corresponding probabilities $p_0$ and $p_1$.

\subsection{Measuring a Random Cbit}\label{sec:ch1:measrc}
A random cbit seems to ``contain more information'' than
		a deterministic one, since its state is defined by the probability vector $p$. This is in the sense that, in order to specify the state of a random cbit, one needs to describe the probability $p_0$ or $p_1$ (since $p_0+p_1=1$), while the state of a deterministic cbit is clearly described by a single binary digit. But how much information can be actually extracted from the observation of a deterministic cbit or a random cbit?
		
		Suppose first that you are handed a deterministic cbit, e.g., a coin resting on one of its two faces. Evidently, a single glance at the coin would reveal its binary value. 
		
		Consider now being given a random cbit -- say a slot machine with a single arm producing either digit $0$ or $1$ with probabilities $p_0$ and $p_1$, respectively. In order to extract the state of the random cbit, that is, its probability vector $p$, a single ``glance'' is not sufficient. Rather, one needs to carry out multiple \textbf{measurements} of the random cbit, producing a number of \emph{independent realizations} of the random cbit, each drawn from probability distribution $p$. From the obtained measurement outputs, one can \emph{estimate} the probability $p_x$, for $x\in\{0,1\}$, by evaluating the fraction of realizations of the random cbit with value $x$. 
		
		In the example of the slot machine, obtaining multiple measurements entails playing the arm of the slot machine several times. Importantly, each independent measurement requires that the system be reset to the original state
		so as to generate a new realization of the same random cbit. Quantitatively, using the outlined fraction-based estimator, by Chebyshev's inequality, one needs $O(1/\epsilon^{2})$ independent measurements of the random cbits in order to produce an estimate with precision
		$\epsilon>0$.

\section{Qubit}

In this section, building on the basic background material presented so far in this chapter, we introduce the notion of a qubit.

\subsection{A Qubit as a Complex Amplitude Vector}\label{sec:ch1:qubitav}

	A \textbf{quantum bit (qubit)} is a two-level quantum system, such as the
		up-down spin of an electron or the vertical-horizontal polarization
		of a photon. In a manner somewhat analogous to a random cbit, whose state is defined by a probability vector $p$ as in (\ref{eq:ch1:rcbit}), the \textbf{state} of a qubit is described by a two-dimensional 
		\textbf{amplitude vector} 
		\begin{equation}\label{eq:ch1:qubit}
		|\psi\rangle=\left[\begin{array}{c}
			\alpha_{0}\\
			\alpha_{1}
		\end{array}\right]=\alpha_{0}|0\rangle+\alpha_{1}|1\rangle. 
		\end{equation} There are two key differences between the state $|\psi\rangle$ in (\ref{eq:ch1:qubit}) of a quantum qubit and the state $p$ of a random cbit: \begin{itemize}
	\item Unlike a probability vector $p$, the amplitude vector has \emph{complex} entries $\alpha_0$ and $\alpha_1$. (Note that complex entries include real-valued entries as a special case.)
\item The qubit state vector (\ref{eq:ch1:qubit}) has the defining property of having \emph{unitary norm}, i.e.,
		\begin{equation}\label{eq:ch1:normq}
		|||\psi\rangle||_{2}^{2}=\langle\psi|\psi\rangle=|\alpha_{0}|^{2}+|\alpha_{1}|^{2}=1.
		\end{equation}This condition is different from the property (\ref{eq:ch1:normp}) satisfied by probability vectors, which stipulates that the sum of the entries of $p$ -- and not the norm of the vector -- equals $1$. We observe that property (\ref{eq:ch1:normq}) is consistent with the convention introduced in the previous section of considering all kets (and bras) to have unitary norm. 
\end{itemize}

While being distinct from the state of a random cbit, the qubit state (\ref{eq:ch1:qubit}) recovers as special cases the two possible states, expressed as one-hot amplitude vectors, of a deterministic cbit. In fact, setting the amplitudes as $\alpha_0=1$ (and hence $\alpha_1=0$), or $\alpha_1=1$ (and hence $\alpha_0=0$), recovers the deterministic cbit states $|0\rangle$ and $|1\rangle$, respectively. Therefore, a qubit that can only assume states $|0\rangle$ and $|1\rangle$ is equivalent to a deterministic cbit.

\begin{figure}
			\begin{center}
				\includegraphics[clip,scale=0.35]{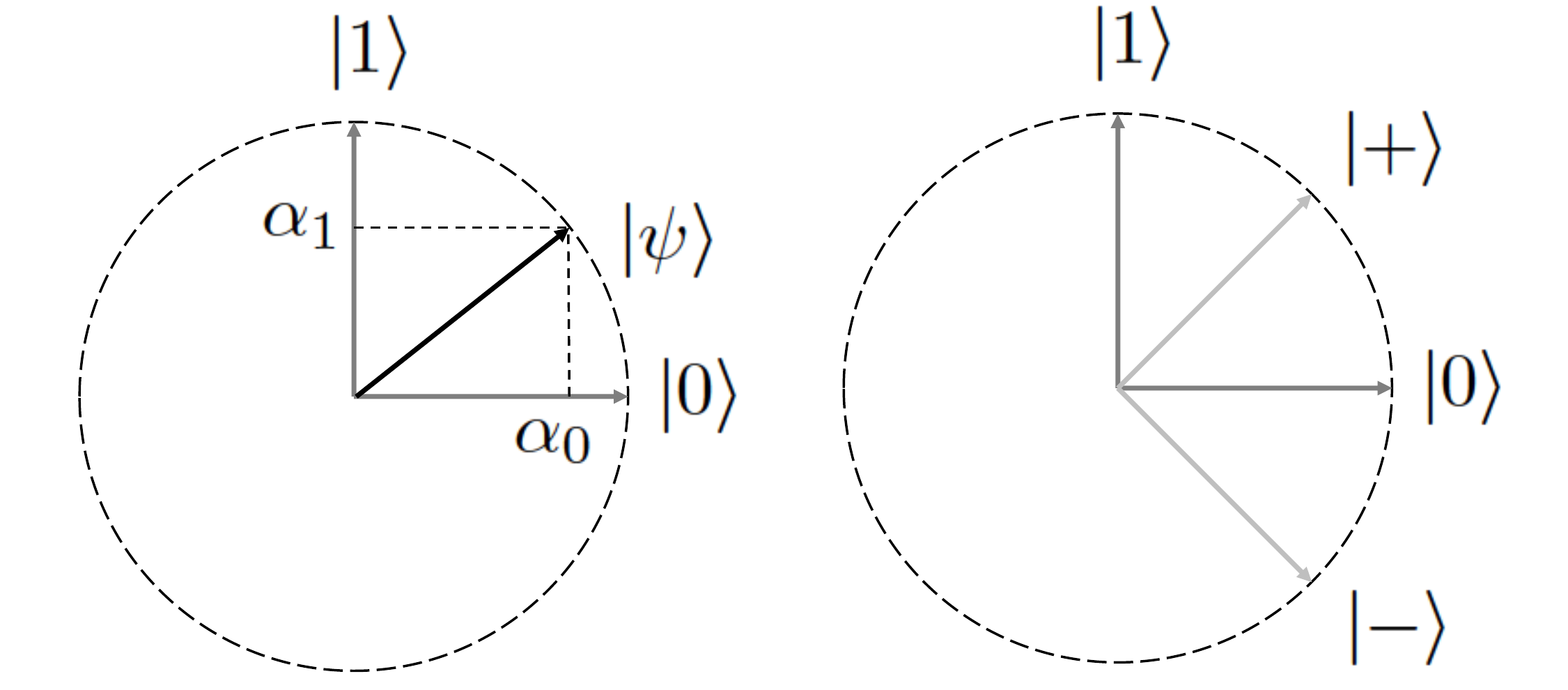}
				\par\end{center}\vspace{-0.4cm}\caption{(left) An illustration of the two-dimensional Hilbert space with the computational basis $\{|0\rangle,|1\rangle\}$ and a given qubit state $|\psi\rangle$ in (\ref{eq:ch1:qubit}); (right) An illustration of computational and diagonal bases. \label{fig:ch1:bases}}
		\end{figure}

By (\ref{eq:ch1:qubit}), we say that the qubit is in a \textbf{superposition} of states $|0\rangle$
		and $|1\rangle$, with respective complex amplitudes $\alpha_{0}$
		and $\alpha_{1}$. Mathematically, this implies that the state of a qubit is a vector that  lies in a two-dimensional
		\emph{complex} linear vector space, referred to as the \textbf{Hilbert space} of
		dimension two. The states $|0\rangle$ and $|1\rangle$ form the so-called \textbf{computational basis} of the Hilbert space. A geometric interpretation of a quantum state, simplified by representing real amplitudes, is provided by the left part of Fig. \ref{fig:ch1:bases}.
		
		Being a two-dimensional vector, the state of the qubit can be equivalently expressed as a superposition of any two orthonormal vectors forming a basis of the Hilbert space. An important example is given by  the so-called \textbf{diagonal basis}, which consists of the two vectors
			\begin{equation}
				|+\rangle=\frac{1}{\sqrt{2}}(|0\rangle+|1\rangle\rangle=\frac{1}{\sqrt{2}}\left[\begin{array}{c}
					1\\
					1
				\end{array}\right]
			\end{equation}and 
			\begin{equation}
				|-\rangle=\frac{1}{\sqrt{2}}(|0\rangle-|1\rangle)=\frac{1}{\sqrt{2}}\left[\begin{array}{c}
					1\\
					-1
				\end{array}\right].
			\end{equation}
			It can be directly checked that vectors $|+\rangle$ and $|-\rangle$ are orthogonal and that they have unitary norm. They are illustrated in the right part of Fig. \ref{fig:ch1:bases}.  The qubit state (\ref{eq:ch1:qubit}) can be expressed as a superposition of diagonal states as
			\begin{equation}
				|\psi\rangle=\left[\begin{array}{c}
					\alpha_{0}\\
					\alpha_{1}
				\end{array}\right]=\frac{1}{\sqrt{2}}(\alpha_{0}+\alpha_{1})|+\rangle+\frac{1}{\sqrt{2}}(\alpha_{0}-\alpha_{1})|-\rangle.
			\end{equation}Therefore, the amplitude of the basis vector $|+\rangle$ is given by the scaled sum $1/\sqrt{2}(\alpha_0+\alpha_1)$, while the amplitude of the basis vector $|-\rangle$ is given by the scaled difference $1/\sqrt{2}(\alpha_0-\alpha_1)$.

\subsection{Measuring a Qubit: von Neumann Measurements}\label{sec:ch1:meas}

In a manner somewhat similar to a random cbit (see Sec. \ref{sec:ch1:measrc}), the state of a qubit is only accessible through
		\textbf{measurements} of the qubit. Note that we will henceforth refer to a qubit and to its state interchangeably. A measurement takes as input a \emph{qubit} in an arbitrary state $|\psi\rangle$, as in (\ref{eq:ch1:qubit}),  and produces a \emph{cbit} as the measurement's output, while leaving the qubit in a generally different state from the original state $|\psi\rangle$.
		
		The most basic type of measurement is known as \textbf{von Neumann measurement in the computational basis}, or \textbf{standard measurement} for short. Given an input qubit state $|\psi\rangle=\alpha_0 |0\rangle + \alpha_1 |1\rangle$, a standard measurement is defined by the following two properties illustrated in Fig. \ref{fig:ch1:figmeas}: \begin{itemize} \item \textbf{Born's rule}: The probability of observing cbit $x\in\{0,1\}$ is
		\begin{equation}\label{eq:ch1:Born}
		\Pr[\textrm{measurement output equals \ensuremath{x\in\{0,1\}}}]=|\alpha_{x}|^{2};
		\end{equation}
\item \textbf{``Collapse'' of the state}: If the measured cbit is $x\in\{0,1\}$, the \textbf{post-measurement state} of the qubit is $|x\rangle$.
\end{itemize}

\begin{figure}\vspace{-0.6cm}
\begin{center}\begin{quantikz}
			\lstick{$\ket{\psi}=\alpha_{0}\ket{0} + \alpha_{1}\ket{1}$} & \meter{$x$}  & \qw  \rstick{$\ket{x}$ w.p. $|\alpha_{x}|^{2}$}
		\end{quantikz}
\end{center}\vspace{-0.4cm}\caption{A von Neumann measurement in the computational basis, also known as a standard measurement, for a single qubit. (The abbreviation ``w.p.'' stands for ``with probability''.)\label{fig:ch1:figmeas}}\vspace{-0.6cm}
\end{figure}

Measurements are typically depicted as shown in Fig. \ref{fig:ch1:figmeas}. Accordingly, a measurement is denoted via a gauge block, with the output cbit $x\in\{0,1\}$ indicated on top of the block and the post-measurement state shown as the output to the right of the block.

By the Born rule (\ref{eq:ch1:Born}), the absolute value squared $|\alpha_x|^{2}$ of the quantum amplitude $\alpha_x$ defines the probability of a measurement outcome $x\in\{0,1\}$. Geometrically, this probability corresponds to the magnitude squared of the projection of the input state $|\psi\rangle$ into the computational-basis vector $|x\rangle$ (see Fig. \ref{fig:ch1:bases}). Note that the quadratic dependence of the measurement probabilities on the amplitudes in the qubit state vector explains the difference between the conditions (\ref{eq:ch1:normp}) and (\ref{eq:ch1:normq}) satisfied by probability vectors and amplitude vectors. One possible way to think of the absolute value squared of the amplitude that appears in Born's rule (\ref{eq:ch1:Born}) is as a measure of intensity, e.g., of a photon beam. 

The Born rule can be equivalently expressed
using the bra-ket notation by noting that the $x$-th amplitude of
qubit state $|\psi\rangle$ can be obtained as
\begin{equation}
	\alpha_{x}=|x\rangle^{\dagger}|\psi\rangle=\langle x|\psi\rangle,
\end{equation}
which implies
\begin{align}\label{eq:ch1:ip}
	|\alpha_{x}|^{2} & =|\langle x|\psi\rangle|^{2}.
\end{align}
It is also useful to note that, by (\ref{eq:ch1:ip}), the probability $|\alpha_{x}|^{2}$
can be expressed as
\begin{align}
	|\alpha_{x}|^{2} & =\langle x|\psi\rangle\langle x|\psi\rangle^{*}=\langle x|\psi\rangle\langle\psi|x\rangle=\langle x|\rho|x\rangle,
\end{align}
where $\rho=|\psi\rangle\langle\psi|$ is the so-called \textbf{density matrix} associated with state $|\psi\rangle$, which will be formally introduced in Chapter 3.

By the ``collapse''-of-the-state property, while the input state $|\psi\rangle$ is generally unknown (that is why one measures it), the post-measurement state is fully determined by the output of the measurement. In fact, the measurement ``collapses'' the input qubit state $|\psi\rangle$ to the computational-basis vector $|x\rangle$ corresponding to the measurement's output cbit $x\in\{0,1\}$.  

As we have discussed in the previous subsection, a qubit that can only take states $|0\rangle$ and $|1\rangle$ behaves like a standard deterministic cbit. As a sanity check, one can directly verify that, by Born's rule, measuring a qubit in state $|x\rangle$, for $x\in\{0,1\}$, returns output $x$ with probability $1$, while leaving the qubit state unchanged.

To conclude this section, it should be mentioned that the interpretation of the ``collapse''-of-the-state property is
much debated in physics and philosophy (and in movies, where the ``many-world''
interpretation provides an easy excuse for a plot twist). 

We will see in Chapter 3, and then again in Chapter 5, that there are different types of measurements; until then, we will always assume standard measurements.

\subsection{Random Cbit vs. Qubit}\label{sec:ch1:rcbitvsqubit}

By Born's rule, if the amplitudes $\{\alpha_{x}\}_{x=0}^{1}$ are real and
		non-negative, we can write the state (\ref{eq:ch1:qubit}) of a qubit as
		\begin{equation}\label{eq:ch1:realqubit}
		|\psi\rangle=\left[\begin{array}{c}
			\sqrt{p_{0}}\\
			\sqrt{p_{1}}
		\end{array}\right],
		\end{equation}
where $p_x$ is the probability (\ref{eq:ch1:Born}) that a measurement of the qubit returns the cbit $x$. Accordingly, there may be a temptation to think of  a qubit state as the square root
		of a probability vector. Even more treacherously, this perspective may lead one to treat a qubit as merely being a random cbit defined by a probability vector $p=[|\alpha_0|^2,|\alpha_1|^2]^{T}$. Accordingly, one would model a qubit in the superposition state (\ref{eq:ch1:qubit}) 
		as having a true, but unknown, classical state $|x\rangle,$ with $x\in\{0,1\}$, which is only
		revealed upon measurement. This temptation should be resisted!
		
	By Born's rule (\ref{eq:ch1:Born}), the viewpoint described in the previous paragraph provides the correct description of the output of the standard measurement of a qubit. However, 	a qubit in state (\ref{eq:ch1:realqubit}) behaves very differently from a random cbit state defined by the  probability vector $p=[|\alpha_0|^2,|\alpha_1|^2]^{T}$  
		in terms of how it evolves over time and of how it interacts with other qubits. Specifically, as we will detail in Sec. \ref{sec:ch1:inference}, the two amplitudes $\alpha_0$ and $\alpha_1$ defining the superposition state $|\psi\rangle$ can  combine over time in ways that produce subsequent measurements that cannot be described by the evolution of a random cbit. (So, Schrodinger's cat is actually neither dead nor alive,
		but it behaves according to a superposition of the two states.)

\section{Single-Qubit Quantum Gates}\label{sec:ch1:sqgates}

In this section, we describe how the state of a qubit evolves in a closed quantum system, introducing the key concept of a quantum gate.

\subsection{Closed Quantum Systems and Unitary Transformations}\label{sec:ch1:unit}

	Consider a \textbf{closed quantum system} consisting of a single qubit that is not subject to measurement. Note that implementing a measurement would require the presence of an instrument connecting the qubit to the outside world of an ``observer''. In a closed system, by the laws of quantum physics, the state of a qubit evolves according
	to \textbf{linear and reversible transformations}.  Linearity may come as a surprise, as one may expect that nature could produce more complex behavior, but it is a  model that has stood the test of time through a large number of experimental validations. Reversibility is a consequence of the principle that a closed system should conserve information.
	
	To elaborate on this last point, irreversible operations imply a loss of information, and deleting information requires 
	energy. By \textbf{Landauer's principle}, it specifically requires $k_{B}T\ln(2)$ joule per bit, where $k_{B}$ is the Boltzmann constant
	and $T$ is the temperature of the heat sink in kelvins.
	 Therefore, a closed system consisting of a single qubit cannot delete
	information, as this would entail the injection of energy from the outside. It follows that transformations in a closed system should be reversible.
	In contrast,  measurements correspond to interactions with external instruments, and are
	not reversible.

	 The only non-trivial linear reversible  operation mapping a deterministic  cbit state to
	a deterministic cbit state is the \textbf{NOT, or bit flip, operation}. The NOT operation is defined by the logical mapping 
	\begin{equation}\label{eq:ch1:NOT1}
	0 \mapsto \bar{0}=1,
	\end{equation}and \begin{equation}\label{eq:ch1:NOT2}
	1 \mapsto \bar{1}=0,
\end{equation}where the bar notation indicates logical negation. The NOT operation can be hence summarized as\begin{equation}\label{eq:ch1:NOTsummary}
x \mapsto \bar{x}=x\oplus 1,
\end{equation} where $\oplus$ indicates the XOR operation. 

The NOT mapping defined by (\ref{eq:ch1:NOTsummary})  can be expressed in terms of the one-hot representation of the state of a cbit by introducing the $2 \times 2$ \textbf{Pauli $X$ matrix}	\begin{equation}\label{eq:ch1:X}
X=\left[\begin{array}{cc}
	0 & 1\\
	1 & 0
\end{array}\right].
\end{equation} In fact, given the input state $|x\rangle$ of a cbit with $x\in\{0,1\}$ (see Table \ref{table:ch1:first}), the state of the cbit at the output of a NOT operation (\ref{eq:ch1:NOT1})-(\ref{eq:ch1:NOT2}) is given by the one-hot amplitude vector
	\begin{equation}\label{eq:ch1:NOTagain}
	X|x\rangle=|\bar{x}\rangle=|x\oplus1\rangle.
	\end{equation}

\begin{figure}\vspace{-0.6cm}
	\begin{center}
		\begin{center}\begin{quantikz}
				\lstick{$\ket{\psi}$ \\ input qubit state} &  \gate{U} & \qw  \rstick{$U\ket{\psi}$\\ output qubit state}
			\end{quantikz}
		\end{center}
		\par\end{center}\vspace{-0.4cm}\caption{The state of a qubit evolves in a closed system according to the product of the input state $|\psi\rangle$ by a unitary matrix $U$, also known as a single-qubit quantum gate.\label{fig:ch1:unitary}}\vspace{-0.2cm}
\end{figure}
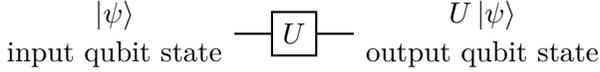

The Pauli $X$ matrix (\ref{eq:ch1:X}), describing a NOT operation, has the following property	\begin{equation}\label{eq:ch1:defuni0}
	XX^{\dagger}=X^{\dagger}X=I,
\end{equation}where $I$ denotes the $2\times 2$ identity matrix \begin{equation} I=\left[\begin{array}{cc}
		1 & 0\\
		0 & 1
	\end{array}\right]. \end{equation} The equalities in (\ref{eq:ch1:defuni0}) are the defining properties of the class of unitary matrices.  

More broadly, any reversible linear transformations mapping a qubit state into a qubit
	state is described by a $2\times 2$ \textbf{unitary matrix} $U$; and, conversely, any unitary matrix $U$ defines a linear reversible transformation between quantum states. Generalizing (\ref{eq:ch1:defuni0}), a unitary matrix $U$ satisfies the equalities 
	\begin{equation}\label{eq:ch1:defuni}
		UU^{\dagger}=U^{\dagger}U=I.
	\end{equation}
Hence, the  inverse of a unitary matrix equals its Hermitian transpose, i.e., we have $U^{-1}=U^{\dagger}$. A unitary matrix maps a qubit state into a qubit state, since it  conserves the  norm of the input vector, i.e., 
\begin{equation}
	||U|\psi\rangle||_{2}^{2}=\langle\psi|U^{\dagger}U|\psi\rangle=|||\psi\rangle||_{2}^{2}=1,
\end{equation}	 where we have used (\ref{eq:ch1:defuni}).

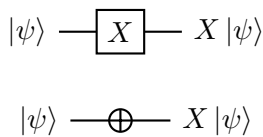
\begin{figure}
\begin{center}\begin{quantikz}
			\lstick{$\ket{\psi}$}  & \gate{X} & \qw  \rstick{$X\ket{\psi}$}
		\end{quantikz}
\end{center}
\begin{center}\begin{quantikz}
			\lstick{$\ket{\psi}$}  & \targ{} & \qw  \rstick{$X\ket{\psi}$}
		\end{quantikz}
\end{center}\vspace{-0.4cm}\caption{Two equivalent representations of a Pauli $X$, or NOT,  gate. \label{fig:ch1:NOT}}\vspace{-0.2cm}
\end{figure}

To summarize, as illustrated in Fig. \ref{fig:ch1:unitary}, in a closed system, a qubit in state $|\psi\rangle$ evolves to the state  
	\begin{equation}
	|\psi'\rangle=U|\psi\rangle
	\end{equation} for some unitary matrix $U$. This transformation is linear and reversible. In fact, by (\ref{eq:ch1:defuni}), one  can return the qubit to the initial state $|\psi\rangle$ by applying the transformation  $U^{\dagger}$ (also unitary by (\ref{eq:ch1:defuni})), i.e.,
\begin{equation}|\psi\rangle = U^{\dagger}|\psi'\rangle.	
\end{equation}  An example of a unitary transformation is given by the NOT, or Pauli $X$, matrix (\ref{eq:ch1:X}), which is typically depicted in one of the two ways shown in Fig. \ref{fig:ch1:NOT}.

\subsection{Quantum Gates}\label{sec:ch1:gates}
By the discussion so far in this subsection, any unitary matrix $U$ is in principle physically realizable as the evolution of a closed quantum system. A unitary matrix operating on a single qubit is referred to as a \textbf{single-qubit quantum gate} in the context of quantum computing. 

\begin{figure}
		\begin{center}\begin{quantikz}
				\lstick{$\ket{\psi}$}  & \gate{U_1} & \gate{U_2} & \cdots & \gate{U_K} &\qw  \rstick{$U_{K} \cdots U_{2}U_{1}\ket{\psi}$}
			\end{quantikz}
		\end{center}\vspace{-0.4cm}\caption{Example of quantum circuit describing the evolution of the state of a qubit in a closed system as a cascade of single-qubit quantum gates implementing unitary matrices $U_1$, $U_2$,..., and $U_K$.\label{fig:ch1:cascade}}\vspace{-0.6cm}
	\end{figure}
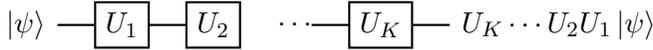

In practice, as illustrated in Fig. \ref{fig:ch1:cascade}, a transformation $U$ of a qubit in a quantum computer is typically implemented via a cascade of basic single-qubit quantum gates selected from a library of transformations  available in the given quantum system. This cascade can be expressed mathematically as the product \begin{equation}\label{eq:ch1:product}
	U=U_K \cdot U_{K-1} \cdots U_1,
\end{equation}where the $2 \times 2$ unitary matrices $U_k$ with $k\in\{1,2,...,K\}$ represent the operation of basic quantum gates. Note that the product of unitary matrices is also unitary (as it can be checked by using (\ref{eq:ch1:defuni})). 

	Accordingly, as in  Fig.  \ref{fig:ch1:cascade}, the evolution of a qubit in a closed quantum system can be generally described by a \textbf{quantum circuit} in which
a wire represents a qubit, and multiple quantum gates are applied in the order from left to right. Note that the order in which the quantum gates are applied to the input state is the inverse of the order in which the corresponding matrices appear in the product (\ref{eq:ch1:product}) when read from left to right.

Examples of basic single-qubit quantum gates implemented in standard quantum computers are given in Table \ref{table:ch1:gates}. These include the four \textbf{Pauli matrices, or Pauli operators}, namely $I$ (the identity matrix), $X$, $Y$, and $Z$; the \textbf{Hadamard gate} $H$; and the \textbf{Pauli $Y$-rotation}  $R_Y(\theta)$. The Pauli matrices $X$, $Y$, and $Z$ are related by the \textbf{cyclic product} properties $XY=iZ$, $YZ=iX$, and $ZX=iY$, where $i$ is the complex unit. They are also \textbf{anti-commuting} in the sense that we have the products $P_1 P_2=-P_2 P_1$ with $P_1,P_2 \in \{X,Y,Z\}$ and $P_1 \neq P_2$.
 
			\begin{table}
		\begin{center}
			\begin{tabular}{cc}
				\toprule 
			name & operator\tabularnewline
				\midrule
				\midrule 
				identity & $I=\left[\begin{array}{cc}
						1 & 0\\
						0 & 1
					\end{array}\right]$\tabularnewline
				\midrule 
			Pauli $X$ & $X=\left[\begin{array}{cc}
						0 & 1\\
						1 & 0
					\end{array}\right]$\tabularnewline
				\midrule 
				Pauli $Z$  & $Z=\left[\begin{array}{cc}
						1 & 0\\
						0 & -1
					\end{array}\right]$\tabularnewline
				\midrule 
			Pauli $Y$  & $Y=iXZ=\left[\begin{array}{cc}
						0 & -i\\
						i & 0
					\end{array}\right]$\tabularnewline
				\midrule 
			Hadamard  & $H=\frac{1}{\sqrt{2}}\left[\begin{array}{cc}
						1 & 1\\
						1 & -1
					\end{array}\right]=\frac{1}{\sqrt{2}}(X+Z)$\tabularnewline
				\midrule 
			Pauli $Y$-rotation & $R_{Y}(\theta)=\left[\begin{array}{cc}
						\cos(\theta/2) & -\sin(\theta/2)\\
						\sin(\theta/2) & \cos(\theta/2)
					\end{array}\right]$\tabularnewline
				\bottomrule
			\end{tabular}
			\par\end{center}\vspace{-0.4cm}\caption{Examples of notable single qubit-quantum gates.\label{table:ch1:gates}}
		\end{table}
	
		In the next two subsections, we describe three useful characterizations of unitary matrices and hence of quantum gates.

	\subsection{Quantum Gates as Change-of-Basis Transformations}\label{sec:ch1:qgcb}

\begin{figure}
		\begin{center}
			\includegraphics[clip,scale=0.35]{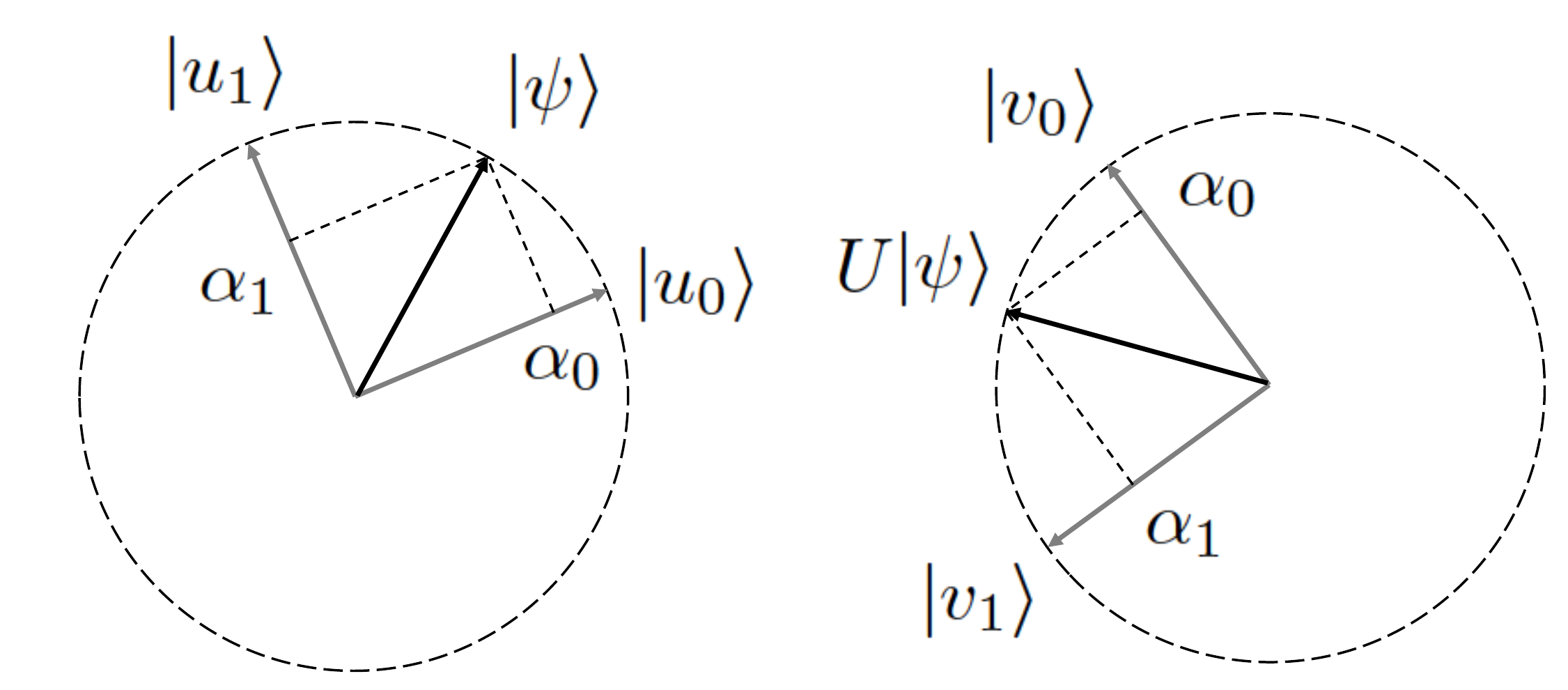}
			\par\end{center}\vspace{-0.4cm}\caption{A geometric illustration of the operation of a unitary matrix as the change-of-basis transformation (\ref{eq:ch1:changeofbasis}). \label{fig:ch1:changeofbasis}}
	\end{figure}

	Any unitary matrix operating on a qubit can be expressed as
	\begin{align}\label{eq:ch1:changeofbasis}
		U & =|v_{0}\rangle\langle u_{0}|+|v_{1}\rangle\langle u_{1}|,
	\end{align}
	where \{$|v_{0}\rangle,|v_{1}\rangle$\} and \{$|u_{0}\rangle,|u_{1}\rangle$\}
	are two orthonormal bases of the two-dimensional Hilbert space. By (\ref{eq:ch1:changeofbasis}), we can interpret a unitary operator as mapping each vector $|u_x\rangle$ from one orthonormal basis to a vector $|v_x\rangle$ in another orthonormal basis for $x\in\{0,1\}$. In fact,  by (\ref{eq:ch1:changeofbasis}), we have the mapping \begin{align}U|u_x\rangle & =(|v_{0}\rangle\langle u_{0}|+|v_{1}\rangle\langle u_{1}|)|u_x\rangle \nonumber \\
		&= |v_{0}\rangle\langle u_{0}|u_x\rangle+|v_{1}\rangle\langle u_{1}|u_x\rangle \nonumber\\
		& = |v_x\rangle
	\end{align} for $x\in\{0,1\}$. Therefore, a unitary matrix (\ref{eq:ch1:changeofbasis}) applies a \textbf{change of basis} from basis \{$|u_{0}\rangle,|u_{1}\rangle$\} to basis \{$|v_{0}\rangle,|v_{1}\rangle$\}. A geometric interpretation is provided in Fig. \ref{fig:ch1:changeofbasis}.

By linearity, once one specifies the operation of a unitary matrix on the two vectors of an orthonormal basis \{$|u_{0}\rangle,|u_{1}\rangle$\}, as in (\ref{eq:ch1:changeofbasis}), the output of the matrix-vector multiplication $U|\psi\rangle$ is  defined for \emph{any} qubit state $|\psi\rangle$. In fact, as discussed in Sec. \ref{sec:ch1:qubitav}, any qubit state $|\psi\rangle$ can be expressed as the superposition $|\psi\rangle =\alpha_0 |u_0\rangle + \alpha_1 |u_1\rangle$  of the two vectors $|u_0\rangle$ and $|u_1\rangle$. Therefore, we have the equality \begin{align}U|\psi\rangle & = \alpha_0 U| u_0\rangle + \alpha_1 U |v_1\rangle \nonumber \\ & = \alpha_0|v_0\rangle + \alpha_1|v_1\rangle.
\end{align}
	
	Table \ref{table:ch1:changeofbasis} reports some examples of single-qubit quantum gates expressed in the form (\ref{eq:ch1:changeofbasis}), which are detailed next.
	
	\noindent $\bullet$ \textbf{Identity gate}: The identity ``gate'' maps any quantum state to itself. Therefore, the form (\ref{eq:ch1:changeofbasis}) applies with \emph{any} orthonormal basis $\{|u_x\rangle=|v_x\rangle\}_{x=0}^{1}$, i.e., we have  \begin{equation}\label{eq:ch1:residentity}
		I=|v_0\rangle\langle v_0|+|v_1\rangle\langle v_1|.
	\end{equation} Condition (\ref{eq:ch1:residentity}) is also known as a \textbf{resolution of the identity}.
	
	\noindent $\bullet$ \textbf{Pauli $X$, or NOT, gate}: The Pauli $X$, or NOT, gate acts as a \textbf{bit flip}, mapping state $|0\rangle$ to $|1\rangle$, and state $|1\rangle$ to $|0\rangle$. By (\ref{eq:ch1:NOTagain}), the Pauli $X$ gate can be also thought of as a \textbf{shift} operator, as it maps each vector $|x\rangle$, with $x\in\{0,1\}$, to the ``shifted'' version $|x \oplus 1\rangle$ with ``shift'' given by $1$.  Given a qubit state in the superposition (\ref{eq:ch1:qubit}), the effect of the Pauli $X$ gate is to assign amplitude $\alpha_0$ to the basis vector $|1\rangle$ and the amplitude $\alpha_1$ to the basis vector $|0\rangle$, i.e., \begin{equation}
		X (\alpha_0 |0\rangle + \alpha_1 |1\rangle)= \alpha_0 |1\rangle + \alpha_1 |0\rangle.
	\end{equation}

	\noindent $\bullet$ \textbf{Pauli Z gate}: While the Pauli $X$ operator swaps the amplitudes of the computational basis vectors, the Pauli $Z$ gate swaps the amplitudes of the vectors in the diagonal basis $\{|+\rangle,|-\rangle\}$ -- an operation known as \textbf{phase flip}. The name is a consequence of the fact that the $Z$ operator flips the phase of the amplitude of the $|1\rangle$ vector for an arbitrary input state (\ref{eq:ch1:qubit}), in the sense that we have  \begin{equation} Z(\alpha_0|0\rangle+\alpha_1|1\rangle)=\alpha_0|0\rangle-\alpha_1|1\rangle. 
	\end{equation}
	
\noindent $\bullet$ \textbf{Hadamard gate}: The Hadamard gate transforms the computational basis into the diagonal
	basis and back, in the sense we have the equalities
	\begin{equation}\label{eq:ch1:Hcomp}
		H|0\rangle=|+\rangle\textrm{ and }H|1\rangle=|-\rangle,
	\end{equation}
	as well as
	\begin{equation}\label{eq:ch1:Hdiag}
		H|+\rangle=|0\rangle\textrm{ and }H|-\rangle=|1\rangle.
	\end{equation}
	The Hadamard gate can be thought of as performing a two-dimensional discrete Fourier transform, as well as the corresponding inverse discrete Fourier transform.  In this interpretation, the vector $|+\rangle$ represents the zero-frequency (i.e., constant) signal, and the vector $|-\rangle$ is the maximum-frequency signal.
	
	\begin{table} 
	\begin{center}
		\begin{tabular}{cc}
			\toprule 
		name & operator\tabularnewline
			\midrule
			\midrule 
			identity & $I=\left[\begin{array}{cc}
					1 & 0\\
					0 & 1
				\end{array}\right]=|v_0\rangle\langle v_0|+|v_1\rangle\langle v_1|$ \textrm{for any orth. basis $\{|v_x\rangle\}_{x=0}^{1}$}\tabularnewline
			\midrule 
			Pauli $X$ & $X=\left[\begin{array}{cc}
					0 & 1\\
					1 & 0
				\end{array}\right]=|0\rangle\langle1|+|1\rangle\langle0|$\tabularnewline
			\midrule 
			Pauli $Z$ & $Z=\left[\begin{array}{cc}
					1 & 0\\
					0 & -1
				\end{array}\right]=|+\rangle\langle-|+|-\rangle\langle+|$\tabularnewline
			\midrule 
		Hadamard  & $H=\frac{1}{\sqrt{2}}\left[\begin{array}{cc}
					1 & 1\\
					1 & -1
				\end{array}\right]=|0\rangle\langle+|+|1\rangle\langle-|=|+\rangle\langle0|+|-\rangle\langle1|$\tabularnewline
			\bottomrule
		\end{tabular}
		\par\end{center}\vspace{-0.4cm}\caption{Examples of single-qubit quantum gates as change-of-basis transformations defined by the decomposition  (\ref{eq:ch1:changeofbasis}).\label{table:ch1:changeofbasis}}
	\end{table}

	\subsection{Quantum Gates as Transformations with Unitary-Magnitude Eigenvalues}\label{sec:ch1:uniteig}
	
To introduce an alternative interpretation of quantum gates, we will need to review first the spectral theorem, which applies to normal matrices.  	A \textbf{normal matrix} $A$ is a square $N\times N$ matrix that satisfies
	the condition	\begin{equation}\label{eq:ch1:normal}
		AA^{\dagger}=A^{\dagger}A.
	\end{equation}
	By the \textbf{spectral theorem}, any normal matrix can be expressed in terms of its \textbf{eigendecomposition}
	\begin{equation}\label{eq:ch1:eigen}
		A=\sum_{x=0}^{N-1}\lambda_{x}|v_{x}\rangle\langle v_{x}|,
	\end{equation}
	where $\{|v_{x}\rangle\}_{x=0}^{N-1}$ are the eigenvectors, which form an orthonormal basis of
	the $N$-dimensional Hilbert space, and $\{\lambda_{x}\}_{x=0}^{N-1}$ are the corresponding
	\textbf{eigenvalues}, which are generally complex. 
	
	Using the eigendecomposition (\ref{eq:ch1:eigen}), given a normal matrix $A$ and a scalar function $f(\cdot)$, we define the matrix function \begin{equation}\label{eq:ch1:matfun}
		f(A)=\sum_{x=0}^{N-1}f(\lambda_{x})|v_{x}\rangle\langle v_{x}|.
	\end{equation}That is, function $f(A)$ is evaluated by applying the scalar function $f(\cdot)$ separately to each eigenvalue of matrix $A$. 
	
	A unitary matrix $U$ is a normal matrix, since it satisfies the condition (\ref{eq:ch1:defuni}) and hence also the equality (\ref{eq:ch1:normal}).  Therefore, a $2\times 2$ unitary matrix can be expressed in terms of its eigendecomposition (\ref{eq:ch1:eigen}) with $N=2$ eigenvectors and eigenvalues, i.e., as 	\begin{equation}\label{eq:ch1:eigenunit}
		U=\lambda_{0}|v_{0}\rangle\langle v_{0}|+\lambda_{1}|v_{1}\rangle\langle v_{1}|, 
	\end{equation} where  $\{|v_{x}\rangle\}_{x=0}^{1}$ is the orthonormal basis of eigenvectors. Furthermore, all the eigenvalues
of unitary matrices have absolute value equal to 1, i.e., $|\lambda_x|=1$ for $x\in\{0,1\}$. To see this, note that, by (\ref{eq:ch1:eigenunit}), the condition (\ref{eq:ch1:defuni}) is equivalent to the equality 	\begin{equation}
|\lambda_{0}|^{2}|v_{0}\rangle\langle v_{0}|+|\lambda_{1}|^{2}|v_{1}\rangle\langle v_{1}|=I,
\end{equation} and we have the resolution-of-identity condition (\ref{eq:ch1:residentity}). 

The change-of-basis representation (\ref{eq:ch1:changeofbasis})  describes a unitary matrix as a  map from a state in one basis to a state in another basis. In contrast, the eigendecomposition (\ref{eq:ch1:eigenunit}) identifies states -- the eigenvectors $\{|v_{x}\rangle\}_{x=0}^{1}$  -- that are left unchanged by the operator except for a scaling by a complex number with a unitary absolute value, namely the eigenvalue $\lambda_x$ for state $|v_x\rangle$. This is in the sense that we have the equalities \begin{equation}
	U|v_x\rangle=\lambda_x |v_x\rangle
\end{equation} for $x\in\{0,1\}$.

Some examples of single-qubit gates expressed in terms of their eigendecompositions can be found in Table \ref{table:ch1:eigen}, where we have defined the so-called \textbf{circular orthonormal basis}
	\begin{align}
		|+i\rangle & =\frac{1}{\sqrt{2}}(|0\rangle+i|1\rangle)\\
		|-i\rangle & =\frac{1}{\sqrt{2}}(|0\rangle-i|1\rangle).
	\end{align}Note that the identity has all eigenvalues equal to $1$, while all other Pauli gates -- $X$, $Y$, and $Z$ -- have one eigenvalue equal to $1$ and the other equal to $-1$.  Given the eigendecompositions in Table \ref{table:ch1:eigen}, the orthonormal bases $\{|+\rangle, |-\rangle\}$, $\{|+i\rangle, |-i\rangle\}$, and $\{|0\rangle, |1\rangle\}$  are also known as $X$, $Y$, and $Z$ bases, respectively.
	
	\begin{table}
	\begin{center}
		\begin{tabular}{cc}
			\toprule 
		name & operator\tabularnewline
			\midrule
			\midrule 
			identity & $I=\left[\begin{array}{cc}
					1 & 0\\
					0 & 1
				\end{array}\right]=|v_0\rangle\langle v_0|+|v_1\rangle\langle v_1| \textrm{ for any orth. basis $\{|v_x\rangle\}_{x=0}^{1}$}$\tabularnewline
			\midrule 
		Pauli $X$ & $X=\left[\begin{array}{cc}
					0 & 1\\
					1 & 0
				\end{array}\right]=|+\rangle\langle+|-|-\rangle\langle-|$\tabularnewline
			\midrule 
			Pauli $Z$ & $Z=\left[\begin{array}{cc}
					1 & 0\\
					0 & -1
				\end{array}\right]=|0\rangle\langle0|-|1\rangle\langle1|$\tabularnewline
			\midrule 
		Pauli $Y$ & $Y=\left[\begin{array}{cc}
					0 & -i\\
					i & 0
				\end{array}\right]=|+i\rangle\langle +i|-|-i\rangle\langle -i|$\tabularnewline
			\bottomrule
		\end{tabular}
		\par\end{center}\vspace{-0.4cm}\caption{Examples of single-qubit quantum gates expressed in terms of their eigendecompositions (\ref{eq:ch1:eigenunit}) with unitary-magnitude eigenvalues.\label{table:ch1:eigen}}
	\end{table}

\subsection{Quantum Gates from Hermitian Generators}\label{sec:ch1:gen}

As we discuss in this subsection, a unitary matrix -- and hence also a quantum gate -- can be expressed as an exponential transformation of a generator matrix. In order to explain this characterization of a unitary matrix, we need to introduce the definition of Hermitian matrices.

A $2\times 2$ matrix $A$ is \textbf{Hermitian} if it satisfies the property \begin{equation}\label{eq:ch1:Herm} A^{\dagger}=A.
\end{equation} Since the equality (\ref{eq:ch1:Herm}) implies (\ref{eq:ch1:normal}), a Hermitian matrix is normal.  Therefore, as a result of the spectral theorem, it has an eigendecomposition (\ref{eq:ch1:eigen}). Furthermore, it can be shown using (\ref{eq:ch1:Herm}) that all eigenvalues of a Hermitian matrix are real.	

Any unitary matrix $U$ can be expressed
in terms of a Hermitian matrix $G$ as
\begin{equation}\label{eq:ch1:gen}
	U=\exp(-iG).
\end{equation}
The Hermitian matrix $G$ is known as the \textbf{generator} -- or, in physics, as the \textbf{Hamiltonian} -- of the unitary
$U$. Recall that a function of a normal matrix is defined as in (\ref{eq:ch1:matfun}).

It can be easily checked that a matrix in the form (\ref{eq:ch1:gen}) is indeed unitary. In fact, the defining property (\ref{eq:ch1:defuni}) of unitary matrices is verified as 
\begin{equation}\label{eq:ch1:UG}
	U^{\dagger}U=\exp(iG^{\dagger})\exp(-iG)=\exp(iG^{\dagger}-iG)=I=UU^{\dagger},
\end{equation} where we have used the property (\ref{eq:ch1:Herm}), i.e., the equality $G^{\dagger}=G$.

As an example of the characterization (\ref{eq:ch1:gen}), the Pauli $Y$-rotation matrix $R_Y(\theta)$ in Table \ref{table:ch1:gates} can be expressed as a function of the Pauli $Y$ matrix as \begin{equation}\label{eq:ch1:RY}R_Y(\theta)=\exp\left(-i\frac{\theta}{2}Y\right),
\end{equation} so that the generator matrix is given by the Hermitian matrix  $G=(\theta/2) Y$. One can similarly define the \textbf{Pauli $X$-rotation matrix}  \begin{equation}R_X(\theta)=\exp\left(-i\frac{\theta}{2}X\right),
\end{equation} as well as the \textbf{Pauli $Z$-rotation matrix}\begin{equation}R_Z(\theta)=\exp\left(-i\frac{\theta}{2}Z\right).
\end{equation} Rotation matrices can be interpreted geometrically by introducing the so-called Bloch sphere (see problems).

	\subsection{Pauli Orthonormal Basis for the Space of Matrices}\label{sec:ch1:Paulibasis}
	
	The Pauli matrices $I,X,Y,$ and $Z$ play a key role in the formalism of quantum theory. One of their useful properties is that they form a basis for the space of $2\times2$ (bounded) matrices $A$ with arbitrary complex entries. This is in the sense that any such matrix $A$ can be written as a linear combination of the Pauli matrices as \begin{equation}\label{eq:ch1:Pauliexp}
	A=a_{0}I+a_{1}X+a_{2}Y+a_{3}Z.
	\end{equation} The coefficients  of the expansion can be computed using the trace operator. The \textbf{trace} $\mathrm{tr}(\cdot)$ of a square matrix is the sum of elements on the main diagonal of the matrix (see Sec. \ref{sec:ch3:trace} for additional information). With this definition, the coefficients in (\ref{eq:ch1:Pauliexp}) are obtained as
	\begin{equation}
	a_{0}=\frac{1}{2}\mathrm{tr}\left(A\right),\textrm{}a_{1}=\frac{1}{2}\mathrm{tr}\left(AX\right),\textrm{ \ensuremath{a_{2}}\ensuremath{=}\ensuremath{\frac{1}{2}}\ensuremath{\mathrm{tr}\left(AY\right)}, \ensuremath{a_{3}}=\ensuremath{\frac{1}{2}}\ensuremath{\mathrm{tr}\left(AZ\right)}}.
	\end{equation}
	 Note that the operation $\textrm{tr}(AB)$ for two square matrices $A$ and $B$ corresponds to the inner product $a^{T} b$ between the vectors $a$ and $b$ obtained by stacking the columns of matrices $A^{T}$ and $B$.
	
	If $A=U$ is unitary, it can
	be shown that the vector of coefficients $[a_{0},a_{1},a_{2},a_{3}]^{T}$ in the decomposition (\ref{eq:ch1:Pauliexp}) has unitary norm. Furthermore, the expansion
	(\ref{eq:ch1:Pauliexp}) can be specialized as 
	\begin{equation}
	U=\exp(i\delta)(\cos(\phi)I+i\sin(\phi)(\lambda_{1}X+\lambda_{2}Y+\lambda_{3}Z))
	\end{equation}
	for some angles $(\delta,\phi)$ and some real numbers $\lambda_{1},\lambda_{2},$ and $\lambda_{3}$. For example, the Pauli $Y$-rotation matrix can be expressed in the form (\ref{eq:ch1:Pauliexp}) as \begin{equation}\label{eq:ch1:PauliYcossin}
		R_Y(\theta)=\cos\left(\frac{\theta}{2}\right)I-i\sin\left(\frac{\theta}{2}\right)Y,
	\end{equation}with the same form applying also to rotations $R_X(\theta)$  and $R_Z(\theta)$ by replacing matrix $Y$ with $X$ and $Z$, respectively. Note that the expression (\ref{eq:ch1:PauliYcossin}) can be derived from (\ref{eq:ch1:RY}) via Euler's formula $\exp(ix)=\cos(x)+i\sin(x)$.
	
	If $A$ is Hermitian, all coefficients $a_{0,},a_{1},a_{2}$, and $a_{3}$ in the expansion (\ref{eq:ch1:Pauliexp}) are real. Therefore, by the characterization (\ref{eq:ch1:gen}), we can write an arbitrary unitary matrix as	\begin{equation}\label{eq:ch1:gen1}
		U=\exp(-i(a_{0}I+a_{1}X+a_{2}Y+a_{3}Z)), 
	\end{equation}for real coefficients $a_{0},a_{1},a_{2}$ and $a_{3}$. For example, the Pauli $Y$-rotation matrix can be expressed in the form (\ref{eq:ch1:gen1}) with $a_0 =a_1 =a_3 =0$ and $a_2 =\theta/2$; and similar characterizations apply to the Pauli $X$- and $Z$-rotations.

\section{Amplitude Diagrams}\label{sec:ch1:amplitude}
In a quantum circuit diagram, such as that in Fig. \ref{fig:ch1:cascade},  a qubit is represented by a wire, and quantum gates are indicated as input-output blocks operating on the qubit with time flowing from left to right. Note that the quantum gates are applied \textbf{in place}, in the sense that  the physical quantum qubit is the same throughout the computation, while its state varies over time. An alternative representation of the operation of a quantum circuit is provided by an \textbf{amplitude diagram}. Amplitude diagrams offer a more detailed description of a unitary by depicting the evolution of the two complex amplitudes defining the qubit state.

To elaborate, let us fix an orthonormal basis. An amplitude diagram contains two wires, one for each of the two amplitudes associated to either basis vector. Adopting the computational basis $|0\rangle$ and $|1\rangle$, each wire reports the evolution of the value of one of the two amplitudes $\alpha_0$ and $\alpha_1$ in the qubit state (\ref{eq:ch1:qubit}). 

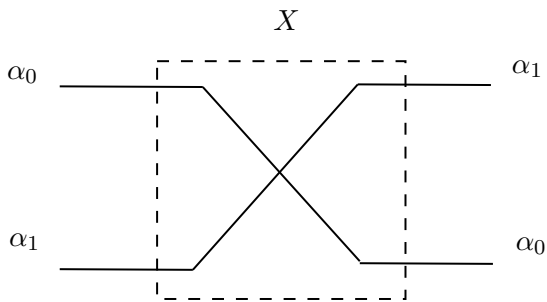
\begin{figure}    

\tikzset{every picture/.style={line width=0.75pt}} 
\begin{center}
\begin{tikzpicture}[x=0.75pt,y=0.75pt,yscale=-1,xscale=1]
	
	\draw    (135,39.67) -- (207,40) ;
	\draw    (207,40) -- (286,127.67) ;
	\draw    (135,131.67) -- (202,132) ;
	\draw    (202,132) -- (285,38.67) ;
	\draw    (285,38.67) -- (352,39) ;
	\draw    (286,128.67) -- (353,129) ;
	\draw  [dash pattern={on 4.5pt off 4.5pt}] (184,27) -- (309,27) -- (309,146.67) -- (184,146.67) -- cycle ;
	
	\draw (120,131) node [anchor=north west][inner sep=0.75pt]   [align=left] {$ $};
	\draw (107,28) node [anchor=north west][inner sep=0.75pt]   [align=left] {$\displaystyle \alpha _{0}$};
	\draw (108,112) node [anchor=north west][inner sep=0.75pt]   [align=left] {$\displaystyle \alpha _{1}$};
	\draw (361,25) node [anchor=north west][inner sep=0.75pt]   [align=left] {$\displaystyle \alpha _{1}$};
	\draw (363,114) node [anchor=north west][inner sep=0.75pt]   [align=left] {$\displaystyle \alpha _{0}$};
	\draw (241,0) node [anchor=north west][inner sep=0.75pt]   [align=left] {$\displaystyle X$};

\end{tikzpicture}\caption{Amplitude diagram describing the operation of a Pauli $X$ gate in the computational basis.\label{fig:ch1:X}}   
\end{center}
\end{figure}

As an example, Fig. \ref{fig:ch1:X} shows the description of a Pauli $X$ gate via an amplitude diagram in the computational basis. As clearly illustrated by the amplitude diagram, the $X$ gate swaps the amplitudes associated to the two vectors $|0\rangle$ and $|1\rangle$ in the computational basis (see Sec.  \ref{sec:ch1:qgcb}). Amplitude diagrams will be used in the next section in order to illustrate the uniquely quantum phenomenon of interference.

\section{Interference}\label{sec:ch1:inference}

The key difference between the behavior of a qubit and that of a random cbit is the
		phenomenon of \textbf{interference}. As the state of a qubit evolves over time, the amplitudes
			$\alpha_{0}$ and $\alpha_{1}$ corresponding to the two computational
			basis state $|0\rangle$ and $|1\rangle$ can combine and ``interfere'' in ways that produce measurement outputs that cannot be described in terms of the evolution of a random cbit. In particular, as we will discuss in this section, it is even possible that the two amplitudes cancel each other
			out, creating destructive interference.

\begin{figure}
	\begin{center}\begin{quantikz}
				\lstick{$\ket{+}$ } & \qw & \meter{$x$}   & \qw &  \rstick{$\ket{x}$\\ w.p. $0.5$}
			\end{quantikz}	\end{center}
		\begin{center}\begin{quantikz}
				\lstick{$\ket{0}$ or $\ket{1}$\\ w.p. $0.5$ } &   \gate{H} \qw & \meter{$x$} & \qw  & \rstick{$\ket{x}$\\ w.p. $0.5$}
			\end{quantikz}
	\end{center}
\begin{center}\begin{quantikz}
		\lstick{$\ket{+}$} &   \gate{H} \qw & \meter{$0$} & \qw  & \rstick{$\ket{0}$\\ w.p. $1$}
	\end{quantikz}

		\end{center}\vspace{-0.4cm}\caption{(top) Measuring directly qubit in the superposition state $|+\rangle$ yields measurement outputs $0$ or $1$ with equal probability; (middle) If the input state is in either state $|0\rangle$ or $|1\rangle$ with equal probability, the measurement output after the Hadamard gate would be $0$ or $1$ with equal probability; (bottom) Measuring a qubit in state $|+\rangle$ after the application of a Hadamard gate yields a measurement output equal to $1$ with probability $1$ owing to the phenomenon of interference.\label{fig:ch1:dirmeas}}
	\end{figure}
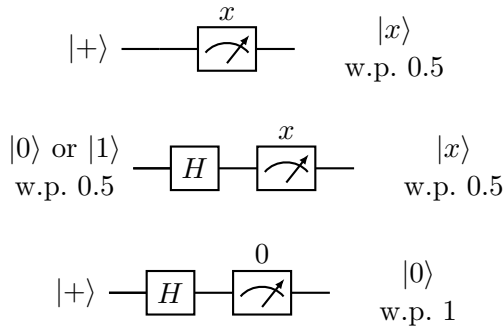

As we detailed in Sec. \ref{sec:ch1:rcbitvsqubit}, if we directly measure a state $|\psi\rangle$ in superposition, it behaves in a manner akin
	to (the square root of) a probability vector. As an example, as seen in the top part of Fig. \ref{fig:ch1:dirmeas}, if we directly measure a qubit in state $|+\rangle=1/\sqrt{2}(|0\rangle+|1\rangle)$, we obtain as measurement output $0$ or $1$ with equal probability $(1/\sqrt{2})^{2}=1/2$. 
However, as anticipated in Sec.  \ref{sec:ch1:rcbitvsqubit},  this cannot be interpreted as indicating that, unbeknownst to us, prior to the measurement, the qubit is in \emph{either} state $|0\rangle$ \emph{or} state $|1\rangle$. This situation would describe the state of a random cbit, for which randomness is of \textbf{epistemic} nature, that is, related to lack of knowledge on the part of the observer making a measurement. For a qubit, uncertainty is of an inherently different nature; and, as we will illustrate next with an example, one needs to describe the state of the qubit as a ``real'' superposition of \emph{both} states $|0\rangle$ \emph{and} state $|1\rangle$.

For reference, let us consider first the situation in which we model,  \emph{incorrectly}, the input state as being equal 
to $|0\rangle$ with probability $1/2$ and $|1\rangle$ with probability
$1/2$ as shown in the middle part of Fig. \ref{fig:ch1:dirmeas}. This would imply that the qubit is equivalent to a random cbit, taking either possible state with equal probability. In this case, after the Hadamard gate in the figure, by (\ref{eq:ch1:Hcomp}), the qubit  would be in state $H |0\rangle = |+\rangle$ with probability $1/2$, and in state $H |1\rangle = |-\rangle$ with probability $1/2$. Measuring each diagonal state $|+\rangle$ and $|-\rangle$  would produce output $0$ or $1$ with equal probability. Therefore, by the law of total probability, the standard measurement in Fig. \ref{fig:ch1:dirmeas} would output $0$ with probability $1/2\cdot 1/2 + 1/2 \cdot 1/2=1/2$, and $1$ with probability $1/2$. 

Let us now consider the situation in the bottom part of Fig. \ref{fig:ch1:dirmeas} in which the input state is given by the superposition $|+\rangle$ of states $|0\rangle$ and $|1\rangle$. Applying the Hadamard gate to the input superposition state $|+\rangle$ gives the qubit state 
	\begin{align}\label{eq:ch1:interf}
		H\left(\frac{1}{\sqrt{2}}(|0\rangle+|1\rangle)\right) & =\frac{1}{\sqrt{2}}(H|0\rangle+H|1\rangle)\nonumber\\
		& =\frac{1}{2}(|0\rangle+|1\rangle+|0\rangle-|1\rangle)=|0\rangle,
	\end{align}where we have used again (\ref{eq:ch1:Hcomp}). Therefore, by the Born rule, measuring the output state produces output $0$ with probability $1$. The calculation (\ref{eq:ch1:interf}) can  be also carried out directly in terms of amplitude vectors by expressing the Hadamard gate as in Table \ref{table:ch1:gates}. With this formalism, the output state can be computed as
	\begin{align}\label{eq:ch1:amplitude}
	\underbrace{\frac{1}{\sqrt{2}}\left[\begin{array}{cc}
			1 & 1\\
			1 & -1
		\end{array}\right]}_{H}\underbrace{\left[\begin{array}{c}
			\frac{1}{\sqrt{2}}\\
			\frac{1}{\sqrt{2}}
		\end{array}\right]}_{|+\rangle}=\underbrace{\left[\begin{array}{c}
			1\\
			0
		\end{array}\right]}_{|0\rangle},
\end{align}confirming (\ref{eq:ch1:interf}). 

\begin{figure}
\begin{center}
	\includegraphics[clip,scale=0.52]{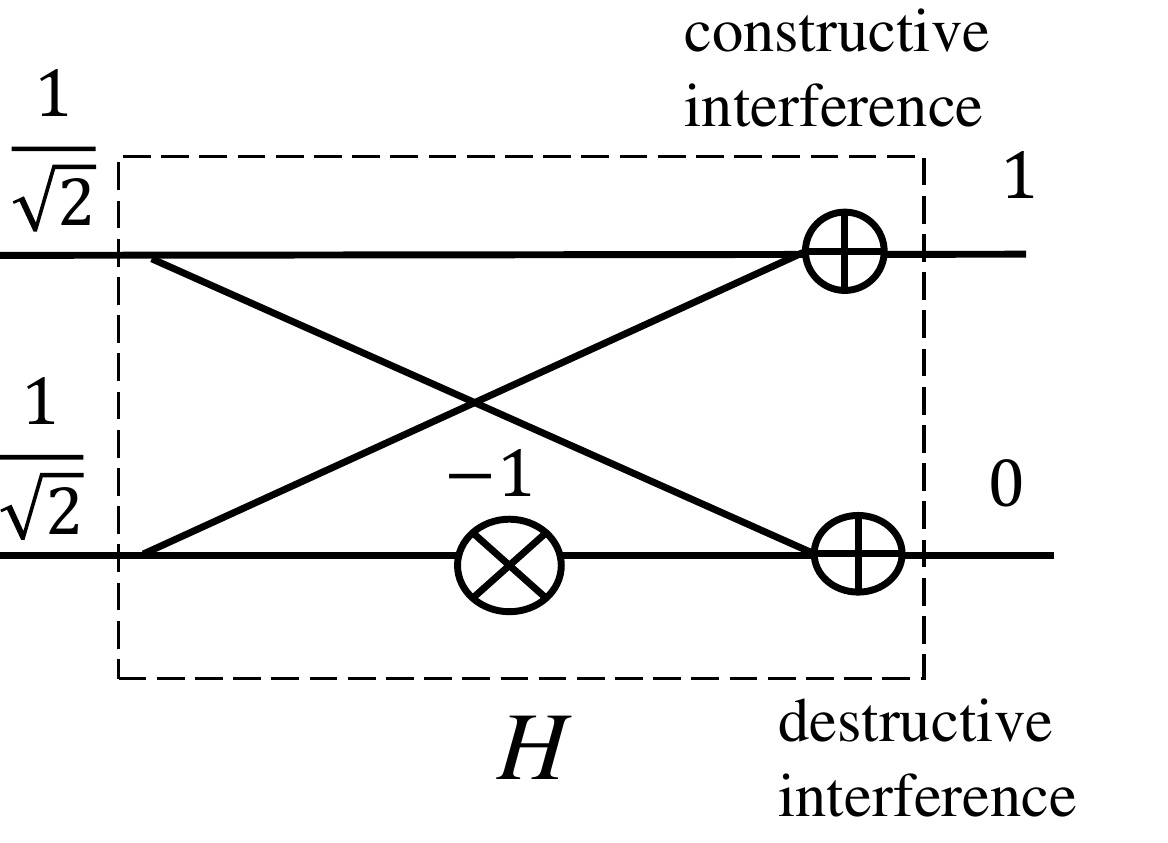}
	\par\end{center}\vspace{-0.4cm}\caption{Amplitude diagram illustrating the effect of interference for the quantum circuit in the bottom part of Fig. \ref{fig:ch1:dirmeas}.\label{fig:ch1:interf}}
\end{figure}

 Graphically, the amplitude vector calculation in (\ref{eq:ch1:amplitude}) can be represented using
	the amplitude diagram shown in Fig. \ref{fig:ch1:interf}. 
	 The amplitude diagram highlights the fact that the amplitudes of states $|0\rangle$
	and $|1\rangle$ interfere with each other as the system evolves through the Hadamard gate, reinforcing the amplitude of state $|0\rangle$ and nulling the amplitude of state $|1\rangle$ after the second Hadamard gate. Accordingly, \textbf{constructive interference} occurs for the state $|0\rangle$ while \textbf{destructive interference} takes place for state $|1\rangle$.
	
	Overall, having observed the significant difference between the output produced by random cbit state in the middle part of Fig.   \ref{fig:ch1:dirmeas} and the superposition qubit state in the bottom part of the figure, we can conclude that a qubit in a superposition state cannot be interpreted as being in either state -- it is, in some precise sense, in both states.
	
\section{Conclusions}
This chapter has introduced the qubit as the basic unit of quantum information and computing. To this end, we have built on a, rather limited, analogy with a random cbit. A qubit can evolve in one of two ways: through multiplication via a unitary transform -- i.e., through a linear reversible norm-preserving  transformation -- or through a measurement. When directly measured, a qubit can be equivalently described as being in a state of epistemic uncertainty. Accordingly, each of the two orthogonal states of the qubit is observed as the post-measurement state with a probability equal to the absolute value squared of the corresponding amplitude. However, if a qubit first evolves through a unitary transformation and is only then measured, describing the distribution of the measurement outputs  requires modelling the state of the qubit as a ``true'' superposition of the two orthogonal states. In the next chapter, we will extend the formalism and concepts introduced here to the case of multiple qubits.

\section{Recommended Resources}
The material covered in this chapter is standard, and recommended references for further reading include \cite{mermin2007quantum}, which makes particularly clear the relationship between cbits and qubits (``Qbits'');  and \cite{lipton2021introduction}, which provides a ``ket-free'' presentation, highlighting the role of linear algebra and offering useful discussions via amplitude diagrams. An extensive and endlessly useful reference is the classical book \cite{nielsen10}.
\section{Problems}

\begin{enumerate}
	\item Prove the equalities $YX=-iZ$, $ZY=-iX$, and $XZ=-iY$.
	\item Describe the output of a standard measurement when a qubit in state $|0\rangle$ is first passed through a Pauli $Y$-rotation $R_Y(\theta)$
	as a function of the angle $\theta$. Explain your results using amplitude-based
	diagrams.
	\item Explain the two-slit experiment (see recommended resources) in terms of interference.
\item Argue that the global phase of the two-dimensional vector describing
	a qubit state is not relevant to describe the outcomes of measurements
	of the qubit. 
	\item Using the Bloch sphere (see recommended references), describe
	the Pauli rotations $R_{X}(\theta)$, $R_{Y}(\theta)$, and $R_{Z}(\theta)$ geometrically, and argue that any unitary transformation can be written
	as the product
	\begin{equation}
		\exp(i\alpha)R_{Z}(\theta_{1})R_{X}(\theta_{2})R_{Z}(\theta_{3})
	\end{equation}
	for suitable angles $\alpha$, $\theta_{1},$ $\theta_{2},$ and $\theta_{3}.$
	Following your argument, demonstrate that any pair of Pauli rotations 
	can be combined to produce an arbitrary unitary transformation. 
\item Show the following equivalence relations\begin{center}
			\begin{quantikz}
				\qw & \gate{H} & \qw & \gate{X} & \qw & \gate{H}
			\end{quantikz}
			=
			\begin{quantikz}
				\qw & \gate{Z} & \qw 
			\end{quantikz}
		\end{center}
		\begin{center}
			\begin{quantikz}
				\qw & \gate{H} & \qw & \gate{Z} & \qw & \gate{H}
			\end{quantikz}
			=
			\begin{quantikz}
				\qw &  \gate{X} & \qw
			\end{quantikz}
		\end{center}
	\begin{center}
		\begin{quantikz}
			\qw & \gate{H} & \qw & \gate{Y} & \qw & \gate{H}
		\end{quantikz}
		=
		\begin{quantikz}
			\qw &  \gate{-Y} & \qw
		\end{quantikz}
	\end{center}

\end{enumerate} 

\chapter{Classical Bits (Cbits) and Quantum Bits (Qubits)}

\section{Introduction}

	In this chapter, we will describe quantum systems with more than one qubit, introducing the formalism used to model states,  
	transformations, and measurements. A key new concept arising from the analysis of multiple-qubit systems is that of entanglement -- a uniquely quantum form of statistical  dependence.

\section{Multiple Random Classical Bits}
As in the previous chapter, we start by discussing a formalism for the description of multiple random cbits, which will then be used as a reference point for the definition of a multi-qubit state. 

\subsection{Classical Bits as Integers and as One-Hot Vectors}\label{sec:ch2:onehot}

To begin, let us consider a system with $n=3$ cbits. As illustrated in Table \ref{table:ch2:mapping}, we can represent the state of a three-cbit system in one of the following ways:\begin{itemize}
	\item \textbf{as a string of $n=3$ bits}\begin{equation}\label{eq:ch2:bitstring0}
		x_{0},x_{1},x_{2},
	\end{equation}with each $k$-th bit denoted as $x_{k}\in\{0,1\}$ for $k\in\{0,1,2\}$; 
	\item \textbf{as the integer $x\in \{0,1,...,7\}$} given as 
	\begin{equation}\label{eq:ch2:msg0}
		x=2^{2} x_0+2x_1+x_2, 
	\end{equation} which converts the bit string (\ref{eq:ch2:bitstring0}) to one of $N=8$ integers by considering  the cbits in the string (\ref{eq:ch2:bitstring0}) as listed from the most significant to the least significant;
	\item \textbf{as a one-hot amplitude vector}, which is a $3\times 1$ vector with all zero elements except for a ``$1$'' digit in the position indexed by integer $x$ in (\ref{eq:ch2:msg0}) when counting from zero and starting from the top of the vector.
\end{itemize}

 As discussed in Sec. \ref{eq:ch1:Dirac}, in quantum theory,  one-hot amplitude vectors are written using Dirac's ket notation. Dirac's notation describes column
vectors with unitary norm. Accordingly, as illustrated in Table  \ref{table:ch2:mapping}, we denote the one-hot vector representing bit string (\ref{eq:ch2:bitstring0}), or equivalently integer (\ref{eq:ch2:msg0}), as $|x_{0}x_{1}x_{2}\rangle$, $|x_{0},x_{1},x_{2}\rangle$, or as $|x\rangle_3$.  The subscript in the ket $|x\rangle_{3}$  indicates the number
of cbits, here $n=3$. The subscript is introduced to avoid ambiguities, and it can be omitted when no confusion
can arise. We will see later in the monograph that the subscript in a ket can also be used also for other purposes.

\begin{table}
	\begin{center}
		\begin{tabular}{ccc}
			\toprule 
			integer $x$ & cbits & one-hot amplitude vector $|x\rangle_{n}$\tabularnewline
			\midrule
			\midrule 
			0 & 000 & $|0\rangle_{3}=|000\rangle=[1,0,0,0,0,0,0,0]^{T}$\tabularnewline
			\midrule 
			1 & 001 & $|1\rangle_{3}=|001\rangle=[0,1,0,0,0,0,0,0]^{T}$\tabularnewline
			\midrule 
			2 & 010 & $|2\rangle_{3}=|010\rangle=[0,0,1,0,0,0,0,0]^{T}$\tabularnewline
			\midrule 
			3 & 011 & $|3\rangle_{3}=|011\rangle=[0,0,0,1,0,0,0,0]^{T}$\tabularnewline
			\midrule 
			4 & 100 & $|4\rangle_{3}=|100\rangle=[0,0,0,0,1,0,0,0]^{T}$\tabularnewline
			\midrule 
			5 & 101 & $|5\rangle_{3}=|101\rangle=[0,0,0,0,0,1,0,0]^{T}$\tabularnewline
			\midrule 
			6 & 110 & $|6\rangle_{3}=|110\rangle=[0,0,0,0,0,0,1,0]^{T}$\tabularnewline
			\midrule 
			7 & 111 & $|7\rangle_{3}=|111\rangle=[0,0,0,0,0,0,0,1]^{T}$\tabularnewline
			\bottomrule
		\end{tabular}
		\par\end{center}\vspace{-0.4cm}\caption{List of cbit strings of $n=3$ cbits with associated integer representation and one-hot amplitude vector. 
		\label{table:ch2:mapping}}
\end{table}

Generalizing the example of $n=3$ qubits in Table \ref{table:ch2:mapping}, a string of $n$ cbits will be denoted throughout the text in one of the following ways: \begin{itemize}
	\item \textbf{as a string of $n$ bits}\begin{equation}\label{eq:ch2:bitstring}
	x_{0},x_{1},...,x_{n-1},
	\end{equation}with each $k$-th bit denoted as $x_{k}\in\{0,1\}$ for $k\in\{0,1,...,n-1\}$; 
\item \textbf{as the integer $x\in \{0,1,...,2^{n}-1\}$} given as 
	\begin{align}\label{eq:ch2:msg}
	x & = 2^{n-1} x_0 + 2^{n-2} x_1 + \cdots + x_{n-1}\nonumber\\
	 & = \sum_{j=0}^{n-1}2^{n-j-1}\cdot x_{j},
	\end{align} which converts the bit string to one of $N=2^n$ integers in the set $\{0,1,...,2^{n}-1\}$ by considering  the cbits in the string (\ref{eq:ch2:bitstring}) as listed from most significant to least significant reading from left to right;
\item \textbf{as a one-hot amplitude vector} denoted as  $|x_{0}x_{1}\cdots x_{n-1}\rangle$, $|x_{0},x_{1},\cdots ,x_{n-1}\rangle$, or $|x\rangle_n$,  which is an $2^{n}\times 1$ vector with all zero elements except for a $1$ digit in the position indexed by integer $x$ in (\ref{eq:ch2:msg}) when counting from zero and starting from the top of the vector.
\end{itemize}

 Representing classical information encoded by $n$ cbits requires $n$ binary
	physical systems, such as on-off switches. The one-hot amplitude  representation introduced in this subsection is exponentially less efficient, since it requires vectors of
	dimension $N=2^{n}$. For example, $n=10$ cbits are encoded into one-hot amplitude vectors of size $N=1024$; and $n=300$
	cbits require amplitude vectors of size $N\simeq10^{100}$ -- which is larger than the number of
	atoms in the universe. 

\subsection{Random Cbits and Probability Vectors}
Despite their inefficiency, one-hot vectors are routinely
		used in machine learning when dealing with discrete random 
		variables. In fact, the probability distribution 
		of a random $n$-cbit string is described by
		the $2^{n}\times1$ probability vector 	\begin{equation}\label{eq:ch2:vecp}
		p=\left[\begin{array}{c}
			p_{0}\\
			p_{1}\\
			\vdots\\
			p_{2^{n}-1}
		\end{array}\right],
		\end{equation} with $p_{x}\geq0$ being the probability of the cbit string being equal to $x\in \{0,1,...,2^{n}-1\}$.  Note again that we refer to a cbit string and to the corresponding integer representation interchangeably. The probability vector (\ref{eq:ch2:vecp}) must satisfy the condition   
	\begin{equation}\label{eq:ch2:sumprob} \sum_{x=0}^{2^{n}-1}p_{x}=1.	\end{equation}

\subsection{From Individual Cbits to Multiple Cbits via the Kronecker Product}

	As we know from the previous chapter, the state of a single (deterministic) cbit can be represented by a $2\times1$ one-hot amplitude
	vector, which may take values
	\begin{equation}
	|0\rangle=\left[\begin{array}{c}
		1\\
		0
	\end{array}\right]\textrm{ or}\textrm{ }|1\rangle=\left[\begin{array}{c}
		0\\
		1
	\end{array}\right].
	\end{equation}
	Furthermore, as introduced in the last subsection, a system of $n$ classical cbits defined by integer $x\in\{0,1,...,2^{n}-1\}$ is represented
	by a $2^{n}\times1$ one-hot amplitude vector $|x\rangle_{n}$.
	 Can we obtain the one-hot amplitude vector $|x\rangle_{n}$ for the overall system from the one-hot amplitude
	vectors $|x_k\rangle$, with $x_k\in\{0,1\}$, for the individual cbits $k\in\{0,1,...,n-1\}$? 
	
	We will see in this subsection that the answer to this question is affirmative: The state of a classical deterministic system can be described as the combination of the states of the individual subsystems. Furthermore, the mathematical tool that enables this combination is the Kronecker product.
	
		Given an $m\times1$ vector 
	\begin{equation}
		a=\left[\begin{array}{c}
			a_{1}\\
			a_{2}\\
			\vdots\\
			a_{m}
		\end{array}\right]
	\end{equation}
	and a $k\times1$ vector $b$, the \textbf{Kronecker product} $a\otimes b$ produces the $mk\times1$
	vector
	\begin{equation}
		a\otimes b=\left[\begin{array}{c}
			a_{1}b\\
			a_{2}b\\
			\vdots\\
			a_{m}b
		\end{array}\right],
	\end{equation}
	where the products $a_k b$ are evaluated element-wise on the entries of vector $b$ for $k\in\{1,2,...,m\}$.

Using this definition,  the $2^{n}\times1$ amplitude vector $|x\rangle_{n}$ for a system of $n$ cbits can be computed from the two-dimensional amplitude vectors $|x_{0}\rangle,|x_{1}\rangle,\dots,|x_{n-1}\rangle$  of the individual cbits via the $n$-fold Kronecker product
	\begin{equation}\label{eq:ch2:kron}
	|x\rangle_{n}=|x_{0}x_{1}...x_{n-1}\rangle=|x_{0}\rangle\otimes|x_{1}\rangle\otimes\cdots\otimes|x_{n-1}\rangle.
	\end{equation} By (\ref{eq:ch2:kron}), the notation $|x_{0}x_{1}...x_{n-1}\rangle$ can be interpreted as describing the Kronecker product of the states indicated within the ket.

To illustrate this operation, let us consider a system with $n=3$ cbits with values $x_{0}=1$,  $x_{1}=1$,
		and $x_{2}=0$. The amplitude vector describing the state of the overall system is given by 
		\begin{align}
			|110\rangle & =|6\rangle_{3}=\left[\begin{array}{c}
				0\\
				0\\
				0\\
				0\\
				0\\
				0\\
				1\\
				0
			\end{array}\right].\end{align} This vector can be expressed via the Kronecker product in (\ref{eq:ch2:kron}) as
			\begin{align}
		& \left[\begin{array}{c}
				\mathbf{0}\\
				\mathbf{1}
			\end{array}\right]\otimes\left[\begin{array}{c}
				0\\
				1
			\end{array}\right]\otimes\left[\begin{array}{c}
				1\\
				0
			\end{array}\right]\nonumber\\
			& =\left[\begin{array}{c}
				\mathbf{0}\cdot0\\
				\begin{array}{c}
					\mathbf{0}\cdot1\end{array}\\
				\mathbf{\mathbf{1}\cdot}0\\
				\mathbf{\mathbf{1}\cdot}1
			\end{array}\right]\otimes\left[\begin{array}{c}
				1\\
				0
			\end{array}\right]\nonumber\\
		&=\left[\begin{array}{c}
				\mathbf{0}\\
				\begin{array}{c}
					\mathbf{0}\end{array}\\
				\mathbf{0}\\
				\mathbf{1}
			\end{array}\right]\otimes\left[\begin{array}{c}
				1\\
				0
			\end{array}\right]=\left[\begin{array}{c}
				\begin{array}{c}
					\mathbf{0}\cdot1\end{array}\\
				\mathbf{0}\cdot0\\
				\begin{array}{c}
					\mathbf{0}\cdot1\end{array}\\
				\mathbf{0}\cdot0\\
				\begin{array}{c}
					\mathbf{0}\cdot1\end{array}\\
				\mathbf{0}\cdot0\\
				\begin{array}{c}
					\mathbf{1}\cdot1\end{array}\\
				\mathbf{1}\cdot0
			\end{array}\right]=\left[\begin{array}{c}
				0\\
				0\\
				0\\
				0\\
				0\\
				0\\
				1\\
				0
			\end{array}\right],
		\end{align} where the bold font has been introduced in order to facilitate the interpretation of the operations at hand, and we have used the associative property of the Kronecker product. The associative property stipulates that Kronecker products can be grouped in pairs in any arbitrary way in order to evaluate a sequence of Kronecker products.

\subsection{From Individual Random Cbits to Multiple Random Cbits?}\label{sec:ch2:ftm}

In this subsection, we address a question analogous to that considered in the previous subsection moving from deterministic to random cbits. To formulate the question, for $n$ random cbits, suppose that we know the $2\times1$ (marginal) 
		probability vectors for all the constituent $n$ cbits. Can we obtain from these
		$n$ vectors the $2^{n}\times1$ probability vector (\ref{eq:ch2:vecp}) of the overall
		system? As we review in this subsection, the answer is negative, unless the random cbits are statistically independent.
		
	To elaborate, let us consider the case with $n=2$ random cbits. The individual probability
		vectors for the two cbits can be written as
		\begin{equation}\label{eq:ch2:marginals}
			p^{A}=\left[\begin{array}{c}
				p_{0}^{A}\\
				p_{1}^{A}
			\end{array}\right]\textrm{ and \ensuremath{p^{B}}=\ensuremath{\left[\begin{array}{c}
						p_{0}^{B}\\
						p_{1}^{B}
					\end{array}\right],}}
		\end{equation} where the superscript $A$ and $B$ identifies the two cbits. In (\ref{eq:ch2:marginals}), the element $p^{A}_x$ is the marginal probability for the first random cbit to take value $x\in\{0,1\}$, and an analogous definition applies for $p^{B}_x$. 
		If the cbits are \textbf{statistically independent}, we can write the probability vector
		of the system of two random cbits as
		\begin{equation}\label{eq:ch2:indcbits}
			p=\left[\begin{array}{c}
				p_{0}^{A}\\
				p_{1}^{A}
			\end{array}\right]\otimes\textrm{\ensuremath{\left[\begin{array}{c}
						p_{0}^{B}\\
						p_{1}^{B}
					\end{array}\right]}}=\left[\begin{array}{c}
				p_{0}^{A}p_{0}^{B}\\
				p_{0}^{A}p_{1}^{B}\\
				p_{1}^{A}p_{0}^{B}\\
				p_{1}^{A}p_{1}^{B}
			\end{array}\right].
		\end{equation} This is because the joint probability of independent random variables is the product of the individual marginal distributions.
	
		For two \textbf{statistically dependent} -- or, somewhat less formally, \textbf{correlated} -- random cbits, the probability vector cannot be expressed in terms of the individual marginal probability vectors, i.e., we have the inequality
		\begin{equation}
			p=\left[\begin{array}{c}
				p_{00}\\
				p_{01}\\
				p_{10}\\
				p_{11}
			\end{array}\right]\neq\left[\begin{array}{c}
				p_{0}^{A}p_{0}^{B}\\
				p_{0}^{A}p_{1}^{B}\\
				p_{1}^{A}p_{0}^{B}\\
				p_{1}^{A}p_{1}^{B}
			\end{array}\right],
		\end{equation}where $p_{xy}$ is the joint probability that the first cbit takes value $x\in\{0,1\}$ and the second cbit takes value $y\in\{0,1\}$. Therefore, unless the random cbits are independent, the  probability vectors for the individual random cbits only provide information about the corresponding marginal distributions, from which one cannot recover the joint distribution. Knowing the individual states of random subsystems does not allow one to reconstruct the state of the overall random system.

\subsection{Probability Vectors as ``Superpositions'' of Basis States}

As we have seen in Sec. \ref{sec:ch1:rcbitprob}, the probability vector
for a single random cbit can be written as
\begin{equation}
	p=\left[\begin{array}{c}
		p_{0}\\
		p_{1}
	\end{array}\right]=p_{0}|0\rangle+p_{1}|1\rangle,
\end{equation}
with $p_{0}+p_{1}=1$. In words, a probability vector $p$ can be thought of as a \textbf{``superposition''} of the computational
basis vectors $|0\rangle$ and $|1\rangle$. In this subsection, we generalize this relationship to any number $n$ of (jointly distributed) random cbits.

To this end, we start by defining the \textbf{computational basis} of the $2^{n}$-dimensional linear vector
space as the set of $2^{n}$ one-hot amplitude vectors 
\begin{equation}\label{eq:ch2:cK}
	|x\rangle=|x_{0},x_{1},\cdots,x_{n-1}\rangle
\end{equation}for $x\in\{0,1,...,2^{n}-1\}$ and $x_k\in\{0,1\}$ with $k\in\{0,1,...,n-1\}.$ Note that we have dropped the subscript $n$ in (\ref{eq:ch2:kron}) in order to simplify the notation. The vectors\begin{equation}\label{eq:ch1:combasis}\{|x\rangle\}_{x=0}^{2^{n}-1}=\{|0\rangle,|1\rangle,....,|2^{n-1}\rangle\}
\end{equation} can be readily proved to be mutually orthogonal, i.e., we have \begin{equation} 
	\langle x|x' \rangle =0 \textrm{ if $x'\neq x$}
\end{equation}		and to have unitary norm, i.e., $\langle x|x \rangle =1$. Therefore, they form an orthonormal basis for the $2^n$-dimensional linear vector space, which is known as the computational basis.

Therefore,  the $2^n$-dimensional probability vector $p$  for a system of $n$
cbits, i.e., (\ref{eq:ch2:vecp}), can be written as 
\begin{equation}\label{eq:ch2:vecp2}
	p=\left[\begin{array}{c}
		p_{0}\\
		p_{1}\\
		\vdots\\
		p_{2^{n}-1}
	\end{array}\right]=\sum_{x=0}^{2^{n}-1}p_{x}|x\rangle,
\end{equation}
with $\sum_{x=0}^{2^{n}-1}p_{x}=1$. Accordingly, probability vector $p$ is the  \textbf{``superposition''} 
of the $2^{n}$ vectors in the computational basis (\ref{eq:ch1:combasis}), with each vector $|x\rangle$ weighted by the corresponding probability $p_x$. It is worth noting that,  vector $p$,  while not generally expressible as a Kronecker product of two-dimensional vectors, can always be written as a linear combination of vectors that admit such decomposition. In fact, each computational-basis vector $|x\rangle$ in (\ref{eq:ch2:vecp2}), by (\ref{eq:ch2:kron}), is a Kronecker product of per-cbit states.

\section{Multiple Qubits}
In this section, we describe systems consisting of multiple qubits.

\subsection{Qubits as Amplitude Vectors}\label{sec:ch2:Bellbasis}

In a manner that is somewhat analogous to the state of a set of $n$ random cbits, described by the probability vector  $p$ in (\ref{eq:ch2:vecp2}), the state of an $n$-qubit system is specified by the $2^{n}\times1$
		vector
		\begin{equation}\label{eq:ch2:multiqubit}
		|\psi\rangle=\left[\begin{array}{c}
			\alpha_{0}\\
			\alpha_{1}\\
			\vdots\\
			\alpha_{2^{n}-1}
		\end{array}\right]=\sum_{x=0}^{2^{n}-1}\alpha_{x}|x\rangle,
		\end{equation}
		with amplitudes $\{\alpha_x\}_{x=0}^{2^{n}-1}$ satisfying the equality
		\begin{equation}\label{eq:ch2:unitnorm}
		\sum_{x=0}^{2^{n}-1}|\alpha_{x}|^{2}=1.
		\end{equation}
The quantum state vector $|\psi\rangle$ is hence an
		element of a $2^{n}$-dimensional \emph{complex} linear vector space space, which is referred to as the $2^{n}$-dimensional \textbf{Hilbert space}. Furthermore, by (\ref{eq:ch2:multiqubit}),  we say that the state $|\psi\rangle$ is a \textbf{superposition}
		of the $N=2^{n}$ vectors in the computational basis $\{|x\rangle\}_{x=0}^{2^{n}-1}$ with respective amplitudes $\{\alpha_x\}_{x=0}^{2^{n}-1}$. 
		
		Generalizing the discussion concerning single qubits in Sec. \ref{sec:ch1:qubitav}, unlike the probability vector $p$, the quantum state (\ref{eq:ch2:multiqubit})  has \emph{complex} entries $\alpha_0$ and $\alpha_1$, and it satisfies the defining property of having \emph{unitary norm} by (\ref{eq:ch2:unitnorm}) (which is different from (\ref{eq:ch2:sumprob})).  
		
		Moreover, while being distinct from the state of $n$ random cbits, the quantum state (\ref{eq:ch2:multiqubit}) recovers as special cases the $2^{n}$ possible states of $n$ deterministic cbits when expressed as one-hot amplitude vectors. In fact, setting the amplitudes as $\alpha_x=1$, and hence $\alpha_x'=0$ for $x'\neq x$, yields the deterministic $n$-cbit state $|x\rangle$. Therefore, a system of $n$ qubits that can only assume the states $\{|x\rangle\}_{x=0}^{2^{n}-1}$ in the computational basis is equivalent to $n$ deterministic cbits.
		
		As an example, for $n=2$, we can write the $N=2^2=4$-dimensional state vector in the following equivalent ways
			\begin{align}\label{eq:ch2:ex}
			|\psi\rangle & =\left[\begin{array}{c}
				\alpha_{0}\\
				\alpha_{1}\\
				\alpha_{2}\\
				\alpha_{3}
			\end{array}\right]=\left[\begin{array}{c}
				\alpha_{00}\\
				\alpha_{01}\\
				\alpha_{10}\\
				\alpha_{11}
			\end{array}\right]\nonumber\\
			& =\alpha_{0}\left[\begin{array}{c}
				1\\
				0\\
				0\\
				0
			\end{array}\right]+\alpha_{1}\left[\begin{array}{c}
				0\\
				1\\
				0\\
				0
			\end{array}\right]+\alpha_{2}\left[\begin{array}{c}
				0\\
				0\\
				1\\
				0
			\end{array}\right]+\alpha_{3}\left[\begin{array}{c}
				0\\
				0\\
				0\\
				1
			\end{array}\right]\nonumber\\
			& =\alpha_{0}|0\rangle+\alpha_{1}|1\rangle+\alpha_{2}|2\rangle+\alpha_{3}|3\rangle\nonumber\\
			& =\alpha_{00}|00\rangle+\alpha_{01}|01\rangle+\alpha_{10}|10\rangle+\alpha_{11}|11\rangle,
		\end{align}where we have indexed the vectors in the computational basis using either the integer or binary string representations. 
	
As highlighted in (\ref{eq:ch2:ex}), the amplitudes defining quantum state (\ref{eq:ch2:multiqubit}) can be indexed by an integer $x\in\{0,1,...,2^n-1\}$ or, equivalently, by an $n$-cbit string. The latter notation formalizes a quantum state as a \textbf{tensor}. A tensor of order $n$ is a multi-dimensional ``table'' whose elements are identified by an $n$-dimensional vector, here of binary numbers.  For the quantum state (\ref{eq:ch2:multiqubit}), the tensor is defined by a table with entries $\alpha_{x_0,x_1,...,x_{n-1}}$, where each $k$-th dimension is indexed by cbit $x_k\in\{0,1\}$. 
	
		The exponential size of the quantum state vector (\ref{eq:ch2:multiqubit}) makes it practically impossible, in general, to use a classical computer to simulate the operation of quantum systems with as low as $n=50$ qubits. In fact, representing a quantum state of $50$ qubits requires around $2^{50}\gtrsim10^{15}$
		bytes to fit into the main memory, which is hard even for the largest supercomputers available today (see also Sec. \ref{sec:ch2:onehot}).

		Being a $2^n$-dimensional vector, the quantum state (\ref{eq:ch2:multiqubit}) can be expressed as a linear combination of the vectors $\{|v_{x}\rangle\}_{x\in\{0,1,...,2^{n}-1\}}$ forming \emph{any} orthonormal basis of the $2^{n}$-dimensional Hilbert space. As an important example, for $n=2$ qubits, an orthonormal basis for the $2^{2}=4$-dimensional
		Hilbert space is given by the so-called \textbf{Bell basis} consisting of the four \textbf{Bell states}
			\begin{align}\label{eq:ch2:Bell}
				|\Phi^{+}\rangle & =\frac{1}{\sqrt{2}}(|00\rangle+|11\rangle)=\frac{1}{\sqrt{2}}\left[1,0,0,1\right]^{T}\\
				|\Psi^{+}\rangle & =\frac{1}{\sqrt{2}}(|01\rangle+|10\rangle)=\frac{1}{\sqrt{2}}\left[0,1,1,0\right]^{T}\\
				|\Phi^{-}\rangle & =\frac{1}{\sqrt{2}}(|00\rangle-|11\rangle)=\frac{1}{\sqrt{2}}\left[1,0,0,-1\right]^{T}\\
				|\Psi^{-}\rangle & =\frac{1}{\sqrt{2}}(|01\rangle-|10\rangle)=\frac{1}{\sqrt{2}}\left[0,1,-1,0\right]^{T}.
			\end{align}As we will see throughout this text, Bell states play an important role in quantum computing as building blocks of routines and protocols. It is useful to know that we can also write the Bell basis in terms of the diagonal basis (see Sec. \ref{sec:ch1:qubitav}) as
			\begin{align}
				|\Phi^{+}\rangle & =\frac{1}{\sqrt{2}}(|++\rangle+|--\rangle)\\
				|\Psi^{+}\rangle & =\frac{1}{\sqrt{2}}(|++\rangle-|--\rangle)\\
				|\Phi^{-}\rangle & =\frac{1}{\sqrt{2}}(|-+\rangle+|+-\rangle)\\
				|\Psi^{-}\rangle & =\frac{1}{\sqrt{2}}(|-+\rangle-|+-\rangle).
			\end{align}

		\subsection{Measuring Qubits: von Neumann Measurements}
		
	A	\textbf{von Neumann measurement in the computational basis}, or \textbf{standard measurement} for short, takes as input $n$ qubits in an arbitrary state $|\psi\rangle$ as in (\ref{eq:ch2:multiqubit}),  and it produces $n$ cbits as the measurement's output, while leaving the qubits in a generally different state from the original state $|\psi\rangle$. Specifically, a standard measurement satisfies the following two properties, which are direct extensions of the properties reviewed in Sec. \ref{sec:ch1:meas} for single-qubit measurements (i.e., for $n=1$): \begin{itemize} \item \textbf{Born's rule}: The probability of observing the $n$-cbit string $x\in\{0,1\}^{n}$ is
			\begin{align}\label{eq:ch2:Born}
				\Pr[\textrm{measurement output equals \ensuremath{x\in\{0,1\}^{n}}}]&=|\alpha_{x}|^{2}\nonumber \\
				&=|\langle x|\psi\rangle|^{2};
			\end{align}
			\item \textbf{``Collapse'' of the state}: If the measured $n$-cbit string is $x$, the \textbf{post-measurement state} of the qubits is $|x\rangle$.
		\end{itemize}
	
Therefore, by (\ref{eq:ch2:Born}), the probability of measuring a cbit string $x$ is given by the absolute value squared of the corresponding amplitude $\langle x|\psi\rangle$. Moreover, a measurement randomly ``collapses'' state $|\psi\rangle$, which is generally in a superposition of computational-basis states, into a single computational-basis vector $|x\rangle$. In this regard,  one can check that a system of $n$ qubits that can only take states $|x\rangle$ in the computational basis behaves like a deterministic cbits. This is in the sense that, by Born's rule, measuring a qubit in state $|x\rangle$, for $x\in\{0,1,...,2^{n}-1\}$, returns output $x$ with probability $1$, while leaving the qubits' state unchanged.
	
\begin{figure}
\begin{center}				
	\begin{quantikz}  	
		\qw & \meter{$x_0$} & \qw \rstick{\ket{x_0}}\\
		\qw & \meter{$x_1$} & \qw \rstick{\ket{x_1}}\\
		& \vdots &\\
		\qw & \meter{$x_{n-1}$} & \qw \rstick{\ket{x_{n-1}}}
		
\end{quantikz}\end{center}\vspace{-0.4cm} \caption{A standard measurement on $n$ qubits can be implemented as $n$ separate standard measurements on the individual qubits. Each $k$-th cbit $x_k$ of the overall $n$-cbit output of the measurement is produced by measuring the $k$-th qubit. \label{fig:ch2:parmeas}}
\end{figure}
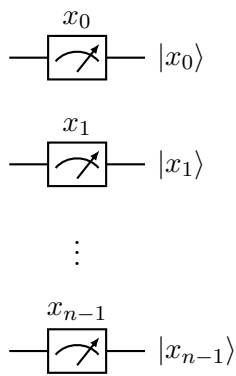

	As we will further clarify in the next chapter, a standard measurement on $n$ qubits can be implemented by applying a standard measurement separately to each individual qubit. This is illustrated in Fig. \ref{fig:ch2:parmeas}, in which a separate measurement block is applied to each qubit. Accordingly, each individual cbit $x_k$ of a measurement output $x_0,x_1,...,x_{n-1}$ corresponding to an integer $x\in\{0,1,...,2^n-1\}$ is produced by measuring the $k$-th qubit. Furthermore, we will see in the next chapter that such per-qubit measurements can be carried out in any order on the qubit, including at the same time, without changing the distribution of the measurement outputs given by Born's rule.

	\section{Quantum Circuits and Local Operations}\label{sec:ch2:singlequbitgates}
	
	Generalizing the discussion in Sec. \ref{sec:ch1:unit}, when not measured, a system of $n$ qubits in state $|\psi\rangle$ evolves according to a linear, norm-preserving, reversible transformation, which is defined by a  $2^n\times 2^n$ unitary matrix $U$. Accordingly, in a closed system, a quantum input state $|\psi\rangle$ evolves into an output state $|\psi'\rangle=U|\psi\rangle$.  In this section, we first introduce the formalism used to describe and design unitary transformations, namely quantum circuits. Then, we elaborate on the simplest type of quantum circuit in which quantum gates are applied in parallel to the qubits.
	
	\subsection{Quantum Circuits}
	
	A unitary matrix satisfies the equalities (\ref{eq:ch1:defuni}), i.e., $UU^{\dagger}=U^{\dagger}U=I$, where $I$ is a $2^n\times 2^n$ identity matrix.	While in theory every unitary matrix $U$ is physically realizable, in practice, transformations $U$ are described and implemented as a cascade of unitary matrices that apply to small subsets of qubits, typically consisting of only one or two qubits. This cascade, and with it the overall matrix $U$, are described by a quantum circuit.

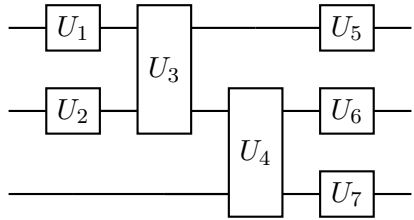
\begin{figure}
		\begin{center} \begin{quantikz} & \gate{U_1} & \gate[wires=2]{U_3} & \qw & \gate{U_5} & \qw \\ & \gate{U_2} & \qw & \gate[wires=2]{U_4}& \gate{U_6}& \qw\\ & \qw & \qw & \qw  & \gate{U_7}& \qw
		\end{quantikz} \end{center}\vspace{-0.4cm}\caption{A quantum circuit representing the operation of an  $8\times 8$ unitary transformation of $n=3$ qubits as a sequence of one- and two-qubit gates (with the former being defined by $2\times2$ matrices, and the latter by $4\times4$ matrices).\label{fig:ch2:cascade}}
	\end{figure}

	A \textbf{quantum circuit} contains $n$ wires -- one per qubit -- and it describes the sequence of unitary matrices applied to the $n$ qubits with time flowing from left to right. An example is shown in Fig. \ref{fig:ch2:cascade}. 	Unitary matrices applied to a subset of $m$ qubits are referred to as  $m$-qubit \textbf{quantum gates}.  For the example in Fig.  \ref{fig:ch2:cascade}, single-qubit gates $U_1$ and $U_2$ are applied to the first two qubits; then a two-qubit gate $U_3$ is applied to the first two qubits; and so on. Existing quantum computers implement a library of \emph{single-qubit} quantum gates and \emph{two-qubit} quantum gates.
	
The operations specified by a quantum circuit are applied \textbf{in place} to the register on $n$ qubits in the order from left to right. Accordingly, quantum gates are applied sequentially to the set of $n$ physical qubits, causing the state of the $n$ qubits to evolve over time via unitary transformations. 
	
	 It is emphasized that, while a quantum circuit contains $n$ wires, a full description of the corresponding quantum transformation applied at each time requires the definition of a $2^n \times 2^n$ unitary matrix $U$. Formally, quantum circuits can be studied within the framework of tensor networks (see recommended resources in Sec. \ref{sec:ch2:rr}).
	
	\subsection{Local Operations}
	
		We will have more to say about  multi-qubit gates in Sec. \ref{sec:ch2:multiqubit}. In this subsection, we consider the simple setting in which the unitary $U$ is composed of separate single-qubit gates applied in \textbf{parallel} to the $n$ qubits. We refer to such operations as being \textbf{local} to each qubit. 
	
 To start, let us generalize the definition of the \textbf{Kronecker product} to matrices. Given an $m\times k$ matrix 
 \begin{equation}
 A=\left[\begin{array}{ccc}
 	a_{11} & a_{12} & \cdots\\
 	a_{21} & a_{22} & \cdots\\
 	\vdots & \vdots\\
 	a_{m1} & a_{m2} & \cdots
 \end{array}\begin{array}{c}
 	a_{1k}\\
 	a_{2k}\\
 	\vdots\\
 	a_{mk}
 \end{array}\right]
 \end{equation}
 and a $l\times r$ matrix $B$, the Kronecker product produces the
 $ml\times kr$ matrix
 \begin{equation}
 A\otimes B=\left[\begin{array}{ccc}
 	a_{11}B & a_{12}B & \cdots\\
 	a_{21}B & a_{22}B & \cdots\\
 	\vdots & \vdots\\
 	a_{m1}B & a_{m2}B & \cdots
 \end{array}\begin{array}{c}
 	a_{1k}B\\
 	a_{2k}B\\
 	\vdots\\
 	a_{mk}B
 \end{array}\right],
 \end{equation}
 where the product $a_{jk} B$ is applied in an element-wise manner to matrix $B$.

	\begin{figure}
\begin{center}
			\begin{quantikz}
			 & \gate{U^{A}} & \qw    \\ 
				& \gate{U^{B}} & \qw  
			\end{quantikz}
	\end{center}\vspace{-0.4cm}\caption{A $4\times4$ unitary matrix obtained as the local application of $2\times 2$ unitary matrices $U^{A}$ and $U^{B}$ to two qubits.\label{fig:ch2:sepgates}}
\end{figure}
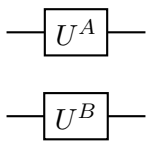

Let us consider first a two-qubit system, i.e., with $n=2$. In it, as illustrated in Fig. \ref{fig:ch2:sepgates},  the $2\times2$ unitary $U^{A}$ is applied to the first qubit, while the $2\times 2$ unitary $U^{B}$ is applied to the second qubit. The overall $4\times 4$ unitary $U$ applied to the system of two qubits can be expressed as the Kronecker product \begin{equation}\label{eq:ch2:kronU}
	U=U^{A} \otimes U^{B}. 
\end{equation} In words, the Kronecker product of $2\times 2$ unitaries describes a transformation in which the corresponding single-qubit quantum gates are applied locally to the qubits. A multi-qubit gate that can be described as the Kronecker product of individual qubit gates is referred to as implementing  \textbf{local operations} on the qubits. 

As an example of local operations, the unitary $I\otimes U$, with $I$ being the $2\times 2$ identity matrix, corresponds to a transformation in which no processing is done on the first qubit and the single-qubit gate $U$ is applied to the second qubit.  

	To understand the effect of the application of local operations, take the example of an input state $|xy\rangle=|x\rangle \otimes |y\rangle$ in the computational basis, where $x,y\in\{0,1\}$. The corresponding output state is given by  \begin{align}\label{eq:ch2:firstkron}
		(U^{A}\otimes U^{B})|xy\rangle &= (U^{A}\otimes U^{B})(|x\rangle \otimes |y\rangle) \nonumber \\ &=U^{A}|x\rangle \otimes U^{B}|y\rangle
	\end{align} for all $x,y\in\{0,1\}$. In the first line of (\ref{eq:ch2:firstkron}), we have used the key property that the multiplication of two Kronecker products
	yields the Kronecker product of the corresponding multiplications. By (\ref{eq:ch2:firstkron}), the output state is obtained by applying the  unitaries $U^A$ and $U^B$  in parallel to the corresponding qubits' states.

	Take now a general two-qubit state (\ref{eq:ch2:multiqubit}), i.e., \begin{equation}\label{eq:ch2:input} |\psi \rangle = \alpha_{00}|00\rangle + \alpha_{01}|01\rangle+ \alpha_{10}|10\rangle+ \alpha_{11}|11\rangle. 
\end{equation} Note that this state cannot be written, apart from special cases, as the Kronecker products of individual states for the two qubits. The output of the circuit in Fig. \ref{fig:ch2:sepgates} is obtained by applying the equality  (\ref{eq:ch2:firstkron}) to each term in the superposition as\begin{align}\label{eq:ch2:output} U|\psi \rangle & = \alpha_{00}U|00\rangle + \alpha_{01}U|01\rangle+ \alpha_{10}U|10\rangle+ \alpha_{11}U|11\rangle \nonumber \\
& = \sum_{x=0}^{1} \sum_{y=0}^{1} \alpha_{xy}(U^A|x\rangle \otimes U^B |y\rangle).
\end{align}

Local operations on $n>2$ qubits are naturally defined as the Kronecker product of $2\times 2$ unitary matrices describing the corresponding qubit gates applied in parallel to the $n$ qubits.

\section{Entanglement}

	 The amplitude vector of an $n$-qubit system
	has some aspects in common with probability vectors describing the state of $n$ random cbits -- including its
	dimension, $2^n$. However, qubits behave in fundamentally different ways as compared to random cbits in terms of their evolution and interactions with other systems. Sec. \ref{sec:ch1:inference} introduced the concept of \textbf{interference}, demonstrating that the behavior exhibited by qubits cannot be explained by  modelling a qubit as a random cbit. 
	In this section, we introduce a new distinguishing feature of quantum systems as compared to classical random systems: \textbf{entanglement}. Entanglement arises in the presence of multiple qubits, and may be thought of as a stronger form
	of statistical dependence that only applies to quantum systems.

	\subsection{Separable and Entangled States}
	 To start, let us define a set of $n$ qubits as \textbf{separable} if the qubits' joint state $|\psi\rangle$, a $2^n$-dimensional vector, can be expressed as the Kronecker product of the individual qubit states. For two arbitrary qubit states,  $|\psi^{A}\rangle=\alpha^{A}_{0}|0\rangle+\alpha^{A}_{1}|1\rangle$  for the first qubit and $|\psi^{B}\rangle=\alpha^{B}_{0}|0\rangle+\alpha^{B}_{1}|1\rangle$ for the second, the general form of a separable state with $n=2$ is \begin{equation}\label{eq:ch2:sep}
	 	|\psi^{A}\rangle \otimes |\psi^{B}\rangle=\left[\begin{array}{c}
	 		\alpha^{A}_0 \cdot 	\alpha^{B}_0\\
	 		\alpha^{A}_0 \cdot \alpha^{B}_1\\
	 		\alpha^{A}_1 \cdot 	\alpha^{B}_0\\
	 		\alpha^{A}_1 \cdot 	\alpha^{B}_1
	 	\end{array}\right].\end{equation} The definition of separable state should call to mind the definition of \emph{independent} random cbits given in (\ref{eq:ch2:indcbits}). In fact, the two definitions are formally equivalent if one replaces quantum amplitudes with probabilities.
 	
 	In terms of notation, a separable state in the form $|\psi^{A}\rangle \otimes |\psi^{B}\rangle$ can  also be written as \begin{equation}|\psi^{A},\psi^{B}\rangle=|\psi^{A}\rangle \otimes |\psi^{B}\rangle=|\psi^{A}\rangle |\psi^{B}\rangle,
 	\end{equation} dropping the Kronecker operation. Note that this notational convention was already used when writing a vector in the computational basis as in (\ref{eq:ch2:kron}).  As a related matter of notation, as in (\ref{eq:ch2:kron}), we will often drop the comma between the identifiers of the states for separate qubits, that is, we may write $|\psi^{A},\psi^{B}\rangle$ as $|\psi^{A}\psi^{B}\rangle$. 
 	
 	As discussed in Sec. \ref{sec:ch2:ftm}, statistically \emph{dependent} random cbits are characterized by probability vectors that cannot be expressed as the Kronecker product of the (marginal) probability vectors of the individual random cbits. In a somewhat analogous manner, a set of qubits is said to be \textbf{entangled} if its state cannot be written
	as the Kronecker product of the states of the individual qubits. As for statistical dependence, there are different degrees of entanglement, with some joint states being more entangled than others. 
	
	
	We will see in the next subsection that the parallel between dependent random cbits and entangled qubits is purely formal, in the sense that measurement outputs of entangled qubits exhibit  statistical behaviors that cannot be explained by means of standard correlations of cbits.

Important examples of entangled states for $n=2$ qubits are given by the Bell states described in Sec.  \ref{sec:ch2:Bellbasis}. For example, the so-called \textbf{Bell pair} 
	\begin{equation}\label{eq:ch2:bellpair}
	|\Phi^{+}\rangle=\frac{1}{\sqrt{2}}(|00\rangle+|11\rangle)=\frac{1}{\sqrt{2}}\left[\begin{array}{c}
		1\\
		0\\
		0\\
		1
	\end{array}\right]
	\end{equation}  cannot be written as the Kronecker product of two individual qubit states. To see this, note that there is no choice for the two individual qubit states $|\psi^{A}\rangle$ and $|\psi^{B}\rangle$, which makes (\ref{eq:ch2:sep}) equal to the Bell state (\ref{eq:ch2:bellpair}). Bell states are \textbf{maximally entangled} pairs of qubits, in the sense that any pair of entangled qubits can be obtained from a Bell pair (or another Bell state) through local operations on the two qubits of the form (\ref{eq:ch2:kronU}). Appendix A in the next chapter describes methods to quantify entanglement in pairs of qubits.

When considering $n\geq 3$ qubits, entanglement may only involve a subset of the qubits. For example, three entangled qubits could be only \textbf{partially entangled} as in a state of the form $|\Phi^+\rangle \otimes |0\rangle$. In it, the first and the second qubits are mutually entangled, forming a Bell pair, while the third qubit is not entangled with the first two qubits. While not separable across the three qubits, and hence entangled according to the definition given here, this state is said to be \emph{biseparable}. More discussion on multipartite entanglement can be found in Appendix B of the next chapter.

A key fact about entanglement is that it does not survive a standard measurement (of all qubits). This is in the sense that, no matter what the quantum state is before the measurement, the post-measurement state is separable, since it is given by one of the vectors in the computational basis as illustrated in Fig. \ref{fig:ch2:parmeas}$  $.

\subsection{A First Example of the Non-Classical Behavior of Entangled Qubits}\label{sec:ch2:nonclass}

	 Consider two qubits in the Bell state $|\Phi^{+}\rangle$. If we directly apply a standard measurements to the two qubits in the Bell pair, by Born's rule, we obtain the output $x=0$, i.e., the cbit string $00$, with probability $1/2$, or $x=3$, i.e., the cbit string $11$, with the same probability. This behavior is equivalent to that of  a classically correlated pair of random cbits that is in state $|00\rangle$ or $|11\rangle$
	 with equal probability, i.e., of random cbits with joint probability
	vector $p=[1/2,0,0,1/2]^{T}.$

	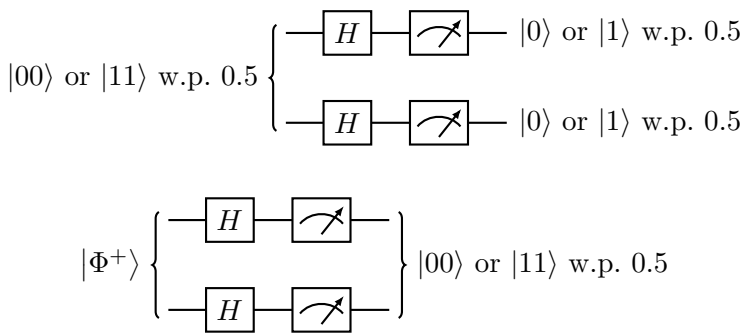
\begin{figure}
			\begin{center}
			\begin{quantikz}
				\lstick[wires=2]{$\ket{00}$ or $\ket{11}$ w.p. $0.5$} & \gate{H} &   \meter{} & \qw  \rstick{$\ket{0}$ or $\ket{1}$ w.p. $0.5$}\\ 
				& \gate{H} &  \meter{} & \qw \rstick{$\ket{0}$ or $\ket{1}$ w.p. $0.5$}
			\end{quantikz}
		\end{center}
		\begin{center}
			\begin{quantikz}
				\lstick[wires=2]{$\ket{\Phi^{+}}$} & \gate{H} &   \meter{} & \qw \rstick[wires=2]{$\ket{00}$ or $\ket{11}$ w.p. $0.5$} \\ 
				& \gate{H} &  \meter{} & \qw
		
		\end{quantikz}
		\end{center}\caption{An example used to illustrate the behavior of entangled qubits: (top) When the input state describes correlated cbits, the outputs of the measurements are independent random cbits; (bottom) When the input state consists of entangled qubits, the measurement outputs are correlated cbits.\label{fig:ch2:ent1}}
	\end{figure}
	 
	 Let us now allow the Bell pair to evolve as described in the bottom part of the quantum circuit in Fig. \ref{fig:ch2:ent1}. Accordingly, a local Hadamard gate is
	 applied separately to each qubit, as described in Sec. \ref{sec:ch2:singlequbitgates}, before a standard measurement is made. We will compare next the behavior of entangled qubits with that of statistically dependent random cbits.
	 
	 Assume first, \emph{incorrectly}, that the input state of the two qubits is state $|00\rangle$ or  state $|11\rangle$
	 	with equal probability, hence being equivalent to two fully correlated random cbits. This is illustrated in the top part of Fig. \ref{fig:ch2:ent1}. By the property
	 	of the Kronecker product (\ref{eq:ch2:firstkron}), under this assumption, the state of the qubits at the output of the two local  Hadamard gates would be \begin{align}
	 		(H\otimes H)|00\rangle & =(H\otimes H)(|0\rangle\otimes|0\rangle)=H|0\rangle\otimes H|0\rangle\nonumber\\
	 		& =|+\rangle\otimes|+\rangle=\frac{1}{2}(|00\rangle+|01\rangle+|10\rangle+|11\rangle)
	 	\end{align} or \begin{align}
	 	(H\otimes H)|11\rangle & =(H\otimes H)(|1\rangle\otimes|1\rangle)=H|1\rangle\otimes H|1\rangle\nonumber\\
	 	& =|-\rangle\otimes|-\rangle=\frac{1}{2}(|00\rangle-|01\rangle-|10\rangle+|11\rangle)
 	\end{align} with equal probability, where we have used the property (\ref{eq:ch1:Hcomp}) of the Hadamard gate. 
 
From the discussion in the previous paragraph, it follows via the law of total probability that a standard measurement would yield outputs  $00$,  $01$, $10$, or $11$ with equal probability. Equivalently, the two output cbits produced by the measurements of the two qubits in Fig. \ref{fig:ch2:ent1} would be independent and would take either value $0$ or $1$ with probability $1/2$. To sum up, as depicted in the top part of Fig. \ref{fig:ch2:ent1}, if the state of the qubits at the input of the circuit was equivalent to that of classical correlated cbits, the presence of the Hadamard gates would destroy the statistical dependence between the two cbits, producing independent cbits at the output of the measurements.
	 
	Consider now the actual situation of interest in which the input quantum state is given by the Bell pair $|\Phi^{+}\rangle$. In this case, the state produced by the circuit in Fig. \ref{fig:ch2:ent1}, prior to the measurement, is  	\begin{align}
	 		(H\otimes H)|\Phi^{+}\rangle & =\frac{1}{\sqrt{2}}(H\otimes H)(|00\rangle+|11\rangle)\nonumber\\
	 		& =\frac{1}{2\sqrt{2}}(|00\rangle+|01\rangle+|10\rangle+|11\rangle+|00\rangle-|01\rangle-|10\rangle+|11\rangle)\nonumber\\
	 		& =|\Phi^{+}\rangle.
	 	\end{align} It follows that the measurement of the two qubits produces the cbit string $00$ or $11$ with equal probability. 	Therefore, the local Hadamard gates leave the state of the system, $|\Phi^{+}\rangle$, unchanged, preserving the correlation
	 	between the two cbits produced by the measurement. (Mathematically, this happens because the state  $|\Phi^{+}\rangle$ is an eigenvector of the operator $H\otimes H$.)
	 
	 This example demonstrates that entangled qubits cannot be treated as correlated random cbits. In particular, we have concluded that  the qubits in the Bell state $|\Phi^{+}\rangle$ cannot be considered as being in either state $|00\rangle$ or $|11\rangle$ with equal probability. This has intriguing implications on the nature of reality, as we will further discuss in Sec. \ref{sec:ch3:spooky}.

\subsection{Entangled States as Superpositions of Separable States}\label{sec:ch2:supsep}

A basic fact about entangled states is that they can be always expressed as superpositions of multiple separable states. This follows straightforwardly from the fact that the $2^n$ vectors (\ref{eq:ch2:kron}) in the computational basis are separable. Therefore, the superposition (\ref{eq:ch2:multiqubit}) is indeed a linear combination of separable states. 

More generally,  given an orthonormal basis $\{|v_{x_{k}}\rangle\}_{x_{k}=0}^{1}$ for each qubit $k\in\{0,1,...,n-1\}$, an  orthonormal basis for the $2^{n}$-dimensional Hilbert space of an $n$-qubit system can be obtained as  the set of all $2^{n}$ states of the form
\begin{equation}
	\{|v_{x_{0}}\rangle\otimes|v_{x_{1}}\rangle\otimes\cdots\otimes|v_{x_{n-1}}\rangle\}.
\end{equation}
It follows that any quantum state for $n$ qubits can be written as the superposition \begin{equation}\label{eq:ch2:supsep0}
	|\psi\rangle=\sum_{x_{0}=0}^{1}\sum_{x_{1}=0}^{1}\cdots\sum_{x_{n-1}=0}^{1}\alpha_{x_{0},x_1....,x_{n-1}}|v_{x_{0}}\rangle\otimes|v_{x_{1}}\rangle\otimes\cdots\otimes|v_{x_{n-1}}\rangle.
\end{equation} 

Even more broadly, we can decompose the set of $n$ qubits into $K$
subsets, each of $n_{k}$ qubits with $\sum_{k=0}^{K-1}n_{k}=n$.
Denoting as $\{|v_{x_{k}}\rangle\}_{x_{k}=0}^{2^{n_{k}}-1}$ 
any orthonormal basis for the $2^{n_k}$-dimensional Hilbert space of the $k$-th subset of qubits,
an orthonormal basis for the $2^{n}$-dimensional Hilbert space for the $n$ qubits is obtained as the set of all $2^n$ states of the form
\begin{equation}
	\{|v_{x_{0}}\rangle\otimes|v_{x_{1}}\rangle\otimes\cdots\otimes|v_{x_{K-1}}\rangle\}.
\end{equation}
Accordingly, we can write the state of the $n$ qubits as the superposition
\begin{equation}\label{eq:ch2:supsep}
	|\psi\rangle=\sum_{x_{0}=0}^{2^{n_0}-1}\sum_{x_{1}=0}^{2^{n_1}-1}\cdots\sum_{x_{K-1}=0}^{2^{n_{K-1}}-1}\alpha_{x_{0},x_1....,x_{K-1}}|v_{x_{0}}\rangle\otimes|v_{x_{1}}\rangle\otimes\cdots\otimes|v_{x_{K-1}}\rangle.
\end{equation}

\section{Multi-Qubit Quantum Gates}\label{sec:ch2:multiqubit}

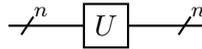
\begin{figure}
	\begin{center}\begin{quantikz} 			 &\qwbundle{n}&   \gate{U}   & \qw & \qwbundle{n} 		\end{quantikz} 	\end{center}\vspace{-0.4cm}\caption{An $n$-qubit system, when not measured, evolves according to a reversible linear transformation described by a $2^n\times 2^n$ unitary matrix $U$.\label{fig:ch2:multiqubit1} }
\end{figure}

As explained in Sec. \ref{sec:ch2:singlequbitgates}, systems of $n$ qubits, when not measured, evolve according to a reversible, norm-preserving, linear transformation described by a $2^n\times 2^n$ unitary matrix $U$. The situation is illustrated in Fig. \ref{fig:ch2:multiqubit1}, in which a bundle of $n$ wires is used to indicate the presence of $n$ wires, one for each qubit.  

While in principle all $2^{n}\times2^{n}$ unitary matrices $U$ are
physically realizable, in practice one is limited to 
transformations that can be implemented by a quantum circuit with a, sufficiently short, sequences of one- and 
two-qubit gates. Specifically, for feasibility, the sequence should consist of a number of gates that grows polynomially with the number of qubits, $n$, and not with the number of amplitudes, $2^n$.  In this section, generalizing the discussion in Sec. \ref{sec:ch1:sqgates}, we cover several ways to interpret and construct unitary matrices.


\subsection{Local Operations and Multi-Qubit Quantum Gates}\label{sec:ch2:Paulibasis}

As introduced in  Sec. \ref{sec:ch2:singlequbitgates}, the simplest type of unitary operating on $n$ qubits is given by the parallel, local, application of single-qubit gates to each of the $n$ individual qubits. This corresponds to a unitary matrix $U$ obtained as the Kronecker product \begin{equation}\label{eq:ch2:sepu}
	U=U_0 \otimes U_1 \otimes \cdots \otimes U_{n-1},
\end{equation} where each $2\times 2$ unitary $U_k$, with $k\in \{0,1,...,n-1\}$, applies to the corresponding $k$-th qubit. 

A particularly important category of local unitary matrices is given by \textbf{strings of $n$ Pauli matrices}. Let us write as $P_s$ with $s\in\{0,1,2,3\}$ a Pauli matrix, where \begin{equation}
	P_{0}=I,\textrm{ \ensuremath{P_{1}=X,\textrm{ \ensuremath{P_{2}=Y,\textrm{ and \ensuremath{P_{3}=Z}}}}}}.
\end{equation} Furthermore, let $s^{n}=[s_0,...,s_{n-1}]$ be a vector of integers $s_k\in\{0,1,2,3\}$ with $k\in\{0,1,...,n-1\}$.  A string of $n$ Pauli matrices $P_{s^{n}}$ is given by the Kronecker product of $n$ Pauli matrices \begin{equation}\label{eq:ch2:stringP} P_{s^{n}}=P_{s_0} \otimes P_{s_1}  \otimes \cdots \otimes P_{s_{n-1}},
\end{equation} where each matrix $P_{s_k}$ is one of the Pauli matrices $\{I,X,Y,Z\}$.

By (\ref{eq:ch2:stringP}), a string of Pauli matrices $P_{s^{n}}$ corresponds to the parallel application of single-qubit gates, with gate $P_{s_k}\in \{I,X,Y,Z\}$ being applied to the $k$-th qubit. Some of the Pauli matrices in (\ref{eq:ch2:stringP}) can be selected as the identity matrix $I$, indicating that non-trivial quantum gates are applied only to a subset of qubits.  The \textbf{weight} of a Pauli string is the number of Pauli operators in it that are different from the identity $I$.

As we have seen in Sec. \ref{sec:ch1:Paulibasis}, an arbitrary $2\times 2$ matrix can be expressed as a linear combination of the four Pauli matrices  $\{I,X,Y,Z\}$. More generally, an arbitrary $2^n\times2^n$ (bounded) matrix $A$, with arbitrary complex entries, can be written as a linear combination of a linear combination of all $4^n$ strings of $n$ Pauli matrices, i.e.,  \begin{equation}\label{eq:ch2:decherm} A=\sum_{s^{n}\in\{0,1,2,3\}^{n}} a_{s^{n}} P_{s^{n}},
\end{equation}with (generally complex) coefficients $a_{s^{n}}$ given as \begin{equation}a_{s^{n}}=\frac{1}{2^n}\textrm{tr}\left(A P_{s^{n}}\right).
\end{equation} For Hermitian matrices, the coefficients are real; while for unitary matrices the vector of coefficients has unitary norm. 

By the decomposition (\ref{eq:ch2:decherm}), in a manner that parallels the relationship between separable and quantum states covered in Sec. \ref{sec:ch2:supsep}, a unitary matrix operating on $n$ qubits can be expressed as a linear combination of local operators.


\subsection{Quantum Gates as Change-of-Basis Transformations}\label{sec:ch2:changeofbasis}

Generalizing the discussion in Sec. \ref{sec:ch1:qgcb}, a transformation is unitary if it can be written in the \textbf{change-of-basis} form 	\begin{equation}\label{eq:ch2:unitarytrans}
	U=\sum_{x=0}^{2^{n}-1}|v_{x}\rangle\langle u_{x}|,
\end{equation} for two orthonormal bases \{$|v_{0}\rangle,|v_{1}\rangle,...,|v_{2^{n}-1}\rangle$\}
and \{$|u_{0}\rangle,|u_{1}\rangle,...,|u_{2^{n}-1}\rangle$\} of the $2^n$-dimensional Hilbert space. Accordingly, the transformation $U$ maps each vector $|u_x\rangle$ in the first set of $2^{n}$  orthogonal
			vectors into a vector $|v_x\rangle$  in the second set of $2^{n}$  orthogonal vectors. Recall that, by linearity, specifying how the operator acts
		on each vector of a basis of $2^n$ orthogonal vectors as in the decomposition (\ref{eq:ch2:unitarytrans}) is sufficient to describe the output of the operator for any input state.

	As a useful special case, a transformation that maps vectors in the computational basis to a \textbf{permutation} of the same set of vectors is unitary. To define such a transformation, let $\pi(x)$ be a permutation of the integers $x\in\{0,1,...,2^{n}-1\}$. Then, the matrix operating as
		\begin{equation}\label{eq:ch2:permu}
		U|x\rangle=|\pi(x)\rangle
		\end{equation} on the computational basis is unitary. 
	
	One way to define a permutation $\pi(x)$ is via a \textbf{cyclic shift} of the $n$-bit binary strings $x$, with shift given by some $n$-cbit string $y$. This yields the permutation $\pi(x)=x\oplus y$, where $\oplus$ represents the element-wise XOR operation and $x$ and $y$ are interpreted as $n$-cbit strings. By (\ref{eq:ch2:permu}), the resulting unitary matrix operates as	\begin{equation}
		U|x\rangle=|x\oplus y\rangle
	\end{equation}on the vectors of the computational basis. By (\ref{eq:ch1:NOTagain}), the Pauli $X$ gate can be interpreted as applying a cyclic shift with $y=1$.

\subsection{Quantum Gates as Transformations with Unitary Magnitude Eigenvalues}

As detailed in Sec. \ref{sec:ch1:uniteig}, unitary matrices are normal, and hence they can be characterized in terms of their eigendecomposition (\ref{eq:ch1:eigen}). Furthermore, all their eigenvalues  have unitary magnitude. 

For instance, a non-trivial Pauli string (\ref{eq:ch2:stringP}), in which at least one of the Pauli matrices is different from the identity, has an equal number, $2^n/2$, of eigenvalues equal to $1$ and to $-1$. This follows from the fact that the set of eigenvalues of a Kronecker product of normal matrices is given by the set of all products of combinations of eigenvalues of the constituent matrices. 

\begin{figure}
	\begin{center}
		\begin{quantikz} 
			j & \ctrl{1} &  \qw \\ 
			k & \gate{U} & \qw 
		\end{quantikz}
	\end{center}\vspace{-0.4cm}\caption{A two-qubit controlled quantum gate, also known as controlled-$U$ gate. The qubit indexed as $j$ is the controlling qubit and the qubit labelled by $k$ is the controlled qubit. \label{fig:ch2:controlled}}
\end{figure}
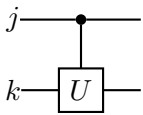

\subsection{Two-Qubit Controlled Quantum Gates}\label{sec:ch2:controlled}

	A common way to construct two-qubit quantum gates is to use one qubit as a ``switch'' controlling whether a given single-qubit gate $U$ is applied to a second qubit. This important construction is described in this subsection.
	
	 Given a single-qubit gate defined by a $2\times 2$ unitary matrix  $U$, a \textbf{two-qubit controlled quantum gate}  is described by a $4\times 4$ matrix $C_{jk}^{U}$, in which the first index $j$ identifies the controlling qubit, while the
	second index $k$ identifies the controlled qubit. The gate $C_{jk}^{U}$ is also referred to as \textbf{controlled-$U$ gate}. The symbol used to represent the controlled-$U$ gate $C_{jk}^{U}$  in a quantum circuit  is shown in Fig. \ref{fig:ch2:controlled}. 

 To describe the operation of the controlled-$U$ gate, let us fix qubit
	$j=0$ as the controlling qubit and qubit $k=1$ as the controlled qubit for simplicity of notation.	The controlled-$U$ quantum gate 	$C_{01}^{U}$, applied to the system of the two qubits, is defined by the unitary matrix
	\begin{equation}\label{eq:ch2:cu}
	C_{01}^{U}=|0\rangle\langle0|\otimes I+|1\rangle\langle1|\otimes U,
	\end{equation}where $I$ is the $2\times2$ identity matrix. 	 One can easily check that this transformation is indeed unitary, since it satisfies the defining condition (\ref{eq:ch1:defuni}), i.e., 
	\begin{align}
		(C_{01}^{U})^{\dagger}C_{01}^{U} & =|0\rangle\langle0|\otimes I+|1\rangle\langle1|\otimes U^{\dagger}U\nonumber\\
		& =|0\rangle\langle0|\otimes I+|1\rangle\langle1|\otimes I\nonumber\\
		& =(|0\rangle\langle0|+|1\rangle\langle1|)\otimes I=I,
	\end{align}
	where we have used the resolution-of-identity equality (cf. (\ref{eq:ch1:residentity})) \begin{equation}|0\rangle\langle0|+|1\rangle\langle1|=I.\end{equation}
	 Similarly, we have the equality $C_{01}^{U}(C_{01}^{U})^{\dagger}=I.$

	 To see what a two-qubit controlled quantum gate does (and to prove in a different way that it is unitary),
	it is sufficient to evaluate the output of the gate when applied  to the four vectors of the computational basis. As we have seen, this follows by linearity since any two-qubit state (\ref{eq:ch2:multiqubit}) can be expressed as a linear combination of computational-basis vectors. We obtain
		\begin{align}\label{eq:ch2:CNOT}
			C_{01}^{U}|00\rangle & =(|0\rangle\langle0|\otimes I+|1\rangle\langle1|\otimes U)(|0\rangle \otimes |0\rangle)\nonumber\\
			& =|0\rangle\langle0|0\rangle\otimes|0\rangle+|1\rangle\langle1|0\rangle\otimes U|0\rangle\nonumber\\
			& =|0\rangle\otimes|0\rangle=|00\rangle \nonumber\\
			C_{01}^{U}|01\rangle & =|01\rangle\nonumber\\
			C_{01}^{U}|10\rangle & =|1\rangle\otimes U|0\rangle\nonumber\\
			C_{01}^{U}|11\rangle & =|1\rangle\otimes U|1\rangle.
		\end{align}
	In words, when applied to vectors in the computational basis, the controlled
	quantum gate $C_{01}^{U}$ applies the single-qubit gate $U$ to the controlled qubit when
	the controlling qubit is in state $|1\rangle$, while no operation is applied to second qubit when
	the controlling qubit is in state $|0\rangle$. The controlling qubit hence acts as a ``switch'' for the application of the single-qubit gate $U$ on the controlled qubit. 
	
	As an example, assume that the input state of the two qubits is given by the separable state $|\psi\rangle = |+\rangle \otimes |\psi\rangle$, where $|\psi\rangle$ is an arbitrary state of the controlled qubit. Then, applying the controlled-$U$ gate yields\begin{equation}\label{eq:ch2:newex}
		C_{01}^{U}(|+\rangle \otimes |\psi\rangle) = \frac{1}{\sqrt{2}} (|0\rangle \otimes |\psi\rangle + |1\rangle \otimes U |\psi\rangle). 
		\end{equation}Therefore, when the controlling qubit is in a superposition state, here state $|+\rangle$, the computation separates into two superimposed \textbf{branches}. In one branch, the controlling qubit is in state $|0\rangle$ and hence the controlled qubit is unchanged; while in the other branch the controlling qubit is in state $|1\rangle$ and hence the controlled qubit undergoes transformation $U$. Note also that, while the input state $|+\rangle \otimes |\psi\rangle$ is separable, the output state (\ref{eq:ch2:newex}) is generally entangled.

 In matrix form, the controlled-$U$ gate $C_{01}^{U}$ gate can be written as	\begin{equation}\label{eq:ch2:controlledU}
	C^{U}_{01}=\left[\begin{array}{cccc}
		1 & 0 & 0 & 0\\
		0 & 1 & 0 & 0\\
		0 & 0 & U_{11} & U_{12}\\
		0 & 0 & U_{21} & U_{22}
	\end{array}\right],
\end{equation}where $U_{lm}$ is the $(l,m)$-th entry of matrix $U$.

\subsection{Examples of Two-Qubit Controlled Gates}

\begin{figure}
	\begin{center}
		\begin{quantikz} 
			& \ctrl{1} &  \qw \\ 
			& \gate{X} & \qw 
		\end{quantikz}
		\begin{quantikz} 
			& \ctrl{1} &  \qw \\ 
			& \targ{} & \qw 
		\end{quantikz}
	\end{center}\vspace{-0.4cm}\caption{A CNOT gate, or controlled-$X$ gate, represented in one of two equivalent ways (cf. Fig. \ref{fig:ch1:NOT}). The representation on the right makes it clear that the CNOT gate, when applied to the computational-basis vectors, reports the XOR of the two states on the controlled qubit. \label{fig:ch2:CNOT}}
\end{figure}

In this subsection, we elaborate on two examples of two-qubit controlled gates. A first important two-qubit controlled gate is  the \textbf{controlled-X (CX) gate}, also known as \textbf{controlled-NOT (CNOT)} gate, which is obtained by choosing the single-qubit gate $U=X$. Given its relevance, the CNOT gate  $C_{01}^{X}$  is commonly denoted as $C_{01}$, hence removing the specification of the single qubit gate $X$. The CNOT gate is represented in a quantum circuit in either one of the two ways shown in Fig. \ref{fig:ch2:CNOT}, mirroring the corresponding representations of the NOT gate in Fig. \ref{fig:ch1:NOT}. 

The representation on the right of Fig. \ref{fig:ch1:NOT} helps one remember that the CNOT gate, when applied to vectors in the computational basis, reports the XOR of the two pre-gate states on the state of the controlled qubit. Indeed, applying the CNOT gate to the states in the computational basis yields the outputs\begin{align}\label{eq:ch2:CNOTmap}
		C_{01}|00\rangle & =|0,0\oplus 0\rangle  = |00\rangle\nonumber\\
		C_{01}|01\rangle & =|0,0\oplus 1\rangle  =|01\rangle\nonumber\\
		C_{01}|10\rangle & =|1,1\oplus 0\rangle  =|11\rangle\nonumber\\
		C_{01}|11\rangle & =|1,1\oplus 1\rangle  =|10\rangle.
	\end{align} Specializing (\ref{eq:ch2:controlledU}), we can also write the CNOT gate as the unitary matrix
\begin{equation}\label{eq:ch2:CNOTgate}
	C_{01}=\left[\begin{array}{cccc}
		1 & 0 & 0 & 0\\
		0 & 1 & 0 & 0\\
		0 & 0 & 0 & 1\\
		0 & 0 & 1 & 0
	\end{array}\right].
\end{equation} 

As a numerical example, consider the application of the CNOT gate to the Bell states $|\Psi^{-}\rangle$, which yields
\begin{equation}
	C_{01}|\Psi^{-}\rangle=\frac{1}{\sqrt{2}}C_{01}(|01\rangle-|10\rangle)=\frac{1}{\sqrt{2}}(|01\rangle-|11\rangle).
\end{equation}

The mapping (\ref{eq:ch2:CNOTmap}) can be also interpreted in terms of  the characterization of unitary matrices discussed in Sec. \ref{sec:ch2:changeofbasis}. In fact, by (\ref{eq:ch2:CNOTmap}), the CNOT gate maps vectors $|x,y\rangle$ in the computational basis to a permuted version of the same vectors as \begin{equation}C_{01}|x,y\rangle=|x,x\oplus y\rangle,
\end{equation} with $x,y\in\{0,1\}.$ The state $|x,y\oplus x\rangle$ is obtained by ``shifting'' by $x$ the state of the second qubit.

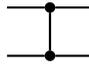
\begin{figure}
	\begin{center}
			\begin{quantikz} 
				& \ctrl{1} &  \qw \\ 
				& \ctrl{-1} & \qw 
			\end{quantikz}
	\end{center}\vspace{-0.4cm} \caption{A controlled-Z (CZ) gate. The symbol reflects the symmetry of the gate with respect to the two qubits.\label{fig:ch2:CZ}}
\end{figure}

A second commonly implemented controlled gate is the \textbf{controlled-Z (CZ) gate}, which is obtained by setting the single qubit gate as $U=Z$. Unlike the CNOT gate, the operation of a CZ gate is  symmetric with respect to the choice of the controlling and controlled qubits, in the sense that we have the equalities
		\begin{equation}
		C_{01}^{Z}=C_{10}^{Z}=\left[\begin{array}{cccc}
			1 & 0 & 0 & 0\\
			0 & 1 & 0 & 0\\
			0 & 0 & 1 & 0\\
			0 & 0 & 0 & -1
		\end{array}\right].
		\end{equation} Reflecting this symmetry, the symbol used to describe a CZ gate in a quantum circuit is as in Fig. \ref{fig:ch2:CZ}. Equivalently, we can describe a CZ gate via its application to the computational-basis vector as \begin{equation}C_{10}^{Z}|x,y\rangle=(-1)^{xy}|x,y\rangle
	\end{equation}  for all
		$x,y\in\{0,1\}.$
	
\subsection{Quantum Gates from Reversible Classical Binary Functions} \label{sec:ch2:reversible}

The characterization (\ref{eq:ch2:CNOTmap}) of the CNOT gate suggests a useful approach to define a larger class of unitary transformations as an extension of \textbf{classical reversible binary functions}. Accordingly, one starts with a reversible binary mapping to define the operation on the computational-basis vectors, and then the unitary is extended by linearity to any quantum input state. 

For example, in the case of the CNOT gate, the classical function is the reversible implementation of the XOR operation given by $(x,y) \mapsto (x,x\oplus y)$ with $x,y\in\{0,1\}$. Note that this function can be readily inverted, since we have the mapping $(x,x\oplus y) \mapsto (x,x\oplus y \oplus x) = (x,y)$.

\subsection{Quantum Gates from Hermitian Generators}

As introduced in Sec. \ref{sec:ch1:gen}, a unitary matrix $U$ can be expressed in terms of a Hermitian generator $G$ via (\ref{eq:ch1:gen}), i.e.,\begin{equation}
	U=\exp(-iG).
\end{equation}
Using the decomposition (\ref{eq:ch2:decherm}) for the generator $G$ allows us to write any $2^n \times 2^n$ unitary matrix $U$ in terms of Pauli strings as \begin{equation}\label{eq:ch2:ext}
	U=\exp(-iG)=\exp\left(-i\left(\sum_{s^{n}\in\{0,1,2,3\}^{n}} a_{s^{n}} P_{s^{n}}\right)\right),
\end{equation}where the coefficients $a_{s^{n}}$ are real. Furthermore, by Euler's formula, if the sum includes a single Pauli string $P$, i.e., if $	U=\exp(-iaP)$ for some real number $a$, we can also write \begin{equation} U=\exp(-iaP)=\cos\left(a\right)I-i\sin\left(a\right)P.
\end{equation} An example for $n=1$ is given by (\ref{eq:ch1:PauliYcossin}). We refer to Appendix A in Chapter 6 for extensions of the characterization (\ref{eq:ch2:ext}).

\subsection{Universal Gates}

We conclude this section by noting that it can be proved that the set of gates \{CNOT, $H$, $T$\}, where
\begin{equation}
	T=\left[\begin{array}{cc}
		1 & 0\\
		0 & \exp(i\frac{\pi}{4})
	\end{array}\right],
\end{equation}
is \textbf{universal}, that is, from these gates one can implement any unitary matrix
on any number of qubits. 

Interestingly, in stark contrast, quantum circuits consisting only of the set of gates \{CNOT, $H$, $S$\}, where
\begin{equation}
	S=\left[\begin{array}{cc}
		1 & 0\\
		0 & i
	\end{array}\right]
\end{equation} is the \textbf{phase gate} (also denoted as $\sqrt{Z}$), can be efficiently simulated on a classical computer. Such circuits implement unitaries that belong to the so-called \textbf{Clifford group}, and the result concerning the feasibility of implementation on classical computers is known as the \textbf{Gottesman–Knill theorem}.

\section{Creating Entanglement}\label{sec:ch2:creating}
In order to create entanglement between two qubits that are initially in a separable state (i.e., not entangled), it is necessary to apply two-qubit gates. A standard circuit that can be used to entangle two qubits is shown in Fig. \ref{fig:ch2:entangling}. For all separable input states of the form $|xy\rangle$ with $x,y\in\{0,1\}$, the outputs of the circuit are given as	\begin{align}
	C_{01}(H\otimes I)|00\rangle & =\frac{1}{\sqrt{2}}(|00\rangle+|11\rangle)=|\Phi^{+}\rangle\nonumber\\
	C_{01}(H\otimes I)|01\rangle & =\frac{1}{\sqrt{2}}(|01\rangle+|10\rangle)=|\Psi^{+}\rangle\nonumber\\
	C_{01}(H\otimes I)|10\rangle & =\frac{1}{\sqrt{2}}(|00\rangle-|11\rangle)=|\Phi^{-}\rangle\nonumber\\
	C_{01}(H\otimes I)|11\rangle & =\frac{1}{\sqrt{2}}(|01\rangle-|10\rangle)=|\Psi^{-}\rangle.
\end{align}
Therefore, the output vectors correspond to entangled pairs of qubits in the Bell basis. We can more concisely write the output of the circuit as the state
\begin{equation}C_{01}(H\otimes I)|xy\rangle=\frac{1}{\sqrt{2}}(|0y\rangle+(-1)^{x}|1\bar{y}\rangle)
\end{equation} with $\bar{y}=y\oplus 1$. 

	
	\begin{figure}
	\begin{center}
			\begin{quantikz} 
				\lstick{\ket{x}} & \gate{H} & \ctrl{1} &  \qw \rstick[wires=2]{$\frac{1}{\sqrt{2}}(|0y\rangle+(-1)^{x}|1\bar{y}\rangle)$}\\ 
				\lstick{\ket{y}} & \qw & \gate{X} & \qw 
			\end{quantikz}
	\end{center}\vspace{-0.4cm}\caption{A CNOT-based circuit that produces entangled pairs in the Bell basis (with $x,y\in\{0,1\}$).\label{fig:ch2:entangling}}
\end{figure}

\begin{figure}
		\begin{center}
			\begin{quantikz} 
				\lstick{\ket{0}} & \gate{H} & \ctrl{1} &  \qw \rstick[wires=2]{$\frac{1}{\sqrt{2}}(|+\rangle|0\rangle+|-\rangle|1\rangle)$}\\ 
				\lstick{\ket{0}} & \gate{H} & \ctrl{-1} & \qw 
			\end{quantikz}
	\end{center}\vspace{-0.4cm}\caption{A CZ-based circuit that produces entangled pairs in the Bell basis (with $x,y\in\{0,1\}$).\label{fig:ch2:entanglingZ}}
\end{figure}
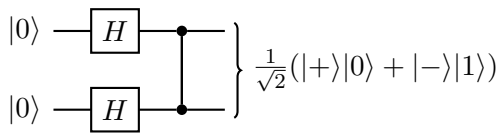

One can also create entanglement via the controlled-Z (CZ)
		gate using the circuit shown in Fig. \ref{fig:ch2:entanglingZ}. It can be readily checked that this circuit yields the entangled state
		\begin{align}
			|\psi\rangle & =\frac{1}{2}\left(|00\rangle+|01\rangle+|10\rangle-|11\rangle\right)\nonumber\\
			& =\frac{1}{\sqrt{2}}(|0\rangle|+\rangle+|1\rangle|-\rangle)\nonumber\\
			& =\frac{1}{\sqrt{2}}(|+\rangle|0\rangle+|-\rangle|1\rangle),
		\end{align}
		which can be converted to the Bell state $|\Phi^{+}\rangle$ by applying
		the Hadamard gate $H$ to the first qubit or to the second qubit.

	\section{Amplitude Diagrams}
As introduced in Sec. \ref{sec:ch1:amplitude}, amplitude diagrams provide an alternative means to visualize and describe the operation of a quantum circuit. Unlike the standard quantum circuit diagram with one wire per qubit, an amplitude diagram has one wire per amplitude, and it illustrate the interaction among the amplitudes as time flows from left to right. As a result, instead of having $n$ wires like a quantum circuit, an amplitude diagram contains $2^n$ wires. This makes amplitude diagrams only applicable, in practice, for systems with a small number of qubits.

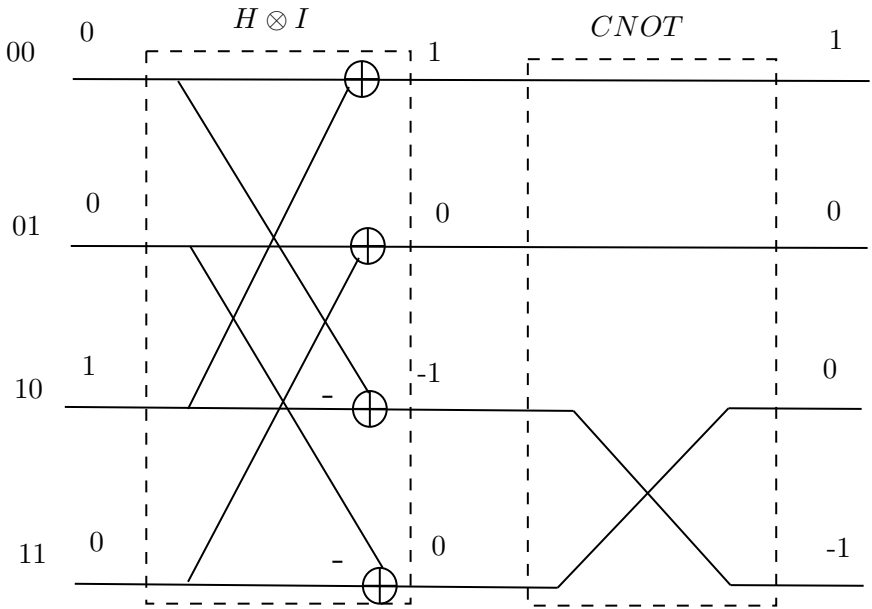
\begin{figure}
	\tikzset{every picture/.style={line width=0.75pt}} 
	
	\begin{tikzpicture}[x=0.75pt,y=0.75pt,yscale=-1,xscale=1]
		
		\draw    (169,198) -- (425,200) ;
		\draw    (425,200) -- (504,287.67) ;
		\draw    (174,287) -- (417,289) ;
		\draw    (417,289) -- (503,198.67) ;
		\draw    (503,198.67) -- (570,199) ;
		\draw    (504,287.67) -- (571,288) ;
		\draw  [dash pattern={on 4.5pt off 4.5pt}] (402,23) -- (527,23) -- (527,298) -- (402,298) -- cycle ;
		\draw    (172,117) -- (573,118) ;
		\draw    (173,33) -- (574,34) ;
		\draw   (310,33) .. controls (310,28.03) and (313.81,24) .. (318.5,24) .. controls (323.19,24) and (327,28.03) .. (327,33) .. controls (327,37.97) and (323.19,42) .. (318.5,42) .. controls (313.81,42) and (310,37.97) .. (310,33) -- cycle ; \draw   (310,33) -- (327,33) ; \draw   (318.5,24) -- (318.5,42) ;
		\draw   (313,117) .. controls (313,112.03) and (316.81,108) .. (321.5,108) .. controls (326.19,108) and (330,112.03) .. (330,117) .. controls (330,121.97) and (326.19,126) .. (321.5,126) .. controls (316.81,126) and (313,121.97) .. (313,117) -- cycle ; \draw   (313,117) -- (330,117) ; \draw   (321.5,108) -- (321.5,126) ;
		\draw   (314,199) .. controls (314,194.03) and (317.81,190) .. (322.5,190) .. controls (327.19,190) and (331,194.03) .. (331,199) .. controls (331,203.97) and (327.19,208) .. (322.5,208) .. controls (317.81,208) and (314,203.97) .. (314,199) -- cycle ; \draw   (314,199) -- (331,199) ; \draw   (322.5,190) -- (322.5,208) ;
		\draw   (319,288) .. controls (319,283.03) and (322.81,279) .. (327.5,279) .. controls (332.19,279) and (336,283.03) .. (336,288) .. controls (336,292.97) and (332.19,297) .. (327.5,297) .. controls (322.81,297) and (319,292.97) .. (319,288) -- cycle ; \draw   (319,288) -- (336,288) ; \draw   (327.5,279) -- (327.5,297) ;
		\draw    (226,34) -- (322,191) ;
		\draw    (232,117) -- (329,279) ;
		\draw    (231,199) -- (312,37) ;
		\draw    (231,286) -- (317,123) ;
		\draw  [dash pattern={on 4.5pt off 4.5pt}] (210,19) -- (343,19) -- (343,297) -- (210,297) -- cycle ;
		
		\draw (338,291) node [anchor=north west][inner sep=0.75pt]   [align=left] {$ $};
		\draw (432,0) node [anchor=north west][inner sep=0.75pt]   [align=left] {$\displaystyle CNOT$};
		\draw (138,13) node [anchor=north west][inner sep=0.75pt]   [align=left] {00};
		\draw (141,101) node [anchor=north west][inner sep=0.75pt]   [align=left] {01};
		\draw (142,183) node [anchor=north west][inner sep=0.75pt]   [align=left] {10};
		\draw (144,266) node [anchor=north west][inner sep=0.75pt]   [align=left] {11};
		\draw (176,171) node [anchor=north west][inner sep=0.75pt]   [align=left] {1};
		\draw (178,89) node [anchor=north west][inner sep=0.75pt]   [align=left] {0};
		\draw (175,3) node [anchor=north west][inner sep=0.75pt]   [align=left] {0};
		\draw (180,260) node [anchor=north west][inner sep=0.75pt]   [align=left] {0};
		\draw (297,191) node [anchor=north west][inner sep=0.75pt]   [align=left] {{\Large -}};
		\draw (302,274) node [anchor=north west][inner sep=0.75pt]   [align=left] {{\Large -}};
		\draw (350,13) node [anchor=north west][inner sep=0.75pt]   [align=left] {1};
		\draw (345,175) node [anchor=north west][inner sep=0.75pt]   [align=left] {\mbox{-}1};
		\draw (354,94) node [anchor=north west][inner sep=0.75pt]   [align=left] {0};
		\draw (352,262) node [anchor=north west][inner sep=0.75pt]   [align=left] {0};
		\draw (552,7) node [anchor=north west][inner sep=0.75pt]   [align=left] {1};
		\draw (551,93) node [anchor=north west][inner sep=0.75pt]   [align=left] {0};
		\draw (551,263) node [anchor=north west][inner sep=0.75pt]   [align=left] {\mbox{-}1};
		\draw (549,173) node [anchor=north west][inner sep=0.75pt]   [align=left] {0};
		\draw (252,-4) node [anchor=north west][inner sep=0.75pt]   [align=left] {$\displaystyle {\displaystyle H\otimes I}$};

	\end{tikzpicture}\vspace{-0.1cm}\caption{An amplitude diagram representing the operation of the entangling circuit in Fig. \ref{fig:ch2:entangling}. The input state is $|10\rangle$, producing as the output state the Bell state $|\Phi^{-}\rangle$. The amplitudes at each step are not normalized for simplicity of visualization.\label{fig:ch2:maze}}
\end{figure}

To illustrate the use of amplitude diagrams, consider the entangling circuit in Fig. \ref{fig:ch2:entangling}. This corresponds to a $4\times 4$ transformation, and hence it requires four wires, each reporting the amplitude of one of the states in the computational basis $\{|00\rangle,|10\rangle,|01\rangle,|11\rangle\}$. The resulting amplitude diagram is shown in Fig. \ref{fig:ch2:maze}. The operation of the Hamadard transformation on the first qubit is expressed by the matrix\begin{equation}
H\otimes I=\frac{1}{\sqrt{2}}\left[\begin{array}{cccc}
	1 & 0 & 1 & 0\\
	0 & 1 & 0 & 1\\
	1 & 0 & -1 & 0\\
	0 & 1 & 0 & -1
\end{array}\right].
\end{equation}Interesting, while the operation of the transformation $H\otimes I$ is local, the transformation applies jointly to all amplitudes, as illustrated by the amplitude diagram. The following CNOT gate implements multiplication with matrix (\ref{eq:ch2:CNOTgate}), which swaps the last two amplitudes. The figure shows the evolution of the amplitudes for the input state $|10\rangle$, producing as the output the Bell state $|\Phi^{-}\rangle$. In the figure, for simplicity, the amplitudes are not normalized.

\section{Superdense Coding}\label{sec:ch2:superdense}

Entanglement provides a new resource that can be leveraged to support communications and computing primitives. In this section, we provide an example of the role that entanglement can play for communications.  In particular, we will see how the availability of an entangled pair of qubits, with a qubit held at a transmitter and the other at a receiver, can increase the communication capacity between the two ends of the link. 

	 To describe the setting of interest, consider a noiseless quantum channel whereby Alice can send one qubit
	(e.g., one photon) to Bob. 
	 With one qubit, it is possible to show that Alice cannot communicate more than a single cbit of information to
	Bob. In fact, it is possible to detect unambiguously only two states of a single
		qubit via a measurement. This is because one can only define two mutually orthogonal states in a  two-dimensional space.  This limitation remains true if Alice and Bob share statistically dependent cbits that are independent of the information cbits. That is, randomizing encoding and decoding strategies, even in a coordinated fashion based on shared randomness, is not useful in increasing
		the capacity between Alice and Bob.
	
Given this context, we ask the question: What if Alice and Bob share a pair of entangled qubits? To avoid trivial answers, we specifically assume that the qubits are  
	prepared before selecting the information cbit and are hence independent
	of the information message.

	To elaborate, assume that Alice and Bob share a Bell pair 
	\begin{equation}
	\frac{1}{\sqrt{2}}(|00\rangle+|11\rangle)=|\Phi^{+}\rangle,
	\end{equation}
	with the first qubit at Alice and the second at Bob.
	 If this pair of qubits behaved as a pair of conventional cbits, the actual state of the system would $|00\rangle$ or  $|11\rangle$ with equal probability. Accordingly, Alice and Bob would share
	one bit of common randomness.
	 As discussed earlier in this section, common randomness cannot help Alice convey more information
	than a single cbit by sending a single qubit.
	 But, as we have argued in Sec. \ref{sec:ch2:nonclass}, entangled qubits do not behave like correlated random cbits.
	This reinforces the question at hand: Can we use the new resource of entanglement to enhance the classical capacity of
	the channel between Alice and Bob?

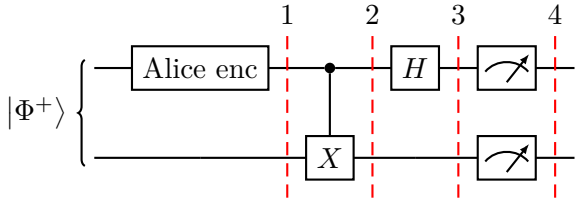
\begin{figure}
	\begin{center}
		\begin{quantikz}[slice all]
			\lstick[wires=2]{$\ket{\Phi^{+}}$} & \gate{\textrm{Alice enc}} & \ctrl{1} &  \gate{H} & \meter{} & \qw  \\ 
			& \qw & \gate{X} & \qw  & \meter{} & \qw
		
	\end{quantikz}
	\end{center}\vspace{-0.4cm} \caption{Operations carried out by Alice and Bob in superdense coding.\label{fig:ch2:super}}
\end{figure}

As we will see next, with a shared Bell pair, Alice
		can communicate \emph{two} cbits to Bob with a single qubit. To this end, we follow the steps illustrated in Fig. \ref{fig:ch2:super}, which are summarized as follows. \begin{itemize}
			\item  Alice selects at random two cbits of information $(x_0,x_1)$, with $x_0,x_1 \in\{0,1\}$. 
			\item Then, Alice encodes this information by applying a one-qubit gate
			to its qubit as indicated in Table \ref{table:ch2:super}.  Mathematically, given two cbits, the transformation applied by Alice is described by the unitary $Z^{x_1}X^{x_0}$. 
			\item After applying this transformation, Alice sends her qubit to Bob. Then, Bob applies the unitary transformation shown in Fig. \ref{fig:ch2:super}
			to both qubits. This transformation comprises the cascade of a CNOT and of a Hadamard gate to the first qubit.  
			\item Finally,  a standard measurement of the two qubits is made at Bob. The output measurement cbits can be shown to equal the message $(x_0,x_1)$  with probability $1$.
		\end{itemize}

\begin{table}
\begin{center}
	\begin{tabular}{cc}
		\toprule 
		message & transformation\tabularnewline
		\midrule
		\midrule 
	$00$ & $I=\left[\begin{array}{cc}
				1 & 0\\
				0 & 1
			\end{array}\right]$\tabularnewline
		\midrule 
		$01$ & $X=\left[\begin{array}{cc}
				0 & 1\\
				1 & 0
			\end{array}\right]$\tabularnewline
		\midrule 
		$10$ & $Z=\left[\begin{array}{cc}
				1 & 0\\
				0 & -1
			\end{array}\right]$\tabularnewline
		\midrule 
		$11$ & $ZX=\left[\begin{array}{cc}
				0 & 1\\
				-1 & 0
			\end{array}\right]$\tabularnewline
		\bottomrule
	\end{tabular}
	\par\end{center}\vspace{-0.4cm}\caption{Mapping between message and transformation applied by Alice to her qubit in superdense coding.\label{table:ch2:super}}
\end{table}

Let us now verify that indeed the superdense coding protocol illustrated in Fig. \ref{fig:ch2:super}, with Alice's encoding mapping in Table \ref{table:ch2:super}, allows Bob to recover Alice's message with probability $1$. To this end, it can be directly checked that the state of the two qubits before the measurement, i.e., 
		at step 3 in Fig. \ref{fig:ch2:super},  is given by
		\begin{align}\label{eq:ch2:super}
			|\psi^{3}\rangle & =|00\rangle\textrm{ if Alice chooses \ensuremath{I}}\nonumber\\
			|\psi^{3}\rangle & =|01\rangle\textrm{ if Alice chooses \ensuremath{X}}\nonumber\\
			|\psi^{3}\rangle & =|10\rangle\textrm{ if Alice chooses \ensuremath{Z}}\nonumber\\
			|\psi^{3}\rangle & =|11\rangle\textrm{ if Alice chooses \ensuremath{ZX}.}
		\end{align}
Therefore, by Born's rule, measuring the two qubits allows Bob to
		recover Alice's message.

To elaborate further on (\ref{eq:ch2:super}), let us consider as an example the case
		in which Alice chooses message $01$ and hence, by Table \ref{table:ch2:super}, it applies the Pauli $X$ gate on her qubit. The two qubits assume the following states at the times numbered in Fig. \ref{fig:ch2:super}: 
		\begin{align}
			|\psi^{1}\rangle & =(X\otimes I)|\Phi^{+}\rangle=\frac{1}{\sqrt{2}}(|10\rangle+|01\rangle)\nonumber\\
			|\psi^{2}\rangle & =C_{AB}|\psi^{1}\rangle=\frac{1}{\sqrt{2}}(|11\rangle+|01\rangle)\nonumber\\
			|\psi^{3}\rangle & =(H\otimes I)|\psi^{2}\rangle=\frac{1}{2}(|01\rangle-|11\rangle+|01\rangle+|11\rangle)=|01\rangle,
		\end{align}where we have used indices $A$ and $B$ to identify the qubits at Alice and Bob, respectively.

	 As we will see in the next chapter, one way to understand why superdense coding works
	is to note that the four 4-dimensional states $|\psi^{1}\rangle$
	obtained upon applying one of the four transformations by Alice are
	orthogonal, and hence distinguishable at Bob.
	 The ``non-classicality'' of this phenomenon is that the four distinguishable
	states are obtained by means of a local operations at Alice, who has only access
	to one of the two entangled qubits.
	 To explain this perspective, in the next chapter, we will generalize the type of
	measurements considered thus far, which apply separately to each qubit, to \emph{joint} measurements across multiple qubits.

\section{Trading Quantum and Classical Resources}

	 Superdense coding illustrates the general concept
	that classical and quantum resources
	can be traded for one another.
	In particular, superdense coding can be interpreted as revealing the
	following \textbf{resource transfer inequality}
	\begin{equation}
	\textrm{1 ebit + 1 qubit \ensuremath{\geq} 2 cbits,}
	\end{equation}
	where ``ebit'' indicates a Bell pair of entangled qubits. 
	 The inequality indicates that having the resources on the left-hand
	side allows one to realize the resources on the right-hand side.
	 Since one qubit can encode 1 cbit of information, one can also think of this resource
	inequality as stating that the initial ebit can be converted into
	one cbit of \emph{potential} information.
	
	 Another classical example of how one can trade classical and quantum resources is \textbf{teleportation}. To describe it, consider a classical channel that Alice can use to share cbits
	with Bob.
	In order to communicate exactly an arbitrary quantum state, Alice
	would need to transmit an infinite number of bits to describe a unitary-norm
	two-dimensional complex vector.
	But what if Alice and Bob share an entangled pair of qubits, which is
	independent of the qubit to the communicated?
	It turns out that in this case it is possible to communicate one qubit 
	in an arbitrary state by communicating only two cbits from Alice
	to Bob. This leads to the resource inequality	\begin{equation}
		\textrm{1 ebit + 2 cbits \ensuremath{\geq} 1 qubits}.
	\end{equation}

	\section{Conclusions}

In this chapter, we have introduced the formalism necessary to describe multi-qubit states. Extending the discussion for single qubits in the previous chapter, we have reviewed the two ways in which a quantum system can evolve, namely through a unitary transformation or through measurements. Unitary transformations are described by quantum circuits that are implemented via sequences of one- and two-qubit gates, with controlled gates playing a key role in the latter category. We have seen how a new phenomenon presents itself in multi-qubit systems: entanglement. Entanglement can be viewed as a strong form of statistical dependence that has no classical counterpart. Entangled qubits provide a novel resource that can be used for computation and communications. The next chapter will cover more implications of entanglement (see Sec. \ref{sec:ch3:spooky}). 

\section{Recommended Resources}\label{sec:ch2:rr}
As for the previous chapter, the reader is referred to \cite{mermin2007quantum} and \cite{lipton2021introduction} for clear and intuitive introductions to the topics covered in this chapter. For a physics-based perspective, useful references include \cite{susskind2014quantum,penrose2005road}. An introduction to tensor networks is provided by \cite{wood2011tensor}.

\section{Problems}
\begin{enumerate}
	\item Show that it is possible to convert one Bell basis vector into
	any other by means of local unitary transformations applied at any
	of the two qubits. For instance, applying local operations on
	the first system as
	\begin{equation}
		(Z^{z}X^{x}\otimes I)|\Phi^{+}\rangle
	\end{equation}
	we have $|\Psi^{+}\rangle$ with $z=0$ and $x=1$; $|\Phi^{-}\rangle$
	with $z=1$ and $x=0$; and $|\Psi^{-}\rangle$ with $z=1$ and $x=1$.
	\item Show that the following four vectors are orthonormal and that they have unitary norm: \begin{align}
		 & \cos(\theta)|00\rangle + \sin(\theta)|11\rangle\nonumber\\
		& \cos(\theta)|01\rangle + \sin(\theta)|10\rangle\nonumber\\
		&\sin(\theta)|00\rangle - \cos(\theta)|11\rangle\nonumber\\
	&\sin(\theta)|01\rangle - \cos(\theta)|10\rangle.
	\end{align}Therefore, for any $\theta$, these four vectors form an orthonormal basis of the $4$-dimensional Hilbert space. 

\item Use the first state in the orthonormal basis identified at the previous problem as the input for the circuit in the bottom part of Fig. \ref{fig:ch2:ent1}. Compute the probability that the two random cbits at the output of the measurement are equal. Argue that, as $\theta$ increases in the interval $[0,\pi/2]$, the two input qubits are increasingly entangled.
	\item Prove that for any $2\times2$ operator $M$, we have the equality $(M\otimes I)||\Phi^{+}\rangle=(I\otimes M^{T})|\Phi^{+}\rangle$
	(this is known as ``transpose trick''). 
	\item Consider the Bell state $|\Phi^+\rangle$ and apply the separable transformation $R_Y(\theta_0) \otimes R_Y(\theta_1)$ followed by a measurement. Demonstrate that the probability that the two cbits produced by the measurement are equal depends on the difference $\theta_0-\theta_1$ of the two rotation angles $\theta_0$ and $\theta_1$.
	\item Describe \textbf{qudits} as an extension of qubits to multi-valued quantum
	states (see recommended resources). 
	\item Prove that superdense coding is private in the sense that an eavesdropper
	intercepting the transmitted qubit would not be able to obtain any
	information about the two bits encoded by Alice.
	\item For superdense coding, show
 that the four 4-dimensional states $|\psi^{1}\rangle$
	obtained upon applying the four transformations by Alice are orthogonal. 
	\item Describe teleportation (see recommended resources).
	\item Prove the equivalence between quantum circuits illustrated in the figures below. \begin{center}
		\begin{quantikz} 				\qw & \qw & \ctrl{1} & \qw & \qw &\qw \\	\qw & \gate{H} & \gate{Z} & \gate{H} & \qw		\end{quantikz} 			= 			\begin{quantikz} 				\qw & \ctrl{1} & \qw\\ \qw & \gate{X} & \qw  			\end{quantikz}
	\end{center}
{\small{}\begin{center}
		\begin{quantikz} 				\qw & \ctrl{1} & \gate{X} & \ctrl{1}  &\qw \\	\qw &\gate{X} & \qw & \gate{X} & \qw &	\end{quantikz} 			= 			\begin{quantikz} 				\qw & \gate{X} & \qw\\ \qw & \gate{X} & \qw  			\end{quantikz}
\end{center}}{\small\par}

{\small{}\begin{center}
		\begin{quantikz} 				\qw & \ctrl{1} & \gate{Z} & \ctrl{1}  &\qw \\	\qw &\gate{X} & \qw & \gate{X} & \qw &	\end{quantikz} 			= 			\begin{quantikz} 				\qw & \gate{Z} & \qw\\ \qw & \qw & \qw  			\end{quantikz}
\end{center}}{\small\par}

{\small{}\begin{center}
		\begin{quantikz} 				\qw & \ctrl{1} & \qw & \ctrl{1}  &\qw \\	\qw &\gate{X} & \gate{X} & \gate{X} & \qw &	\end{quantikz} 			= 			\begin{quantikz} 				\qw & \qw & \qw\\ \qw & \gate{X} & \qw  			\end{quantikz}
\end{center}}{\small\par}

{\small{}\begin{center}
		\begin{quantikz} 				\qw & \ctrl{1} & \qw & \ctrl{1}  &\qw \\	\qw &\gate{X} & \gate{Z} & \gate{X} & \qw &	\end{quantikz} 			= 			\begin{quantikz} 				\qw & \gate{Z} & \qw\\ \qw & \gate{Z} & \qw  			\end{quantikz}
\end{center}}{\small\par}
\end{enumerate}

\chapter{Generalizing Quantum Measurements (Part I)}

\section{Introduction}

	In this chapter, we generalize quantum measurements in two important ways:\begin{itemize} \item We introduce von Neumann measurements in \emph{any} orthonormal basis -- not limited to the computational
		basis;
		\item  and we discuss situations in which only a \emph{subset} of the qubits is measured.
		\end{itemize}The first generalization will allow us to define \emph{joint} measurements across multiple qubits, which go beyond the separate, per-qubit, measurements considered thus far. The second will leads us to define the key concepts of density state, decoherence, and partial trace.

\section{Measurements in an Arbitrary Orthonormal Basis}\label{sec:ch3:any}

%

	 So far, we have considered the most common type of
		von Neumann measurements, namely measurements in the computational
		basis $\{|x\rangle\}_{x=0}^{2^{n}-1}$, also known as \textbf{standard measurements}. As described in the previous chapter and illustrated in Fig. \ref{fig:ch2:parmeas}, these amount to separate measurements
		of each qubit. In this section, we generalize von Neumann measurements by allowing for measurements in an arbitrary  orthonormal basis. As we will see, this generalized class of von Neumann measurements includes also joint measurements across multiple qubits.
		
	As a point of notation,	we will frequently use interchangeably the integer notation $x\in\{0,1,...,2^{n}-1\}$ and the $n$-cbit string notation $x_{0},x_{1},...,x_{n-1}$ 
		with $x_{k}\in\{0,1\}$, which are mutually related via (\ref{eq:ch2:msg}). Specifically, we write $x \in \{0,1\}^{n}$ to denote either an integer in the range $\{0,1,...,2^{n}-1\}$ or the corresponding $n$-cbit string. Given the one-to-one mapping (\ref{eq:ch2:msg}),  conflating the two quantities is well justified and notationally convenient.
		
		\subsection{Reviewing Standard Measurements}

		For a system of $n$ qubits in pre-measurement state $|\psi\rangle$, by Born's rule (\ref{eq:ch2:Born}), a standard measurement return $x\in\{0,1\}^n$ with probability 
			\begin{align}\label{eq:ch3:prob}
				|\langle x|\psi\rangle|^{2},
			\end{align}leaving the system in the  post-measurement state
			equal the computational basis vector $|x\rangle$. The probability (\ref{eq:ch3:prob})	corresponds to the magnitude squared of the projection of state $|\psi\rangle$ onto the computational-basis vector $|x\rangle$. As a result of the measurement, the quantum state $|\psi\rangle$ ``collapses'' into one of the vectors in the computational basis. Recall that each vector $|x\rangle=|x_0,x_1,...,x_{n-1}\rangle$ in the computational basis corresponds to a separable state in which each qubit $k$ has an individual state $|x_k\rangle$. Therefore, a standard measurement destroys any entanglement that may exist in the pre-measurement state $|\psi\rangle$.

		\subsection{Defining Measurements in an Arbitrary Orthonormal Basis}

 To specify the more general form of von Neumann measurement of interest in this section, we fix a set of $2^{n}$ orthonormal basis vectors $|v_{0}\rangle,...,|v_{2^{n}-1}\rangle$,
	each of dimension $2^{n}\times1$. Choosing the vectors in the computational basis, i.e., setting  $|v_{x}\rangle=|x\rangle$
	for all $x\in\{0,1\}^n$, recovers the standard 
	measurement. More generally, the vectors $|v_{x}\rangle$ may be chosen in an arbitrary way, as long as they form an orthonormal basis. In particularly, they may be separable or entangled. We will see that separable states describe local, per-qubit, measurements, while entangled states specify joint measurements across multiple qubits.

\begin{figure}
\begin{center}
		\begin{quantikz}
			\lstick{$\ket{\psi}$} & \qwbundle{n} & \meter{$x$} & \qwbundle{n}  \rstick{$\ket{v_{x}}$ w.p. $|\langle v_{x}|\psi\rangle|^{2}$}
	
	\end{quantikz}
\end{center}\vspace{-0.4cm} \caption{Illustration of a von Neumann measurement in an arbitrary orthonormal basis $\{|v_{x}\rangle\}_{x=0}^{2^{n}-1}$. Note that in this figure we make an exception to the conventional used in the rest of the text of using the  gauge block to represent a measurement in the computational basis. \label{fig:ch3:measany}}
\end{figure}
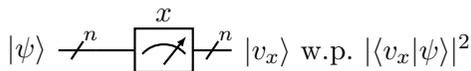

	Given an input state $|\psi\rangle$, a measurement in the basis $\{|v_{0}\rangle,...,|v_{2^{n}-1}\rangle\}$ is defined by the following two properties, which are illustrated in Fig. \ref{fig:ch3:measany}. \begin{itemize} \item \textbf{Born's rule}: The probability of observing the $n$-cbit string $x\in\{0,1\}^{n}$ is
		\begin{equation}\label{eq:ch3:Born}
			\Pr[\textrm{measurement output equals \ensuremath{x\in\{0,1\}^{n}}}]=|\langle v_{x}|\psi\rangle|^{2};
		\end{equation}
\item \textbf{``Collapse'' of the state}: If the measured $n$-cbit string is $x$, the \textbf{post-measurement state} of the qubits is $|v_x\rangle$.
\end{itemize}

As anticipated, these properties recover the definition of a standard measurement given in the previous chapter when we set $|v_{x}\rangle=|x\rangle$
for all  $x\in\{0,1\}^n$. Note, in particular, that the probability (\ref{eq:ch3:Born}) corresponds, geometrically, to the projection of the quantum state $|\psi\rangle$ onto the basis vector $|v_x\rangle$. It is also useful to observe that, unlike the special case of standard measurements, the post-measurement state $|v_x\rangle$ need not be separable. 

To illustrate the impact of the choice of the measurement basis, let us consider two examples. First, for a single-qubit setting, let us adopt the diagonal basis $\{|+\rangle,|-\rangle\}$. With it, the output of a measurement of a qubit in state $|+\rangle$ is equal to $0$ with probability $1$, leaving the state $|+\rangle$ unchanged. In contrast, a standard measurement of state $|+\rangle$ would output $0$ or $1$ with equal probability,  ``collapsing'' the state to either state $|0\rangle$ or $|1\rangle$.

As a second example,  for $n=2$ qubits, one could consider the  measurement of two qubits in the Bell basis  $\{|\Phi^{+}\rangle,|\Psi^{+}\rangle,|\Phi^{-}\rangle,|\Psi^{+}\rangle\}$ for the $2^n=2^2=4$-dimensional Hilbert space. Note that these vectors correspond to entangled states. Given an input Bell state $|\Phi^{+}\rangle$, this measurement produces output $0$ with probability $1$, leaving the state unchanged. This is in clear contrast to the case of a standard measurement of state $|\Phi^{+}\rangle$, which would yield outputs $00$ or $11$ with equal probability, leaving the qubits in the corresponding separable state $|00\rangle$ or $|11\rangle$.

\subsection{Implementing Measurements in an Arbitrary  Orthonormal Basis}\label{sec:ch3:implany}

\begin{figure}
\begin{center}
		\begin{quantikz}
			\lstick{$\ket{\psi}$} & \qwbundle{n} & \gate{U_{v_{x}\rightarrow x}} & \meter{$x$} & \qwbundle{n} & \gate{U_{x\rightarrow v_{x}}} &  \qw \rstick{$\ket{v_{x}}$ w.p. $|\langle v_{x}|\psi\rangle|^{2}$}
	
	\end{quantikz}
\end{center}\vspace{-0.4cm} \caption{Implementation of a measurement in an arbitrary orthonormal basis $\{|v_{x}\rangle\}_{x=0}^{2^{n}-1}$ via standard measurements and unitary pre- and post-measurement transformations. Note that here, and in the rest of the text, the gauge block represents a standard measurement. \label{fig:ch3:implany}}
\end{figure}
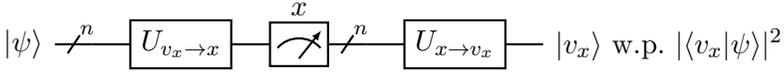

	As we discuss in this subsection, a measurement in an arbitrary orthonormal basis $\{|v_{x}\rangle\}_{x=0}^{2^{n}-1}$ can
	be equivalently realized via standard measurements (in the computational basis) by applying suitable unitary transformations before and after the measurement as illustrated in Fig. \ref{fig:ch3:implany}. While standard measurements are always separable across qubits, the pre- and post-measurement unitaries are generally not separable. In particular,  measurements defined by orthonormal bases with entangled state will be seen to require unitary transformations that apply across multiple qubits. 
	
	The unitary transformation to be applied before the measurement maps each vector $|v_{x}\rangle$ into the computational-basis
	vector $|x\rangle$. Intuitively, by applying such a  change-of-basis mapping, one can use a standard measurement to measure in the basis $\{|v_{x}\rangle\}_{x=0}^{2^{n}-1}$. Following the discussion in Sec. \ref{sec:ch2:changeofbasis}, this transformation is given by the unitary matrix 
	\begin{equation}\label{eq:ch3:cbmeas}
	U_{v_{x}\rightarrow x}=\sum_{x=0}^{2^{n}-1}|x\rangle\langle v_{x}|.
	\end{equation} The post-measurement transformation is given by the inverse of the pre-measurement transformation $U_{v_{x}\rightarrow x}$, i.e., 
	\begin{equation}
	U_{v_{x}\rightarrow x}^{-1}=U_{v_{x}\rightarrow x}^{\dagger}=U_{x\rightarrow v_{x}}=\sum_{x=0}^{2^{n}-1}|v_{x}\rangle\langle x|, 
	\end{equation} which maps the computational-basis
vector $|x\rangle$ into  vector $|v_{x}\rangle$. 
 
 To see that the circuit in Fig. \ref{fig:ch3:implany} implements a von Neumann measurement in basis $\{|v_{x}\rangle\}_{x=0}^{2^{n}-1}$, let us evaluate the probability of observing outcome $x$ given an input state $|\psi\rangle$.  By the Born rule, since the state entering the standard measurement is $U_{v_{x}\rightarrow x}|\psi\rangle$, this probability is computed as
		\begin{align}
			|\langle x|U_{v_{x}\rightarrow x}|\psi\rangle|^{2} & =|(U_{x\rightarrow v_{x}}|x\rangle)^{\dagger}|\psi\rangle|^{2}=|\langle v_{x}|\psi\rangle|^{2}, 
		\end{align} which coincides with (\ref{eq:ch3:Born}). Furthermore, given measurement output $x$, after the second transformation, the qubits are in state $U_{x\rightarrow v_{x}}|x\rangle=|v_x\rangle$.
	
To illustrate the implementation of the circuit in Fig. \ref{fig:ch3:implany}, let us consider the two examples described in the previous subsection. Measuring a single qubit in the diagonal basis $\{|+\rangle, |-\rangle\}$ requires implementing the pre-measurement unitary (\ref{eq:ch3:cbmeas}) that maps the diagonal basis into the computational basis. As discussed in Sec. \ref{sec:ch1:qgcb}, this can be done via the Hadamard transform $H$, which maps state $|+\rangle$ to $|0\rangle$ and state $|-\rangle$ to $|1\rangle$. Alternatively, it can also be implemented via the Pauli $Y$-rotation $R_Y(\pi/2)$, which maps  state $|-\rangle$ to $|0\rangle$ and $|+\rangle$ to $|1\rangle$. 

\begin{figure}
	\begin{center}
		\begin{quantikz}
			\lstick[wires=2]{$\frac{1}{\sqrt{2}}(|0y\rangle+(-1)^{x}|1\bar{y}\rangle)$} & \qw  & \ctrl{1} &    \gate{H} & \qw \rstick{$\ket{x}$} \\ 
			& \qw & \gate{X} & \qw  & \qw \rstick{$\ket{y}$}
		\end{quantikz}
	\end{center}\vspace{-0.4cm} \caption{Unitary transformation $U_{\textrm{Bell}\rightarrow \textrm{Comp}}$ between Bell-basis vectors and computational-basis vectors.\label{fig:ch3:Belltocomp}}
\end{figure}
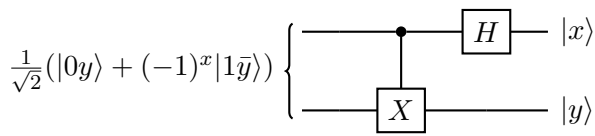

The second example is given by the measurement of $n=2$ qubits in the Bell basis. The Bell basis measurement can be implemented by means of the general circuit in Fig.  \ref{fig:ch3:implany} by employing the transformation $U_{v_{x}\rightarrow x}$ between Bell-basis vectors and computational-basis vectors and its inverse. The change-of-basis transformation $U_{v_{x}\rightarrow x}$ at hand -- which we will denote as $U_{\textrm{Bell}\rightarrow \textrm{Comp}}$ -- can be implemented as in Fig. \ref{fig:ch3:Belltocomp} via the cascade of a CNOT gate and a Hadamard gate on the first qubit. In fact, we have the mapping \begin{align}\label{eq:ch3:BtoC}
	U_{\textrm{Bell}\rightarrow \textrm{Comp}}|\Phi^{+}\rangle=|00\rangle     \nonumber\\
	U_{\textrm{Bell}\rightarrow \textrm{Comp}}|\Psi^{+}\rangle=|01\rangle     \nonumber\\
	U_{\textrm{Bell}\rightarrow \textrm{Comp}}|\Phi^{-}\rangle=|10\rangle     \nonumber\\
	U_{\textrm{Bell}\rightarrow \textrm{Comp}}|\Psi^{-}\rangle=|11\rangle, 
\end{align} as summarized in Fig. \ref{fig:ch3:Belltocomp}. The inverse transformation,  \begin{equation}U_{\textrm{Comp}\rightarrow \textrm{Bell}}=U^{\dagger}_{\textrm{Bell}\rightarrow \textrm{Comp}},
\end{equation} is implemented by running the circuit in Fig. \ref{fig:ch3:Belltocomp}  backwards, which corresponds to the entangling circuit in Fig. \ref{fig:ch2:entangling}. 


	The architecture in Fig. \ref{fig:ch3:implany} suggests  that the complexity of performing measurements in arbitrary
basis functions depends on the degree to which the unitary $U_{v_{x}\rightarrow x}$
can be efficiently described by one or two-qubit gates. As suggested by the examples above and as can be readily proved, if all basis vectors $|v_x\rangle$ are separable, the change-of-basis transformation  $U_{v_{x}\rightarrow x}$ in (\ref{eq:ch3:cbmeas}) amounts to local operations across the $n$ qubits, and so does its inverse $U_{x\rightarrow v_{x}}$. Therefore, in this case, the circuit in Fig. \ref{fig:ch3:implany} can be realized via the parallel application of single-qubit gates. In contrast, if the orthonormal bases contain entangled states, implementing the  transformation  $U_{v_{x}\rightarrow x}$ requires the use of multi-qubit gates. 

\subsection{Revisiting Superdense Coding}

As an application of the generalized measurements introduced in this section, here we revisit the operation of superdense coding, which is illustrated in Fig. \ref{fig:ch2:super}. As discussed in Sec. \ref{sec:ch2:superdense}, by applying local unitaries as in Table \ref{table:ch2:super}, Alice can ensure that the pair of entangled qubits -- one at Alice, the other at Bob -- is in one of the four Bell basis states. After transferring her bit to Bob, Alice has thus guaranteed that the four messages consisting of $2$ cbits of information are encoded into orthogonal states of the two qubits, now both at Bob's side. How can Bob detect which of the four orthogonal states was produced by Alice's encoding operation?

This can be readily done by implementing a joint measurement in the Bell basis as described in the previous subsection. For instance, if Alice encodes message $01$, the joint state of the two qubits after Alice's encoding operation is $|\Psi^{+}\rangle$ (see Sec. \ref{sec:ch2:superdense}). Therefore,  by the Born rule (\ref{eq:ch3:Born}) and given the mapping (\ref{eq:ch3:BtoC}), a Bell basis measurement returns output $01$ with probability $1$.

Based on this observation, we can now interpret the operation carried out at Bob after Alice's encoding  in Fig.  \ref{fig:ch2:super} as a measurement in Bell basis. In fact, the circuit consisting of CNOT gate followed by the Hadamard gate implements the  Bell-basis-to-computational basis transformation $U_{\textrm{Bell}\rightarrow \textrm{Comp}}$ in Fig. \ref{fig:ch3:Belltocomp}. Therefore, the final measurement in the computational basis in Fig. \ref{fig:ch2:super} effectively produces the output of a joint measurement in the Bell basis.

\subsection{Computing the Norm of the Inner Product of Two Vectors}\label{sec:ch3:spiking}

 As another example of the usefulness of measuring in an arbitrary
		basis, consider the seemingly unrelated problem of computing the squared absolute value 
		of the inner product between two vectors $|w\rangle$ and $|\psi\rangle$ of equal dimension $2^n$, 
		i.e., $|\langle w|\psi\rangle|^{2}$. This is known as the \textbf{fidelity} between states $|w\rangle$ and $|\psi\rangle$. The inner product plays a key role in many signal processing
		and machine learning applications, including \textbf{kernel methods}. An application will be discussed later in this subsection. We will see next that we can estimate the quantity $|\langle w|\psi\rangle|^{2}$ by making
		a measurement of a state vector $|\psi\rangle$ in an orthonormal basis that includes vector
		$|w\rangle$ -- along with other $2^n-1$ arbitrary mutually orthogonal vectors of unitary norm.

By Born's rule, for such a measurement, the desired quantity $|\langle w|\psi\rangle|^{2}$ is the probability
		of observing the output corresponding to vector $|w\rangle$.
	 Using the architecture in Fig. \ref{fig:ch3:implany}, this measurement can be implemented by constructing a unitary 
		$U_{w}$ that maps the basis at hand to the computational basis. Without loss of generality, one may assume that this transformation  satisfies the equality $U_{w}|w\rangle=|0\rangle$, mapping state $|w\rangle$ into state $|0\rangle$. This way, the probability of observing output $0$ from a standard measurement
		of the state $U_{w}|\psi\rangle$ equals $|\langle w|\psi\rangle|^{2}.$ Therefore, the quantity $|\langle w|\psi\rangle|^{2}$ can be estimated by repeating the described measurement multiple times and by evaluating the fraction of measured outcomes equal to $0$.
		
		The circuit outlined in the previous paragraph can be  used to design a \textbf{stochastic binary ``quantum neuron''}, in which the state $|w\rangle$ plays the role of the weight vector. To this end, given an input state $|\psi\rangle$, we apply the unitary $U_w$ and then make a standard measurement. We take the output of the neuron to be ``active'' if the measurement returns $0$, and to be ``inactive'' otherwise. As we have seen, the probability of the neuron being ``active'' is $|\langle w|\psi\rangle|^{2}$, and hence we can think of the function $|\cdot|^{2}$ as an activation function for the neuron. Note that with only $n$ qubits, this type of neuron can operate on vectors of exponentially large size $2^n$. Caution should, however, be exercised in interpreting this conclusion, since preparing state $|\psi\rangle$ based on a classical data vector of size $2^n$ generally entails an exponential complexity in $n$.

\section{Partial Measurements}\label{sec:ch3:partial}
 
In this section, we study a further extension of quantum measurements whereby only a subset of qubits are measured (in some orthonormal basis). We are interested in generalizing the Born rule, identifying the probability distribution over the measurement outputs; as well as in describing the post-measurement state of all qubits, both measured and not measured.

	 Measurements of a subset of qubits are of interest for the
	following reasons, among others: \begin{itemize} \item 
Many quantum protocols, such as teleportation, entanglement swapping,
		and quantum error correction, rely on measurements of a subset of qubits to determine subsequent quantum operations to be applied on the system.
\item		 Open quantum systems interact, and become entangled with, qubits in
		the environment -- a process that can be modelled as a partial measurement applied  by the environment
		on the quantum system.
		\item Partial measurements can be used to implement projective measurements, which find many applications, including quantum error correction (see Sec. \ref{sec:ch5:projmeas}).
		\end{itemize} 
	
	In this section, we describe partial measurements, and we also cover a ```spooky'' implication of the formalism introduced here concerning the behavior of entangled qubits.

\subsection{Introducing Partial, Single-Qubit, Measurements}

	 To describe partial measurements, we start by considering a quantum system of $n$ qubits in which
	we measure only the first qubit. The description provided in this subsection applies directly to the measurement of \emph{any} one of the $n$ qubits  in the system by properly reordering the operations involved. We focus first on measurements in the computational basis.

To begin,	observe that the state $|\psi\rangle_{n}$ of the overall system of $n$ qubits
	can always be expressed as 
	\begin{equation}\label{eq:ch3:partialdec}
	|\psi\rangle_{n}=\alpha_{0}|0\rangle|\phi_{0}\rangle_{n-1}+\alpha_{1}|1\rangle|\phi_{1}\rangle_{n-1},
	\end{equation}
	where $|\phi_{0}\rangle_{n-1}$ and $|\phi_{1}\rangle_{n-1}$ are
	(unitary norm) $2^{n-1}\times1$ state vectors, and the amplitudes $\alpha_0$ and $\alpha_1$ satisfy the equality  $|\alpha_{0}|^{2}+|\alpha_{1}|^{2}=1$. The decomposition  (\ref{eq:ch3:partialdec}) expresses the $n$-qubit state $|\psi\rangle_n$ as a superposition of two states in which the first qubit is in either state $|0\rangle$ or state  $|1\rangle$. 	We will see in the next subsection that the amplitudes can be chosen without loss of generality as real numbers. 
	  Note that, throughout this section, the subscript on the ket and bra states is used to indicate the number of qubits. 
	  
	  As a numerical example of the decomposition in (\ref{eq:ch3:partialdec}), consider the state of $n=3$ qubits \begin{equation}\label{eq:ch3:exstate}
	  	|\psi\rangle_{3}=  \frac{i}{2}|000\rangle+\frac{12+5i}{26}|001\rangle-\frac{1}{2}|101\rangle+\frac{3}{10}|110\rangle-\frac{2i}{5}|111\rangle,
	  \end{equation}which can be written in the form (\ref{eq:ch3:partialdec}) as
	  \begin{align}\label{eq:ch3:stateex}
	  	|\psi\rangle_{3}
	  	= & |0\rangle\left(\frac{i}{2}|00\rangle+\frac{12+5i}{2}|01\rangle\right)+|1\rangle\left(\frac{1}{2}|01\rangle+\frac{3}{10}|10\rangle-\frac{2i}{5}|11\rangle\right)\nonumber\\
	  	= & \frac{1}{\sqrt{2}}|0\rangle\underbrace{\left(\frac{i}{\sqrt{2}}|00\rangle+\sqrt{2}\frac{12+5i}{26}|01\rangle\right)}_{|\phi_{0}\rangle}\nonumber\\
	  	& +\frac{1}{\sqrt{2}}|1\rangle\underbrace{\left(\frac{1}{\sqrt{2}}|01\rangle+\frac{3}{10}\sqrt{2}|10\rangle-\sqrt{2}\frac{2i}{5}|11\rangle\right)}_{|\phi_{1}\rangle}.
	  \end{align}

	  In the decomposition (\ref{eq:ch3:partialdec}), the vectors $|\phi_{0}\rangle_{n-1}$ and $|\phi_{1}\rangle_{n-1}$
	  are not necessarily orthogonal. This can be readily checked in the example above. In fact, the two vectors $|\phi_{0}\rangle_{n-1}$ and $|\phi_{1}\rangle_{n-1}$ can even be equal, i.e.,
	$|\phi_{0}\rangle_{n-1}=|\phi_{1}\rangle_{n-1}=|\phi\rangle_{n-1}$,
	in which case the first qubit is not entangled with the other $n-1$
	qubits, since the joint state is separable: 
	\begin{equation}\label{eq:ch3:sep}
	|\psi\rangle_{n}=(\alpha_{0}|0\rangle+\alpha_{1}|1\rangle)\otimes|\phi\rangle_{n-1}.
	\end{equation}

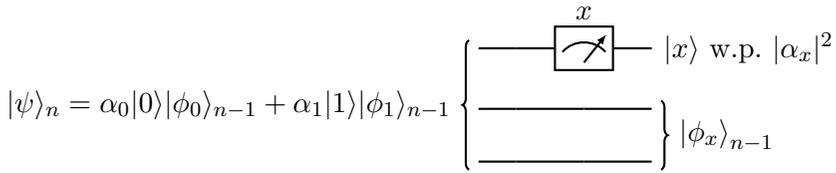
\begin{figure}
\begin{center}
	\begin{quantikz}
		\lstick[wires=3]{$|\psi\rangle_{n}=\alpha_{0}|0\rangle|\phi_{0}\rangle_{n-1}+\alpha_{1}|1\rangle|\phi_{1}\rangle_{n-1}$} & \qw & \meter{$x$} & \qw  \rstick{$\ket{x}$ w.p. $|\alpha_x|^{2}$}  \\ 
		& \qw & \qw & \qw \rstick[wires=2]{$\ket{\phi_x}_{n-1}$} \\[0.2cm] 
		& \qw & \qw & \qw      
	\end{quantikz}
\end{center}\vspace{-0.4cm} \caption{Illustration of a  partial measurement for a system of $n=3$ qubits in which a standard measurement is applied to the first qubit.\label{fig:ch3:partialmeas} }
\end{figure}

Having introduced the decomposition (\ref{eq:ch3:partialdec}), we are now ready to define the operation of a single-qubit measurement on the first qubit. As illustrated in Fig. \ref{fig:ch3:partialmeas}, a partial measurement in the computational basis for the first qubit of a system in state $|\psi\rangle$ is  described as follows.
\begin{itemize}
	\item \textbf{Generalized Born's rule}: The probability of observing $x\in\{0,1\}$ is
	\begin{align}\label{eq:ch3:genBorn0}
		\Pr[\textrm{measurement output equals \ensuremath{x}}]&=|\alpha_x|^2;
	\end{align} 
	\item \textbf{``Partial collapse'' of the state}: If the measured cbit is $x\in\{0,1\}$, the post-measurement state is $|x\rangle|\phi_x\rangle_{n-1}$. 
\end{itemize}
It can be readily seen that a partial measurement coincides with a standard measurement when $n=1$. 


	After a partial measurement of the first qubit, the post-measurement state $|x\rangle|\phi_x\rangle_{n-1}$ is such that the first qubit is not
entangled with the remaining $n-1$ qubits, irrespective of whether it was entangled under the original state $|\psi\rangle_n$. The unmeasured qubits can instead by in an entangled state. We may say that the ``collapse'' of the state is partial, in the sense that only the state of the first qubit reduces to one of the vectors, $|x\rangle$, in the computational basis. In contrast, the state of the remaining qubits, namely $|\phi_{x}\rangle_{n-1}$, still depends on the amplitudes of the original state $|\psi\rangle$. Therefore, an observer obtaining measurement outcome $x$ would only know the state of the first qubit, namely $|x\rangle$, while specifying the state $|\phi_{x}\rangle_{n-1}$ of the remaining qubits would generally require knowledge of the amplitudes of the original state $|\psi\rangle$.

To illustrate the effect of a partial measurement, consider again the state (\ref{eq:ch3:stateex}). 	 Making a standard measurement on the first qubit, the probability
of observing cbit $0$ is $1/2$, and the corresponding
post-measurement state is $|0\rangle\left(\frac{i}{\sqrt{2}}|00\rangle+\sqrt{2}\frac{12+5i}{26}|01\rangle\right)$.
After the measurement, the first qubit is no longer entangled with the rest
of the system, but the last two qubits are mutually entangled.

\subsection{Defining Partial, Single-Qubit, Measurements}

In this subsection, we provide a more formal description of partial measurements of a single qubit introduced in the previous subsection. To this end, we start by observing that the decomposition (\ref{eq:ch3:partialdec}) can be constructed as follows: \begin{enumerate}
\item Evaluate the unnormalized states 
\begin{equation}
\label{eq:ch3:unphi}
	|\tilde{\phi}_{x}\rangle_{n-1}=(\langle x|\otimes I)|\psi\rangle_{n}
\end{equation}
for $x\in \{0,1\}$, where the identity matrix is of dimension $2^{n-1}\times 2^{n-1}$.
\item Set the amplitudes as 
\begin{equation}
	\alpha_{x}=\sqrt{\langle\tilde{\phi}_{x}|_{n-1}|\tilde{\phi}_{x}\rangle_{n-1}}=|||\tilde{\phi}_{x}\rangle_{n-1}||_2
\end{equation} for $x\in \{0,1\}$.
\item Compute the normalized states 
\begin{equation}
	|\phi_{x}\rangle_{n-1}=\frac{1}{\alpha_{x}}|\tilde{\phi}_{x}\rangle_{n-1}
\end{equation}for $x\in \{0,1\}$.
\end{enumerate}
Note that this construction demonstrates that the amplitudes $\{\alpha_{x}\}_{x=0}^{1}$
can be taken without loss of generality to be real and positive. 

As an example, for the state in (\ref{eq:ch3:exstate}), we have the unnormalized states as
\begin{equation}
|\tilde{\phi}_{0}\rangle_{2}=(\langle0|\otimes I)|\psi\rangle_{3}=\frac{i}{2}|00\rangle+\frac{12+5i}{2}|01\rangle
\end{equation}
and 
\begin{equation}
|\tilde{\phi}_{1}\rangle_{2}=(\langle1|\otimes I)|\psi\rangle_{3}=\frac{1}{2}|01\rangle+\frac{3}{10}|10\rangle-\frac{2i}{5}|11\rangle, 
\end{equation}from which we can obtain (\ref{eq:ch3:stateex}) by following steps 2 and 3 in the procedure outlined in the previous paragraph. 

With these definitions, a partial measurement in the computational basis for the first qubit of a quantum system in state $|\psi\rangle$ is described as follows.
		\begin{itemize}
			\item \textbf{Generalized Born's rule}: The probability of observing $x\in\{0,1\}$ is
			\begin{align}\label{eq:ch3:genBorn}
				\Pr[\textrm{measurement output equals \ensuremath{x}}]&=(\langle x|\otimes I)|\psi\rangle_n\langle\psi|_n(|x\rangle\otimes I);
			\end{align} 
			\item \textbf{``Partial collapse'' of the state}: If the measured cbit is $x\in\{0,1\}$, the post-measurement state is  $|x\rangle|\phi_{x}\rangle_{n-1}$ with \begin{equation}\label{eq:ch3:genBorn1}|\phi_x\rangle_{n-1}=\frac{(\langle x|\otimes I)|\psi\rangle_n}{\sqrt{(\langle x|\otimes I)|\psi\rangle_n\langle\psi|_n(|x\rangle\otimes I)}}.
			\end{equation}
		\end{itemize}
	This description coincides with that given in the previous subsection, and has the advantage of depending directly on the state $|\psi\rangle$. Furthermore, it is straightforward to extend it to a measurement of the first qubit in any orthonormal basis $\{|v_x\rangle\}_{x=0}^{1}$ by replacing $x$ with $v_x$ in (\ref{eq:ch3:genBorn}) and (\ref{eq:ch3:genBorn1}). This extension is elaborated on in Sec. \ref{sec:ch3:multimeas}.


	\subsection{Spooky Action at a Distance}\label{sec:ch3:spooky}
	
	As an example of the use of partial measurements, let us consider  a  situation in which two qubits in the Bell pair $|\Phi^+\rangle=1/\sqrt{2}(|00\rangle+|11\rangle)$ are physically separated by a large distance. Recall that Bell states describe entangled qubits. The first qubit is at an agent, conventionally called Alice, while the second qubit is at a second agent called Bob. Alice may be in London, while Bob may be arbitrarily far, say on the moon. Alice carries  out a standard measurement of her qubit, while Bob does not apply any operation or measurement on his qubit. 
	
	By the generalized Born rule described in the previous two subsections, Alice will observe measurement outputs $0$ or $1$ with equal probability $1/2$. Furthermore, the post-measurement state is $|x,x\rangle$ when the measurement output at Alice is $x\in\{0,1\}$. In words, after Alice's measurement, the qubits are no longer entangled,  and they assume the same state $|x\rangle$, where $x\in\{0,1\}$ is the measurement output. What is strange about this situation?
	
	As we have discussed in Sec. \ref{sec:ch2:nonclass}, prior to Alice's measurement, the two qubits are \emph{not} in either state $|00\rangle$ or state $|11\rangle$. They \emph{must} instead be described as being in the superposition state $|\Phi^+\rangle=1/\sqrt{2}(|00\rangle+|11\rangle)$. Therefore, Alice's measurement has somehow caused a change in the joint state of the two qubits, which are no longer in state $|\Phi^+\rangle$, but rather in either state $|00\rangle$ or state $|11\rangle$ depending on the measurement output. This is surprising because, as mentioned,  Bob's qubit may be in principle arbitrarily far from Alice's. 
	
	Overall, it appears that there has been an \textbf{instantaneous ``action''} from
	one qubit to the other -- a physical impossibility due to limit imposed by the speed of light. 
	This is what Einstein called \textbf{``spooky action at a distance''}. But has anything measurable actually been transferred instantaneously from Alice to Bob?
	
	It is true that Alice, based on her measurement, can predict, with certainty, the output of a standard measurement of Bob's qubit carried out after Alice's measurement. But from Bob's perspective nothing has changed. Since Bob does not know Alice's measurement, he should view his qubit as being in either state $|0\rangle$ or $|1\rangle$, corresponding to the two possible measurement outputs. Therefore, a measurement of Bob's qubit would return $0$ or $1$ with equal probability, as if Alice had not carried out her measurement. So, no information has actually been transferred from Alice to Bob through Alice's measurement, and there has been no violation of the speed of light limit.
	
	That said, if we wished to simulate the evolution of the state of this two-qubit system using a classical distributed system, 
	information about Alice's measurement output should indeed reach instantaneously the classical system at Bob's end  in order to ensure a correct update of the joint quantum state.
	
 More discussion on entanglement and partial measurements can be found in Appendix B at the end of this chapter for systems involving more than two qubits.
	
	\subsection{Defining Partial, Multi-Qubit, Measurements}\label{sec:ch3:multimeas}

In this subsection, we extend the partial measurement formalism introduced in this section to systems
		of  $n$ qubits in which we measure any subset of $n'\leq n$ qubits in
		any orthonormal basis $\{|v_{x}\rangle_{n'}\}_{x=0}^{2^{n'}-1}$. 		 To this end, without loss of generality, let us order the qubits so that the first $n'$ qubits are measured. Then, given an input state $|\psi\rangle$, the generalized Born rule (\ref{eq:ch3:genBorn}) and the post-measurement state (\ref{eq:ch3:genBorn1}) are directly extended as follows. 	\begin{itemize}
			\item \textbf{Generalized Born's rule}: The probability of observing $x\in\{0,1\}^{n'}$ is
			\begin{align}\label{eq:ch3:genBornproj}
				\Pr[\textrm{measurement output equals \ensuremath{x}}]&=(\langle v_x|_{n'}\otimes I)|\psi\rangle_n\langle\psi|_n(|v_x\rangle_{n'}\otimes I),	\end{align}   where the identity matrix $I$ is of dimension $2^{n-n'}\times2^{n-n'}$; 
			\item \textbf{``Partial collapse'' of the state}: If the measured cbits are given by string $x\in\{0,1\}^{n'}$, the post-measurement state is $|v_x\rangle_{n'} |\phi_x\rangle_{n-n'}$, with \begin{equation}\label{eq:ch3:collgen}
				|\phi_x\rangle_{n-n'}=\frac{(\langle v_x|_{n'}\otimes I)|\psi\rangle_n}{\sqrt{(\langle v_x|_{n'}\otimes I)|\psi\rangle_n\langle\psi|_n(|v_x\rangle_{n'}\otimes I)}}.
			\end{equation}
		\end{itemize}

\section{Non-Selective Partial Measurements and Decoherence}\label{sec:ch3:decoh}
\begin{figure}
\begin{center}
	\begin{quantikz}
		\lstick{$A$} \qw & \meter{} & \trash{\text{trash}} \\
		\lstick{$B$} \qw & \qw & \qw  
	\end{quantikz}
\end{center}\vspace{-0.4cm} \caption{A non-selective partial measurement on subsystem $A$.\label{fig:ch3:nonsel}}
\end{figure}
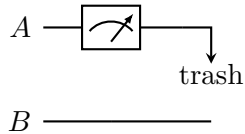

	 What happens if, as illustrated in Fig. \ref{fig:ch3:nonsel}, we measure the qubits of one subsystem, say $A$, and we ``throw
	away'' the result of this measurement? By ``throwing away'' we mean that the result of the measurement is not retained, and hence there is \textbf{epistemic uncertainty} about the outcome of the measurement and about the corresponding post-measurement state. Intuitively, discarding the outcome of the measurement introduces \textbf{classical randomness} in the evolution of the system. We are interested in describing the state
	of the remaining subsystem $B$, so as to be able to describe how subsystem $B$ may evolve after the measurement of subsystem $A$. A  measurement of the type illustrated in Fig. \ref{fig:ch3:nonsel} is known as a \textbf{non-selective
	measurement}, and is the subject of this section.

\subsection{Defining Non-Selective Partial Measurements}

 To elaborate, let us consider again the example in Fig. \ref{fig:ch3:partialmeas}, in which the measurement of the first qubit is discarded. In this case,  the post-measurement state of the subsystem $B$ consisting of the second set of $n-1$ qubits is $|\phi_{0}\rangle_{n-1}$ with
probability $|\alpha_{0}|^{2}$ and $|\phi_{1}\rangle_{n-1}$
with probability $|\alpha_{1}|^{2}$. 	 This is a situation  characterized by classical randomness, in that there 
is epistemic uncertainty on the state of the $n-1$ qubits.

As a result, the state of the subsystem $B$ cannot be described by using the ket state
formulation, since the outlined situation is not equivalent to having the
$n-1$ qubits in the superposition state $\alpha_{0}|\phi_{0}\rangle_{n-1}+\alpha_{1}|\phi_{1}\rangle_{n-1}$.  	Instead, we can only say that subsystem $B$ is in the \textbf{ensemble state} 	\begin{equation}\label{eq:ch3:ensemble0}
	\{(|\phi_{0}\rangle_{n-1},|\alpha_{0}|^2),(|\phi_{1}\rangle_{n-1},|\alpha_{1}|^2)\},
\end{equation} in which subsystem $B$ is in state  $|\phi_{x}\rangle_{n-1}$ with
probability $|\alpha_{x}|^{2}$. To describe such quantum states, we will need to introduce a new formalism, that of
density matrices. This is the subject of the next section. 

Importantly, the random behavior highlighted in the previous paragraph only occurs when the  $n-1$ qubits in subsystem $B$ are \emph{entangled} with the first qubit being measured in a non-selective manner. To see this, consider the case in which the two subsystem are not entangled, and hence the state of the system is given by (\ref{eq:ch3:sep}). In this case, the ensemble (\ref{eq:ch3:ensemble0}) reduces to the coherent state $|\phi\rangle_{n-1}$.	

The definition of non-selective measurements given in this subsection can be readily generalized to any partial measurement of $n'$ qubits as described in Sec. \ref{sec:ch3:multimeas}.

\subsection{Decoherence}

A ket state is said to be \textbf{coherent}, while a process producing an ensemble state such as (\ref{eq:ch3:ensemble0}) from a coherent state is said to cause \textbf{decoherence}. Decoherence entails that the produced quantum state is characterized by classical randomness, so that there is epistemic uncertainty regarding the ket state describing the system. The discussion in the previous subsection has revealed that decoherence occurs when a non-selective measurement is made on a part of an entangled state. 

	 Non-selective measurements, and hence decoherence, may be intentionally added in a quantum circuit as a resource
to implement probabilistic computing. In this type of situations, the randomization caused by a measurement can be harnessed to carry out probabilistic computations. 

Decoherence also happens naturally in an \textbf{open quantum system} in which
	a subsystem -- subsystem $B$ in Fig. \ref{fig:ch3:nonsel} -- gets entangled with its environment -- subsystem $A$. This is a typical situation in quantum computers, whose qubits inevitably become entangled with the environment after some (short) period of time.  	In this setting, the environment effectively acts
	as an observer, whose measurements are not available to the original subsystem $B$, e.g., to the quantum computer.

%

%
\section{Density Matrices}

In this section, we introduce density matrices as a means to describe a ``noisy'' ket state. This corresponds to the situation outlined in the previous section in which there is classical uncertainty on the ket state the system is in. As such, density matrices provide a more general formalism in which to define the state of a quantum state as compared to ket vectors.


\subsection{Describing Ensemble States as Density Matrices}

	 Consider a general situation in which the state of a quantum system
	of $n$ qubits is classically uncertain among $M$ possible states
	$\{|\phi_{m}\rangle\}_{m=0}^{M-1}$, with each state $|\phi_{m}\rangle$
	having probability $p_{m}.$ The states $\{|\phi_{m}\rangle\}_{m=0}^{M-1}$
	need not be orthogonal. Note that we will henceforth drop the subscript indicating the number of qubits from  ket and bra vectors. 
	 This setting can be described by the \textbf{ensemble state}
	\begin{equation}\label{eq:ch3:ensemble}
	\{(|\phi_{0}\rangle,p_{0}),(|\phi_{1}\rangle,p_{1}),...,(|\phi_{M-1}\rangle,p_{M-1})\}.
	\end{equation} If we have $p_{m}=1$ for some $m\in\{0,1,...,M-1\}$, we say that we have a \textbf{pure, or
coherent, state}; otherwise we have a \textbf{mixed state}.
	
	 The probabilities $p_{m}$, also known as \textbf{epistemic probabilities},
	are conceptually distinct from the squared absolute values of the amplitudes that
	give the measurement outputs' probabilities for pure quantum states, as
	they reflect a state of classical uncertainty. This state of uncertainty may reflect ignorance about the way in which the state is prepared,
		 or about the result of a measurement of another subsystem as in the situation studied in the previous section.

	 Suppose that we make a measurement of the system described by the ensemble state (\ref{eq:ch3:ensemble}). For generality, we consider a measurement in some arbitrary orthonormal basis $\{|v_{x}\rangle\}_{x=0}^{2^{n}-1}.$
	Partial measurements can be also similarly studied. How can we describe the probability distribution of the output of the measurement?
	
	If the state of the system is $|\phi_m\rangle$, by Born's rule, the output of the measurement is  $x\in\{0,1\}^n$ with probability $|\langle v_{x}|\phi_{m}\rangle|^{2}$. Since, under the ensemble (\ref{eq:ch3:ensemble}), the state is given by  $|\phi_m\rangle$ with probability $p_m$, by the law of total probability we have
	\begin{align}\label{eq:ch3:derdens}
	\Pr[\textrm{measurement output equals \ensuremath{x}}] & =	\sum_{m=0}^{M-1}p_{m}|\langle v_{x}|\phi_{m}\rangle|^{2} \nonumber \\ & =\sum_{m=0}^{M-1}p_{m}\langle v_{x}|\phi_{m}\rangle\langle\phi_{m}|v_{x}\rangle \nonumber\\
		& =\langle v_{x}|\left(\underbrace{\sum_{m=0}^{M-1}p_{m}|\phi_{m}\rangle\langle\phi_{m}|}_{\rho}\right)|v_{x}\rangle,
	\end{align}
	where \begin{equation}\label{eq:ch3:derdens1}
		\rho=\sum_{m=0}^{M-1}p_{m}|\phi_{m}\rangle\langle\phi_{m}|
		\end{equation} is known as the \textbf{density matrix} associated with the ensemble state (\ref{eq:ch3:ensemble}).

 The derivation (\ref{eq:ch3:derdens}) shows that we can describe the probability distribution of measurement outcomes on a noisy
	quantum state as a function of the density matrix $\rho$. The density matrix
	$\rho$ in (\ref{eq:ch3:derdens1}) hence fully represents the state of the system (\ref{eq:ch3:ensemble}) as it pertains the description of measurable outputs. 
	
	By the definition in (\ref{eq:ch3:derdens1}), the density matrix satisfies two conditions:\begin{itemize}
		\item it is \textbf{positive semidefinite}, i.e., it is Hermitian and it has non-negative real eigenvalues;
		\item and it has \textbf{unitary trace}, i.e., \begin{equation}\label{eq:ch3:untrace}
			\textrm{tr}(\rho)=1.
		\end{equation}
	\end{itemize}	
		To validate the second property, we can compute the trace of the density matrix in (\ref{eq:ch3:derdens}) as 
			\begin{align}
				\textrm{tr}(\rho) & =\sum_{x=0}^{2^{n}-1}\langle x|\rho|x\rangle=\sum_{x=0}^{2^{n}-1}\langle x|\left(\underbrace{\sum_{m=0}^{M-1}p_{m}|\phi_{m}\rangle\langle\phi_{m}|}_{\rho}\right)|x\rangle \nonumber\\
				& =\sum_{m=0}^{M-1}p_{m}\sum_{x=0}^{2^{n}-1}\langle x|\phi_{m}\rangle\langle\phi_{m}|x\rangle=\sum_{m=0}^{M-1}p_{m}=1,
			\end{align}
		where the first equality follows from the definition of the trace as the sum of the elements on the diagonal of a square matrix (see also Sec. \ref{sec:ch3:trace}).

		\subsection{From Density Matrices to Ensemble States}
		
		As discussed in the previous subsection, an ensemble state, in which there is epistemic uncertainty about the ket state the system is in, can be described by a density matrix. We now verify that  \emph{any} positive semidefinite matrix $\rho$ with unitary trace is a valid density matrix, in the sense that it describes an ensemble state.

		The key observation is that any positive semidefinite $2^n \times 2^n$ matrix $\rho$ can be written
		in terms of its eigendecomposition as
		\begin{equation}\label{eq:ch3:eigen}
			\rho=\sum_{x=0}^{2^{n}-1}p_{x}|u_{x}\rangle\langle u_{x}|,
		\end{equation}
		where $\{|u_{x}\rangle\}_{x=0}^{2^{n}-1}$ is the orthonormal basis
		of eigenvectors, and the corresponding eigenvalues $\{p_x\}_{x=0}^{2^{n}-1}$ are non-negative.
		Furthermore, by the unitary trace condition (\ref{eq:ch3:untrace}) one can directly check (see also Sec. \ref{sec:ch3:trace}) that we have 
		\begin{equation}
			\textrm{tr}(\rho)=\sum_{x=0}^{2^{n}-1}p_{x}=1.
		\end{equation}
		Therefore, comparing (\ref{eq:ch3:eigen}) with (\ref{eq:ch3:derdens1}), we conclude that the  matrix $\rho$ describes the ensemble state 	\begin{equation}\label{eq:ch3:ensembleex}
			\{(|u_{0}\rangle,p_{0}),(|u_{1}\rangle,p_{1}),...,(|u_{2^{n}-1}\rangle,p_{2^{n}-1})\}.
		\end{equation} This corresponds to a noisy ket state of the type studied in the previous subsection, in which the system is in pure state $|u_{x}\rangle$ with probability $p_{x}$. Note that, if the some of the eigenvalues are repeated (i.e., if they have multiplicity larger than $1$), the decomposition (\ref{eq:ch3:eigen}), and hence the resulting ensemble (\ref{eq:ch3:ensembleex}), is not unique.

		\subsection{Density Matrix as a General Description of a Quantum State}
	
	Density matrices provide a general description of quantum states that includes as special cases pure, i.e., ket, states, as well as random and deterministic cbits. Therefore, one can use a density matrix to specify an arbitrary classical state involving deterministic or random cbits, as well as a pure or mixed quantum state. In other words, density matrices account for epistemic uncertainty, as well as for quantum uncertainty related to the randomness of quantum measurements.  
	
	 To elaborate on this point, we first observe that a quantum coherent state, described by a ket $|\psi\rangle$, is described by the rank-1 density
	 matrix 
	 \begin{equation}\label{eq:ch3:rank1}
	 	\rho=|\psi\rangle\langle\psi|.
	 \end{equation} 	More general mixed quantum states, accounting for ensembles of pure states, are described by density matrices with
 rank larger than $1$. An interesting observation is that multiplying the pure state $|\psi\rangle$ by a complex number of the form $\exp(i\theta)$ for some angle $\theta$ does not change the corresponding density matrix (\ref{eq:ch3:rank1}). This reflects the fact that, by Born's rule, the global phase of a state $|\psi\rangle$ does not affect the distribution of the output of measurements of state $|\psi\rangle$.

	Second, we note that density matrices can also be used to describe classical random cbits, and hence also deterministic cbits. In fact, a random cbit taking value $0$  with probability $p_{0}$
	and $1$ with probability $p_{1}=1-p_0$ can be described by the ensemble $\{(|0\rangle,p_0),(|1\rangle,p_1)\}$. By the definition (\ref{eq:ch3:derdens1}), this ensemble state is described by the 
	$2\times2$ diagonal density matrix
	\begin{equation}\label{eq:ch3:rcbit}
		\rho=p_0 |0\rangle \langle 0| + p_1 |1\rangle \langle 1| =\left[\begin{array}{cc}
			p_{0} & 0\\
			0 & p_{1}
		\end{array}\right].
	\end{equation}
	This is a valid density matrix since it is positive semidefinite and it satisfies
	the condition $\textrm{tr}(\rho)=p_{0}+p_{1}=1$.
	
	More generally, a  random $n$-cbit string with probability vector 
	$p=[p_{0},...,p_{2^{n}-1}]^T$ is described by the diagonal density matrix
	\begin{equation}
		\rho=\sum_{x=0}^{2^{n}-1} p_x |x\rangle \langle x|= \left[\begin{array}{cccc}
			p_{0} & 0 & \cdots & 0\\
			0 & p_{1} & \ddots & \vdots\\
			\vdots & \ddots & \ddots & 0\\
			0 & \cdots  & 0 & p_{2^{n}-1}
		\end{array}\right],
	\end{equation}
	which is positive semidefinite and satisfies the condition
	$\textrm{tr}(\rho)=\sum_{x=0}^{2^{n}-1}p_{x}=1$.
		
		\subsection{Evolution and Measurement of Density States}\label{sec:ch3:densityform}
		
	We have seen in the previous subsection that a density matrix $\rho$ provides a general description of quantum  states. It is, therefore, important to extend the definition of unitary, i.e., closed-system, evolutions and of measurements from pure to density states.
	
 	When not measured, density states undergo unitary evolutions for the same reasons explained in Sec. \ref{sec:ch1:unit}. Generalizing the unitary evolution of a coherent state, given a unitary transformation $U$ and an input density state $\rho$, the density
 matrix obtained at the output of the transformation is given as
 \begin{equation}
 	\rho'=U\rho U^{\dagger}.
 \end{equation}One can directly check that this reduces to the standard unitary evolution of coherent states by considering rank-1 matrices.

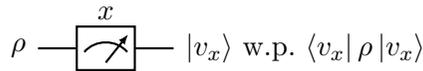
\begin{figure}
			\begin{center}
				\begin{quantikz}
					\lstick{$\rho$} & \meter{$x$} & \qw  \rstick{$\ket{v_x}$ w.p.  $\bra{v_x}\rho\ket{v_x}$} 
				\end{quantikz}
		\end{center}\vspace{-0.2cm} \caption{Measurement in an arbitrary orthonormal basis $\{|v_{x}\rangle\}_{x=0}^{2^{n}-1}$ for a density state $\rho$.\label{fig:ch3:densmes}}
	\end{figure}

Let us now turn to measurements. Following (\ref{eq:ch3:derdens}), a von Neummann measurement
			in the orthonormal basis $\{|v_{x}\rangle\}_{x=0}^{2^{n}-1}$ for an $n$-qubit density state $\rho$ operates as illustrated in Fig. \ref{fig:ch3:densmes} and detailed as follows.	\begin{itemize}
				\item \textbf{Born's rule}: The probability of observing $x\in\{0,1\}^{n}$ is
				\begin{align}\label{eq:ch3:genBorndens}
					\Pr[\textrm{measurement output equals \ensuremath{x}}]&=\langle v_x|\rho|v_x\rangle;	\end{align}   
				\item \textbf{``Collapse'' of the state}: If the measured cbits are given by string $x\in\{0,1\}^{n}$, the post-measurement state is given by the rank-1 density matrix $|v_x \rangle \langle v_x|$.
			\end{itemize} 
		
The definition of measurement can be extended to account for the partial measurement of $n'$ qubits as defined in Sec. \ref{sec:ch3:partial}. We refer the reader to Sec. \ref{sec:ch5:projmeas} for further discussion on this point in the context of projective measurements.


\subsection{An Application of Ensemble States: Quantum Encryption}

As an application of the formalism of density matrices, consider the problem of \textbf{quantum encryption}. In it, we have an encoder, which holds some quantum system in state $|\psi\rangle$, an intended decoder, and, possibly, an eavesdropper. The eavesdropper may intercept the quantum system while in transit between encoder and decoder. The goal of quantum encryption is twofold. First, we would like to encode the quantum state  in such a way that the intended decoder 
 can recover it exactly, assuming that the state is not intercepted by the eavesdropper. Second, should the eavesdropper get hold of the quantum system, we would like the eavesdropper to be unable to obtain useful information about the state  $|\psi\rangle$ from it. 

In a symmetric-key encryption strategy, encoder and decoder share a key $\kappa\in\{0,1,...,K-1\}$, which is randomly selected from the set of $K$ possible keys. While encoder and decoder know the selected key $\kappa$, the eavesdropper must model the key as being a uniformly distributed random variable in the set $\{0,1,...,K-1\}$.

 The encoder applies a unitary transformation $U_{\kappa}$ dependent on the key $\kappa$, which is also known to the decoder, to the ``plaintext'', i.e., to the quantum state  $|\psi\rangle$ to be encrypted. From the point of view of the decoder, which knows the key, the resulting state is $U_{\kappa}|\psi\rangle$. From it, the decoder can readily recover $|\psi\rangle$ by applying the inverse of  the unitary $U_{\kappa}$.
 
  In contrast, as far as the eavesdropper is concerned, the encrypted system is the mixed state $K^{-1}\sum_{\kappa=0}^{K-1} U_\kappa |\psi\rangle \langle \psi | U^{\dagger}_\kappa$, since the key is not available at the eavesdropper. Encryption can be considered as effective if, for any two ``plaintexts'' $|\psi\rangle$ and $|\psi'\rangle$, the two corresponding encrypted versions $K^{-1}\sum_{\kappa=0}^{K-1} U_\kappa |\psi\rangle \langle \psi | U^{\dagger}_\kappa$ and $K^{-1}\sum_{\kappa=0}^{K-1} U_\kappa |\psi'\rangle \langle \psi'| U^{\dagger}_\kappa$ are hard to distinguish, through measurement, by the eavesdropper (see Sec. \ref{sec:ch5:povmd} for discussion on quantum state detection).

\section{Partial Trace}\label{sec:ch3:partialtrace}

In Sec.  \ref{sec:ch3:decoh}, we have seen that a non-selective partial measurement leaves the subsystem that is not measured in an ensemble state when the two subsystems are entangled. The previous section has then shown how an ensemble state can be described by a density matrix. In this section, we describe a useful operation -- the partial trace -- that allows one to relate the density state resulting from a non-selective partial measurement to the pre-measurement state.

\subsection{Trace as a Linear Operator}\label{sec:ch3:trace}

Let us start by first reviewing the trace operation $\textrm{tr}(\cdot)$ by highlighting its operation as a linear function of the input matrix. Recall that the trace returns the sum of the elements on the main diagonal of a  square, $N \times N$,  matrix. We are interested in values of $N$ that can be written as $N=2^n$ for some integer $n$.

To proceed, we need the simple observation that any $N\times N$ square matrix $A$ can be written as \begin{equation}\label{eq:ch3:AA}
	A=\sum_{x=0}^{N-1} \sum_{y=0}^{N-1} [A]_{(x,y)} |x\rangle \langle y|,
\end{equation} where $[A]_{(x,y)}$ is the $(x,y)$-th element of the matrix, while $\{|x\rangle\}_{x=0}^{N-1}$ and $\{|y\rangle\}_{x=0}^{N-1}$ are the computational basis vectors. This follows immediately by noting that the outer product $|x\rangle \langle y|$ yields an $N \times N$ matrix with all zero entries except for a $1$ in the $(x,y)$-th position.  Furthermore, the same decomposition -- with generally different coefficients, say $[A']_{(x,y)}$ --  applies with any choice of the orthonormal basis $\{|v_x\rangle\}_{x=0}^{N-1}$, i.e., \begin{equation}\label{eq:ch3:AAB}
A=\sum_{x=0}^{N-1} \sum_{y=0}^{N-1} [A']_{(x,y)} |v_x\rangle \langle v_y|.
\end{equation}

The trace is a linear operator, and hence, by the decomposition (\ref{eq:ch3:AAB}) for an arbitrary square matrix, it can be described by its operation on an 
arbitrary outer product $|v\rangle\langle u|$ of two $N \times 1$ vectors $|v\rangle$ and $|u\rangle$. With this in mind, we can characterize the trace as the linear operator that acts on outer products of vectors as 
\begin{align}\label{eq:ch3:tracedef0}
	\textrm{tr}(|v\rangle\langle u|) & =\langle v|u\rangle.
\end{align}
In words, the trace ``converts'' an outer product of vectors into the corresponding inner product.

Using this characterization, it can be verified that the trace
can be equivalently written as
\begin{equation}\label{eq:ch3:tracedef}
	\textrm{tr}(\rho)=\sum_{x=0}^{2^{n}-1}\langle v_{x}|\rho|v_{x}\rangle
\end{equation}
for \emph{any} orthonormal basis $\{|v_{x}\rangle\}_{x=0}^{2^{n}-1}.$ In fact, the operation in (\ref{eq:ch3:tracedef}) satisfies (\ref{eq:ch3:tracedef0}).


\subsection{Partial Trace via Non-Selective Partial Measurements}\label{sec:ch3:nonsel}

	 Let us return to the motivating setting of a non-selective partial measurement by focusing on the system of $n$ cbits in a general pure state as studied in Sec. \ref{sec:ch3:multimeas}. We 	measure the first set of $n'$ qubits in some basis $\{|v_{x}\rangle_{n'}\}_{x=0}^{2^{n'}-1}.$ As illustrated in Fig. \ref{fig:ch3:nonsel}, we refer to the measured subsystem of $n'$ qubits as $A$ and to the rest of the $n-n'$ qubits as subsystem $B$.
	 
	 Following the discussion in Sec. \ref{sec:ch3:partial}, we can always write the joint state as (cf. (\ref{eq:ch3:partialdec}))
	\begin{equation}\label{eq:ch3:AB}
	|\psi\rangle_{AB}=\sum_{x=0}^{2^{n'}-1}\text{\ensuremath{\alpha{}_{x}}}|v_{x}\rangle_A|\phi_{x}\rangle_B,
	\end{equation}
	where $\{|\phi_{x}\rangle_B\}_{x=0}^{2^{n'}-1}$ are states for subsystem $B$ (generally not orthogonal). Note that in (\ref{eq:ch3:AB}),  we have used subscripts to indicate the subsystem corresponding to each state. This is a commonly used approach. 
	
	 The density matrix for subsystem $B$ upon a non-selective measurement of subsystem $A$ is given as (\ref{eq:ch3:derdens1}), i.e., 
	\begin{equation}\label{eq:ch3:rhob}
	\rho_{B}=\sum_{x=0}^{2^{n'}-1}|\alpha_{x}|^{2}|\phi_{x}\rangle_{B}\langle\phi_{x}|_{B}.
	\end{equation}Note again the use of the subscript in $\rho_B$ to emphasize the identity of the subsystem corresponding to the given quantum state. How can we express the relationship between the density state $\rho_{B}$ and the original density state $\rho_{AB}=|\psi\rangle_{AB}\langle\psi|_{AB}$ of the overall system?

	 A convenient way to do this is by means of an operation that generalizes the trace (\ref{eq:ch3:tracedef}): the partial trace. The \textbf{partial trace} of the density $\rho_{AB}$  over subsystem $A$ -- i.e., over the system being measured and discarded -- is defined as\begin{align}\label{eq:ch3:ptrace}
	 	\rho_{B}&=\textrm{tr}_{A}(\rho_{AB}) \nonumber\\
	 	&=\sum_{x=0}^{2^{n'}-1}(\langle v_{x}|_{A}\otimes I)\rho_{AB}(|v_{x}\rangle_{A}\otimes I),
	 \end{align}
where $I$ is the $2^{n-n'} \times 2^{n-n'}$ identity matrix. Note that the subscript in $\textrm{tr}_{A}(\cdot)$ indicates which system is being ``traced over''.  The equality between the partial trace output (\ref{eq:ch3:ptrace}) and the density matrix (\ref{eq:ch3:rhob}) can be directly checked by plugging $\rho_{AB}=|\psi\rangle_{AB}\langle\psi|_{AB}$ with (\ref{eq:ch3:AB}) into (\ref{eq:ch3:ptrace}).

By comparison with (\ref{eq:ch3:tracedef}), the trace corresponds to a partial trace applied to the entire system, which is obtained for $n=n'$. Importantly, the partial trace over a subsystem $A$ of $n'$ qubits returns a density of size $2^{n-n'}\times 2^{n-n'}$ over $n-n'$ qubits, while the trace produces a scalar (i.e., a $1 \times 1$ matrix).  As for the trace, the choice of the orthonormal basis $\{|v_x\rangle_A\}_{x=0}^{2^{n'}-1}$ is immaterial in the definition of the partial trace (\ref{eq:ch3:ptrace}). That is, the output of the partial trace operation is the same irrespective of the choice of the orthonormal basis for subsystem $A$. 

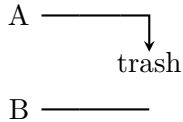
\begin{figure}
	\begin{center}
		\begin{quantikz}
			\lstick{A} \qw & \qw & \trash{\text{trash}} \\
			\lstick{B} \qw & \qw & \qw  
		\end{quantikz}
	\end{center}\vspace{-0.4cm} \caption{Since the density state of subsystem $B$ in the situation depicted in Fig. \ref{fig:ch3:nonsel1} does not depend on the specific measurement made on subsystem $A$, one can illustrate the setting from the perspective of subsystem $B$ as in this figure by removing the specification of a measurement on subsystem $A$.\label{fig:ch3:nonsel1}}
\end{figure}

Interpreting this last result in terms of partial non-selective measurements, we conclude that the state of subsystem $B$ when subsystem $A$ is measured in a non-selective fashion does not depend on the specific basis in which subsystem $A$ is measured. Therefore, from the perspective of subsystem $B$, the situation in Fig. \ref{fig:ch3:nonsel} can be also illustrated as in Fig. \ref{fig:ch3:nonsel1} by removing the specification of a measurement on subsystem $A$. The fact that any unterminated wire in a quantum circuit can be terminated with a measurement is known as the \textbf{principle of implicit measurement}.

Some further discussion on the implications of entanglement on the output of non-selective measurements can be found in Appendix B.

\subsection{Partial Trace as a Linear Operator}
Like the trace, the partial trace operation (\ref{eq:ch3:ptrace}) is also linear. We will now generalize the definition of the trace given in (\ref{eq:ch3:tracedef0}), which relies on the linearity of the trace, to the partial trace. 

From (\ref{eq:ch3:AAB}), it follows that any matrix square $2^n\times 2^n$ matrix can be expressed as a linear combination of matrices for the form 
	\begin{equation}
	|v,u\rangle_{AB}\langle v',u'|_{AB}=|v\rangle_{A}\langle v'|_{A}\otimes|u\rangle_{B}\langle u'|_{B},
	\end{equation}where $|v\rangle_A$ and $|v'\rangle_A$ states for subsystem $A$, and  $|u\rangle_{B}$ and $|u'\rangle_{B}$ are states for subsystem $B$. The partial trace can be hence equivalently described, generalizing the trace definition (\ref{eq:ch3:tracedef0}), as the linear operator satisfying the equalities
	\begin{align}\label{eq:ch3:ptrace1}
		\textrm{tr}_{A}(|v,u\rangle_{AB}\langle v',u'|_{AB}) & =\textrm{tr}_{A}(|v\rangle_{A}\langle v'|_{A}\otimes|u\rangle_{B}\langle u'|_{B})\nonumber\\
		& =\textrm{tr}(|v\rangle_{A}\langle v'|_{A})\cdot|u\rangle_{B}\langle u'|_{B}\nonumber\\
		& =\langle v|_{A}|v'\rangle_{A}\cdot|u\rangle_{B}\langle u'|_{B}.
	\end{align} In words, the partial trace ``converts'' outer products into inner products only for the system that is being traced over, namely subsystem $A$. 

The equivalent definitions (\ref{eq:ch3:ptrace}) and (\ref{eq:ch3:ptrace1}) can be directly generalized to the partial measurement of any subset of qubits in a system. Accordingly, one can straightforwardly extend the partial trace operation to trace over any subset of qubits.

\subsection{Partial Trace and Classical Marginalization}

The partial trace can be interpreted as a form of ``quantum marginalization''. To explain, consider a set of $n=2$ random cbits with general probability vector $p=[p_{00},p_{01},p_{10},p_{11}]^{T}$. This state can be described by the density matrix
		\begin{align}
			\rho_{AB} & =\left[\begin{array}{cccc}
				p_{00} & 0 & 0 & 0\\
				0 & p_{01} & 0 & 0\\
				0 & 0 & p_{10} & 0\\
				0 & 0 & 0 & p_{11}
			\end{array}\right]\nonumber\\
			& =p_{00}|00\rangle\langle00|+p_{01}|01\rangle\langle01|+p_{10}|10\rangle\langle10|+p_{11}|11\rangle\langle11|.
		\end{align}
	By (\ref{eq:ch3:ptrace1}), the partial
		trace over subsystem $A$, i.e., over the first qubit (encoding the first cbit), returns
		\begin{align}
			\rho_{B} & =\textrm{tr}_{A}(\rho_{AB})=p_{00}|0\rangle\langle0|+p_{10}|0\rangle\langle0|+p_{01}|1\rangle\langle1|+p_{11}|1\rangle\langle1|\nonumber\\
			& =\left[\begin{array}{cc}
				p_{00}+p_{10} & 0\\
				0 & p_{01}+p_{11}
			\end{array}\right].
		\end{align}
This is the density matrix to a random cbit with distribution equal to the marginal distribution of
		the second cbit. So, as anticipated, the partial trace generalizes the operation of classical
		marginalization.

\section{Conclusions}
In this chapter, we have generalized the quantum measurements  in two important directions. First, we have defined measurements in any orthonormal basis, which encompass joint measurements across multiple qubits. We have seen that such measurements can be implemented via standard measurements (in the computational basis) via suitable pre- and post-measurement unitary operations. Then, we have discussed situations in which only a subset of qubits is measured, and the output of the measurement is either retained or discarded. In the latter case, we have concluded that, if there is entanglement between the qubits being measured and the rest of the system,  the state of the subsystem that is not subject to measurement must be described by a density matrix. The density matrix of the remaining subsystem can be related to the original density matrix via the partial trace operation.

\section{Recommended Resources}
 For a basic introduction to partial measurements, the reader is referred to \cite{mermin2007quantum}. Non-selective measurements and quantum noise are clearly presented in \cite{nielsen10,wilde2013quantum,watrous2018theory}. For discussions on the implementations of measurements using quantum circuits, a useful reference is \cite{kaye2006introduction}.

\section{Problems}

	\begin{enumerate}
		\item For a two-qubit system show that measuring qubits sequentially
	in the computational basis gives the same distribution of the outputs
	as a joint measurement in the computational basis. Generalize to any
	number of qubits.
	\item Describe mixed states as vectors within the Bloch sphere (see recommended resources).
\item  Describe the von Neumann entropy and show that for entangled states
	the entropy of a whole can be smaller than the entropy of individual
	subsystems (see Appendix A).
	\item Contrast the coherent superpositon
	\begin{equation}
	\sum_{x}\alpha_{x}|\psi_{x}\rangle
	\end{equation}
	with the non-coherent superposition
	\begin{equation}
	\sum_{x}p_{x}|\psi_{x}\rangle\langle\psi_{x}|
	\end{equation}
	by demonstrating quantum circuits in which these two states as inputs
	yield different measurement probabilities. Argue that for the former
	state the relative phases between the component states $|\psi_{x}\rangle$ matter,
	while this is not the case for the latter.
\item For the circuit below, describe the state of the first three qubits
	when a measurement on the last qubit is discarded. Note
	that the last qubit is initialized in the ground state $|0\rangle$.
	Consider first the eight vectors in the computational basis for the first
	three qubits, and then describe what happens to an arbitrary superposition
	state for the first three qubits. Relate your conclusions to the concept
	of parity (see also Chapter 5).\begin{center}
			\begin{quantikz}
				\qw & \ctrl{3} & \qw & \qw & \qw &  \qw\\
				\qw & \qw & \ctrl{2} & \qw & \qw &  \qw\\
				\qw & \qw & \qw & \ctrl{1} & \qw &  \qw\\
				\lstick{$\ket{0}$} & \gate{X} & \gate{X} & \gate{X} & \meter{} &  \trash{\text{trash}}
			\end{quantikz}
	\end{center}
\end{enumerate}

%
%
%
%
%

\section*{Appendix A: Quantifying Bipartite Entanglement}

 In this appendix, we focus on bipartite pure states $|\psi\rangle_{AB}$ of $n$ qubits that are partitioned into a subsystem $A$ encompassing
		$n_{A}$ qubits and a subsystem $B$ including $n_{B}=n-n_{A}$ qubits
		(see Sec. \ref{sec:ch3:nonsel}). We describe two
		ways of quantifying entanglement of the two systems, namely the Schmidt
		rank and the entropy of entanglement. 
		
		\subsection*{Schmidt Decomposition and Schmidt Rank}
		
		 Any pure state $|\psi\rangle_{AB}$ can be expressed as the amplitude vector
		\begin{align}
			|\psi\rangle_{AB} & =\sum_{x=0}^{2^{n_{A}}-1}\sum_{y=0}^{2^{n_{B}}-1}\alpha_{x,y}|x\rangle_{A}\otimes|y\rangle_{B}\nonumber \\
			& =\left[\begin{array}{ccccccccc}
				\alpha_{0,0} &  \cdots & \alpha_{0,2^{n_{B}}-1} & \cdots & \alpha_{2^{n_{A}}-1,0} &  \cdots & \alpha_{2^{n_{A}}-1,2^{n_{B}}-1}\end{array}\right]^{T},
		\end{align}
		where $\{|x\rangle_{A}\}_{x=0}^{2^{n_{A}}-1}$ and $\{|y\rangle_{B}\}_{x=0}^{2^{n_{B}}-1}$
		are the computational bases for subsystems $A$ and $B$. 
		 Equivalently, we can write the state as
		\begin{equation}
		|\psi\rangle_{AB}=\textrm{vec}(\text{\ensuremath{\Omega_{AB}}})
		\end{equation}
		in terms of the $2^{n_{B}}\times2^{n_{A}}$ \textbf{amplitude matrix}
	\begin{equation}\label{eq:ch3:amplmat}
		\text{\ensuremath{\Omega}}_{AB}=\left[\begin{array}{cccc}
			\alpha_{0,0} & \alpha_{1,0} & \cdots & \alpha_{2^{n_{A}}-1,0}\\
			\alpha_{0,1} & \alpha_{1,1} & \cdots & \alpha_{2^{n_{A}}-1,1}\\
			\vdots & \vdots & \vdots & \vdots\\
			\alpha_{0,2^{n_{B}}-1} & \alpha_{1,2^{n_{B}}-1} & \cdots & \alpha_{2^{n_{A}}-1,2^{n_{B}}-1}
		\end{array}\right],
	\end{equation}
		where the operator $\textrm{vec}(\cdot)$ stacks the columns of the
		argument matrix.
		 The amplitude matrix (\ref{eq:ch3:amplmat}) provides an equivalent characterization of the
		state $|\psi\rangle_{AB}$ that highlights the assumed partition into
		the subsystems $A$ and $B$. 
		
		 Consider now the singular value decomposition (SVD) of the amplitude
		matrix
	\begin{equation}\label{eq:ch3:Svd}
		\text{\ensuremath{\Omega}}_{AB}=\sum_{d=0}^{r-1}\lambda_{d}|u_{d}\rangle_{B}\langle v_{d}|_{A},
	\end{equation}
		where the sets $\{|v_{d}\rangle_{A}\}_{d=0}^{r-1}$ and $\{|u_{d}\rangle_{B}\}_{d=0}^{r-1}$
		comprise the right and left singular vectors, respectively, associated
		with the $r$ positive singular values $\{\lambda_{d}\}_{d=0}^{r-1}$.
		Recall that the $2^{n_{A}}\times1$ vectors $\{|v_{d}\rangle_{A}\}_{d=0}^{r-1}$
		are orthonormal, and so are the $2^{n_{B}}\times1$ vectors $\{|u_{d}\rangle_{B}\}_{d=0}^{r-1}.$ 
		 Furthermore, the number of positive singular values $r$ satisfies
		the inequality
		\begin{equation}\label{eq:ch3:Src}
		r\leq\min(2^{n_{A}},2^{n_{B}}).
		\end{equation}

		 Using the SVD (\ref{eq:ch3:Svd}), we then have
		\begin{equation}\label{eq:ch3:svdsvd}
		|\psi\rangle_{AB}=\textrm{vec}(\Omega_{AB})=\sum_{d=0}^{r-1}\lambda_{d}|v_{d}\rangle_{A}^{*}\otimes|u_{d}\rangle_{B},
	\end{equation}
		where $|v_{d}\rangle_{A}^{*}$ represents the complex conjugate of
		vector $|v_{d}\rangle_{A}$. Note that the vectors $\{|v_{d}\rangle_{A}^{*}\}_{d=0}^{r-1}$
		are still orthonormal.
		 The expression (\ref{eq:ch3:svdsvd}) is known as the Schmidt decomposition of state $|\psi\rangle_{AB}$ 
		as it pertains its partition into systems $A$ and $B$, and the Schmidt
		rank is the number $r$ of positive singular values. 
		
		 Restating this result, given a state $|\psi\rangle_{AB}$ of a system partitioned into subsystems $A$ and $B$ of $n_A$ and $n_B$ qubits, respectively, there exist an orthonormal
		set of $r$  vectors $\{|a_{d}\rangle_{A}\}_{d=0}^{r-1}$ of dimension $2^{n_{A}}\times1$ and an orthonormal set 
		of $r$  vectors $\{|b_{d}\rangle_{B}\}_{d=0}^{r-1}$ of dimension $2^{n_{B}}\times1$, such that the \textbf{Schmidt decomposition}
		\begin{equation}\label{eq:ch3:SD}
		|\psi\rangle_{AB}=\sum_{d=0}^{r-1}\lambda_{d}|a_{d}\rangle_{A}\otimes|b_{d}\rangle_{B}
	\end{equation} holds. Note that we have $|a_{d}\rangle_{A}=|v_{d}\rangle^{*}_{A}$ and $|b_{d}\rangle_{B}=|u_{d}\rangle_{B}$ in the derivation above.
		Accordingly, any pure state $|\psi\rangle_{AB}$ can be expressed as a linear combination of $r\leq\min(2^{n_{A}},2^{n_{B}})$ separable states $\{|a_{d}\rangle_{A}\otimes|b_{d}\rangle_{B}\}_{d=0}^{r-1}$
		with positive amplitudes $\{\lambda_{d}\}_{d=0}^{r-1}$ satisfying
		the normalization condition
			\begin{equation}
			\sum_{d=0}^{r-1}\lambda_{d}^{2}=1.
			\end{equation}
	
		 An interesting aspect of the Schmidt decomposition is that, if one
		of the two systems comprises a small number of qubits, the Schmidt
		rank $r$ is accordingly small due to the inequality (\ref{eq:ch3:Src}).
		 Another useful property of the Schmidt decomposition has to do with
		the quantification of entanglement between subsystems $A$ and $B$. This is the main subject of this appendix.
		
The key result in this regard is that the two subsystems $A$ and $B$ are entangled if and
		only if the Schmidt rank $r$ of state $|\psi\rangle_{AB}$ is larger than 1. Conversely, if and only if $r=1,$
		by (\ref{eq:ch3:SD}), the state 
		of the two subsystems is separable (and is given by $|\psi\rangle_{AB}=|a_{0}\rangle_{A}\otimes|b_{0}\rangle_{B}$). Therefore, the Schmidt rank
		$r$ can be used to assess the degree of entanglement between the two
		systems.
		
		The significance of the Schmidt rank as a measure of entanglement can  be also understood by studying the setting shown in Sec.   \ref{sec:ch3:nonsel} in which one of the two subsystems is discarded. From (\ref{eq:ch3:SD}), if we make a measurement of subsystem $A$ in a non-selective
		way, we obtain the density matrix
		\begin{align}\label{eq:ch3:rhoB}
			\rho_{B} & =\mathrm{tr}_{A}(|\psi\rangle_{AB}\langle\psi|_{AB})=\sum_{d=0}^{r-1}\lambda_{d}^{2}|b_{d}\rangle_{B}\langle b_{d}|_{B},
		\end{align}
		while measuring subsystem $B$ in a non-selective way yields
		\begin{align}\label{eq:ch3:rhoA}
			\rho_{A} & =\mathrm{tr}_{B}(|\psi\rangle_{AB}\langle\psi|_{AB})=\sum_{d=0}^{r-1}\lambda_{d}^{2}|a_{d}\rangle_{A}\langle a_{d}|_{A}.
		\end{align}
		Therefore, the Schmidt rank corresponds to the rank of the density
		matrix of one subsystems when the other is discarded. Accordingly, if the two systems
		are not entangled, the density matrices describe pure states, i.e.,
		they have rank $r=1$; while entanglement is manifested in both subsystems having
		mixed states with rank $r>1$. 
		
		\subsection*{Entropy of Entanglement}

		 As we have seen in the previous subsection, entanglement between two subsystems causes non-selective
		measurements of one subsystem to yield a (non-trivial) mixed, and
		hence noisy, state for the other subsystem. Taking this perspective further, entanglement
		can be more precisely quantified by the degree of classical randomness
		of the reduced states (\ref{eq:ch3:rhoB}) and (\ref{eq:ch3:rhoA}). Accordingly, the more random the reduced states $\rho_A$ and $\rho_B$ 
		are, the more entangled the original pure state $|\psi\rangle_{AB}$
	 can be said to be. 
		
		 To quantify the classical randomness associated with a density state,
		a standard measure is the \textbf{von Neumann entropy}. Given a density
		state $\rho$ with eigendecomposition $\rho=\sum_{x=0}^{r-1}p_{x}|u_{x}\rangle\langle u_{x}|,$
		the von Neumann entropy is the classical, Shannon, entropy of the probability 
		vector $[p_{0},p_{1},...,p_{r-1}]$, i.e., 
		\begin{equation}
		H(\rho)=-\sum_{x=0}^{r-1}p_{x}\log_{2}(p_{x}),
		\end{equation}
		which is measured in bits. 
		 For a system of $n$ qubits, the von Neumann entropy satisfies the
		inequalities
		\begin{equation}\label{eq:ch3:entropyin}
		0\leq H(\rho)\leq n.
		\end{equation}
		The lower bound is attained if and only if the state $\rho$ is pure (i.e., with 
		rank 1); while the upper bound is achieved if and only if the density state is maximally mixed, i.e., if it corresponds to uniformly distributed random cbits with density matrix $\rho=2^{-n} I$. 
		
		 Returning to the density states (\ref{eq:ch3:rhoB}) and (\ref{eq:ch3:rhoA}), their von Neumann entropy
		is equal to 
		\begin{equation}\label{eq:ch3:enten}
		H(\rho_{A})=H(\rho_{B})=-\sum_{x=0}^{r-1}\lambda_{x}^{2}\log_{2}(\lambda_{x}^{2}).
		\end{equation}
		 By the discussion in the previous subsection, the von Neumann entropy
		of the reduced states is equal to zero if and only if the state
		$|\psi\rangle_{AB}$ is separable. Furthermore,  it is positive if and only
		if the state $|\psi\rangle_{AB}$ is entangled, and thus the Schmidt
		rank is larger than 1. Finally, by (\ref{eq:ch3:entropyin}), the maximum value of the von Neumann entropy
		for the reduced states  is $\min(n_{A},n_{B}),$ which
		corresponds to a \textbf{maximally entangled} state. 
		
		In light of the properties described in the last paragraph, given a pure bipartite state $|\psi\rangle_{AB}$, the von Neumann entropy of the reduced states, i.e., \begin{equation}
			H(\textrm{tr}_{A}(|\psi\rangle_{AB}\langle \psi|_{AB}))=H(\textrm{tr}_{B}(|\psi\rangle_{AB}\langle \psi|_{AB}))
		\end{equation}
		is known as the \textbf{entropy of entanglement} of the state $|\psi\rangle_{AB}$. The entropy of entanglement is zero if and only if the two subsystems $A$ and $B$ are not entangled, and it reaches the maximum value of $\min(n_{A},n_{B})$ when the two subsystems are maximally entangled.  For example, any Bell state can be checked to have the maximum entropy
		of entanglement of $\min(n_{A},n_{B})=1$  bit. Therefore, Bell pairs are maximally entangled.
		
		 As a final note, we observe that,  when the two
		 subsystems are entangled, the von Neumann entropy (\ref{eq:ch3:enten}) of the individual subsystems
		 is larger than the von Neumann entropy of the overall system $H(|\psi\rangle_{AB}\langle \psi|_{AB})=0$ -- a phenomenon that has no classical counterpart. (It would be as if the individual pages of a book contained random gibberish, and all the information enclosed in book could only be revealed by looking at all pages at once.)

\section*{Appendix B: On Multipartite Entanglement}

In this appendix, we briefly discuss entanglement involving more than
		two subsystems. We specifically elaborate on the case of $n=3$ qubits, each corresponding to a distinct subsystem. As we have seen in Appendix A, with two subsystems, entanglement can be readily quantified by the Schmidt rank or by the entropy of entanglement. As we will discuss here, multipartite entanglement exhibits a more complex behavior that is not as easily quantified. Multipartite entanglement is central to many quantum computing and communication primitives, such as on-demand Bell state generation between pairs of nodes.
		
		For $n=3$, there are two main types of entanglement, which are exemplified
		by the  \textbf{Greenberger--Horne--Zeilinger (GHZ) state}
		\begin{equation}
		|\textrm{GHZ\ensuremath{\rangle=\frac{1}{\sqrt{2}}(|000\rangle+|111\rangle)}}
		\end{equation}
		and by the \textbf{W state} (named after Wolfgang D\"{u}r)
		\begin{equation}\label{eq:ch3:Wstate}
		|\textrm{W\ensuremath{\rangle=\frac{1}{\sqrt{3}}(|001\rangle+|010\rangle+|100\rangle)}}.
		\end{equation} Note that the GHZ is a natural generalization of the Bell state $|\Phi^{+}\rangle=1/\sqrt{2}(|00\rangle+|11\rangle)$
		from $n=2$ to $n=3$ qubits; while the W state is a natural generalization
		of the Bell state $|\Psi^{+}\rangle=1/\sqrt{2}(|01\rangle+|10\rangle)$
		from $n=2$ to $n=3$ qubits. 
		
		The three qubits are entangled under both GHZ and W states, since
		the respective state vectors cannot be written as the Kronecker product of individual qubit
		states. More precisely, they are \emph{fully} entangled in the sense that they are not even \emph{biseparable}. A biseparable state can be partitioned into two subsystems such that the Kronecker product of the individual subsystems' states gives the overall state of the system.

 To elaborate on the different behaviors exhibited by systems in the
	 GHZ and W states, let us consider what happens if we observe one of the
		three qubits, say the first, and we are interested in the resulting
		state of the other two qubits. We specifically focus on non-selective
		measurements as described in Sec. \ref{sec:ch3:nonsel} and also considered in Appendix A.  By the generalized Born rule (\ref{eq:ch3:genBorn})-(\ref{eq:ch3:genBorn1}),
		for the GHZ state, a non-selective measurement of the first qubit
		yields the ensemble state
		\begin{equation}
		\{(|00\rangle,1/2),(|11\rangle,1/2)\}
		\end{equation}
		for the last two qubits. This ensemble is described by the density
		matrix
		\begin{equation}
		\rho=\frac{1}{2}|00\rangle\langle00|+\frac{1}{2}|11\rangle\langle11|=\frac{1}{2}I.
		\end{equation} We conclude that the two remaining qubits are no longer entangled.
		Rather, they behave like classical correlated cbits: With probability
		$1/2$ they take state $|00\rangle$, and with probability $1/2$ their
		state is $|11\rangle$. So, in a manner similar to the Bell states for $n=2$ qubits, the
		measurement of a single qubit destroys entanglement for the GHZ state. 

Let us now consider the W state. In order to apply the generalized
		Born rule, we rewrite (\ref{eq:ch3:Wstate}) equivalently as 
			\begin{equation}
			|\textrm{W\ensuremath{\rangle=\sqrt{\frac{2}{3}}|0\rangle|\Psi^{+}\rangle+\frac{1}{\sqrt{3}}|1\rangle|00\rangle,}}
			\end{equation}
			where $|\Psi^{+}\rangle=1/\sqrt{2}(|01\rangle+|10\rangle)$.Therefore, following the same steps used above for the GHZ
			state, upon measuring the first qubit in a non-selective way, the
			last two qubits are in the mixed state
		\begin{equation}
		\rho=\frac{2}{3}|\Psi^{+}\rangle\langle\Psi^{+}|+\frac{1}{3}|00\rangle\langle00|.
		\end{equation} It follows that the two qubits are still partially entangled, albeit not
		maximally so. In fact, with probability $2/3$ they are in the (maximally
		entangled) Bell state $|\Psi^{+}\rangle$, while with probability
		$1/3$ they are in the separable state $|00\rangle$.

Overall, the discussion in this appendix suggests that GHZ states
		exhibit a larger degree of entanglement, which is, however, completely lost
		upon the measurement of any of the three qubits. In contrast, the
		``connectivity'' among the three qubits under the W state is less
		strong, as the measurement of one qubit preserves some degree of entanglement
		between the other two qubits. That said, the two types of entanglement are distinct in the sense
		that it is not possible to transform either state into the other via
		local operations (see also Problem 1 in Chapter 2 for the case $n=2$).

\section*{Appendix C: More on Density Matrices and on the Partial Trace}

This appendix reports some useful additional facts about density matrices and partial trace.

As we have seen in Sec. \ref{sec:ch1:uniteig}, the notation $f(\rho)$ describes the matrix
	\begin{equation}
		f(\rho)=\sum_{x=0}^{2^{n}-1}f(p_{x})|u_{x}\rangle\langle u_{x}|,
	\end{equation}where we have used the eigendecomposition (\ref{eq:ch3:eigen}).
	and this definition extends more generally to all normal operators.  Using this definition, a pure state can be readily seen to satisfy the equality
	\begin{equation}
		\textrm{tr}(\rho^{2})=1,
	\end{equation}
	while, more generally, we have the inequalities
	\begin{equation}
		1/2^{n}\leq\textrm{tr}(\rho^{2})\leq1.
	\end{equation}
	The quantity $\textrm{tr}(\rho^{2}),$ known as \textbf{purity}, decreases
	with the ``mixedness'' of the state.
	
	Following Sec. \ref{sec:ch1:Paulibasis}, for a single qubit, the density matrix can also be represented using
	the Pauli operators as
	\begin{equation}
	\rho=\frac{1}{2}(I+r_{x}X+r_{y}Y+r_{z}Z),
	\end{equation}where the coefficients are computed as $r_{x}=\textrm{tr}(\rho X),$ \textrm{$r_{y}=\textrm{tr}(\rho Y)$,
	and $r_{z}=\textrm{tr}(\rho Z)$.}
	 Furthermore, the so-called Bloch vector $r=[r_{x},r_{y},r_{z}]^{T}$ can be shown to satisfy
	the inequality $||r||\leq1$ and the two eigenvalues can be computed as $1/2(1\pm||r||_{2}).$

	 If we write the joint density matrix of a system of two qubits ($n=2$) in the block form
	\begin{equation}
	\rho_{AB}=\left[\begin{array}{cc}
		P & Q\\
		R & S
	\end{array}\right]
	\end{equation}
	where the size of $P$ is $2\times2$ and the size of S is $2\times2$,
	respectively, then we have
	\begin{equation}
	\text{\textrm{tr}}_{A}(\rho_{AB})=P+S
	\end{equation}
	and
	\begin{equation}
	\text{\textrm{tr}}_{B}(\rho_{AB})=\left[\begin{array}{cc}
		\textrm{tr}(P) & \textrm{tr}(Q)\\
		\textrm{tr}(R) & \textrm{tr}(S)
	\end{array}\right].
	\end{equation}


\chapter{Quantum Computing}

\section{Introduction}

This chapter provides a brief introduction to the approach traditionally adopted for the design of quantum algorithms based on quantum circuits. To this end, we first review the gate-based model of quantum computation specified by a quantum circuit. Then, we describe basic building blocks of traditional quantum algorithms aimed at computing binary functions and at implementing a form of quantum memory. This background material will provide us with the necessary tools to study  Deutsch's algorithm, the first demonstration of the benefits
of quantum computation. Finally, we discuss the no cloning theorem, which imposes fundamental limitations on quantum
computing (as well as quantum communications). Along the way, we will present several key concepts such as quantum parallelism, basis-copying circuits, and phase kick-back.  

\section{Gate-Based Model of Quantum Computation}\label{sec:ch4:gatebased}

As described in Sec. \ref{sec:ch2:singlequbitgates}, a quantum algorithm is specified by a quantum circuit. A quantum circuit, in turn, consists of a sequence
of quantum gates (unitaries) and measurements that are applied\emph{ in place} to
a register of qubits. In a quantum circuit, we have a wire for each qubit, and \emph{double lines} represent a wire that carries cbits. 

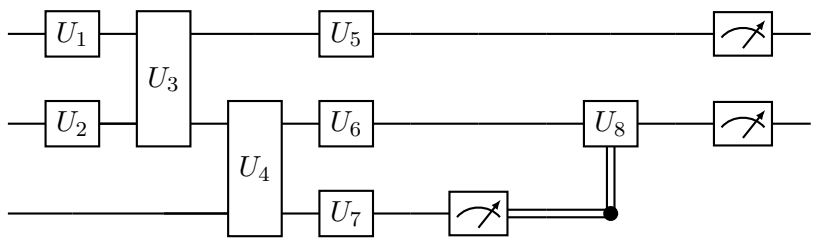
\begin{figure}
		 	\begin{center} \begin{quantikz} & \gate{U_1} & \gate[wires=2]{U_3} & \qw & \gate{U_5} & \qw &\qw & \qw & \qw &\qw & \meter{} & \qw\\ & \gate{U_2} & \qw & \gate[wires=2]{U_4}& \gate{U_6}& \qw &\qw & \qw & \gate{U_8} &\qw & \meter{} & \qw \\ & \qw & \qw & \qw  & \gate{U_7}& \qw & \meter{} & \cw & \cwbend{-1}
		 	\end{quantikz} \end{center}\vspace{-0.4cm}\caption{An illustration of the gate-based model of quantum computing.\label{fig:ch4:gatebased}}
		 \end{figure}

In the example of Fig. \ref{fig:ch4:gatebased}, we have a sequence of quantum gates $U_m$ with $m\in\{1,...,7\}$, acting on one or two qubits. The cbit at the output of the measurement on the third qubits controls the gate $U_8$ applied to the second qubit. For instance, the gate $U_8$ may be the identity $I$ if the controlling cbit is equal to 0 and an $X$ gate if the cbit is equal to 1. It is recalled that measurements are irreversible operations, and hence the outlined operation is different from a controlled gate as defined in Sec. \ref{sec:ch2:controlled}, which is reversible.  That said, it can be proved that all measurements can be always moved to the end of the circuit by replacing measurement-controlled gates with controlled gates -- this is the so-called \textbf{principle of deferred measurement}.  We will see a use of measurement-controlled unitaries in the next chapter in the context of quantum error correction (see Sec. \ref{sec:ch5:qec}).  Finally, the output of the computation is given by the result of the measurement of the first two qubits, as further discussed next. 

The output of a quantum algorithm can be defined in one of two ways. \begin{itemize}
		 \item \textbf{Probabilistic quantum computing}: The quantum algorithm returns a measurement output obtained from
		a \emph{single run} of the quantum circuit. Due to \textbf{shot noise}, i.e., to the inherent stochasticity of quantum measurements, the output is generally a random cbit string. Therefore, this approach implements a form of probabilistic computing.
		\item  \textbf{Deterministic quantum computing}: The quantum algorithm returns the average  of several measurement outputs that are 
		obtained from \emph{multiple runs} of the quantum circuit. This type of output is typically formulated in terms of expected values of observables, which will be introduced in the next chapter (see Sec. \ref{sec:ch5:obs}).
		\end{itemize}

\section{Computing Binary Functions and Quantum RAM}\label{sec:ch4:cso}

Having introduced the general model of gate-based quantum computing, in this section we describe a useful building blocks for quantum algorithms: the controlled shift operator. We show how this gate can be used to evaluate binary functions and to implement a quantum random-access memory (RAM).
	
	\subsection{Shift Operators}
	
\begin{figure}
\begin{center}
		\begin{quantikz}
			\qw & \gate{X^{a_0}} & \qw \\
			\qw & \gate{X^{a_1}} & \qw\\
			& \vdots &\\
			\qw & \gate{X^{a_{m-1}}} & \qw
		\end{quantikz}
\end{center}\vspace{-0.4cm} \caption{The shift operator $X^{\otimes a}$ on $m$ qubits, defined in (\ref{eq:ch4:shift1}) or equivalently in (\ref{eq:ch4:shift}), is a Pauli string that applies  single-qubit Pauli $X$ gates or the identity $I$ in a manner controlled by the cbit string $a=a_{0},...,a_{m-1}\in\{0,1\}^{m}$ as in (\ref{eq:ch4:shift1}).  \label{fig:ch4:shift}}
\end{figure}
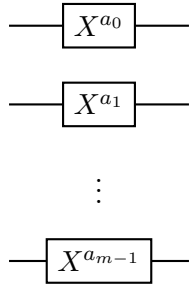

Given a binary string $a=a_{0},...,a_{m-1}\in\{0,1\}^{m}$, the \textbf{shift operator} $X^{\otimes a}$ is a separable quantum gate that applies either the identity gate $I$ or the  Pauli $X$ gate to each $k$-th qubit, with $k\in\{0,1,...,m-1\}$. Specifically, the Pauli $X$ gate is applied to qubit $k$ if the corresponding cbit $a_{k}$ equals $a_{k}=1$, and no operation is applied to qubit $k$ if
$a_{k}=0$. Mathematically, as illustrated in Fig. \ref{fig:ch4:shift}, the shift operator $X^{\otimes a}$ with shift cbit string $a$ can be expressed as the Kronecker product
	\begin{align}\label{eq:ch4:shift1}
		X^{\otimes a} =X^{a_{0}}\otimes X^{a_{1}}\otimes\cdots\otimes X^{a_{m-1}}.
	\end{align}Note that the shift operator $X^{\otimes a}$ is a Pauli string as defined in  Sec. \ref{sec:ch2:Paulibasis}.
	
	As an example, with $m=2$ qubits and the shift bit string $a=01,$ the shift operator is given as
	\begin{align}\label{eq:ch4:IX}
		X^{\otimes 01} =I\otimes X.
	\end{align}Therefore, the operator $X^{\otimes 01}$ applies a Pauli $X$ gate to the second qubit only.

The name ``shift'' operator arises from the  change-of-basis form of the  unitary (\ref{eq:ch4:shift1}). To describe this interpretation, for two binary strings $a=a_{0},...,a_{m-1}\in\{0,1\}^{m}$ and  $b=b_{0},...,b_{m-1}\in\{0,1\}^{m}$, we   write as $a\oplus b$ the $m$-cbit string obtained as the bit-wise XOR of the two cbit strings, i.e., \begin{equation}
	a\oplus b=a_{0}\oplus b_{0},...,a_{m-1} \oplus b_{m-1}.
\end{equation} Following the characterization of unitary matrices in Sec. \ref{sec:ch2:changeofbasis}, we can then express the shift operator (\ref{eq:ch4:shift1}) as \begin{align}\label{eq:ch4:shift}
	X^{\otimes a} & =\sum_{y=0}^{2^{m}-1}|y\oplus a\rangle\langle y|.
\end{align}  Accordingly, the shift operator maps each vector $|y\rangle$ in the
computational basis to the ``shifted'' version $|y\oplus a \rangle$. This can be readily verified by applying (\ref{eq:ch4:shift1}) to a vector $|y\rangle$ in the computational basis, which yields\begin{equation}
	X^{\otimes a}|y\rangle=|y\oplus a\rangle.
\end{equation} 
	
	As an example, the shift operator (\ref{eq:ch4:IX}) can be expressed in the change-of-basis form \begin{equation}
	X^{\otimes 01} = 	|01\rangle\langle00|+|00\rangle\langle01|+|11\rangle\langle10|+|10\rangle\langle11|.
		\end{equation}
	


	\subsection{Controlled Shift Operators}
	 Consider now a partition of the set of qubits into two subsystems,  with $n$ qubits in the first subsystem and $m$ in the second subsystem.  In this subsection, we will introduce the controlled shift operator, in which the first set of qubits determines, in a sense to be specified below, the shift to be applied to the second set of qubits. We will see in the following subsections that this operator allows the computation of binary functions on a quantum computer, as well as the implementation of a form of quantum memory.

	 To start, let us study the effect of applying the shift operator $X^{\otimes a}$ to the
	second subsystem of $m$ qubits within the larger system encompassing also the first $n$ qubits. This operation amounts to the application of the unitary  $I\otimes X^{\otimes a}$, with the identity matrix $I$ being of size $2^{n} \times 2^{n}$, i.e., to the transformation
	\begin{equation}\label{eq:ch4:um}
	(I\otimes X^{\otimes a})|x,y\rangle=|x,y\oplus a\rangle,
	\end{equation}
	where $\{|x\rangle\}_{x=0}^{2^{n}-1}$ and $\{|y\rangle\}_{y=0}^{2^{m}-1}$ are the computational
	bases for the Hilbert spaces of the two subsystems. In a manner similar to (\ref{eq:ch4:shift1}) and (\ref{eq:ch4:shift}), we can write the unitary matrix (\ref{eq:ch4:um}) as 
	\begin{align}\label{eq:ch4:shiftsep}
		I\otimes X^{\otimes a} & =\sum_{x=0}^{2^{n}-1}\sum_{y=0}^{2^{m}-1}|x,y\oplus a\rangle\langle x,y|,
	\end{align}
	where $I$ is the $2^{n}\times2^{n}$ identity matrix.
	
	\begin{figure}
		\begin{center}
			\begin{quantikz}
				\lstick{$\ket{x}$} & \qwbundle{n} & \ctrl{1} & \qwbundle{n} \rstick{$\ket{x}$}\\
				\lstick{$\ket{y}$} & \qwbundle{m} & \gate{X^{\otimes f(x)}} & \qwbundle{m} \rstick{$\ket{y\oplus f(x)}$}
			\end{quantikz}
		\end{center}\vspace{-0.4cm}\caption{The controlled shift operator $CX^{\otimes f}$ is a controlled qubit gate in which the first $m$ qubits control the shift operator $X^{\otimes f(x)}$ applied to last $n$ qubits. \label{fig:ch4:cshiftc}}
	\end{figure}
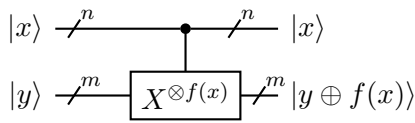

	Consider now a \textbf{binary function} \begin{equation}\label{eq:ch4:binfun}f(x)=f_{0}(x),...,f_{m-1}(x)\in\{0,1\}^{m}
		\end{equation}
taking as input the $n$-cbit string $x\in\{0,1\}^{n}$ and producing as output the $m$-cbit string $f(x)$. Given a binary function $f(\cdot)$, the \textbf{controlled shift operator} is the unitary  transformation \begin{align}\label{eq:ch4:controlledshiftU}
	CX^{\otimes f} & = \sum_{x=0}^{2^{n}-1}|x\rangle\langle x|\otimes(X^{f_{0}(x)}\otimes\cdots\otimes X^{f_{m-1}(x)}).
\end{align} In a manner that directly extends the discussion about two-qubit controlled gates in Sec. \ref{sec:ch2:controlled}, if the first subset of qubits is in some state $|x\rangle$,  the effect of the unitary $CX^{\otimes f}$ is to apply the shift operator $X^{\otimes f(x)}$ to the second set of qubits. More generally, the unitary operates on $2^n$ separate computational branches, applying the shift operator $X^{\otimes f(x)}$ on the computational branch in which the first subset of $n$ qubits assumes state $|x\rangle$. Note that, unlike the simpler controlled qubit gates studied in Sec. \ref{sec:ch2:controlled}, here a different non-trivial transformation may be applied on all computational branches. 

The unitary $CX^{\otimes f}$ generally depends on all the values of the function $f(\cdot)$. Furthermore, it recovers the (two-qubit) CNOT gate by setting $m=n=1$ and $f(x)=x$ for $x\in\{0,1\}$. As such, unlike the shift operator $X^{\otimes a}$, which is separable across the qubits, the controlled shift applies jointly to all qubits, and can create entanglement (see Sec. \ref{sec:ch2:creating}).

\begin{figure}\begin{center}
			\begin{quantikz}
				\lstick[wires=2]{$\ket{x}$} &[2mm] \qw & \ctrl{2} & \qw & \qw & \qw \rstick[wires=2]{$\ket{x}$} \\
				&[2mm] \qw & \qw & \ctrl{2} & \qw & \qw \\
				\lstick[wires=2]{$\ket{y}$} & \qw & \gate{X} & \qw & \qw & \qw  \rstick[wires=2]{$\ket{y\oplus x}$}\\
				& \qw & \qw & \gate{X} & \qw & \qw  
			\end{quantikz}
	\end{center}\vspace{-0.4cm}\caption{An implementation of the controlled shift operator $CX^{\otimes f}$ with $m=n=2$ and $f(x)=x$.  \label{fig:ch4:Ufid}}
\end{figure}
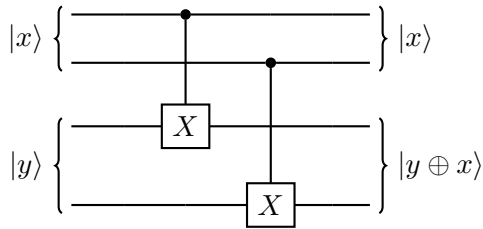

Let us now demonstrate an implementation of the controlled shift operator.	For $m=n=2$ and given the identity function $f(x)=x,$ the circuit implementing the unitary $CX^{\otimes f}$ is illustrated in Fig. \ref{fig:ch4:Ufid}. This implementation follows directly from the characterization in (\ref{eq:ch4:controlledshiftU}). In fact, as shown in the figure, when applied to vectors $|x,y\rangle$ in the computational basis, the two CNOT gates compute the XOR between the states of the first qubit and of the third and between the second and fourth qubits' states, producing the desired output $|x,y\oplus x\rangle$. 



\subsection{Computing a Binary Function}

In this subsection, we show how to use the controlled shift operator  $CX^{\otimes f}$ to evaluate the binary function $f(\cdot)$. As anticipated in Sec. \ref{sec:ch2:reversible}, quantum circuits can be used to implement reversible binary functions when the inputs are restricted to computational-basis vectors. Accordingly, we will show in this subsection that the controlled shift operator $CX^{\otimes f}$ implements function $f(\cdot)$ in a reversible manner on the computational basis.

\begin{figure}
\begin{center}
			\begin{quantikz}
				\lstick{$\ket{x}$} & \qwbundle{n} & \gate[wires=2]{CX^{\otimes f}} & \qwbundle{n}  \rstick{$\ket{x}$}\\
				\lstick{$\ket{y}$} & \qwbundle{m} & & \qwbundle{m}  \rstick{$\ket{y\oplus f(x)}$}
			\end{quantikz}
			\begin{quantikz}
				\lstick{$\ket{x}$} & \qwbundle{n} & \gate[wires=2]{CX^{\otimes f}} & \qwbundle{n}  \rstick{$\ket{x}$}\\
				\lstick{$\ket{0}$} & \qwbundle{m} & & \qwbundle{m}  \rstick{$\ket{f(x)}$}
			\end{quantikz}
		\end{center}
	\vspace{-0.4cm}\caption{(left) Illustration of the operation of the controlled shift operator $CX^{\otimes f}$ on an arbitrary vector $|x,y\rangle$ in the computational basis; (right) Illustration of the use of the controlled shift operator to compute function $f(x)$ for some input  $x\in\{0,1\}^{n}$, with the input $x$ being  encoded in the state of the first subsystem of $n$ qubits and the output $f(x)\in\{0,1\}^{m}$ being encoded in the state of the second subsystem of $m$ qubits. \label{fig:ch4:Uf}}\end{figure}
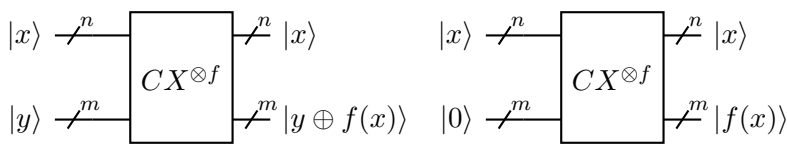

To this end, we first observe, in a manner similar to (\ref{eq:ch4:um}), that the unitary $CX^{\otimes f}$  maps any vector $|x,y\rangle$ in the computational basis to vector $|x,y\oplus f(x)\rangle$, where $\{|x\rangle\}_{x=0}^{2^{n}-1}$ and $\{|y\rangle\}_{y=0}^{2^{m}-1}$ are the computational
bases for the Hilbert spaces of the two subsystems, that is, we have \begin{equation}\label{eq:ch4:Uf}
	CX^{\otimes f} |x,y\rangle=|x,y\oplus f(x)\rangle.\end{equation} Accordingly, the controlled shift operator can be expressed in the change-of-basis form
\begin{align}\label{eq:ch4:Uf1}
	CX^{\otimes f} & =\sum_{x=0}^{2^{n}-1}\sum_{y=0}^{2^{m}-1}|x,y\oplus f(x)\rangle\langle x,y|,
	\end{align}as illustrated in Fig.  \ref{fig:ch4:Uf}-(left).  Importantly, the mapping  $(x,y) \mapsto (x,y \oplus f(x))$ implemented by the circuit is reversible for any binary function $f(\cdot)$.

 Based on this observation, the controlled shift operator can be used to compute function $f(x)$ using the approach depicted in Fig. \ref{fig:ch4:Uf}-(right). In it, one sets the first set of $n$ qubits to state $|x\rangle$ for the given bit string $x$ of interest. This can be done starting from the ground state $|0\rangle$ by applying an $X$ gate to all qubits corresponding to values 1 in the cbit string $x$. The second set of $m$ qubits is initialized in state $|0\rangle$. From (\ref{eq:ch4:Uf}), the output of the controlled shift operator is given by a separable state in which the first set of qubits is in state $|x\rangle$, while the second set of qubits is in state $|f(x)\rangle$. One can then measure the output qubits, i.e., the second set of qubits, in order to determine the $m$-cbit string $f(x)$ with  probability $1$.



	
		
\subsection{Quantum Random Access Memory (QRAM)}

	 As another important application of controlled shift operators, in this subsection we discuss the implementation of a \textbf{quantum random access memory (QRAM)}.  To start,  we view $x\in\{0,1,...,2^{n}-1\}$ as an index pointing to
	an entry $f(x)$ in a data base $\{f(0),f(1),....,f(2^{n}-1)\}$ of
	$m$-cbit strings. A classical RAM would return the entry $f(x)$ when queried at  index value $x$. In contrast, as we will show next,  querying a QRAM with an input in superposition retrieves a superposition
	of multiple, possibly all, entries in the data base. 
	
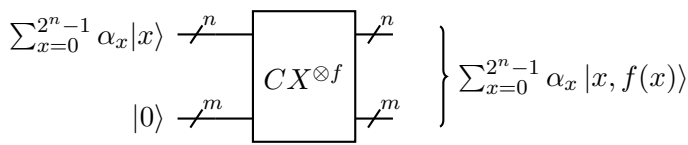
\begin{figure}
\begin{center}
		\begin{quantikz}
			\lstick{$\sum_{x=0}^{2^{n}-1}\alpha_x |x\rangle$} & \qwbundle{n} & \gate[wires=2]{CX^{\otimes f}} & \qwbundle{n} & \rstick[wires=2]{$\sum_{x=0}^{2^{n}-1}\alpha_x \ket{x,f(x)}$}\\
			\lstick{$\ket{0}$} & \qwbundle{m} & & \qwbundle{m}& 
		\end{quantikz}
\end{center}\vspace{-0.4cm}\caption{Querying a QRAM $CX^{\otimes f}$ with an input in superposition produces an entangled state encoding the  superposition
of all entries in the data base.  \label{fig:ch4:QRAM}}
\end{figure}

	A QRAM is defined by a  controlled shift gate $CX^{\otimes f}$ operating on $n+m$ qubits, with the first set of $n$ qubits encoding the query and the second set of $m$ qubits encoding the output of the QRAM. Assume that the input state for the first $n$ qubits is the superposition $\sum_{x=0}^{2^{n}-1}\alpha_x |x\rangle$, while the second subset of qubits are in the ground state $|0\rangle$. By (\ref{eq:ch4:Uf}), as illustrated in Fig. \ref{fig:ch4:QRAM}, the output state is then given by the superposition \begin{equation}\label{eq:ch4:supQRAM}
		\sum_{x=0}^{2^{n}-1}\alpha_x \ket{x,f(x)}.
	\end{equation}After the application of the QRAM gate, the qubits are in the entangled state (\ref{eq:ch4:supQRAM}) that encodes all input and output values corresponding to non-zero amplitudes $\alpha_x$. As a result, a single query to a QRAM can potentially encode information about \emph{all} entries in the data base.

Crucially, while the output state (\ref{eq:ch4:supQRAM}) depends on all the entries of the data base, measuring the output qubits would ``collapse'' the state to one of the states  $|x,f(x)\rangle$ with probability $|\alpha_x|^2$. Therefore, a measurement would produce the same result as a classical RAM fed with a random query $x$ with probability $|\alpha_x|^2$. We will discuss in the next section a (more useful!) application of the outlined parallelism of QRAMs.

\section{Deutsch's Problem and Quantum Parallelism}\label{sec:ch4:deutschprob}

The design of quantum algorithms revolves around the definition of mechanisms that leverage the unique quantum properties of superposition and entanglement. Deutsch's algorithm, the first demonstration of the possibility of a \textbf{quantum speed-up}, takes advantage of the property of a QRAM illustrated in Fig. \ref{fig:ch4:QRAM}. This property is an example of \textbf{quantum parallelism}: With a single application of the controlled shift operator $CX^{\otimes f}$, one can compute the superposition state (\ref{eq:ch4:supQRAM}), which can encode all the input-output values of function $f(\cdot)$.

To focus on Deutsch's problem, consider a binary function
		\begin{equation}
		f(x)\in\{0,1\}\text{ with \ensuremath{x\in\{0,1\}}}
		\end{equation}
		taking one bit as input and producing one bit as the output, i.e., having $n=1$ and $m=1$. There are only four such binary functions $f(\cdot)$, which are listed in Table \ref{table:ch4:binfun}. Suppose that we only wanted to know whether $f(\cdot)$ is constant -- i.e., if it is one of the functions
			($f_{0},f_{3}$) -- or not -- i.e., if it is one of the functions ($f_{1},f_{2}$). In a classical system, we would need to query the value of function $f(\cdot)$
			at both values of $x\in\{0,1\}$ to find out. By Deutsch's algorithm, quantum computing requires only one application of the function-computing
			operator $U_{f}$, demonstrating a quantum speed-up. How can this be accomplished?

		\begin{table}
		\begin{center}
			{\small{}}%
			\begin{tabular}{ccc}
				\toprule 
				& {\small{}$x=0$} & {\small{}$x=1$}\tabularnewline
				\midrule
				\midrule 
				{\small{}$f_{0}$} & {\small{}$0$} & {\small{}$0$}\tabularnewline
				\midrule 
				{\small{}$f_{1}$} & {\small{}$0$} & {\small{}$1$}\tabularnewline
				\midrule 
				{\small{}$f_{2}$} & {\small{}$1$} & {\small{}$0$}\tabularnewline
				\midrule 
				{\small{}$f_{3}$} & {\small{}$1$} & {\small{}$1$}\tabularnewline
				\bottomrule
			\end{tabular}{\small\par}
			\par\end{center}
		\vspace{-0.4cm} 
		\caption{The four possible binary functions with a single cbit as the input and a single cbit as the output ($m=n=1$). The table lists the outputs as a function of the input (column value). \label{table:ch4:binfun}}
		\end{table}

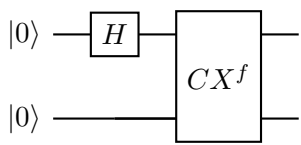
\begin{figure}
\begin{center}
		\begin{quantikz}
			\lstick{$\ket{0}$} & \gate{H} & \gate[wires=2]{CX^{f}} & \qw & \\
			\lstick{$\ket{0}$}  & \qw & \qw &  \qw
		\end{quantikz}
	\end{center}\vspace{-0.4cm}\caption{An application of  the QRAM circuit in Fig. \ref{fig:ch4:QRAM} for the special case $m=1$ and $n=1$. The Hadamard gate puts the first qubit in the superposition state $|+\rangle$. \label{fig:ch4:deutsch0}}
\end{figure}

	As discussed in the previous section,	with a single use of the controlled shift operator $CX^{f}$, we can compute the superposition of both input-output values of function $f(\cdot)$. Note that we have dropped the Kronecker product from the notation  $CX^{\otimes f}$ to indicate that we have a single controlled qubit in this example. Specifically, this can be done using the circuit in Fig. \ref{fig:ch4:deutsch0}, which specializes the QRAM circuit in Fig. \ref{fig:ch4:QRAM} for $m=1$ and $n=1$. Note that the use of the Hadamard gate on the first qubit produces  of the equal superposition $|+\rangle = 1/\sqrt{2}(|0\rangle + |1\rangle)$, corresponding to the case  $\alpha_x = 1/\sqrt{2}$ for $x=0,1$ in Fig. \ref{fig:ch4:QRAM}. By the general result summarized in Fig. \ref{fig:ch4:QRAM}, after the $CX^{f}$ gate in Fig. \ref{fig:ch4:deutsch0}, we have the entangled state	\begin{align}
		|\psi\rangle & =CX^{f}|+,0\rangle=\frac{1}{\sqrt{2}}\sum_{x=0}^{1}|x,f(x)\rangle.
	\end{align}

	While this state encodes all input-output values of the function $f(x)$, it is not clear how to extract useful information from it. In fact, if we measure the first  qubit in the computational basis, by the generalized Born rule (see Sec. \ref{sec:ch3:partial}), 
	we  end up with  the post-measurement state $|x,f(x)\rangle$, where $x=0$ or $x=1$ with equal probability $1/2$. 
	This amounts to observing the value $f(x)$ of the function at a single, randomly selected, input $x\in\{0,1\}$. 
	
	Therefore, we cannot learn anything more than a single value of function $f(x)$ from a single measurement of the superposition
	state $|\psi\rangle$.
	The hope is that, based on a single, properly designed, measurement of the superposition
	state $|\psi\rangle$, we can instead learn something about the \emph{relation} among values
	of $f(x)$ across different values of $x$. Deutsch's algorithm shows that indeed a single measurement is sufficient to determine whether function $f(\cdot)$ is constant or not.
		
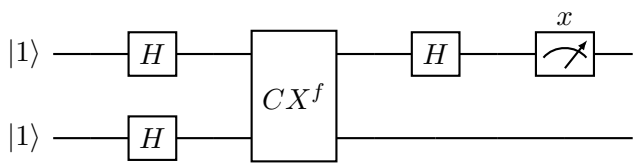
\begin{figure}
\begin{center}
		\begin{quantikz}
			\lstick{$\ket{1}$} & \qw & \gate{H} & \qw & \gate[wires=2]{CX^{f}} & \qw & \gate{H} & \qw & \meter{$x$} 
			& \qw \\
			\lstick{$\ket{1}$} & \qw & \gate{H} & \qw &  & \qw & \qw & \qw & \qw 
			&  \qw
			
		\end{quantikz}
\end{center}\vspace{-0.4cm}\caption{Deutsch's algorithm.\label{fig:ch4:deutsch1}}
\end{figure}

Deutsch's algorithm is described by the quantum circuit shown in Fig. \ref{fig:ch4:deutsch1}. In it, we can recognize two elements that we have encountered before. First, the top Hadamard gate has the same role explained in the context of Fig.  \ref{fig:ch4:QRAM} of probing the function $f(\cdot)$ via a superposition (here given by $H|1\rangle=|-\rangle=1/\sqrt{2}(|0\rangle-|1\rangle)$). Second, the cascade of Hadamard gate 
and measurement applied to the top qubit after the controlled shift operator implements a measurement in the diagonal basis $\{|+\rangle,|-\rangle\}$ (see Sec. \ref{sec:ch3:implany}).

We will demonstrate in Sec. \ref{sec:ch4:validityD} that, if $f(\cdot)$ is constant, i.e., if it is function $f_0$ or $f_3$ in Table \ref{table:ch4:binfun}, measuring the first qubit as in Fig. \ref{fig:ch4:deutsch1} gives $x=1$
with probability $1$; while, if $f(\cdot)$ is not constant, i.e., if it is function $f_1$ or $f_2$, 
 the measurement returns $x=0$ with probability $1$. Therefore, a single measurement is sufficient to settle the question of whether function $f(\cdot)$ is constant or not. To prove this result, which demonstrates a quantum speed-up for the problem at hand, we will introduce the key idea of phase kick-back, and return to Deutsch's algorithm in Sec. \ref{sec:ch4:validityD}.

\section{Phase Kick-Back}
In this section, we describe the key idea that underlies the design of Deutsch's algorithm: phase kick-back. The same principle underpins also other important quantum algorithms, such as phase estimation. As we will see in the next section, phase kick-back implies that the controlled shift operator in Deutsch's algorithm (see Fig. \ref{fig:ch4:deutsch1}) encodes the value of function $f(x)$ in the \emph{relative phase} of the \emph{first} qubit (rather than in the state of the second qubit as in Fig. \ref{fig:ch4:Uf}). This property will be the key to proving the validity of Deutsch's algorithm.

	To start, consider a controlled two-qubit gate. As described in Sec. \ref{sec:ch2:controlled}, a controlled-$U$ gate with qubit 0 as the control qubit and qubit 1 as the controlled qubit can be expressed as   
	\begin{equation}\label{eq:ch4:controlledU}
	C_{01}^{U}=|0\rangle\langle0|\otimes I+|1\rangle\langle1|\otimes U.
	\end{equation}  When applied on vectors of the computational basis, this operator acts with gate $U$ on qubit 1 if qubit 0 is in state $|0\rangle$; otherwise, if qubit 0 is in state $|1\rangle$, no change is applied to qubit 1. Note that, as in (\ref{eq:ch4:controlledshiftU}), the definition of a controlled gate can be generalized to gates that apply a non-trivial transformations to qubit 1 also in the computational branch in which qubit 0 is in state $|0\rangle$. In this section we will focus on controlled gates of the form (\ref{eq:ch4:controlledU}) for simplicity of explanation, and the general case will be considered in the next section.

The key observation in this section is the following. While it may seem that the control bit 0 cannot be changed by the application
of  the controlled gate (\ref{eq:ch4:controlledU}), this is not the case. In fact, the application of the two-qubit gate $C_{01}^{U}$
can change the relative phase of the amplitudes of the control qubit
-- a phenomenon known as \textbf{phase kick-back}.

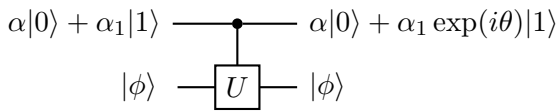
\begin{figure}
	\begin{center}
	\begin{quantikz} 
		\lstick{$\alpha|0\rangle+\alpha_1|1\rangle$} & \ctrl{1} &  \qw  \rstick{$\alpha|0\rangle+\alpha_1\exp(i\theta)|1\rangle$} \\ 
		\lstick{$\ket{\phi}$}\ & \gate{U} & \qw  \rstick{$\ket{\phi}$}
	\end{quantikz}
\end{center}\vspace{-0.4cm} \caption{The phase-kick phenomenon: When the state of the controlled qubit is an eigenvector $|\phi\rangle$ of the single-qubit gate $U$ with eigenvalue $\exp(i\theta)$, the net effect is that the state of the first qubit acquires a relative phase term $\exp(i\theta)$.\label{fig:ch4:kick}}
\end{figure}

	 
To proceed, define as $|\phi\rangle$ an eigenvector of the single-qubit gate $U$ with eigenvalue $\exp(i\theta)$.
Recall that all eigenvalues of unitary matrices have unitary magnitude (see Sec. \ref{sec:ch1:uniteig}). Consider now the operation of the controlled-$U$ gate on a separable state $|\psi,\phi\rangle$, where the state of the controlling qubit 0 is a generic superposition $|\psi\rangle=\alpha_0|0\rangle+\alpha_1|1\rangle$. This situation is illustrated in Fig. \ref{fig:ch4:kick}.  Accordingly, we have two branches of computation -- in the first the controlling qubit is in state $|0\rangle$, and in the second the controlling qubit is in state $|1\rangle$.

With the mentioned input $|\psi,\phi\rangle =\alpha_0|0,\phi\rangle+\alpha_1|1,\phi\rangle$, the state of the controlled qubit is unchanged in the first branch, since the controlling qubit is in state $|0\rangle$; and is merely multiplied by the  phase term $\exp(i\theta)$ in the second branch, in which the controlling qubit is in state $|1\rangle$.  As summarized in Fig. \ref{fig:ch4:kick} and derived next, this phase term can be equivalently thought of as being ``kicked
back'' to the first qubit, modifying the relative phase between the complex amplitudes of the computational basis states $|0\rangle$ and $|1\rangle$.

	To elaborate, let us evaluate the output of the circuit in Fig. \ref{fig:ch4:kick} as
	\begin{align}
		C_{01}^{U}(|\psi\rangle\otimes|\phi\rangle) & =(|0\rangle\langle0|\otimes I)(|\psi\rangle\otimes|\phi\rangle)+(|1\rangle\langle1|\otimes U)(|\psi\rangle\otimes|\phi\rangle)\nonumber \\
		& =\alpha_0|0\rangle|\phi\rangle+\alpha_1|1\rangle\exp(i\theta)|\phi\rangle\nonumber \\
		& =(\alpha_0|0\rangle+\alpha_1\exp(i\theta)|1\rangle)|\phi\rangle,
	\end{align} where in the first line we have used the definition (\ref{eq:ch4:controlledU}) of controlled-$U$ gate.  This calculation confirms that the net effect of the application of the controlled-$U$ gate is to leave the second qubit in the eigenstate $|\phi\rangle$ of the gate $U$, while changing the relative phase of the amplitudes of the first qubit.

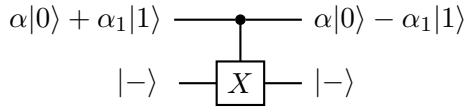
\begin{figure}
	\begin{center}
			\begin{quantikz} 
				\lstick{$\alpha|0\rangle+\alpha_1|1\rangle$} & \ctrl{1} &  \qw  \rstick{$\alpha|0\rangle -\alpha_1|1\rangle$} \\ 
				\lstick{$\ket{-}$}\ & \gate{X} & \qw  \rstick{$\ket{-}$}
			\end{quantikz}
	\end{center}\vspace{-0.4cm} \caption{The phase kick-back effect for the CNOT gate. \label{fig:ch4:CNOTkick}}
\end{figure}

	As an example, as depicted in Fig. \ref{fig:ch4:CNOTkick}, consider the CNOT gate as the controlled gate. The Pauli $X$ gate has eigenvectors $|+\rangle$ and $|-\rangle$ with respective eigenvalues $1$ and $-1$. Therefore, the phase kick-back effect is observed by setting the controlled qubit to state  $|-\rangle$, in which case the relative phase of the first qubit is modified by $\pi$.  

We conclude this section by noting that the phase kick-back effect may also lead to undesired effects in the presence of quantum errors (see problems).

\section{Validity of Deutsch's Algorithm}\label{sec:ch4:validityD}

	In this section, we show that Deutsch's algorithm, which is detailed in Fig. \ref{fig:ch4:deutsch1}, is valid, in the sense that, as explained in Sec. \ref{sec:ch4:deutschprob}, it returns the correct solution to the problem of identifying constant or non-constant binary functions with probability 1. We will see that this goal is accomplished by leveraging the phase kick-back effect described in the previous section to encode the value of the function $f(\cdot)$ in the relative phase of the first qubit in Fig. \ref{fig:ch4:deutsch1}.
	
	
	To start, we recall from Sec. \ref{sec:ch4:cso} that the controlled shift operator $CX^{f}$ is the controlled qubit gate (\ref{eq:ch4:controlledshiftU}), which can be specialized for $m=n=1$ as
	\begin{align}\label{eq:ch4:shiftc}
		CX^{f} & =|0\rangle\langle0|\otimes X^{f(0)}+|1\rangle\langle1|\otimes X^{f(1)}.
	\end{align} In words, for input states in the computational basis, the controlled shift operator applies the unitary $X^{f(x)}$ to the second qubit if the first qubit is in state $|x\rangle$ for $x\in\{0,1\}$. Note again that this form of controlled gate is more general than in (\ref{eq:ch4:controlledU}) in that a non-trivial gate is applied also when the controlling qubit is in state $|0\rangle$.



\begin{figure}
	\begin{center}
	\begin{quantikz}
	\lstick{$\ket{-}$}  & \qw & \ctrl{1} & \qw &  \meter{in $|\pm\rangle$ basis} 
	&\qw \\
	\lstick{$\ket{-}$}   & \qw & \gate{X^{f(x)}} & \qw & \qw 
	&  \qw \rstick{$\ket{-}$}
\end{quantikz}\end{center} \vspace{-0.4cm}\caption{ An equivalent description of the quantum circuit describing the Deutsch algorithm in Fig. \ref{fig:ch4:deutsch1}. Note that the gauge block here is used to implement a measurement in the diagonal basis. \label{fig:ch4:eqDeutsch}}
\end{figure}
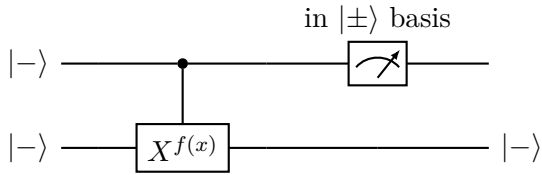

Now, recalling that the cascade of the Hadamard and standard measurement implements a measurement in the diagonal basis (see Sec. \ref{sec:ch3:implany}), we can represent the quantum circuit describing the Deutsch algorithm in Fig. \ref{fig:ch4:deutsch1} as in Fig. \ref{fig:ch4:eqDeutsch}. In this circuit, we have used a controlled gate representation for the controlled shift operator $CX^{f}$ in order to highlight the description (\ref{eq:ch4:shiftc}).

\begin{figure}
\begin{center}
		\begin{quantikz}
			\lstick{$\ket{-}$} & \gate{X^{f(x)}} & \qw  \rstick{$(-1)^{f(x)}|-\rangle$}
		\end{quantikz}
\end{center}\vspace{-0.4cm} \caption{The state $|-\rangle$ is an eigenvector of the unitary $U_{\oplus f(x)}=X^{f(x)}$ with eigenvalue $(-1)^{f(x)}$.\label{fig:ch4:eigD}}
\end{figure}

	The key observation at this point is that the state $|-\rangle$ is an eigenvector of the operator $X^{f(x)}$
	for both $x=0$ and $x=1$. In fact, we have the equalities 
	\begin{equation}
		X^{f(x)}|-\rangle=\frac{1}{\sqrt{2}}(|f(x)\rangle-|f(x)\oplus1\rangle)=(-1)^{f(x)}|-\rangle, 
	\end{equation}
	showing that the associated eigenvalue is $(-1)^{f(x)}$ (see Fig. \ref{fig:ch4:eigD}). It follows that, due to phase kick-back introduced in the previous section, after the controlled shift gate and prior to the measurement in the diagonal basis in the quantum circuit of Fig. \ref{fig:ch4:eqDeutsch}, we have the state
	\begin{align}
		CX^{f}|-,-\rangle & =\frac{1}{\sqrt{2}}CX^{f}(|0,-\rangle-|1,-\rangle)\nonumber\\
		& =\frac{1}{\sqrt{2}}(|0\rangle X^{f(0)}|-\rangle-|1\rangle X^{f(1)}|-\rangle)\nonumber\\
		& =\frac{1}{\sqrt{2}}((-1)^{f(0)}|0,-\rangle-(-1)^{f(1)}|1,-\rangle)\nonumber\\
		& =\frac{1}{\sqrt{2}}((-1)^{f(0)}|0\rangle-(-1)^{f(1)}|1\rangle)|-\rangle. 
	\end{align} 
Therefore, the state of the first qubit, $\frac{1}{\sqrt{2}}((-1)^{f(0)}|0\rangle-(-1)^{f(1)}|1\rangle),$
	is equal to \textrm{
		\begin{equation}
		|-\rangle\textrm{ (or \ensuremath{-|-\rangle)\textrm{ if the function is constant}}}
		\end{equation}
		and 
		\begin{equation}
		|+\rangle\textrm{ (or \ensuremath{-|+\rangle)\textrm{ if the function is not constant. }}}
		\end{equation}
	}Note that the negative sign does not change the quantum state, since quantum states are unaffected by global phases. 

The upshot of this calculation is that, as anticipated, the controlled shift operator in Fig. \ref{fig:ch4:eqDeutsch} encodes the values of the function in the relative phase of the state of the first qubit.

	 This completes the proof of validity of Deutsch's algorithm. In fact, measuring the first qubit in the diagonal basis as in Fig. \ref{fig:ch4:eqDeutsch} returns $1$ (corresponding to state $|-\rangle$) with probability $1$ if the function is constant; and it returns $0$ (corresponding to state $|+\rangle$) with probability $1$ if the function is not constant. Note that Deutsch's algorithm produces a deterministic output despite requiring a single run of the quantum circuit.

\section{No Cloning Theorem}

As discussed in Sec. \ref{sec:ch4:gatebased}, for many quantum algorithms, including the most common quantum machine learning methods to be discussed in Chapter 6,  the output of the algorithm is obtained by averaging the outputs of measurements across multiple runs of the circuit. One may wonder why it is indeed necessary to run the entire circuit anew when the goal is to measure again the same output quantum state produced by the circuit. Couldn't we simply run the circuit once, produce the quantum state, ``copy'' it to multiple quantum systems so that all systems are in the same state, and finally measure separately such systems? Unfortunately, this procedure is physically impossible to implement -- a result known as the no cloning theorem. 

\begin{figure}
	\begin{center}
		\begin{quantikz}
			\lstick{$\ket{\psi}$} &   \gate[wires=2]{U?} & \qw \rstick{$\ket{\psi}$}\\ \lstick{$\ket{0}$} &   & \qw \rstick{$\ket{\psi}$}
		\end{quantikz}
	\end{center}\vspace{-0.4cm}\caption{No, there is no unitary $U$ that can implement the cloning operation in the figure for any arbitrary quantum state $|\psi\rangle$.\label{fig:ch4:nocloning}}
\end{figure}
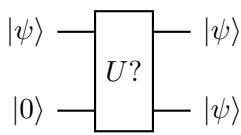

	 The \textbf{no cloning theorem} says that there is no unitary transformation
	that can copy an arbitrary, unknown, state $|\psi\rangle$ of $n$ qubits into another quantum system of $n$ qubits. Such a unitary would operate as shown in Fig. \ref{fig:ch4:nocloning}. Accordingly, given an arbitrary state $|\psi\rangle$ and a set of $n$ additional qubits -- typically referred to as \textbf{ancillas} -- initially in the ground state $|0\rangle$, the hypothetical cloning circuit would produce the separable state $|\psi,\psi\rangle=|\psi\rangle \otimes |\psi\rangle$. In it, both the original qubit and the ancillas are in the unknown state $|\psi\rangle$.

	 The proof of the non-existence of cloning unitaries is straightforward and follows directly from
	the linearity of the operator $U$. To see this, suppose that we could devise a unitary with the properties
		\begin{equation}
		U|\psi,0\rangle=|\psi,\psi\rangle
		\end{equation}
		and 
			\begin{equation}
		U|\psi',0\rangle=|\psi',\psi'\rangle
	\end{equation}
		for two distinct states $|\psi\rangle$ and $|\psi'\rangle$. If we applied such a unitary to the superposition state $\alpha|\psi\rangle+\beta|\psi'\rangle$, we would obtain
		\begin{align}
			U((\alpha|\psi\rangle+\beta|\psi'\rangle)\otimes|0\rangle) & =U(\alpha|\psi,0\rangle+\beta|\psi',0\rangle)\nonumber\\
			& =\alpha|\psi,\psi\rangle+\beta|\psi',\psi'\rangle,
		\end{align}
		which is different from the desired state $(\alpha|\psi\rangle+\beta|\psi'\rangle)\otimes(\alpha|\psi\rangle+\beta|\psi'\rangle)$ containing a clone of the input superposition  state. This yields a contradiction, and hence it concludes the proof.

	 The constraint on no cloning has several key implications, such as the following:\begin{itemize} \item \textbf{Quantum state accessibility}: One can generally only obtain partial
		information about a quantum state, since, once the state is measured, it is no longer
		accessible;
		\item \textbf{Quantum error correction}: Standard error correction schemes such
		as repetition coding are not directly applicable to fight quantum noise
		such as decoherence, and novel solutions are required (see Sec. \ref{sec:ch5:qec}); 
		\item \textbf{No superliminal communications}: Should quantum cloning be allowed,
		it would be possible to communicate instantaneously via a variant
		of superdense coding, violating relativity theory;
\item		\textbf{Quantum crypotography}: Unconditionally secure communication schemes
		can be designed since an eavesdropper cannot copy quantum information. Any unauthorized measurement of a quantum system would modify the state of the system, potentially revealing the presence of an eavesdropper.
	\end{itemize}

\section{Classical Cloning: Basis-Copying Gate}\label{sec:ch4:BC}

As we have seen in the previous section, the no cloning theorem states that one cannot design a quantum circuit that is able to copy \emph{any arbitrary} state. However, this does not rule out the possibility to design quantum circuits that copy \emph{specific} quantum states. A useful and standard example is a circuit that copies only 
		vectors in the computational basis. The fact that this type of transformation is feasible should be clear from the fact that we can, of course, copy
		classical (deterministic) cbits. We may hence refer to this transformation as a \textbf{classical cloning operator}.
		
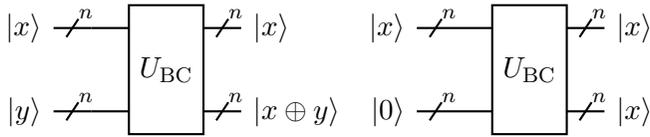
\begin{figure}
\begin{center}
		\begin{quantikz}
			\lstick{$\ket{x}$} &  \qwbundle{n} & \gate[wires=2]{U_{\textrm{BC}}} & \qwbundle{n} \rstick{$\ket{x}$}\\ \lstick{$\ket{y}$}  &  \qwbundle{n} &   & \qwbundle{n} \rstick{$\ket{x\oplus y}$}
		\end{quantikz}
	\begin{quantikz}
		\lstick{$\ket{x}$} &  \qwbundle{n} & \gate[wires=2]{U_{\textrm{BC}}} & \qwbundle{n} \rstick{$\ket{x}$}\\ \lstick{$\ket{0}$}  &  \qwbundle{n} &   & \qwbundle{n} \rstick{$\ket{x}$}
	\end{quantikz}
\end{center}\vspace{-0.4cm}\caption{The input-output operation of a basis-copying circuit\label{fig:ch4:basiscopying}}
\end{figure}

		The classical cloning operator, or \textbf{basis-copying unitary},  $U_{\textrm{BC}}$ operates as illustrated in Fig. \ref{fig:ch4:basiscopying}. To describe it, let us consider $n$ qubits of interest, along with $n$ ancilla qubits initially in the ground state $|0\rangle$. Given an input state $|x,0\rangle$, for any quantum state $|x\rangle$ in the computational basis, the basis-copying unitary should produce at the output the state $|x,x\rangle$ as shown in Fig. \ref{fig:ch4:basiscopying}-(right). 

	As can be observed by comparing Fig. \ref{fig:ch4:basiscopying}-(left) with Fig. \ref{fig:ch4:Uf}, the basis-copying circuit amounts to the controlled shift unitary
	$CX^{\otimes f}$ with function $f(x)=x$ being the identity matrix. Therefore, specializing (\ref{eq:ch4:controlledshiftU}), we can write the basis-copying unitary as
	\begin{align}
		U_{\textrm{BC}} = \sum_{x=0}^{2^{n}-1}|x\rangle\langle x|\otimes(X^{x_{0}}\otimes\cdots\otimes X^{x_{n-1}}).
	\end{align} 

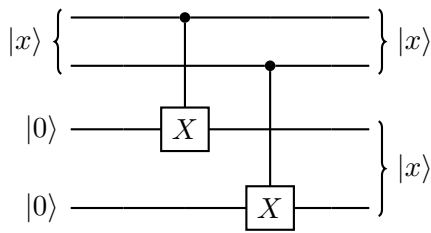
\begin{figure}
\begin{center}
			\begin{quantikz}
				\lstick[wires=2]{$\ket{x}$} &[2mm] \qw & \ctrl{2} & \qw & \qw & \qw \rstick[wires=2]{$\ket{x}$} \\
				&[2mm] \qw & \qw & \ctrl{2} & \qw & \qw \\
				\lstick{$\ket{0}$} & \qw & \gate{X} & \qw & \qw & \qw  \rstick[wires=2]{$\ket{x}$}\\
				\lstick{$\ket{0}$} & \qw & \qw & \gate{X} & \qw & \qw  
			\end{quantikz}
	\end{center}\vspace{-0.4cm}\caption{Implementation of the basis-copying unitary $U_{\textrm{BC}}$ with $n=2$.\label{fig:ch4:ubc}}
\end{figure}

Moreover, the basis-copying unitary can be implemented using 	CNOT gates as illustrated in Fig. \ref{fig:ch4:Ufid} for $n=2$. This circuit is reproduced in Fig. \ref{fig:ch4:ubc} with the input of the ancillas set to the ground state $|0\rangle$ to emphasize its operation as a basis-copying circuit. For any number $n$ of qubits, each of the $n$ top qubits acts as the controlling qubit of a CNOT gate that controls a distinct ancilla qubit. One can readily check that this circuit implements the transformation described in Fig. \ref{fig:ch4:basiscopying}-(left).

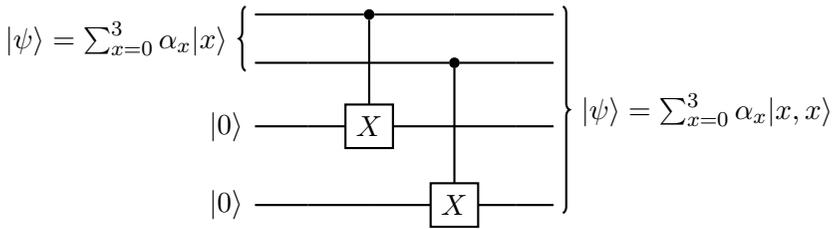
\begin{figure}
\begin{center}
		\begin{quantikz}
			\lstick[wires=2]{$|\psi\rangle=\sum_{x=0}^{3}\alpha_{x}|x\rangle$} &[2mm] \qw & \ctrl{2} & \qw & \qw & \qw \rstick[wires=4]{$|\psi\rangle=\sum_{x=0}^{3}\alpha_{x}|x,x\rangle$} \\
			&[2mm] \qw & \qw & \ctrl{2} & \qw & \qw \\
			\lstick{$\ket{0}$} & \qw & \gate{X} & \qw & \qw & \qw  \\
			\lstick{$\ket{0}$} & \qw & \qw & \gate{X} & \qw & \qw  
		\end{quantikz}
\end{center}\vspace{-0.4cm} \caption{The basis-copying unitary is not a general quantum cloning machine.\label{fig:ch4:basisno}}
\end{figure}

It is important to emphasize that the basis-copying circuit is not a general cloning circuit -- it cannot be by the no cloning theorem. To see this, consider the situation in Fig. \ref{fig:ch4:basisno}, in which the $n=2$ qubits are in a general superposition state  $|\psi\rangle=\sum_{x=0}^{3}\alpha_{x}|x\rangle$. The output of the basis cloning circuit is given by the generally entangled state $|\psi\rangle=\sum_{x=0}^{3}\alpha_{x}|x,x\rangle$, which is different from the ideal output $|\psi\rangle \otimes |\psi\rangle$ of a hypothetical cloning circuit.

\section{Conclusions}
In this chapter, we have introduced some key elements of traditional quantum computing, including the gate-based architecture, phase kick-back, and the constraints imposed by the no cloning theorem. In the next chapter, we will broaden the scope of the quantum primitives at our disposal by further generalizing quantum measurements.

\section{Recommended Resources}

Recommended references on quantum computing include \cite{lipton2021introduction,kaye2006introduction,nielsen10}. Hardware implementation aspects are discussed in \cite{bergou2021quantum,hidary2019quantum}.

\section{Problems}
\begin{enumerate}
\item  Provide a quantum circuit that implements a general controlled shift operator defined by any
	function $f(x)$ from $m$ to $n$ bits. {[}Assume that you have access
	to a classical data base of $2^{m}$ binary strings of $n$ bits defining
	function $f(x)$.{]}
\item Describe a situation in which a phase flip  occurring on a qubit causes a phase flip on another qubit when the two qubits interact via a CNOT gate. Recall that a phase flip error is described by the application of a Pauli $Z$ operator. [Hint: Use phase kick-back.]
\item  Describe the Deutsch-Jozsa algorithm that extends Deutsch algorithm
	to any $n$. Describe its operation in terms of phase kick-back (see
	recommended resources).
	\item Show that, for any function $f(x)$ from an $n$-cbit string $x$ to a single cbit ($m=1$), the $2^n \times 2^n$ matrix $U$ defined as $U|x\rangle=(-1)^{f(x)}|x\rangle$ is unitary. Relate this observation to Deutsch's algorithm.
\item Show that if cloning was possible, it would also be feasible
	to communicate information faster than light.
\end{enumerate}

\chapter{Generalizing Quantum Measurements (Part II)}

\section{Introduction}

The most general type of measurement studied so far consist of partial measurements in which a subset of qubits is measured in an arbitrary basis (see Sec. \ref{sec:ch3:multimeas}). In this chapter, we provide two successive extensions of such measurements. First, we introduce projective measurements, and, with it, the important concept of quantum observables. As an application of projective measurements, we also briefly discuss quantum error correction. Then, we further extend projective measurements to  positive operator-valued measurements (POVMs), which provide the most general form of quantum measurements. As an application of POVMs, we cover the problem of unambiguous state detection. Finally, the chapter derives quantum channels from non-selective measurements. Along with POVMs, quantum channels exhaust the set of all possible operations applicable on a quantum system.

	\section{Projective Measurements}\label{sec:ch5:projmeas}
	
	In this section, we introduce projective measurements.
	
	\subsection{Complete vs. Incomplete Measurements}
		 A von Neumann measurement of $n$ qubits in an arbitrary orthonormal basis $\{|v_{x}\rangle\}_{x=0}^{2^{n}-1}$
	(see Sec. \ref{sec:ch3:any}) is said to be \textbf{complete}, reflecting the fact that the post-measurement state is fully described by the  measurement output
			 $x\in \{0,1,...,2^{n}-1\}$. Indeed, when the measurement output is $x$, the post-measurement state is   given by vector $|v_{x}\rangle$.  Note that the post-measurement state $|v_{x}\rangle$ is known even when, as is usually the case, the pre-measurement state is not known. In contrast, partial measurements of a subset of qubits	(see Sec. \ref{sec:ch3:partial}) are said to be \textbf{incomplete}, since the post-measurement state is not fully specified by the measurement output. Rather, the state of the qubits that are not measured generally retains its dependence on the pre-measurement state.

		As we will introduce in this section, projective measurements provide a more general framework to define incomplete measurements. While a complete measurement necessarily causes a ``collapse'' of the waveform, unless the pre-measurement state is one of the vectors in the measurement basis $\{|v_{x}\rangle\}_{x=0}^{2^{n}-1}$, projective measurements allow for \textbf{``gentler'' measurements} that leave unchanged pre-measurement states belonging to specific (non-trivial) subspaces. As we will see, this property has useful practical implications.
		

\subsection{Projection Matrices}
	
	Projective measurements are defined by projection matrices.  
		 A \textbf{projection matrix} $\Pi$ is a (square) positive semidefinite matrix with eigenvalues
		equal to 0 or 1.  Specifically,	an $N\times N$ projection matrix $\Pi$ with rank  $r\in\{1,...,N\}$ has $r$ eigenvalues equal to $1$ and all other, $N-r$,  eigenvalues equal to $0$. Therefore, by the spectral theorem (see Sec. \ref{sec:ch1:uniteig}), it can be expressed as \begin{equation}\label{eq:ch5:eigP}
			\Pi=\sum_{x=0}^{r-1} |v_x\rangle \langle v_x|,		\end{equation} where $\{|v_x\rangle\}_{x=0}^{r-1}$ are the eigenvectors associated with eigenvalue $1$. Note that the eigendecomposition (\ref{eq:ch5:eigP}) is not unique, in the sense that any other orthonormal basis of the $r$-dimensional subspace spanned by vectors $\{|v_x\rangle \}_{x=0}^{r-1}$ can be used in (\ref{eq:ch5:eigP}) in lieu of vectors $\{|v_x\rangle\}_{x=0}^{r-1}$.

		 Geometrically, the operation $\Pi |a\rangle$ for some $N \times 1$ state vector $|a\rangle$ corresponds to the \textbf{orthogonal projection} of vector $|a\rangle$ onto the subspace spanned by the $r$ eigenvectors $\{|v_x\rangle \}_{x=0}^{r-1}$ of matrix  $\Pi$ that are associated with eigenvalues equal to $1$. We will refer to 
		this  $r$-dimensional subspace as the \textbf{subspace spanned by projection matrix $\Pi$}.	Note that, by this geometric interpretation, unless all $N$ eigenvalues are equal to $1$, a projection matrix $\Pi$ is not unitary, since it does not preserve the norm of an arbitrary input vector.
		
		In the decomposition (\ref{eq:ch5:eigP}) an important role is played by \textbf{rank-1 projection matrices} \begin{equation}\label{eq:ch5:rank1}
		v_x=	|v_x\rangle \langle v_x|.
		\end{equation} Geometrically, the effect of a rank-1 projection matrix $v_x$ on an input vector $|a\rangle$ is to compute the projection $v_x |a\rangle$ of the latter in the direction specified by vector $|v_x\rangle$.

			\subsection{Reviewing von Neumann Measurements}
			
			For reference, we review here von Neumann measurements by highlighting the role of rank-1 projection matrices in the definition of measurement probabilities and post-measurement states.
			
			As discussed in Sec. \ref{sec:ch3:densityform}, a von Neumann measurement
is specified by an orthonormal
	basis $\{|v_{0}\rangle,|v_{1}\rangle,...,|v_{N-1}\rangle\}$ of $N=2^{n}$ vectors,
	each of dimension $N\times1$. According to Born's rule, the probability of observing output  $x\in\{0,1,...,N-1\}$
		is
		\begin{align}
			\Pr[\textrm{measurement output equals }x] & =\langle v_{x}|\rho|v_{x}\rangle \nonumber\\ &=\textrm{tr}(|v_{x}\rangle\langle v_{x}|\rho)=\textrm{tr}(v_{x}\rho),
		\end{align}
		where we have used the definition (\ref{eq:ch5:rank1}) of rank-1 projection matrices $v_x$ with  $x\in\{0,1,...,N-1\}$.	Furthermore, the post-measurement state, expressed as a density matrix,
		is 
		\begin{equation}\label{eq:ch5:postm}
		|v_{x}\rangle\langle v_{x}|=\frac{v_{x}\rho v_{x}}{\langle v_{x}|\rho|v_{x}\rangle},
		\end{equation}
		 as it can be proved by noting that we can always write the density matrix $\rho$ as 
		\begin{equation}
		\rho=\sum_{x=0}^{N-1}\sum_{x'=0}^{N-1}\lambda_{x,x'}|v_{x}\rangle\langle v_{x'}|,
		\end{equation}
		with $\lambda_{x,x'}=\langle v_{x}|\rho|v_{x'}\rangle$.

		 From the derivations in the previous paragraph, von Neumann measurements are defined by \emph{rank-1} projection matrices
		$v_{x}=|v_{x}\rangle\langle v_{x}|$ for $x\in\{0,1,...,N-1\}$. Importantly, these matrices satisfy the resolution-
		of-identity condition 
		\begin{equation}\label{eq:ch5:resid}
		\sum_{x=0}^{N-1}v_{x}=I,
		\end{equation}where $I$ is the $N\times N$ identity matrix. 
		 To see that the equality (\ref{eq:ch5:resid}) holds, it is sufficient to note that we
		have the equality 
		\begin{equation}
		\left(\sum_{x=0}^{N-1}v_{x}\right)|\psi\rangle=|\psi\rangle
		\end{equation}
		for any input state $|\psi\rangle$, since the latter can always be expressed 
		as the superposition $|\psi\rangle=\sum_{x=0}^{N-1}\alpha_{x}|v_{x}\rangle$
		and we have 
		\begin{equation}
		v_{x}|v_{x'}\rangle=|v_{x}\rangle\langle v_{x}|v_{x'}\rangle,
		\end{equation}
		with $\langle v_{x}|v_{x'}\rangle=0$ for $x\neq x'$ and $\langle v_{x}|v_{x}\rangle=1$.

	\subsection{Defining Projective Measurements}\label{sec:ch5:defproj}
	
		Generalizing the description given in the previous subsection of von Neumann measurements, projective measurements are defined as follows. A projective measurement is specified by a set 
		of $N'\leq N$ projection matrices $\{\Pi_{y}\}_{y=0}^{N'-1}$, with rank possibly
		larger than 1 (when $N'<N$), such that the resolution-of-identity
		condition 
		\begin{equation}\label{eq:ch5:residg}
		\sum_{y=0}^{N'-1}\Pi_{y}=I
		\end{equation}
		is satisfied. Geometrically, condition (\ref{eq:ch5:residg}) indicates that the subspace spanned by the union of the eigenvectors of all projection matrices $\{\Pi_{y}\}_{y=0}^{N'-1}$ equals the entire $N$-dimensional Hilbert space.
		
		 A projective measurement defined by projection matrices $\{\Pi_{y}\}_{y=0}^{N'-1}$ satisfying the resolution-of-identity condition (\ref{eq:ch5:residg}) operates on an input quantum state $\rho$ as follows. \begin{itemize}
		 	\item \textbf{Born's rule}: The probability of observing $y\in\{0,1,...,N'-1\}$ is
		\begin{align}\label{eq:ch5:Born}
			\Pr[\textrm{measurement output equals }y]=\textrm{tr}(\Pi_{y}\rho),
		\end{align}which, for a pure state  $\rho=|\psi\rangle\langle\psi|$ simplifies as \begin{align}\label{eq:ch5:Bornpure}
		\Pr[\textrm{measurement output equals  }y]=\textrm{tr}(\Pi_{y}\rho)=\langle\psi|\Pi_{y}|\psi\rangle;
	\end{align}
	\item \textbf{``Incomplete collapse'' of the state}:	The post-measurement state is
		\begin{equation}\label{eq:ch5:postmeas}
		\frac{\Pi_{y}\rho\Pi_{y}}{\textrm{tr}(\Pi_{y}\rho)}, 
		\end{equation}which, for a pure state  $\rho=|\psi\rangle\langle\psi|$ simplifies as\begin{equation}\label{eq:ch5:postmeaspure}
		\frac{\Pi_{y}|\psi\rangle}{\sqrt{\langle\psi|\Pi_{y}|\psi\rangle}}.
	\end{equation}
\end{itemize}

With rank-1 projection matrices, these rules reduce to the von Neumann
measurement reviewed in the previous subsection, which are complete.	More generally, by the post-measurement state (\ref{eq:ch5:postmeaspure}), projective measurements are incomplete.  To see this, assume that projection matrix $\Pi_y$ has rank larger than $1$ and that the input state $|\psi\rangle$ is an unknown coherent state. Then, upon measuring output $y$, the post-measurement state (\ref{eq:ch5:postmeaspure}) can be generally only determined as being any state vector in the subspace spanned by projection matrix $\Pi_y$.

A related observation concerns the pre-measurement states that are left \emph{unchanged} by a projective measurement. A von Neumann measurement in basis $\{|v_{0}\rangle,|v_{1}\rangle,...,|v_{N-1}\rangle\}$
leaves a pre-measurement state $|\psi\rangle$ unchanged if and only if it is one of the vectors $|v_{x}\rangle$
in the basis, i.e., if $|\psi\rangle=|v_{x}\rangle$ for some $x\in\{0,1,...,N-1\}$. In contrast,  a projective measurement leaves a state $|\psi\rangle$ unchanged if
it is \emph{any} vector  in the subspace spanned by one of the projection matrices 
$\Pi_{y}$. This is because, in this case, the projective measurement returns $y$ with probability $1$, and the post-measurement state (\ref{eq:ch5:postmeaspure}) equals the pre-measurement state. This property is at the core of the application of projective measurements for quantum error correction (see Sec. \ref{sec:ch5:qec}).

\subsection{Projective Measurements Generalize von Neumann Measurements and Partial Measurements}\label{sec:ch5:long}

As mentioned at the beginning of this chapter, projective measurements include the most general type of measurement studied prior to this chapter, i.e., partial measurements in an arbitrary basis, which were detailed in Sec. \ref{sec:ch3:multimeas}. 

To see this, consider as in Sec. \ref{sec:ch3:multimeas} systems
of  $n$ qubits in which we measure any subset of $n'\leq n$ qubits in
any orthonormal basis $\{|v_{y}\rangle\}_{y=0}^{2^{n'}-1}$. Without loss of generality, we order the qubits so that the first $n'$ qubits are measured. Now, define the projection matrices \begin{equation}\label{eq:ch5:partial}
	\Pi_y = v_y \otimes I,
	\end{equation}where $v_y=|v_y\rangle \langle v_y|$ is a rank-1 projection matrix and the identity matrix is of size $2^{n-n'}\times2^{n-n'}$. It can be now readily checked that, for any given pre-measurement state $|\psi\rangle$, the generalized Born rule (\ref{eq:ch3:genBornproj}) and the post-measurement state (\ref{eq:ch3:collgen}) are obtained from the projective measurement rules (\ref{eq:ch5:Bornpure}) and (\ref{eq:ch5:postmeaspure}), respectively.

	\subsection{Local vs. Joint Measurements}

As discussed in Sec. \ref{sec:ch3:any}, some measurements, like standard measurements, can be carried out separately on each qubit, while others require joint measurements across multiple qubits. The formalism of projective measurements makes it easy to distinguish these two situations. 

To elaborate, consider a quantum system in which we identify a number of subsystems, e.g., the individual qubits. A local measurement across the subsystems is specified by projection matrices that can be expressed as the Kronecker product of individual projection matrices, one for each of the subsystems. 

An example of a local measurement is given by the partial measurements studied in the previous subsection. In fact, the projection matrix in  (\ref{eq:ch5:partial}) is written as the Kronecker product $v_y \otimes I$ of a  projection matrix, $v_y$, operating on the subsystem being measured and of the degenerate projection matrix given by the identity matrix $I$. The latter indicates that no measurement is carried out on the second subset of qubits.
	
		More generally, when the projection matrices cannot be expressed in Kronecker product form, the measurement needs to be carried out jointly across the involved subsystems.

		\subsection{How to Construct a Projective Measurement}\label{eq:ch5:howto}
	
		 The projection matrices $\{\Pi_{y}\}_{y=0}^{N'-1}$ defining a projective
		measurement can be always obtained from some orthonormal basis $\{|v_{0}\rangle,|v_{1}\rangle,...,|v_{N-1}\rangle\}$, with $N=2^n$, 
		by: (\emph{i}) partitioning vectors $\{|v_{0}\rangle,|v_{1}\rangle,...,|v_{N-1}\rangle\}$ into $N'$ subsets; and (\emph{ii})  defining $\Pi_{y}$ as the projection matrix onto the subspace spanned by the $y$-th subset of basis vectors. 
		
		More formally, in order to define a projective measurement, let us fix:\begin{itemize}
			\item an orthonormal basis $\{|v_{0}\rangle,|v_{1}\rangle,...,|v_{N-1}\rangle\}$; 
			\item and a partition of the index set $\{0,1,...,N-1\}$ into $N'$ disjoint subsets  $\mathcal{X}_{y}\subseteq \{0,1,...,N-1\}$ with $y\in\{0,1,...,N'-1\}$, such that their union equals the overall set $\{0,1,...,N-1\}$. 
		\end{itemize} Then, each matrix $\Pi_{y}$ is defined as the projection matrix onto the
		subspace spanned by the $y$-th subset of vectors, i.e., 
		\begin{equation}
		\Pi_{y}=\sum_{x\in\mathcal{X}_{y}}|v_{x}\rangle\langle v_{x}|.
		\end{equation} If each set  $\mathcal{X}_{y}$ contains a single vector, and hence all projection matrices have rank 1, this construction recovers the von Neumann measurement in the orthonormal basis $\{|v_{0}\rangle,|v_{1}\rangle,...,|v_{N-1}\rangle\}$.

\subsection{An Example: Parity Measurements}\label{sec:ch5:paritymeas}
	
		 To illustrate the operation of a projective measurement, let us consider  $n=2$ qubits (i.e., $N=4$) and fix the computational
		basis $\{|00\rangle$,$|01\rangle$, $|10\rangle$, $|11\rangle\}.$ A projective measurement with $N'=2$ possible outcomes can be defined by  partitioning these vectors into \textbf{even-parity states} $\{|00\rangle,|11\rangle \}$, for which the number of ones in the defining cbit string is an even number; 
		and \textbf{odd-parity states} $\{|01\rangle,|10\rangle \}$ in which the number of ones is odd. Note that this corresponds to the subsets  $\mathcal{X}_{0}=\{00,11\}$ and $\mathcal{X}_{1}=\{01,10\}$. 
		
		Accordingly,  the \textbf{parity projective measurement} is defined by the rank-2 projection matrices
		\begin{equation}\label{eq:ch5:projmat0}
		\Pi_{0}=|00\rangle\langle00|+|11\rangle\langle11|
		\end{equation} and 
		\begin{equation}\label{eq:ch5:projmat1}
		\Pi_{1}=|01\rangle\langle01|+|10\rangle\langle10|.
		\end{equation} Note that this is a \textbf{joint} measurement since the projection matrices $\Pi_0$ and $\Pi_1$ cannot be expressed as the Kronecker products of $2\times 2$ projection matrices.
	
Let us now study the effect of the parity projective measurements on two qubits in an arbitrary pure state
			\begin{equation}\label{eq:ch5:genstate}
			|\psi\rangle=\alpha_{00}|00\rangle+\alpha_{01}|01\rangle+\alpha_{10}|10\rangle+\alpha_{11}|11\rangle.
			\end{equation} By the Born rule (\ref{eq:ch5:Bornpure}), 
			the output of the measurement is $y=0$ with probability
			\begin{align}\label{eq:ch5:parityp1}
				\langle\psi|\Pi_{0}|\psi\rangle=|\alpha_{00}|^{2}+|\alpha_{11}|^{2},
			\end{align} and $y=1$ with probability 	
		\begin{align}\label{eq:ch5:parityp2}
		\langle\psi|\Pi_{1}|\psi\rangle=|\alpha_{01}|^{2}+|\alpha_{10}|^{2}.
	\end{align}
		Furthermore,	the post-measurement state (\ref{eq:ch5:postmeaspure}) is 
			\begin{equation}\label{eq:ch5:postex}
			\frac{\Pi_{0}|\psi\rangle}{\sqrt{\langle\psi|\Pi_{0}|\psi\rangle}}=\frac{1}{\sqrt{|\alpha_{00}|^{2}+|\alpha_{11}|^{2}}}(\alpha_{00}|00\rangle+\alpha_{11}|11\rangle)
			\end{equation} when the measurement output is $y=0$; while it is \begin{equation}\label{eq:ch5:postex1}
			\frac{\Pi_{1}|\psi\rangle}{\sqrt{\langle\psi|\Pi_{1}|\psi\rangle}}=\frac{1}{\sqrt{|\alpha_{01}|^{2}+|\alpha_{10}|^{2}}}(\alpha_{01}|01\rangle+\alpha_{10}|10\rangle)
		\end{equation} when the measurement output is $y=1$.

		Therefore, if the measurement output is $y\in\{0,1\}$, the post-measurement state (\ref{eq:ch5:postex})  is in the subspace spanned
			by $\Pi_{y}$, i.e., by the even- or odd-parity computational-basis vectors. Apart from this information,
			nothing else can be inferred from the output of the measurement, unless one knows  the amplitudes $\alpha_{00},$ $\alpha_{01}$,  $\alpha_{10}$, and  $\alpha_{11}$ of the pre-measurement state. This makes the measurement incomplete. 
		  Note the difference with a von Neumann measurement
			in the computational basis, in which the post-measurement state is
			the computational-basis vector $|x_{0},x_{1}\rangle$ corresponding to the measurement's
			output $(x_{0},x_{1})\in\{0,1\}^2$.
	
Consider now the situation in which the measured  state $|\psi\rangle$ is in the subspace spanned by $\Pi_0$, and hence it can be written as  $|\psi\rangle=\alpha_{00}|00\rangle+\alpha_{11}|11\rangle.$ Since this state is the superposition of even-parity computational basis vectors, it is also said to have an \textbf{even parity}. When applied to such an even-parity state, the
	parity measurement produces output $y=0$ with probability 1, and
	the post-measurement state equals the pre-measurement state. 
	Similarly, if the state $|\psi\rangle$ has an \textbf{odd parity}, i.e., if it is given as $|\psi\rangle=\alpha_{01}|01\rangle+\alpha_{10}|10\rangle,$ lying in the subspace spanned by projection matrix $\Pi_1$, the
	parity measurement produces output $y=1$ with probability 1, and
	the post-measurement state coincides with the pre-measurement state.
	
		We finally observe that the projection matrices defining the parity projective measurement can be also written  in terms of orthonormal bases other than the computational basis. This is because the have rank larger than $1$. As a  useful example, it can be directly checked that the projection matrices in (\ref{eq:ch5:eigZ2}) can be expressed in terms of the Bell basis (\ref{eq:ch2:Bell}) as	\begin{equation}
		\Pi_{0}=|\Phi^{+}\rangle\langle\Phi^{+}|+|\Phi^{-}\rangle\langle \Phi^{-}|
	\end{equation} and 
	\begin{equation}
		\Pi_{1}=|\Psi^{+}\rangle\langle\Psi^{+}|+|\Psi^{-}\rangle\langle \Psi^{-}|.
	\end{equation}

	\section{Observables}\label{sec:ch5:obs}

	Observables model numerical quantities that can be extracted from a quantum system through measurements. While the measurements presented so far always return cbit strings, observables allow the description of measurements that output real  numerical values. In this section, we define observables and their expectations; provide examples of single- and multi-qubit observables; and finally introduce the concept of compatible observables.
		
		\subsection{Defining Observables}
		
		An observable $O$ assigns a numerical, real, value $o_{y}$ to each
		output $y$ of a projective measurement $\{\Pi_{y}\}_{y=0}^{N'-1}$. Therefore, by the Born rule (\ref{eq:ch5:Born}), measuring the observable $O$
	 returns the numerical value $o_{y}$ with probability $\textrm{tr}(\Pi_{y}\rho)$ 
		when the input quantum state is $\rho$. We can express this defining property as the probability \begin{equation}\label{eq:ch5:probobs}  \Pr[\textrm{measurement of $O$ equals  }o_y]=\textrm{tr}(\Pi_{y}\rho),
		\end{equation} for $y\in\{0,1,...,N'-1\}$, which reduces to \begin{equation}\label{eq:ch5:probobsA}  \Pr[\textrm{measurement of $O$ equals  }o_y]=\langle\psi|\Pi_{y}|\psi\rangle
	\end{equation}for a pure pre-measurement state $|\psi\rangle$. 

Note that, when using the formalism of observables in quantum computing, one is often interested only  in the numerical value produced by the measurement and not in the post-measurement state. We will take this viewpoint here, although one can readily define also the post-measurement state via (\ref{eq:ch5:postmeas}) given the underlying projective measurement $\{\Pi_{y}\}_{y=0}^{N'-1}$.

		The output of the measurement of an observable is a random variable with distribution (\ref{eq:ch5:probobs}).  
		An important role in many quantum algorithms is played by the \textbf{expectation of an observable} (see Sec. \ref{sec:ch4:gatebased}). By (\ref{eq:ch5:probobs}), given an input quantum state
		$\rho$, the expectation of observable $O$, denoted as $\langle O\rangle_{\rho}$, can be computed as 
		\begin{equation}\label{eq:ch5:expobs}
		\langle O\rangle_{\rho}=\sum_{y=0}^{N'-1}o_{y}\textrm{tr}(\Pi_{y}\rho)=\textrm{tr}(O\rho),
		\end{equation}
		where we have defined the matrix
		\begin{equation}\label{eq:ch5:expobsmat}
		O=\sum_{y=0}^{N'-1}o_{y}\Pi_{y}.
		\end{equation} Matrix $O$ is Hermitian (since  projection matrices are Hermitian);
		but not necessarily positive semidefinite, since the values $o_{y}$
		can be negative.  	
		
		Specializing (\ref{eq:ch5:expobs}) to a pure input quantum state $|\psi\rangle$, the expectation of  observable $O$, denoted as $\langle O\rangle_{|\psi\rangle}$, can be written
		as 
		\begin{equation}\label{eq:ch5:expobspure}
			\langle O\rangle_{|\psi\rangle}=\textrm{tr}(O|\psi\rangle\langle\psi|)=\langle\psi|O|\psi\rangle.
		\end{equation}
		
		
%
		
		The results derived in the last two paragraphs highlight the important role played by matrix $O$ in the formalism of observables. We now further elaborate on this point by demonstrating that \emph{any} Hermitian matrix $O$ fully describes an observable in terms of both the probabilities (\ref{eq:ch5:probobs}) of all possible outputs $\{o_y\}_{y=0}^{N'-1}$ and of the expectation (\ref{eq:ch5:expobs}). Accordingly, we will conclude that an observable $O$ can be specified by an Hermitian matrix $O$, justifying the use of the same notation $O$ for both concepts.
				
		 To proceed, let us fix an arbitrary Hermitian matrix $O$. By the spectral theorem, matrix $O$ has eigendecomposition
		\begin{equation}\label{eq:ch5:obseig}
		O=\sum_{x=0}^{N-1}\lambda_{x}|v_{x}\rangle\langle v_{x}|,
		\end{equation}
		where the eigenvectors $\{|v_{0}\rangle,|v_{1}\rangle,...,|v_{N-1}\rangle\}$ form an
		orthonormal basis and the associated eigenvalues $\{\lambda_{0},\lambda_{1},...,\lambda_{N-1}\}$
		are real. The eigenvectors  $\{|v_{0}\rangle,|v_{1}\rangle,...,|v_{N-1}\rangle\}$  define the \textbf{eigenbasis} of observable $O$. 
		
		To see how to use the eigendecomposition (\ref{eq:ch5:obseig}) to define an observable, let us now write as $o_{y}$, with $y\in\{0,1,...,N'-1\}$, the $N'$ distinct  eigenvalues. Note that we have the inequality $N' \leq N$. The projective measurement associated with observable $O$ is then defined by the projection matrices \begin{equation}\label{eq:ch5:decproj}\Pi_{y}=\sum_{x:\textrm{ }\lambda_{x}=o_{y}}|v_{x}\rangle\langle v_{x}|
		\end{equation} for $y\in\{0,1,...,N'-1\}$. The sum in (\ref{eq:ch5:decproj}) is over all eigenvectors corresponding to the same eigenvalue $o_y$. That is, each matrix $\Pi_y$ projects onto the subspace spanned by all the eigenvectors with eigenvalues equal to $o_y$. With this choice, the observable (\ref{eq:ch5:obseig}) can be written as in (\ref{eq:ch5:expobsmat}).
	
	The derivation outlined in the previous paragraph shows that, given a Hermitian matrix $O$, one can identify the possible numerical outputs of the observable $O$ as the eigenvalues of matrix $O$. Furthermore, the subspaces spanned by the corresponding eigenvectors associated with each distinct eigenvalue define the projection matrices of the underlying projective measurement. With this information, based on Hermitian matrix $O$, the observable $O$ is fully specified in terms of probabilities (\ref{eq:ch5:probobs}) and expectation (\ref{eq:ch5:expobs}). 
	

	\subsection{Single-Qubit Observables}
	In this subsection, we study two examples of observables for single qubits $(n=1)$. The first is the \textbf{$Z$ observable}, whose defining Hermitian matrix is given by the Pauli $Z$ matrix. For this matrix, the eigendecomposition (\ref{eq:ch5:obseig}) specializes to 
		\begin{equation}\label{eq:ch5:Zeig}
		Z=|0\rangle\langle0|-|1\rangle\langle1|, 
		\end{equation}and hence the computational basis is the eigenbasis of the $Z$ observable.
		  By the discussion in the previous subsection, measuring the $Z$ observable corresponds to a standard von Neumann measurement in the
		computational basis, in which the measurement output 0 is associated
		the numerical value $o_0=1$ and the measurement output 1 is associated the numerical value 
		$o_1=-1$. Accordingly, for any pure state $|\psi\rangle=\alpha_{0}|0\rangle+\alpha_{1}|1\rangle$, by (\ref{eq:ch5:expobspure}), 
		we have the expectation
		\begin{equation}
		\langle Z\rangle_{|\psi\rangle}=\langle\psi|Z|\psi\rangle=(+1)|\langle0|\psi\rangle|^{2}+(-1)|\langle1|\psi\rangle|^{2}=|\alpha_{0}|^{2}-|\alpha_{1}|^{2}.
		\end{equation}

		 Consider now the \textbf{$X$ observable}. 
		 This is also a valid observable because the Pauli matrix $X$ is Hermitian, and it can be expressed
		with its eigendecomposition as
		\begin{equation}
		X=|+\rangle\langle+|-|-\rangle\langle-|.
		\end{equation}
		 This corresponds to a measurement in the diagonal, $\{|+\rangle,|-\rangle\}$, 
		basis -- the eigenbasis of observable $X$ -- in which the measurement output 0 is associated with numerical value 
		$o_0=1$ and the measurement output 1 is associated with $o_1=-1$.
	 	Moreover, for any pure state $|\psi\rangle=\alpha_{0}|0\rangle+\alpha_{1}|1\rangle$,
		we have the expectation
		\begin{align}
			\langle X\rangle_{|\psi\rangle} & =\langle\psi|X|\psi\rangle=(+1)|\langle+|\psi\rangle|^{2}+(-1)|\langle-|\psi\rangle|^{2}\nonumber\\
			& =\frac{|\alpha_{0}+\alpha_{1}|^{2}}{2}-\frac{|\alpha_{0}-\alpha_{1}|^{2}}{2}.
		\end{align}
	
	\subsection{Local Multi-Qubit Observables}\label{sec:ch5:multiqobs}

		Let us now consider observables on any number $n$ of qubits. We start by studying the important class of observables defined by Pauli strings. These provide an important example of local observables, whose numerical outputs can be produced via separate measurements applied on each qubit.
		
		 Using the notation from Sec. \ref{sec:ch2:Paulibasis}, let us write as $P_s$ with $s\in\{0,1,2,3\}$ a Pauli matrix, with \begin{equation}
			P_{0}=I,\textrm{ \ensuremath{P_{1}=X,\textrm{ \ensuremath{P_{2}=Y,\textrm{ and \ensuremath{P_{3}=Z}}}}}}.
		\end{equation} Furthermore, let $s^{n}=[s_0,...,s_{n-1}]$ be a vector of integers $s_k\in\{0,1,2,3\}$ with $k\in\{0,1,...,n-1\}$.  A \textbf{string of $n$ Pauli matrices}, $P_{s^{n}}$, is given by the Kronecker product of $n$ Pauli matrices \begin{equation}\label{eq:ch5:Paulistring}P_{s^{n}}=P_{s_0} \otimes P_{s_2}  \otimes \cdots \otimes P_{s_{n-1}},
		\end{equation} where each matrix $P_{s_k}$ is one of the Pauli matrices $\{I,X,Y,Z\}$. It can be readily checked that any Pauli string (\ref{eq:ch5:Paulistring}) is a valid observable.
		

		To illustrate the process of measuring local  observables such as a Pauli string, we study the following example. 	Consider $n=2$ qubits and the observable given by the Pauli string $O=Z\otimes Z$. 
		By using the eigendecomposition (\ref{eq:ch5:Zeig}), this observable can be written as
		\begin{align}\label{eq:ch5:eigZ2}
			O & =(|0\rangle\langle0|-|1\rangle\langle1|)\otimes(|0\rangle\langle0|-|1\rangle\langle1|)\nonumber\\
			& =(\underbrace{|00\rangle\langle00|+|11\rangle\langle11|}_{\Pi_{0}})-(\underbrace{|01\rangle\langle01|+|10\rangle\langle10|}_{\Pi_{1}}).
		\end{align}
		The expression (\ref{eq:ch5:eigZ2}) reveals that the eigendecomposition of the observable $O$ is characterized by two eigenvalues, $+1$ and $-1$, each with an associated subspace of dimension 2. Specifically, the eigenvalue $o_0=1$ is associated with the subspace spanned by the even-parity computational basis vectors, i.e., by projection matrix $\Pi_0$; while the eigenvalue $o_1=-1$ is associated with the subspace spanned by the odd-parity computational basis vectors, i.e., by projection matrix $\Pi_1$. 
		
		From the discussion in Sec. \ref{sec:ch5:obs}, it follows that  a measurement of the observable $O=Z\otimes Z$ corresponds
		to the parity projective measurement in which the numerical output $o_0=1$ is
		assigned to an even parity output, and numerical output $o_1=-1$ is obtained with an odd-parity
		output. Therefore,  the numerical output equals $1$ with probability  (\ref{eq:ch5:parityp1}), and it equals $-1$ with probability (\ref{eq:ch5:parityp2}).
		
		
		
		While the parity measurement is not local (see Sec. \ref{sec:ch5:paritymeas}),  if one is only interested in the numerical output of the measurement of the observable $O=Z\otimes Z$ (and
		not also in the post-measurement state), it is possible to implement the measurement of $O$ separately across the two qubits. To this end, we measure
		the $Z$ observable at each of two qubits, and then multiply the numerical 
		outputs obtained from the two measurements. 
		
	Let us further elaborate on the measurement of observable $O=Z\otimes Z$. 	Measuring the $Z$ observable separately at the two qubits amount to a standard measurement of the two qubits. Therefore, given an input state (\ref{eq:ch5:genstate}),  this measurement yields the cbits 
		$(1,1)$ with probability $|\alpha_{00}|^{2}$; $(-1,-1)$ with probability
		$|\alpha_{11}|^{2}$; $(1,-1)$ with probability $|\alpha_{01}|^{2}$;
		and $(-1,1)$ with probability $|\alpha_{10}|^{2}$.
		It follows that the product of the two observations equals $1$ with probability
		$|\alpha_{00}|^{2}+|\alpha_{11}|^{2}$, and it equals $-1$ with probability
		$|\alpha_{01}|^{2}+|\alpha_{10}|^{2}$, as in the parity projective measurement. This calculation confirms that the numerical value of observable $O=Z\otimes Z$ can be obtained via local measurements of the two qubits in the $Z$ basis.

		\subsection{Joint Multi-Qubit Observables}
		
		 Not all observables are local. However, following Sec. \ref{sec:ch2:Paulibasis}, any observable on $n$ qubits, i.e., any Hermitian matrix of dimension
		$N\times N$ with $N=2^{n}$, can be expressed
		as a linear combination with real coefficients of $n$-qubit Pauli strings. Therefore, any observable on $n$ qubits can be written as\begin{equation}\label{eq:ch5:obsPauli} O=\sum_{s^{n}\in\{0,1,2,3\}^{n}} a_{s^{n}} P_{s^{n}} 
	\end{equation} for some real coefficients $\{a_{s^{n}}\}$. 

This decomposition does not directly provide a way to measure arbitrary observables via local measurements. However, it suggests an approach to evaluate \emph{expectations} of such observables via the separate measurements of observables of Pauli strings. In fact, by (\ref{eq:ch5:obsPauli}), the expectation $\langle O\rangle_{\rho} $ of any observable $O$ can be expressed as the weighted sum, with weights $\{a_{s^{n}}\}$, of the expectations $\langle P_{s^{n}}\rangle_{\rho} $ of the $4^n$ Pauli observables as \begin{equation}\label{eq:ch5:expPauli}\langle O \rangle_{\rho}=\sum_{s^{n}\in\{0,1,2,3\}^{n}} a_{s^{n}} \langle{P_{s^{n}}}\rangle_{\rho}.
\end{equation}

	\subsection{Compatible Observables}\label{sec:ch5:compatible}

			 Suppose that we are interested in measuring multiple observables --
			say $O_{1}$ and $O_{2}$ -- at the same time. Can we do this with
			a single measurement? If so, we say that the two observables are compatible. 	When two observables are not compatible, measuring one observable would produce a post-measurement state for which a measurement of the second observable would no longer have the same distribution as under the original, pre-measurement, state. 
			

			 It can be proved that two observables $O_{1}$ and $O_{2}$ are \textbf{compatible} if and only if they share a common eigenbasis.		 Mathematically, let 
			\begin{equation}
			O_{1}=\sum_{x=0}^{N-1}\lambda_{1,x}|v_{x}\rangle\langle v_{x}|\textrm{ and \ensuremath{O_{2}=}\ensuremath{\sum_{x=0}^{N-1}\lambda_{2,x}}|\ensuremath{v_{x}\rangle\langle v_{x}}|}
			\end{equation}
		be	two  observables. They are compatible since they share the same eigenbasis
			$\{|v_{x}\rangle\}_{x=0}^{N-1}$. Note that they have  generally different
			eigenvalues. 
			 Measuring both observables simultaneously is possible by carrying
			out a measurement in the eigenbasis $\{|v_{x}\rangle\}_{x=0}^{N-1}$.
			In particular, when obtaining the output $x$, the numerical output for observable
			$O_{1}$ is the eigenvalue $\lambda_{1,x}$, while for observable $O_{2}$
			we have the numerical output $\lambda_{2,x}.$ 
			
			As an alternative, equivalent, characterization, two observables $O_{1}$ and $O_{2}$ are compatible if and only if
			they \textbf{commute}, that is, if and only if we have the equality
			\begin{equation}
				O_{1}O_{2}=O_{2}O_{1}.
			\end{equation}
			This condition can provide an efficient way to check whether two observables
			are compatible.

		As a first example, single-qubit observables applied to different qubits -- e.g., $O_1=X\otimes I$ and $O_2=I\otimes Z$ -- are compatible, since they commute -- e.g., $O_1 O_2= O_2 O_1 = X\otimes Z$. In fact, they can be measured via separate measurements on the qubits. 
		
			 As another example, consider the observables \textrm{$O_{1}=Z\otimes I$}
			and $O_{2}=Z\otimes Z$. The two observables commute, i.e., 
			\begin{equation}
				(Z\otimes I)(Z\otimes Z)=I\otimes Z=(Z\otimes Z)(Z\otimes I),
			\end{equation}
			and hence they are compatible. They can be both measured simultaneously via
			a standard measurement in the computational basis. 
			 
			 In contrast, the observables $O_{1}=Z$ and $O_{2}=X$ on a single
			qubit are not compatible since they do not commute, given that we have the equality
			\begin{equation}
				XZ=iY=-ZX.
			\end{equation}
			 It follows that they cannot be measured simultaneously. Rather, one needs
			to prepare the pre-measurement state twice,  and implement two separate
			measurements for the two observables.

			 The existence of incompatible observables is at the core of \textbf{Heisenberg
			uncertainty principle}. This fundamental result indicates that, for a given input quantum
			state $|\psi\rangle$, the product of the standard deviations of the
			observations of two measurements is lower bounded by a quantity that
			depends on the degree to which the two observables commute.

\section{Implementing Projective Measurements}\label{sec:ch5:balanced}
	
		 As we have seen in Sec. \ref{sec:ch3:implany}, any complete measurement in an arbitrary basis $\{|v_{0}\rangle,|v_{1}\rangle,...,|v_{N-1}\rangle\}$
		can be implemented as illustrated in Fig. \ref{fig:ch3:implany} by means of standard measurements in the computational basis. This is done by adding pre- and post-processing unitary transformations
		$U_{v_{x}\rightarrow x}=\sum_{x=0}^{N-1}|x\rangle\langle v_{x}|$ and $U_{v_{x}\rightarrow x}^{\dagger}=U_{x\rightarrow v_{x}}$ 
		that convert the given basis into the computational basis $\{|x\rangle\}_{x=0}^{N-1}$ and back, respectively.
		  As we discuss in this section, a similar approach can be used to implement more general projective
		measurements for an important special case. 
		
		\subsection{Balanced Projective Measurements}
		
		 We specifically focus on projective measurements on $n$ qubits that satisfy the following two conditions: \begin{itemize}
			\item the number of measurement outcomes, and hence the number of projection matrices, is $N'=2^{n'}$ for some integer for $n'\leq n$; and 
			\item  all the projection matrices $\Pi_y$, with $y\in\{0,1,...,2^{n'}-1\}$, have the same rank $N/N'=2^{n-n'}$ with $N=2^n$.
		\end{itemize} We will refer to such projective measurements as being \textbf{balanced}, since the subspaces corresponding to the projection matrices defining it have the same dimension. Balanced measurements include as special cases the parity measurement introduced in Sec. \ref{sec:ch5:paritymeas}, as well as standard partial measurements of $n'$ qubits (see Sec. \ref{sec:ch5:long}).

	\subsection{Implementing Balanced Projective Measurements via Partial Measurements}\label{sec:ch5:balancedcomp}
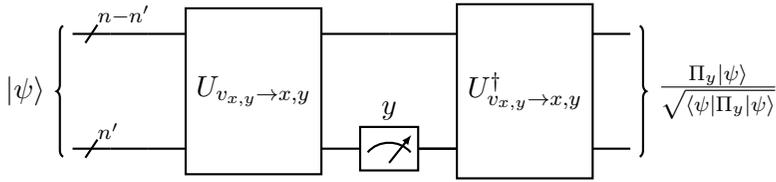
\begin{figure}
\begin{center}
			\begin{quantikz}
				\lstick[wires=2]{$\ket{\psi}$}& \qwbundle{n-n'} & \qw & \gate[wires=2]{U_{v_{x,y}\rightarrow x,y}}& \qw & \gate[wires=2]{U^{\dagger}_{v_{x,y}\rightarrow x,y}}& \qw  \rstick[wires=2]{$\frac{\Pi_{y}|\psi\rangle}{\sqrt{\langle\psi|\Pi_{y}|\psi\rangle}}$}\\
				& \qwbundle{n'} &\qw & & \meter{$y$} & \qw & \qw 
			\end{quantikz}
	\end{center}\vspace{-0.4cm}\caption{Architecture implementing a general balanced projective measurement via a partial measurement of a subset of qubits. \label{fig:ch5:partialproj}}
	\end{figure}

		In this subsection, we show that a balanced projective measurement can be carried out via a partial measurement (in the computational basis) of $n'$ qubit. As illustrated in Fig. \ref{fig:ch5:partialproj}, this is done by adding suitable pre- and post-measurement unitaries that  perform a specific change-of-basis transformation. This transformation, as well as the overall system depicted in Fig.  \ref{fig:ch5:partialproj}, generalize the approach detailed in Fig.  \ref{fig:ch3:implany} for the special case of von Neumann measurements (in which $n'=n$). 
		
		  Following Sec. \ref{eq:ch5:howto}, any balanced projective measurement can be expressed in terms of an
		orthonormal basis $\{|v_{x'}\rangle\}_{x'=0}^{N-1}$ by partitioning the basis 
		into $N'=2^{n'}$ subsets, each with $N/N'=2^{n-n'}$
		elements. This partition can be defined without loss of generality in such a way  that the $n$-cbit string $x'$ is expressed as the concatenation of two cbit strings as $x'=(x,y)$, where \begin{itemize}
			\item the $n'$-cbit string $y$ indicates the subset to which the vector $|v_{x'}\rangle$ belongs, i.e., we have $x'\in\mathcal{X}_{y}$ (recall the notation in Sec. \ref{eq:ch5:howto}); and 
			\item the $(n-n')$-cbit string $x$ identifies the specific vector $|v_{x'}\rangle$ within the $y$-th subset.
		\end{itemize}
	
	With the partition described in the previous paragraph, a balanced projective measurement is described by projection matrices \begin{equation}\label{eq:ch5:Pro}
		\Pi_y=\sum_{x=0}^{N/N'-1} |v_{x,y}\rangle \langle v_{x,y}|
	\end{equation} with $y\in \{0,1,...,N'-1\}$. 
		
We now introduce the change-of-basis unitary $U_{v_{x,y}\rightarrow x,y}$ that maps each vector $|v_{x,y}\rangle$ in the orthonormal basis used to describe the projective measurement to the computational-basis vector $|x,y\rangle$, i.e., \begin{equation}
		U_{v_{x,y}\rightarrow x,y}|v_{x,y}\rangle= |x,y\rangle. 
	\end{equation} Note that this transformation reduces to the unitary $U_{v_x\rightarrow x}$ used in Fig.  \ref{fig:ch3:implany} for the special case of von Neumann measurements ($n'=n$).

	With this transformation, the balanced projective measurement defined by projection matrices (\ref{eq:ch5:Pro}) can be implemented as in Fig. \ref{fig:ch5:partialproj}. In this circuit, the change-of-basis transformation $U_{v_{x,y}\rightarrow x,y}$ is first applied to the input state $|\psi\rangle$ to be measured. Then,  the second set of $n'$ qubits is measured using a standard measurement to produce output $y\in\{0,1,...,2^{n'}-1\}$ with the desired probability (\ref{eq:ch5:Bornpure}). This can be readily checked by using the generalized Born rule (see problems).  	Furthermore, after the second unitary transformation $U^{\dagger}_{v_{x,y}\rightarrow x,y}$, one recovers the post-measurement state (\ref{eq:ch5:postmeaspure}).


	\subsection{An Example: Parity Measurements}\label{sec:ch5:partialprojpar}
		As an example, in this subsection, we specialize the architecture in Fig. \ref{fig:ch5:partialproj} to the parity measurement  with projection matrices $\{\Pi_{0},\Pi_{1}\}$ in  (\ref{eq:ch5:projmat0})-(\ref{eq:ch5:projmat1}). Note that we have $n'=1$ and that both projection matrices have rank equal to 2. Therefore, the parity projective measurement is a balanced projective measurement as defined in this section.
		 
		 For the parity measurement, the decomposition (\ref{eq:ch5:Pro}) holds with 	$|v_{00}\rangle=|00\rangle$,  $|v_{10}\rangle=|11\rangle$, $|v_{01}\rangle=|01\rangle$, and $|v_{11}\rangle=|10\rangle$. Note that in each vector $|v_{x,y}\rangle$, the bit $y$ identifies the projector, while the bit $x$ determines the specific eigenvector with the $y$-th subspace. Accordingly,   the transformation $U_{v_{x,y}\rightarrow x,y}$ described in the previous subsection maps states
		$|v_{00}\rangle=|00\rangle$ and $|v_{10}\rangle=|11\rangle$ to $|00\rangle$ and $|10\rangle$,
		respectively; and states $|v_{01}\rangle=|01\rangle$ and $|v_{11}\rangle=|10\rangle$ to $|01\rangle$
		and $|11\rangle$, respectively. 
		
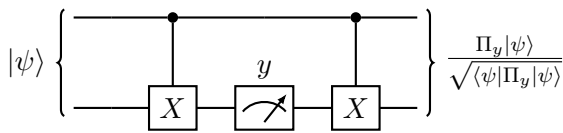
\begin{figure}
\begin{center}
			\begin{quantikz}
				\lstick[wires=2]{$\ket{\psi}$} & \qw & \ctrl{1}& \qw & \ctrl{1}& \qw  \rstick[wires=2]{$\frac{\Pi_{y}|\psi\rangle}{\sqrt{\langle\psi|\Pi_{y}|\psi\rangle}}$}\\
				& \qw & \gate{X}& \meter{$y$} & \gate{X} & \qw 
			\end{quantikz}
	\end{center}\vspace{-0.4cm}\caption{A quantum circuit implementing a parity measurement on two qubits. \label{fig:ch5:partialprojpar}}
	\end{figure}

		The transformation $U_{v_{x,y}\rightarrow x,y}$ can be readily seen to amount to a single CNOT gate with the first qubit controlling the second. Accordingly,  a parity measurement can be carried out by specializing the architecture in Fig. \ref{fig:ch5:partialproj} as depicted in Fig. \ref{fig:ch5:partialprojpar}. 
		
		To verify that the circuit in Fig. \ref{fig:ch5:partialprojpar} indeed produces the desired outcome, consider an arbitrary pure state (\ref{eq:ch5:genstate}) as the input. 
		After the first CNOT  in Fig. \ref{fig:ch5:partialprojpar}, we have
		\begin{equation}
			C_{01}|\psi\rangle=\alpha_{00}|00\rangle+\alpha_{01}|01\rangle+\alpha_{10}|11\rangle+\alpha_{11}|10\rangle.
		\end{equation} Note that for each vector in the computational basis, i.e., every input vector of the form $|x,y\rangle$ with $x,y\in\{0,1\}$, the second qubit now reports the parity of the corresponding input state. 
		By the generalized Born rule, measuring the second qubit produces the output $y=0$ 	with probability $|\alpha_{00}|^{2}+|\alpha_{11}|^{2}$, which coincides with (\ref{eq:ch5:parityp1}), as desired. Furthermore, when $y=0$, the second CNOT transforms the post-measurement state
		\begin{equation}
			\frac{1}{\sqrt{|\alpha_{00}|^{2}+|\alpha_{11}|^{2}}}(\alpha_{00}|00\rangle+\alpha_{11}|10\rangle)
		\end{equation} into the desired post-measurement state (\ref{eq:ch5:postex}) for the parity measurement, i.e., 
		\begin{equation}
			\frac{\Pi_{0}|\psi\rangle}{\sqrt{\langle\psi|\Pi_{0}|\psi\rangle}}=\frac{1}{\sqrt{|\alpha_{00}|^{2}+|\alpha_{11}|^{2}}}(\alpha_{00}|00\rangle+\alpha_{11}|11\rangle).
		\end{equation} A similar calculation applies for the case in which the measurement output is $y=1$, recovering probability (\ref{eq:ch5:parityp2}) and post-measurement state (\ref{eq:ch5:postex1}).
	
\section{Quantum Error Correction}\label{sec:ch5:qec}
	
		 As an application of projective measurements, in this section, we briefly introduce the
		problem of quantum error correction. To this end, suppose that we have a noisy quantum  ``channel'', which may represent, for instance, a communication link or noisy
		hardware. We would like to design an error correction scheme that is able to recover the original state of a qubit irrespective of the noise on the channel. Classically, the problem could be solved via redundancy: By copying and transmitting the information of interest multiple times through the channel, one can ensure some level of protection against channel errors. However, by the no cloning theorem, we cannot duplicate an unknown quantum
		state in order to increase robustness to noise.

		 Notwithstanding the outlined limitation imposed by no cloning, is some form of error correction possible? We will see in this section that the answer to this question is affirmative, and that quantum error correction hinges on the implementation of projective measurements. It is emphasized that quantum error correction is a vast area of research, and that this section is only meant to illustrate some basic ideas pertaining the connection with projective measurements.
	
	\subsection{Bit-Flip Channel}
		 In order to illustrate the process of quantum error correction,
		we consider a simple channel characterized by a ``bit flip'' noise. Given three qubits with a given input state, the channel may ``flip'' at most one out of three qubits. Formally, this means that, if we have three qubits in a
			general coherent state 
			\begin{equation}
			|\psi_{in}\rangle=\sum_{x=0}^{1}\sum_{y=0}^{1}\sum_{z=0}^{1}\alpha_{x,y,z}|x,y,z\rangle
			\end{equation}
			before the application of the channel, the state $|\psi_{out}\rangle$
			after the application of the channel may be
			\begin{align}
			|\psi_{out}\rangle &=	 |\psi_{in}\rangle, \textrm{ or}\nonumber\\
				|\psi_{out}\rangle &=\textrm{ \ensuremath{(X\otimes I\otimes I)|\psi_{in}\rangle=}}\sum_{x=0}^{1}\sum_{y=0}^{1}\sum_{z=0}^{1}\alpha_{x,y,z}|\bar{x},y,z\rangle, \textrm{ or}\nonumber\\
				|\psi_{out}\rangle &= (I\otimes X\otimes I)|\psi_{in}\rangle=\sum_{x=0}^{1}\sum_{y=0}^{1}\sum_{z=0}^{1}\alpha_{x,y,z}|x,\bar{y},z\rangle, \textrm{ or}\nonumber\\
				|\psi_{out}\rangle &= (I\otimes I\otimes X)|\psi_{in}\rangle=\sum_{x=0}^{1}\sum_{y=0}^{1}\sum_{z=0}^{1}\alpha_{x,y,z}\ensuremath{|x,y,\bar{z}\rangle},
			\end{align}
			where $\bar{x}=1\oplus x$, $\bar{y}=1\oplus y$, and $\bar{z}=1\oplus z$. The first state output $|\psi_{out}\rangle$ corresponds to the case in which no bit flip is applied by the channel; the second to the case in which the first qubit is flipped; and so on. Note that a bit flip corresponds to a Pauli string with a single $X$ operator in the position corresponding to the qubit affected by the flip.
			
			We refer to Sec. \ref{sec:ch5:quantumch} for more discussion on quantum channels.

		\subsection{Quantum Coding}
\begin{figure}
\begin{center}
			\begin{quantikz}
				\lstick[wires=1]{$|\psi\rangle=\alpha_{0} \ket{0} + \alpha_{1} \ket{1}$} &[2mm] \qw & \ctrl{1} & \ctrl{2} & \qw & \qw  \rstick[wires=3]{$\alpha_{0} \ket{000} + \alpha_{1} \ket{111}$}\\
				\lstick{$\ket{0}$} & \qw & \gate{X} & \qw & \qw & \qw  \\
				\lstick{$\ket{0}$} & \qw & \qw & \gate{X} & \qw & \qw  
			\end{quantikz}
	\end{center}\vspace{-0.4cm}\caption{Quantum encoding for a quantum error correction scheme designed to correct at most one bit flip.\label{fig:ch5:enc}}
	\end{figure}
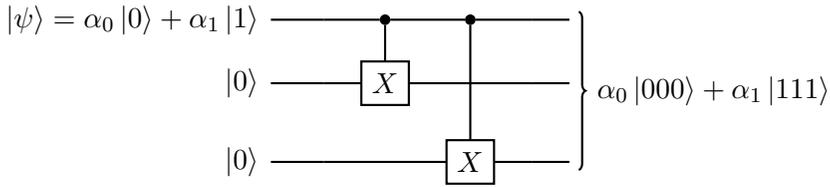

		 Let us assume that we have a qubit with an unknown state $|\psi\rangle=\alpha_{0}|0\rangle+\alpha_{1}|1\rangle$, which we would like to protect against the bit-flip channel described in the previous subsection. 
		 Specifically, we would like to encode this qubit in such a way that errors due to
		the channel can be corrected.
		 While duplication (cloning) is not possible, we can use the ``classical
		cloning'', or basis-copying, circuit introduced in Sec. \ref{sec:ch4:BC}.	The corresponding encoding circuit is shown in Fig. \ref{fig:ch5:enc}. Importantly, the resulting encoded, entangled, state\begin{equation}
			\alpha_{0}|000\rangle+\alpha_{1}|111\rangle
		\end{equation}is different from a simple replica of the quantum state $|\psi\rangle$, which, by the no cloning theorem, cannot be realized for any input state $|\psi\rangle$. 
	
		 With this encoding circuit, the channel at hand can output one of the following
		states $|\psi_{out}\rangle$
		\begin{align*}
		|\psi_{out}\rangle &=	\alpha_{0}|000\rangle+\alpha_{1}|111\rangle & \textrm{ if there is no error}\\
		|\psi_{out}\rangle &=	\alpha_{0}|100\rangle+\alpha_{1}|011\rangle & \textrm{ if the first qubit is flipped}\\
		|\psi_{out}\rangle &=	\alpha_{0}|010\rangle+\alpha_{1}|101\rangle & \textrm{ if the second qubit is flipped}\\
		|\psi_{out}\rangle &=\alpha_{0}|001\rangle+\alpha_{1}|110\rangle & \textrm{ if the last qubit is flipped.}
		\end{align*}
	
\subsection{Quantum Decoding via a Projective Measurement}\label{sec:ch5:qecdec}
		 Based on the discussion in the previous subsection, if there is no error, the state lies in the two-dimensional
		space spanned by vectors $|000\rangle$ and $|111\rangle$. The projection matrix onto this subspace is given as 
		\begin{equation}\label{eq:ch5:qecdec1}
		\Pi_{00}=|000\rangle\langle000|+|111\rangle\langle111|.
		\end{equation}
		 If the first qubit is flipped, the state lies instead in the two-dimensional
		space spanned by vectors $|100\rangle$ and $|011\rangle$, whose corresponding projection matrix is
		\begin{equation}
		\Pi_{10}=|100\rangle\langle100|+|011\rangle\langle011|.
		\end{equation} Similarly, if the second qubit is flipped, the state lies in the two-dimensional
		space spanned by vectors $|010\rangle$ and $|101\rangle$ with projection matrix 
		\begin{equation}
		\Pi_{11}=|010\rangle\langle010|+|101\rangle\langle101|; 
		\end{equation}
		 and, if the third qubit is flipped, the state lies in the two-dimensional
		space spanned by vectors $|001\rangle$ and $|110\rangle$ with projection matrix 
		\begin{equation}\label{eq:ch5:qecdec2}
		\Pi_{01}=|001\rangle\langle001|+|110\rangle\langle110|.
		\end{equation}
		 The reason for the specific choice of the  numbering of the projection matrices given above will be made clear later.

		 Based on the observations in the previous paragraph, applying the projective measurement $\{\Pi_{00},\Pi_{01},\Pi_{10},\Pi_{11}\}$ to the output of the quantum channel $|\psi_{out}\rangle$ allows one to correct one bit flip error by applying a NOT gate, i.e., a Pauli $X$ operator, to the
		first, second, or third qubit depending on the output of the measurement. Specifically, if the output of the measurement is $y=00$, no operation should be applied; if it is  $y=10$, a NOT gate should be applied to the first qubit; and so on. 
		
		Crucially, the success of this operation in correcting a bit flip hinges on the use of a projective measurement. In fact, with a complete measurement, the state
		of the qubits would ``collapse'', destroying the encoded superposition state 
		$\alpha_{0}|000\rangle+\alpha_{1}|111\rangle$ and hence losing information about the original qubit state $|\psi\rangle=\alpha_{0}|0\rangle+\alpha_{1}|1\rangle$. In contrast, as we have seen in Sec. \ref{sec:ch5:projmeas}, the projective measurement leaves unchanged quantum states that lie within the subspace corresponding to each projection matrix.
		 
		 As detailed in this subsection, in order to apply error correction, one should implement a Pauli $X$ gate depending on the output of the projective measurement $\{\Pi_{0},\Pi_{01},\Pi_{10},\Pi_{11}\}$. To make this possible, it is necessary to have both measurement output and post-measurement state 
		 simultaneously encoded in the state of the quantum system. This way, at the next computation step, the measurement output can be used to determine the operation to be applied to the post-measurement state.  This is typically done by adding qubits, whose state encodes the output $y$ of the projective measurement. This will be discussed in the next section.

	\subsection{Key Principles of Quantum Error Correction}
		 To sum up the discussion so far,
		it is useful to reiterate and generalize the principle underlying
		quantum error correction.
		 Quantum error correction aims at protecting an unknown state of $k$
		qubits -- $k=1$ in the example above -- by encoding such state
		into the state of $n$ qubits -- $n=3$ in the example. Note that the unknown original state lies in a Hilbert space of dimension $2^k$. 
		 As a result of encoding, the state of the encoded qubit hence occupies a $2^{k}$-dimensional
		subspace in the $2^{n}$ Hilbert space corresponding to the $n$ encoded
		qubits.
		
		Encoding is designed in such a way that any possible distinct error causes the state
		of the encoded qubits to lie in an orthogonal subspace of dimension
		$2^{k}$.  A projective measurement defined by projectors onto these orthogonal
		subspaces can thus distinguish among the different types of errors, while leaving the state
		within each subspace undisturbed.
		
		If there are $m$ different errors -- $m=3$ in the example -- the design requirement outlined in the previous paragraph calls for a number of encoded qubits, $n$,  satisfying the condition \begin{equation}\label{eq:ch5:Hamming}2^{n}\geq2^{k}(m+1).
		\end{equation} In fact, the Hilbert space of the encoded qubits should include $m+1$ orthogonal subspaces, each of dimension $2^k$, with one subspace corresponding to the absence of errors on the channel and the others accounting for each of the $m$ types of errors.  
	
	In the example studied in this section, we have $k=1$, $n=3$, and $m=3$, which satisfy this condition with equality as  $2^{3}=2^{1}(1+3)=8.$ When further elaborated on by accounting for all possible error types on single qubits, the inequality (\ref{eq:ch5:Hamming}) yields the so called \textbf{quantum Hamming bound}. 
		 
		 \subsection{Quantum Error Detection}
		
		As a less powerful alternative to quantum error correction, it is also possible to design \textbf{quantum error detection}
		schemes, whereby errors are only detected and not corrected. With error detection, one can determine the presence of an error, without necessarily identifying, and correcting, the specific error. With reference to the discussion in the previous subsection, for quantum error detection, there is no need for the $m$ subspaces corresponding
		to different errors to be orthogonal. In fact, it is sufficient that all types of errors cause the post-channel state to lie in the \emph{same} subspace, as long as the latter is orthogonal to that encompassing the encoded, noiseless, state.
		
		 To implement a quantum error detector, one can define an observable that assigns the $+1$ eigenvalue to the ``no-error'' projection matrix $\Pi_0$ describing the subspace spanned by the code, and a $-1$ eigenvalue to the subspace $\Pi_1$ orthogonal to $\Pi_0$. This way, if a measurement of the observable returns $-1$, one can conclude with certainty that an error has occurred, while, if the output is $+1$, the state can be concluded to be unaffected by errors.

		 

%
\section{Implementing Projective Measurements with Ancillas}
	
	As we have discussed in the previous section, for some applications, including quantum error correction, it is useful to have the output of the measurement and the post-measurement state simultaneously encoded in the state of separate quantum subsystems. This requires adding to the system being measured a set of ancilla qubits, whose state encodes the measurement output. We will hence refer to such measurements as \textbf{measurements with ancillas}. In this section, we describe how to implement measurements with ancillas, starting with the case of complete measurements, and then generalizing the presentation to projective measurements.
	
	\subsection{Complete Measurements with Ancillas}
\begin{figure}
			\begin{center}
				\begin{quantikz}
					\lstick{$\ket{\psi}$} & \qwbundle{n} & \gate{U_{v_{x}\rightarrow x}} & \gate[wires=2]{U_{\textrm{BC}}} & \qw & \qw & \gate{U_{x\rightarrow v_{x}}} & \qw \rstick{$\ket{v_x}$}\\
					\lstick{$\ket{0}$} & \qwbundle{n} & \qw &  & \qw & \meter{$x$} & \qw & \qw \rstick{$\ket{x}$}
				\end{quantikz}
		\end{center}\vspace{-0.4cm}\caption{von Neumann measurement in an arbitrary orthonormal basis $\{|v_{0}\rangle,|v_{1}\rangle,...,|v_{N-1}\rangle\}$ with ancillas. The ancillas qubits are the second set of $n$ qubits, which are  initialized to the ground state $|0\rangle$. Unlike the circuit in Fig. \ref{fig:ch3:implany}, this architecture produces both the post-measurement state $|v_x\rangle$ for the first $n$ qubits and the output of the measurement, $x\in \{0,1,...,2^{n}-1\}$, with the latter encoded in the post-measurement state $|x\rangle$ of the $n$ ancilla qubits. \label{fig:ch5:ancilla}}
	\end{figure}
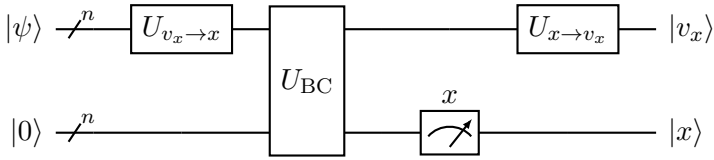

		As we demonstrate in this subsection, we can implement a von
		Neumann measurement with ancillas in any orthonormal basis $\{|v_{0}\rangle,|v_{1}\rangle,...,|v_{N-1}\rangle\}$ for a system of $n$ qubits ($N=2^n$) via the circuit shown in Fig. \ref{fig:ch5:ancilla}. The circuit operates on the $n$ qubits being measured -- in an arbitrary state $|\psi\rangle$ -- as well as on $n$ ancilla qubits initialized in the ground state $|0\rangle$. 
		
	The circuit in Fig. \ref{fig:ch5:ancilla} first applies the change-of-basis unitary $U_{v_x \rightarrow x}$ to the $n$ qubits being measured as in Fig. \ref{fig:ch3:implany}. As detailed in Sec. \ref{sec:ch3:implany}, this unitary maps each vector $|v_{x}\rangle$ of the basis defining the measurement to the vector $|x\rangle$ in the computational basis.  Unlike the circuit in Fig. \ref{fig:ch3:implany}, the first set of $n$ qubits are not directly measured. Rather, they are ``copied'' via a basis-copying circuit to the second set of $n$ qubits, i.e., to the ancilla qubits. The ancilla qubits are measured, and finally  the inverse change-of-basis transformation  $U_{x \rightarrow v_x}$ is applied to the first $n$ qubits. We recall from Sec. \ref{sec:ch4:BC} that the basis-copying transformation is defined as
	\begin{equation}
		U_{\textrm{BC}}=\sum_{x=0}^{N-1}\sum_{y=0}^{N-1}|x,x\oplus y\rangle\langle x,y|,
	\end{equation}
	where $\{|y\rangle\}_{y=0}^{N-1}$ represents the computational basis
	for the Hilbert space of the ancillas. The unitary $U_{\textrm{BC}}$  has the effect of ``cloning'' the vectors in the computational basis from the first $n$ qubits to the ancillas.

For an arbitrary pure state $|\psi\rangle=\sum_{x=0}^{N-1}\alpha_{x}|v_{x}\rangle$
			of the $n$ qubits being measured, and hence a global state $\sum_{x=0}^{N-1}\alpha_{x}|x,0\rangle$
			after the change-of-basis transformation $U_{v_{x}\rightarrow x}$, the basis-copying
			unitary operator yields the state
			\begin{equation}
			U_{\textrm{BC}}(U_{v_{x}\rightarrow x}\otimes I)\sum_{x=0}^{N-1}\alpha_{x}|v_{x},0\rangle=\sum_{x=0}^{N-1}\alpha_{x}|x,x\rangle.
			\end{equation} Therefore, by the generalized Born rule, measuring the ancillas yields
			the post-measurement state $|x,x\rangle$ with probability
			$|\alpha_{x}|^{2}.$ Finally, this produces the state $|v_{x},x\rangle$ after
			the inverse change-of-basis transformation $U_{x\rightarrow v_{x}}$. 
			
			This calculation demonstrates that the architecture in Fig. \ref{fig:ch5:ancilla} implements a von Neumann measurement in the basis $\{|v_{0}\rangle,|v_{1}\rangle,...,|v_{N-1}\rangle\}$.  In fact, the circuit in the figure leaves the first subsystem of $n$ qubits in the post-measurement state $|v_x\rangle$ with probability $|\alpha_{x}|^{2}=|\langle \psi |x \rangle |^2$. Furthermore, the output of the measurement, $x\in \{0,1,...,2^{n}-1\}$ is encoded in the post-measurement state $|x\rangle$ of the $n$ ancilla qubits.

\subsection{Projective Measurements with Ancillas}
	
We now generalize the architecture in Fig. \ref{fig:ch5:ancilla} to the balanced projective measurements introduced in Sec. \ref{sec:ch5:balanced}. To proceed, we focus here on the class of partial measurements of $n'$ qubits in the computational basis, since, as seen in Sec. \ref{sec:ch5:balancedcomp}, by adding suitable pre- and post-measurement transformations, one can implement a general balanced measurement from such partial measurements.  We will provide an example of this procedure at the end of this subsection.

A partial measurement of $n'$ qubits in the computational basis can be implemented via a generalization of the architecture introduced in the previous subsection whereby only the second set of $n'$ qubits, whose state is to be measured, are ``copied'' via a basis-copying circuit to the ancillas. 

Specifically, the measurement involves the introduction of $n'$ ancilla qubits, which are used to encode the output index $y \in \{0,1,...,2^{n'}-1\}$.  A basis-copying unitary 
``copies'' only the second set of $n'$ qubits as controlling qubits onto the $n'$  ancillas. This amounts to a unitary transformation that maps each vector $|x\rangle \otimes |y\rangle \otimes |0\rangle=|x,y,0\rangle$ -- with $x\in\{0,1,...,2^{n-n'}-1\}$, $y\in\{0,1,...,2^{n'}-1\}$, and $|0\rangle$ being the initial ground state of the $n'$ ancillas -- into the state $|x,y,y\rangle$.

\begin{figure}
	\begin{center}
			\begin{quantikz}
				\lstick[wires=2]{$\ket{\psi}$} &[2mm] \qw & \ctrl{1} & \qw & \qw & \qw & \ctrl{1}& \qw \rstick[wires=2]{$\frac{\Pi_{y}|\psi\rangle}{\sqrt{\langle\psi|\Pi_{y}|\psi\rangle}}$} \\
				&[2mm] \qw & \gate{X} & \ctrl{1} & \qw & \qw &\gate{X}&\qw\\
				\lstick{$\ket{0}$} & \qw & \qw & \gate{X} & \qw & \meter{$y$} &\qw& \qw \rstick{$\ket{y}$}
			\end{quantikz}
	\end{center}\vspace{-0.4cm}\caption{An ancilla-based implementation of the parity measurement introduced in Sec. \ref{sec:ch5:paritymeas}.  \label{fig:ch5:parityanc}}
\end{figure}
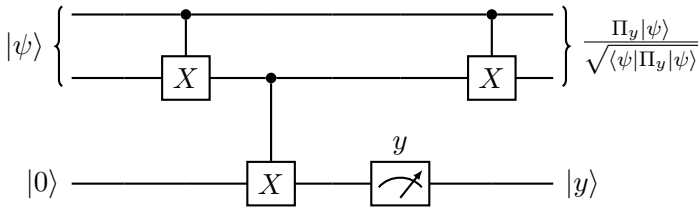

		As an example, consider again the parity measurement $\{\Pi_{0},\Pi_{1}\}$ introduced in Sec. \ref{sec:ch5:paritymeas}. Since
		there are only two values for the measurement ($n'=1$), it is sufficient
		add a single ancilla qubit. As detailed in Sec. \ref{sec:ch5:partialprojpar}, in order to implement this measurement,  one can pre- and post-process the state to be measured with the change-of-basis unitaries illustrated in Fig. \ref{fig:ch5:partialprojpar}. Accordingly, as depicted in Fig. \ref{fig:ch5:parityanc}, the first step is to apply a CNOT gate to the two qubits being measured, which are initially in state $|\psi\rangle$. Then, one applies a basis-copying circuit from the second qubit to the ancilla, followed by a measurement of the ancilla. Note that the basis-copying circuit also amounts to a CNOT gate.  Finally, the initial change-of-basis unitary is inverted with a last CNOT gate. It can be directly checked that the circuit in Fig. \ref{fig:ch5:parityanc} indeed produces the output $y$ and corresponding post-measurement state with the probability (\ref{eq:ch5:parityp1})-(\ref{eq:ch5:parityp2}). 

\begin{figure}
\begin{center}
			\begin{quantikz}
				\lstick[wires=2]{$\ket{\psi}$} &[2mm] \qw & \ctrl{2} & \qw & \qw & \qw \rstick[wires=2]{$\frac{\Pi_{y}|\psi\rangle}{\sqrt{\langle\psi|\Pi_{y}|\psi\rangle}}$} \\
				&[2mm] \qw & \qw & \ctrl{1} & \qw & \qw \\
				\lstick{$\ket{0}$} & \qw & \gate{X} & \gate{X} & \meter{$y$} & \qw  \rstick{$\ket{y}$}
			\end{quantikz}
	\end{center}\vspace{-0.4cm}\caption{An equivalent implementation of the parity measurement with ancilla. Unlike the equivalent circuit in Fig. \ref{fig:ch5:parityanc}, here the ancilla qubit -- initialized in the ground state -- computes the parity of the two qubits being measured, separately for each computational basis state, via the two CNOTs. \label{fig:ch5:paritys}}
\end{figure}
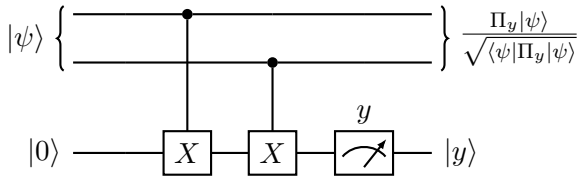

		The outlined general approach to implement projective measurements with ancillas may not necessarily provide the most efficient circuit. For example, for the parity measurement discussed in the last paragraph, the circuit in Fig. \ref{fig:ch5:parityanc}  can be seen to be equivalent to that in Fig. \ref{fig:ch5:paritys}, which requires only two CNOT gates. The operation of the circuit in Fig. \ref{fig:ch5:paritys} has the interesting interpretation that the ancilla qubit computes the parity -- separately on each 
		computational basis state -- of the two qubits being measured via the two CNOT gates. In fact, if the two qubits being measured are in state $|x,y\rangle$, with $x,y\in\{0,1\}$, the state of the ancilla qubit after the two CNOT gates is $|x\oplus y\rangle$.
\subsection{Application to Quantum Error Correction}
	
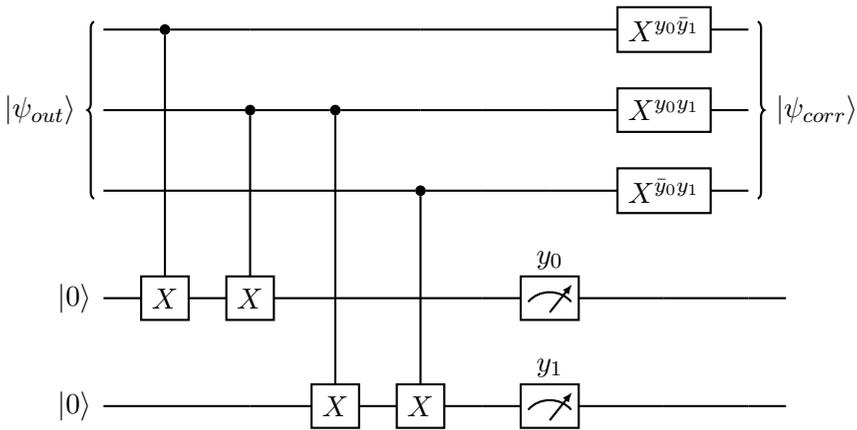
\begin{figure}
			\begin{center}
				\begin{quantikz}
					\lstick[wires=3]{$\ket{\psi_{out}}$} \qw & \ctrl{3} & \qw & \qw & \qw & \qw & \qw & \gate{X^{y_0\bar{y}_1}} &\qw \rstick[wires=3]{$\ket{\psi_{corr}}$}\\
					\qw & \qw & \ctrl{2} & \ctrl{3} & \qw  & \qw & \qw & \gate{X^{y_0y_1}} &\qw\\
					\qw & \qw & \qw & \qw & \ctrl{2}  & \qw & \qw &\gate{X^{\bar{y}_0y_1}} &\qw\\
					\lstick{$\ket{0}$}\qw & \gate{X} & \gate{X} & \qw & \qw  & \qw & \meter{$y_0$}& \qw & \qw & \qw\\
					\lstick{$\ket{0}$}\qw & \qw & \qw & \gate{X} & \gate{X}  & \qw & \meter{$y_1$}& \qw & \qw & \qw
				\end{quantikz}
		\end{center}\vspace{-0.4cm} \caption{Decoder for the quantum error correction code described in Sec. \ref{sec:ch5:qec}. The decoder uses the output of the measurement encoded in the state of the two ancilla qubits (the last two qubits) in order to determine the operation to be applied on the encoded qubits (the first three qubits) at the next computation step.\label{fig:ch5:qdec}}
	\end{figure}

Let us now return to the problem of designing a decoding circuit for the quantum error correction code introduced in Sec. \ref{sec:ch5:qec}. As we have seen in Sec. \ref{sec:ch5:qecdec},  the decoder carries out the projective measurement $\{\Pi_{00},\Pi_{01},\Pi_{10},\Pi_{11}\}$ with projection matrices in (\ref{eq:ch5:qecdec1})-(\ref{eq:ch5:qecdec2}). According to the discussion in this section, this measurement can be implemented by introducing two ancilla qubits ($n'=2$), since there are four possible outcomes. The resulting circuit is illustrated in Fig. \ref{fig:ch5:qdec}, and is detailed next.

Generalizing the circuit in Fig. \ref{fig:ch5:paritys}, the decoder takes as input the post-channel state, and computes the parity -- in the computational basis -- of the first and second pair of qubits in the post-channel state. Specifically, the first ancilla qubit measures the parity of the first pair, and the second ancilla qubit of the second pair. The outputs $(y_0, y_1)\in \{0,1\}^{2}$ of the measurement of the ancillas (in the computational basis) are then used to determine whether an $X$ gate should be applied to the any of the three input qubits. Since it is used to ``diagnose'' the type of error that may have occurred on the encoded qubits, the output of the measurement of the ancillas is referred to as the \textbf{error syndrome}. 

The advantage of introducing the ancillas in the circuit of Fig. \ref{fig:ch5:qdec} is that their post-measurement state encodes the transformation to be applied to the first three qubits to ensure error correction. In this regard, note that the shortcut notation $X^{ab}$ in Fig. \ref{fig:ch5:qdec} indicates that an $X$ gate is applied when $a=b=1$, recovering the decoding rule described in Sec. \ref{sec:ch5:qec}. Accordingly, as long as the channel applies at most one bit flip, the corrected state $|\psi_{corr}\rangle$ equals the encoded state $|\psi_{in}\rangle=\alpha_{0}|000\rangle+\alpha_{1}|111\rangle$.

	In closing this subsection, we note that it is also possible to implement quantum error correction
			without an explicit measurement of the ancillas through the use of
			three-qubit gates, namely Toffoli gates. Furthermore, when the goal is recovering the original qubit, and not the encoded state, it is also possible to avoid using ancilla qubits.

\section{Positive Operator-Valued Measurements}\label{sec:ch5:povm}
	
		 As we have studied in the previous section, general projective measurements can be implemented
		by adding ancillas in ground state $|0\rangle$, applying an entangling operator,
		namely the basis-copying circuit, and then measuring the ancillas. In this section, we address the following question: 
		 What if we applied a more general entangling operator, i.e., an arbitrary
		unitary matrix, to the overall system including the ancillas, and
		then measured the ancillas (in a selective way)? As we will detail, this leads to the most general form of (selective) measurements, namely \textbf{positive operator-valued measurement (POVM)}.

\subsection{Deriving POVM Measurements}
	
\begin{figure}
		\begin{center}
				\begin{quantikz}  				\lstick{$\ket{0}$} & \gate[wires=2]{U}\qwbundle{} &   \meter{$y$}  & \qw\rstick{w.p. $\textrm{tr}(M_{y}\rho)$} \\ 				\lstick{$\rho$} &  \qwbundle{} & \qwbundle{} &	\qw		\end{quantikz}
		\end{center}\vspace{-0.4cm}\caption{A POVM describes a situation in which the system of interest interact through a unitary $U$ with ancilla qubits, initially in the ground state $|0\rangle$, and then the ancillas are measured (in a selective manner). \label{fig:ch5:povm}}
	\end{figure}
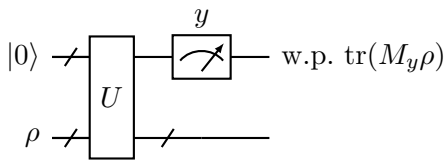

		 To derive POVMs, let us consider a system of $n$ qubits with the addition of  $n'$
		ancillas. The ancillas are initially in the ground state $|0\rangle.$
		 If the system of $n$ qubit is in state $|\psi\rangle$, the overall
		system is hence in the state $|0,\psi\rangle=|0\rangle\otimes|\psi\rangle.$
		Note that, unlike what we have done in the rest of this chapter, here we append the ancillas in front of the $n$-qubit state in order to simplify the derivations that
		follow.  The setting under study is illustrated in Fig. \ref{fig:ch5:povm}.

		Applying a unitary matrix $U$ to the overall system results in the state 
		\begin{equation}
		U|0,\psi\rangle=\left[\begin{array}{cccc}
			U_{00} & U_{01} & \cdots & U_{0,N'-1}\\
			U_{10} & U_{11} & \cdots & U_{1,N'-1}\\
			\vdots & \vdots &  & \vdots\\
			U_{N'-1,0} & U_{N'-1,1} & \cdots & U_{N'-1,N'-1}
		\end{array}\right]\left[\begin{array}{c}
			|\psi\rangle\\
			0\\
			\vdots\\
			0
		\end{array}\right],
		\end{equation}
		where the matrices $U_{km}$, with $k,m\in\{0,1,...,N'-1\}$, are of size $N\times N$, with $N=2^{n}$, and we set $N'=2^{n'}.$ It follows that the output of the unitary operation is
		\begin{equation}\label{eq:ch5:povmfc}
		U|0,\psi\rangle=\left[\begin{array}{c}
			U_{00}|\psi\rangle\\
			U_{10}|\psi\rangle\\
			\vdots\\
			U_{N'-1,0}|\psi\rangle
		\end{array}\right]=\sum_{y=0}^{N'-1}|y\rangle\otimes U_{y0}|\psi\rangle.
		\end{equation}
		 Furthermore, by the unitarity of matrix $U$, i.e., by the equality
		$U^{\dagger}U=I$, we have the condition 
		\begin{equation}\label{eq:ch5:povm1}
		\sum_{y=0}^{N'-1}U_{y0}^{\dagger}U_{y0}=I,
		\end{equation}where $I$ is the $N\times N$ identity matrix.

		 Now, if we measure the ancillas in the computational basis, by the
		generalized Born rule, we get 
		\begin{align}\label{eq:ch5:povmprob}
		\Pr[\textrm{measurement output equals  }y]&=\langle\psi|U_{y0}^{\dagger}U_{y0}|\psi\rangle \nonumber\\&=\textrm{tr}(M_{y}\rho),
		\end{align}
		where we have defined matrix $M_{y}=U_{y0}^{\dagger}U_{y0}$ and the density matrix 
		$\rho=|\psi\rangle\langle\psi|$. 
		 Furthermore, the post-measurement (density) state is 
		\begin{equation}
		\frac{U_{y0}\rho U_{y0}^{\dagger}}{\textrm{tr}(M_{y}\rho)}.
		\end{equation}

\subsection{Defining a POVM}
	
		 Generalizing the derivation above, a POVM is defined by a set of $N\times N$ positive
		semidefinite matrices $M_{y}$ with $y\in\{0,1,...,N'-1\}$, with
		$N'$ generally different from $N$ and possibly larger than $N$,  such that the following condition -- from (\ref{eq:ch5:povm1}) -- applies
		\begin{equation}\label{eq:ch5:povm2}
		\sum_{y=0}^{N'-1}M_{y}=I.
		\end{equation}
		 A POVM stipulates that output $y$ is observed with probability 
		\begin{equation}\label{eq:ch5:povmform}
		\Pr[\textrm{measurement output equals }y]=\textrm{tr}(M_{y}\rho).
		\end{equation}

		 The post-measurement state is not uniquely specified by a
		POVM defined by matrices $\{M_{y}\}_{y=0}^{N'-1}$. 
		 However, by taking the positive semidefinite square root of $M_{y}$, i.e.,
		\begin{equation}
		M_{y}^{1/2}=\sum_{x=0}^{N-1} \sqrt{\lambda_{x,y}} |v_{x,y}\rangle \langle v_{x,y}| 
		\end{equation}
		where $M_{y}=\sum_{x=0}^{N-1} \lambda_{x,y} |v_{x,y}\rangle \langle v_{x,y}| $ is the eigendecomposition
		of $M_{y}$, one typically sets the post-measurement state to the density matrix
		\begin{equation}\label{eq:ch5:POVMpm}
		\frac{M_{y}^{1/2}\rho M_{y}^{1/2}}{\textrm{tr}(M_{y}\rho)}.
		\end{equation}
		
		 POVMs generalize projective measurements, which are obtained by choosing the positive semidefinite matrices $M_y$ as projection matrices, i.e., by setting 
		$M_{y}=\Pi_{y}$ (note that we have $\Pi_{y}^{1/2}=\Pi_{y})$. We observe that, in a projective measurement, the matrices $M_{y}=\Pi_{y}$  are mutually  orthogonal, in the sense that we have the equalities  $\Pi_{y}^{\dagger}\Pi_{y'}=\Pi_{y}\Pi_{y'}=0$, where $0$ is an all-zero matrix, for all $y'\neq y$; and the are idempotent as they satisfy the condition  $\Pi_{y}\Pi_{y}=\Pi_y$. In contrast, in a more general POVM, matrices $M_y$ are only constrained to be positive semidefinite and to satisfy condition (\ref{eq:ch5:povm2}). Therefore, they may not be mutually orthogonal or individually idempotent.

\subsection{POVM for Unambigous Quantum State Discrimination}\label{sec:ch5:povmd}
	
		In this subsection, we study an application of POVMs by focusing on quantum state detection. To this end, assume that, unbeknownst to us, the state of a quantum system can be either   $|\psi_{0}\rangle$ or  $|\psi_{1}\rangle$. 
		 If the vectors $|\psi_{0}\rangle$ and $|\psi_{1}\rangle$ are orthogonal,
		it is possible to distinguish them exactly (with probability 1) by
		using the standard von Neumann measurement defined by the projection matrices
		\begin{align}
			\Pi_{0} & =|\psi_{0}\rangle\langle\psi_{0}|\nonumber\\
			\Pi_{1} & =|\psi_{1}\rangle\langle\psi_{1}|\nonumber\\
			\Pi & _{2}=I-\Pi_{0}-\Pi_{1}.
		\end{align}Note that if the system consists of a single qubit, the third projection matrix $\Pi_2$ is not necessary.
		 If the true state is $|\psi_{y}\rangle$, with $y\in\{0,1\}$, this measurement
		indeed returns $y$ with probability $\textrm{tr}(\Pi_{y}|\psi_{y}\rangle\langle\psi_{y}|)=1$. What if states  $|\psi_{0}\rangle$ and $|\psi_{1}\rangle$ are not
		orthogonal?

		 In this case, we can design a POVM such that, whenever a decision
		is made, the decision is correct; but it is allowed for the measurement
		to return a ``don't know'' decision.
		 To this end, consider the POVM
		\begin{align}
			M_{0} & =a(I-|\psi_{1}\rangle\langle\psi_{1}|)\nonumber\\
			M_{1} & =a(I-|\psi_{0}\rangle\langle\psi_{0}|)\nonumber\\
			M & _{2}=I-M_{0}-M_{1},
		\end{align}
		where $a=(1+|\langle\psi_{0}|\psi_{1}\rangle|)^{-1}$.

		By (\ref{eq:ch5:povmform}), if the true state is $|\psi_{0}\rangle$, the measurement returns
		$y=0$ with probability
		\begin{align}
			\textrm{tr}(M_{0}|\psi_{0}\rangle\langle\psi_{0}|) & =a\langle\psi_{0}|(I-|\psi_{1}\rangle\langle\psi_{1}|)|\psi_{0}\rangle\nonumber\\
			& =a\cdot(1-|\langle\psi_{0}|\psi_{1}\rangle|^{2})=1-|\langle\psi_{0}|\psi_{1}\rangle|;
		\end{align}
		it returns $y=1$ with probability 
		\begin{align}
			\textrm{tr}(M_{1}|\psi_{0}\rangle\langle\psi_{0}|) & =a\langle\psi_{0}|(I-|\psi_{0}\rangle\langle\psi_{0}|)|\psi_{0}\rangle\nonumber\\
			& =a\cdot(1-1)=0;
		\end{align}
		and ``don't know'', i.e., $y=2$, with probability
		\begin{align}
			\textrm{tr}(M_{2}|\psi_{0}\rangle\langle\psi_{0}|) & =|\langle\psi_{0}|\psi_{1}\rangle|.
		\end{align} A similar situation occurs if the true state is $|\psi_{1}\rangle.$
		
		We conclude that, with the designed POVM,  there is no ambiguous detection, but a ``declaration of ignorance''
		with probability \textrm{$|\langle\psi_{0}|\psi_{1}\rangle|$:} The 
		more aligned the two vectors are, the more likely it is that the detector
		will not be able to make a definite decision.
		 
		 Importantly, in order to obtain unambiguous state discrimination as
		described in this section, it is necessary to use a POVM and projective measurements
		are not sufficient. To see this, consider the special case of single-qubit states $|\psi_0\rangle$ and $|\psi_1\rangle$. A projective measurement in this setting could only produce one of two outputs, ruling out the possibility to include a ``don't know'' decision.
		
		\section{Quantum Channels}\label{sec:ch5:quantumch}
		
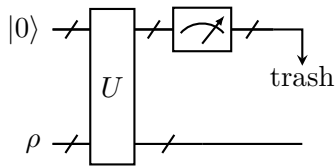
\begin{figure}
			\begin{center}
				\begin{quantikz} 
					\lstick{$\ket{0}$}& \gate[wires=2]{U}\qwbundle{} & \meter{} \qwbundle{}& \trash{\text{trash}} \qwbundle{}\\
					\lstick{$\rho$} & \qwbundle{} &  \qwbundle{} & \qw 
				\end{quantikz}
			\end{center}\vspace{-0.4cm} \caption{A quantum channel can be described as a setting in which the quantum system of interest, consisting of $n$ qubits, interacts with $n'$ ancilla, or ``environment'', qubits via a unitary $U$. The fact that the system does not have access to the environment is modelled via a non-selective measurement of the ancillas (i.e., the output of the measurement is discarded).\label{fig:ch5:qch}}
		\end{figure}

		In the previous section we have introduced POVMs -- the most general form of quantum measurement -- by studying the setting in Fig. \ref{fig:ch5:povm}. In it, the quantum system of interest, encompassing $n$ qubits in some state $\rho$, interacts with $n'$ ancilla qubits through a general unitary transformation $U$, and then a \emph{selective} measurement is made on the ancillas. Following the discussion in Sec. \ref{sec:ch3:decoh}, the ancilla qubits may model the environment in which the quantum computer of $n$ qubits operates. In this case, the unitary $U$ accounts for the unwanted, entangling, interactions with the environment. Furthermore, the fact that the quantum computer does not have access to the environment can be accounted for by assuming that the ancillas, rather than being measured selectively as in Fig. \ref{fig:ch5:povm}, are measured \textbf{non-selectively} as shown in Fig. \ref{fig:ch5:qch}.
		
		The setting in Fig. \ref{fig:ch5:qch} -- in which the system of interest interacts with ``environment'' qubits, initially in the ground state $|0\rangle$, through a unitary and then the environment qubits are measured non-selectively -- defines a \textbf{quantum channel}. Quantum channels are used to model the impact of noise due to decoherence in quantum computers (see Sec. \ref{sec:ch3:decoh}). More broadly, as we will further discuss in this section, quantum channels represent general quantum operations on a quantum state $\rho$ that do not involve selective measurements.

		\subsection{Defining Quantum Channels}
		From Sec. \ref{sec:ch5:povm}, we know that the state at the output of the unitary $U$ in Fig. \ref{fig:ch5:qch} is given by (\ref{eq:ch5:povmfc}), which we rewrite here for convenience as \begin{equation}\label{eq:ch5:qc1} \sum_{y=0}^{N'-1}|y\rangle\otimes K_{y}|\psi\rangle,
		\end{equation} where we have $N'=2^{n'}$ and the $N\times N$ matrices $K_y$ satisfy condition (\ref{eq:ch5:povm1}), i.e., 	\begin{equation}\label{eq:ch5:povmfcx}
			\sum_{y=0}^{N'-1}K_{y}^{\dagger}K_{y}=I.
		\end{equation} Unlike Sec. \ref{sec:ch5:povm}, we now perform a \textbf{non-selective} measurement of the ancillas. The outcome of this operation on the $n$ qubits of interest can be formally described by the partial trace operation introduced in Sec. \ref{sec:ch3:partialtrace}.
		
		Before applying the partial trace, let us follow an equivalent and more direct argument by assuming that the ancillas are measured in the computational basis $\{|y\rangle\}_{y=0}^{N'-1}$. Recall from Sec. \ref{sec:ch3:partialtrace} that the impact of a non-selective measurement on the rest of the system does not depend on the specific choice of the orthonormal basis. By the generalized Born rule, from (\ref{eq:ch5:qc1}), we obtain the post-measurement state $K_{y}|\psi\rangle/\sqrt{p_y}$ with probability $p_y= \langle \psi | K^\dagger_{y} K_{y}|\psi\rangle$ for the system of $n$ qubits. It follows that, from the viewpoint of the system of interest, which does not have access to the measurement outcome $y$, the state of the system is described by the density matrix	\begin{align}
			\rho_{out} & = \sum_{y=0}^{2^{n'}-1} p_y \frac{K_{y}|\psi\rangle \langle \psi |K_{y}^{\dagger}}{p_y}\nonumber \\
			& =\sum_{y=0}^{2^{n'}-1}K_{y}\rho K_{y}^{\dagger},
		\end{align}with $\rho=|\psi\rangle \langle \psi|$.

		This result can also be obtained by tracing out the ancillas -- denoted as $A$ -- to account for the non-selective measurement
		of the ancillas in Fig. \ref{fig:ch5:qch}. In fact, using the definition (\ref{eq:ch3:ptrace}) of partial trace, we compute the output state as
		\begin{align}
			\rho_{out} & =\textrm{tr}_{A}(U(|0\rangle\langle0|\otimes\rho)U^{\dagger})\nonumber\\
			& =\sum_{y=0}^{2^{n'}-1}(\langle y|\otimes I)(U(|0\rangle\langle0|\otimes\rho)U^{\dagger})(|y\rangle \otimes I)\\
			& =\sum_{y=0}^{2^{n'}-1}K_{y}\rho K_{y}^{\dagger},
		\end{align}where we have used the equality $K_y=U_{y0}=(\langle y |\otimes I) U$, which follows from (\ref{eq:ch5:povmfc}).

		To summarize, a \textbf{quantum channel} operating on $n$ qubits is defined by a set of $2^n \times 2^n$ matrices $\{K_{y}\}_{y=0}^{r-1}$, for some integer $r$, which satisfy the property   (\ref{eq:ch5:povmfc}), restated here as
		\begin{equation}\label{eq:ch5:povmfc1}
			\sum_{y=0}^{r-1}K_{y}^{\dagger}K_{y}=I.
		\end{equation}The matrices $K_y$ are known as \textbf{Kraus matrices}. 
		Given an input state $\rho$, the output state produced by a quantum channel defined by Kraus matrices $\{K_{y}\}_{y=0}^{r-1}$ is given as 
		\begin{equation}\label{eq:ch5:qchinout}
			\rho_{out}=\sum_{y=0}^{r-1}K_{y}\rho K_{y}^{\dagger}=\mathcal{N}(\rho_{in}),
		\end{equation}where the notation $\mathcal{N}(\cdot)$ is often used to describe the operation of the channel.
		
		By (\ref{eq:ch5:qchinout}), one can intuitively think of a quantum channel as a quantum operation that modifies the input
		density $\rho$ to the (unnormalized)
		density $K_{y}\rho K_{y}^{\dagger}$ with probability $p_y$. Therefore, each Kraus matrix $K_{y}$ can be interpreted as a different type of error that the channel causes on the quantum state $\rho$.
		Unless we have $r=1,$ quantum channels are generally \textbf{irreversible}, as they
		describe situations in which the system gets entangled with  the ``environment'' causing information to be irretrievably lost.
		
		%
		%
		
		\subsection{Quantum Channels as CPTP Operators}
		An equivalent way to define a quantum channels $\rho_{out}=\mathcal{N}(\rho)$ is as follows. A quantum channel is a mapping $\rho_{out}=\mathcal{N}(\rho)$ from a density matrix $\rho$ to a matrix $\rho_{out}$ that is \textbf{linear, completely positive (CP), and 			 trace preserving (TP)}. Therefore, quantum channels are also referred to as \textbf{CPTP operators}. Linearity is evident from the characterization (\ref{eq:ch5:qchinout}). The TP property requires that the output state has unitary trace, i.e., $\textrm{tr}(\rho_{out})=1$ -- a necessary condition since the output $\rho_{out}$ must be a density matrix. This property can be directly proved based on the condition (\ref{eq:ch5:povmfc1}).  Finally, the CP condition means that, when applied on a subset of a larger		system, the quantum channel yields a valid density, and hence a positive semidefinite matrix, for the whole system. This is a necessary condition to guarantee that, if the channel is applied to only to some qubits of a system, the resulting output matrix is still a valid density state.
		
		%
		%
		%
		%

		
		\subsection{Examples of Quantum Channels}
		
		Quantum channels, i.e., CPTP operators, include all quantum operations that do not involve
		a selective measurement, including trace and partial trace. In fact, CPTPs and POVMs -- with the latter used to describe selective measurements -- exhaust the set of \emph{all} possible operations that can be applied to a quantum system. 	Examples of quantum channels include the following.

		\noindent $\bullet$ \textbf{Quantum state preparation}: This is the process of preparing a quantum ensemble $\{p_{x},|v_{x}\rangle\},$
		i.e., a density matrix $\rho_{out}=\sum_{x}p_{x}|v_{x}\rangle\langle v_{x}|$. It can be described as a quantum channel with input $\rho=1$ and Kraus operators $K_{x}=\sqrt{p_{x}}|v_{x}\rangle$,
		yielding the output state \textrm{$\rho_{out}=\sum_{x} p_{x}|v_{x}\rangle\langle v_{x}|$}.
		
		\noindent $\bullet$ \textbf{Pauli channels}: Pauli channels are single-qubit channels
		that have Kraus matrices
		\begin{equation}
			K_{0}=\sqrt{p_{0}}I,\textrm{}K_{1}=\sqrt{p_{1}}X,\textrm{}K_{2}=\sqrt{p_{2}}Y,\textrm{}K_{3}=\sqrt{p_{3}}Z,
		\end{equation}for $p_{k}\geq0$ with $\sum_{k=0}^{3}p_{k}=1$, 
		yielding the output state
		\begin{equation}
			\rho_{out}=p_{0}\rho+p_{1}X\rho X+p_{2}Y\rho Y+p_{3}Z\rho Z.
		\end{equation}
		Single-qubit channels are practically important since most error models
		in quantum computing assume independent errors across qubits.
		Pauli channels make particularly clear the interpretation mentioned above of individual Kraus operators as
		describing the different types of errors that can occur on a qubit. 
		In particular, in a Pauli channel, the different types of errors correspond to the application
		of $X,$ $Y,$ or $Z$ operators to the qubit. 
		
		Pauli channels specialize to some important subclasses of channels, such as the bit-flip channel and the dephasing channel.  The \textbf{bit-flip channel} applies $X$ errors with some probability $p$, i.e., 
		\begin{equation}
			\rho_{out}=(1-p)\rho +pX\rho X,
		\end{equation}
		where we can assume without loss of generality $p\leq0.5$.  (Otherwise,
		we can apply an $X$ operator to $\rho$ to guarantee this condition.) In contrast, \textbf{dephasing channels} apply  $Z$ errors with probability $p\leq0.5$, i.e., 
		\begin{equation}
			\rho_{out}=(1-p)\rho +pZ\rho Z.
		\end{equation}

\section{Conclusions}
In this chapter, we have extended the formalism of quantum measurements to projective measurements and to POVMs, with the latter providing the most general form of quantum measurement. Throughout, we have emphasized quantum circuits that can implement such measurements via standard measurements in the computational basis. We have also highlighted two important applications: quantum error correction for projective measurements and quantum detection for POVMs. Finally, we have studied a general form of non-selective measurements, leading to the definition of quantum channels. 

\section{Recommended Resources}
Projective measurements and POVMs are well explained in \cite{nielsen10,wilde2013quantum}. Quantum error correction is a vast field, and the books \cite{mermin2007quantum,nielsen10} provide excellent introductions. Quantum communications and networking are covered by \cite{wilde2013quantum,rohde2021quantum}.
\section{Problems}
	
	\begin{enumerate}
		
		\item Argue that, in order to implement a ``gentle'' projective measurement that leaves non-trivial subspaces of pre-measurement states unchanged, it is necessary to have $n\geq 2$ qubits.
		
		\item For an observable $O$ and two states $|\psi\rangle=\sum_{x}\sqrt{p_{x}}|x\rangle$
		and $\rho=\sum_{x}p_{x}|x\rangle\langle x|$, show the relationship
		\begin{equation}
			\langle O\rangle_{|\psi\rangle}=\langle O\rangle_{\rho}+\sum_{x,y:x\neq y}\sqrt{p_{x}p_{y}}\langle x|O|y\rangle.\end{equation} Interpret this equality.
		
		\item Consider the observables $X\otimes X$, $Y\otimes Y$, and $Z\otimes Z$,
		and show that the numerical outputs of their observations can be obtained
		by using either local measurements or joint measurements in the
		Bell basis.
		\item  Derive the projective measurement that
		has the minimum error probability for the problem of distinguishing between
		two density states $\rho_0$ and $\rho_1$.
		\item Describe the measurement-based definition of the quantum entropy (see recommended resources).
		\item Derive a necessary condition on the numbers $k$, $n$, and $m$ of information qubits, encoded qubits, and types of errors, that is required for quantum error detection.
		\item Describe a circuit that measures the observable $X \otimes X$ in a system with two qubits. Consider the implementation involving only the two qubits, as well as the implementation that includes ancilla qubits. 
		\item Describe a quantum error correction code, including encoder and decoder, for the channel that may flip the phase of at most one out of three qubits. A phase flip corresponds to the application of a Pauli $Z$ operator. [Hint: Phase flips correspond to bit flips with a suitable change of basis.]
		\item Prove that the circuit in Fig. \ref{fig:ch5:partialproj} implements a balanced projective measurement by calculating the probability of obtaining output $y$, as well as the corresponding post-measurement state.
		\item Given a state $a |00\rangle + b |11\rangle$ with real numbers $a\geq b >0$ such that the equality $a^2+b^2=1$ holds, show that with probability $2b^2$ the POVM described by positive semidefinite matrices $M_0=b^2/a^2 |0\rangle \langle 0|+|1\rangle \langle 1|$ and $M_1=I-M_0$ produces as post-measurement state (\ref{eq:ch5:POVMpm}) a fully entangled state.
		\item Show that repeating a projective measurements returns the same result of the first measurement with probability $1$ (when conditioned on the output of the first measurement), while this is not the case for more general POVMs.
		\item Prove that a quantum channel is trace preserving.
		\item Explain the following statement: A noiseless quantum channel is given by the unitary transformation $U=\sum_{x=0}^{2^n-1}|x\rangle\langle x|$, while a noiseless classical channel is a noisy quantum channel with Kraus operators $K_{x}=|x\rangle\langle x|$ for $x\in\{0,1,...,2^{n}-1\}$.
	\end{enumerate}

\chapter{Quantum Machine Learning}

\section{Introduction}

In this chapter, we introduce the topic of quantum machine learning. We  start by presenting a taxonomy of approaches, and proceed by reviewing the key concepts of parametrized quantum circuit (PQC) and ansatz. We then describe the general form of cost functions optimized in quantum machine learning, which depend on the expected values of observables. With this background, as a preliminary step, we describe the variational quantum eigensolver (VQE), which is presented as a tool to address a specific class of combinatorial problems. VQEs share with quantum machine learning techniques the main driving idea of using classical optimization to minimize the expectation of an  observable as the cost function. We then overview probabilistic and deterministic machine learning models obtained from PQCs, with applications to unsupervised generative learning, as well as to supervised learning for regression and classification tasks. We conclude with further discussions regarding the choice of the PQC and the problem of data encoding onto quantum states, ending with some considerations on future directions for research.

\section{What is Quantum Machine Learning?}

Quantum computing algorithms, including the techniques reviewed in Chapter 4, have been traditionally designed by hand without imposing restrictions on the number and on the reliability of qubits, quantum gates, and quantum measurements. In practice, current  quantum computers support a few tens of qubits, and projections are that in the near term the number of qubits will not exceed around 1000. Furthermore, quantum state preparation, quantum gates, and quantum measurements in real quantum computers are subject to decoherence and noise (see Sec. \ref{sec:ch3:decoh} and Sec. \ref{sec:ch5:quantumch}). This makes it necessary to implement forms of quantum error correction in order to emulate the noiseless operation assumed by formally designed quantum algorithms. And this, in turn, further increases the number of necessary physical qubits to ensure the desired level of reliability (see Sec. \ref{sec:ch5:qec}).

\begin{figure}
\begin{center}
	\includegraphics[clip,scale=0.25]{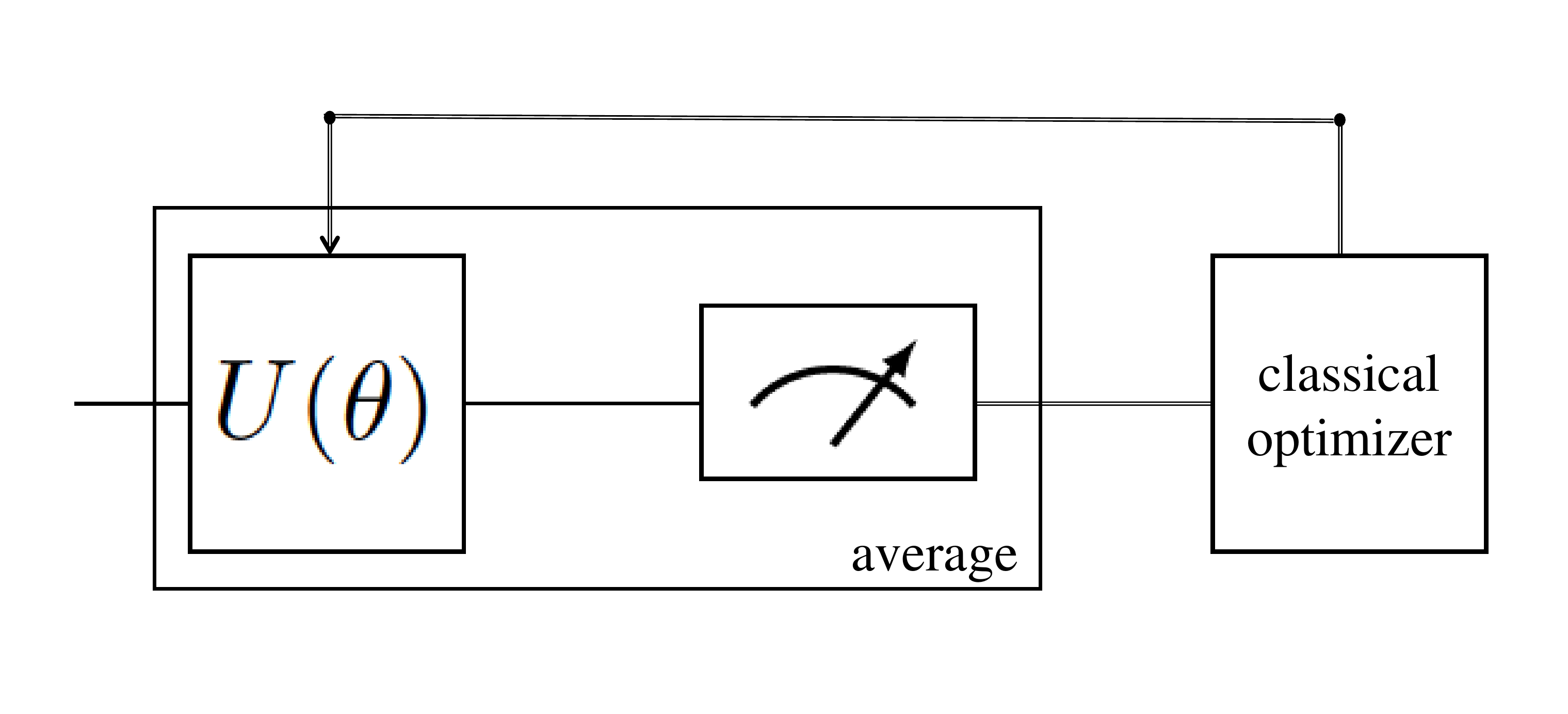}
	\par\end{center}\vspace{-0.4cm}\caption{A high-level description of the quantum machine learning design methodology. A parametric quantum circuit (PQC) implementing a unitary matrix $U(\theta)$ is optimized via its vector of parameters, $\theta$,  based on measurements of the output of the PQC. \label{fig:ch6:qmlintro}}
\end{figure}

Against this backdrop, \textbf{quantum machine learning} refers to an alternative design paradigm, illustrated in Fig. \ref{fig:ch6:qmlintro}, based on the  integration of a small-scale quantum circuit with a classical optimizer. Like classical machine learning, quantum machine learning follows an \textbf{inductive} methodology based on the following two-step approach: \begin{itemize}
	\item  \textbf{Selection of the architecture of a parametric quantum circuit (PQC), also known as ansatz}: The designer first selects the architecture of a  PQC by specifying a sequence of parametrized quantum gates. Parametrized quantum gates may, for instance, include Pauli rotations whose angles are treated as free parameters that can be optimized.  Overall, the operation of the PQC is defined by a unitary matrix $U(\theta)$, which is dependent on a vector of free parameters $\theta$.	The choice of the PQC should be ideally dictated by knowledge available to the designer about quantum algorithmic architectures that are well suited for the problem at hand. The step of selecting a PQC architecture mirrors the problem of choosing a neural network architecture in classical machine learning.
	\item \textbf{Parametric optimization}: The PQC implementing the unitary $U(\theta)$ is connected to a classical optimizer. The optimizer is fed measurements of the quantum state produced by the PQC, typically in the form of estimated expectations of observables; and it produces updates to the parameter vector $\theta$. The updates are aimed at minimizing some cost function, which may depend also on training data. This parametric optimization step is common to the classical machine learning methodology, in which the model parameters $\theta$ are optimized via gradient descent-based methods by targeting the minimization of the training loss.
\end{itemize}



As mentioned in the Preface, the quantum machine learning framework depicted in Fig. \ref{fig:ch6:qmlintro} has a number of potential advantages over the standard approach of handcrafting quantum algorithms assuming fault-tolerant quantum computers: \begin{itemize} 
	\item By keeping the quantum computer in the loop, the classical optimizer 
can directly account for the non-idealities and limitations of the available quantum resources.
\item If the PQC $U(\theta)$ is sufficiently flexible and the classical optimizer sufficiently effective, the approach may automatically design well-performing quantum algorithms that would have been hard to conceive  ``by hand'' through a formal design methodology.
\end{itemize}

\section{A Taxonomy of Quantum Machine Learning}

Unlike classical machine learning, in which data and processing are classical, in quantum machine learning
data and/or processing are quantum. The four possible scenarios are summarized in Table \ref{table:ch6:table1}, and are reviewed in this section.

\begin{table}
\begin{center}
	\begin{tabular}{cccc}
		&  & \textbf{data} & \textbf{processing}\tabularnewline
		&  & C & Q\tabularnewline
		\cmidrule{3-4} \cmidrule{4-4} 
		\textbf{data} & C & CC & CQ\tabularnewline
		\cmidrule{3-4} \cmidrule{4-4} 
		\textbf{generation} & Q & QC & QQ\tabularnewline
		\cmidrule{3-4} \cmidrule{4-4} 
	\end{tabular}
	\par\end{center}\vspace{-0.4cm}\caption{Taxonomy of machine learning settings as a function of the type of data and processing (C=classical, Q=quantum). \label{table:ch6:table1}}
\end{table}

\begin{figure}
\begin{center}
	\includegraphics[clip,scale=0.35]{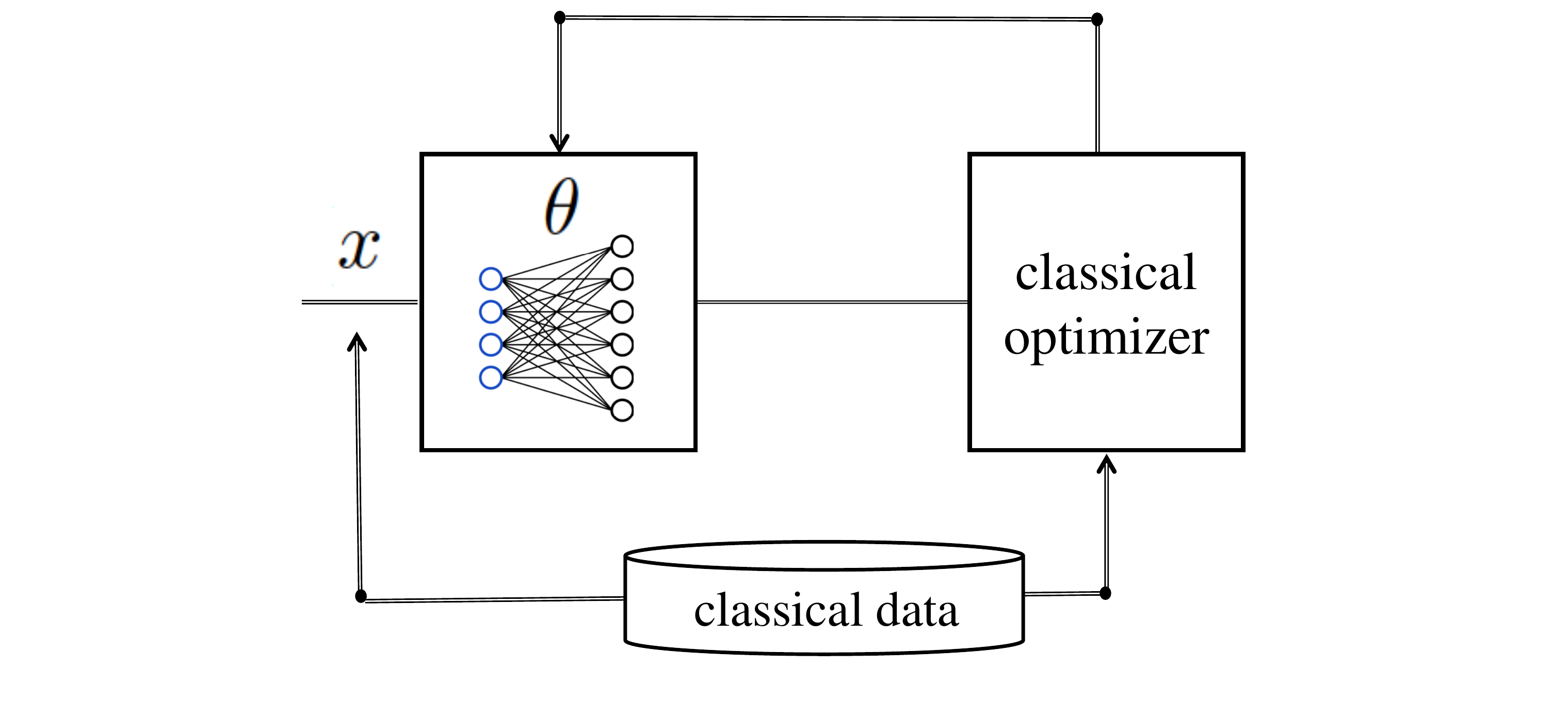}
	\par\end{center}\vspace{-0.4cm} \caption{An illustration of classical machine learning, in which both data and processing are classical. \label{fig:ch6:cc}}
\end{figure}

The \textbf{classical data and classical processing (CC)} case prescribes the
optimization of a standard machine learning model, e.g., of a neural
network, via a classical optimizer.
As illustrated in Fig. \ref{fig:ch6:cc}, a classical machine learning model implements
a parametrized function $h(x|\theta)$ mapping classical input data
$x$ to an output. The classical optimizer minimizes a cost function dependent on classical data over vector $\theta$. For example, in supervised learning, classical data consist of pairs of inputs $x$ and corresponding desired outputs $y$. Moreover, the cost function is given by the training loss, which measures how well the predictions of the outputs produced by function $h(x|\theta)$ match the desired outputs $y$ in the available data. 

\begin{figure}
\begin{center}
	\includegraphics[clip,scale=0.35]{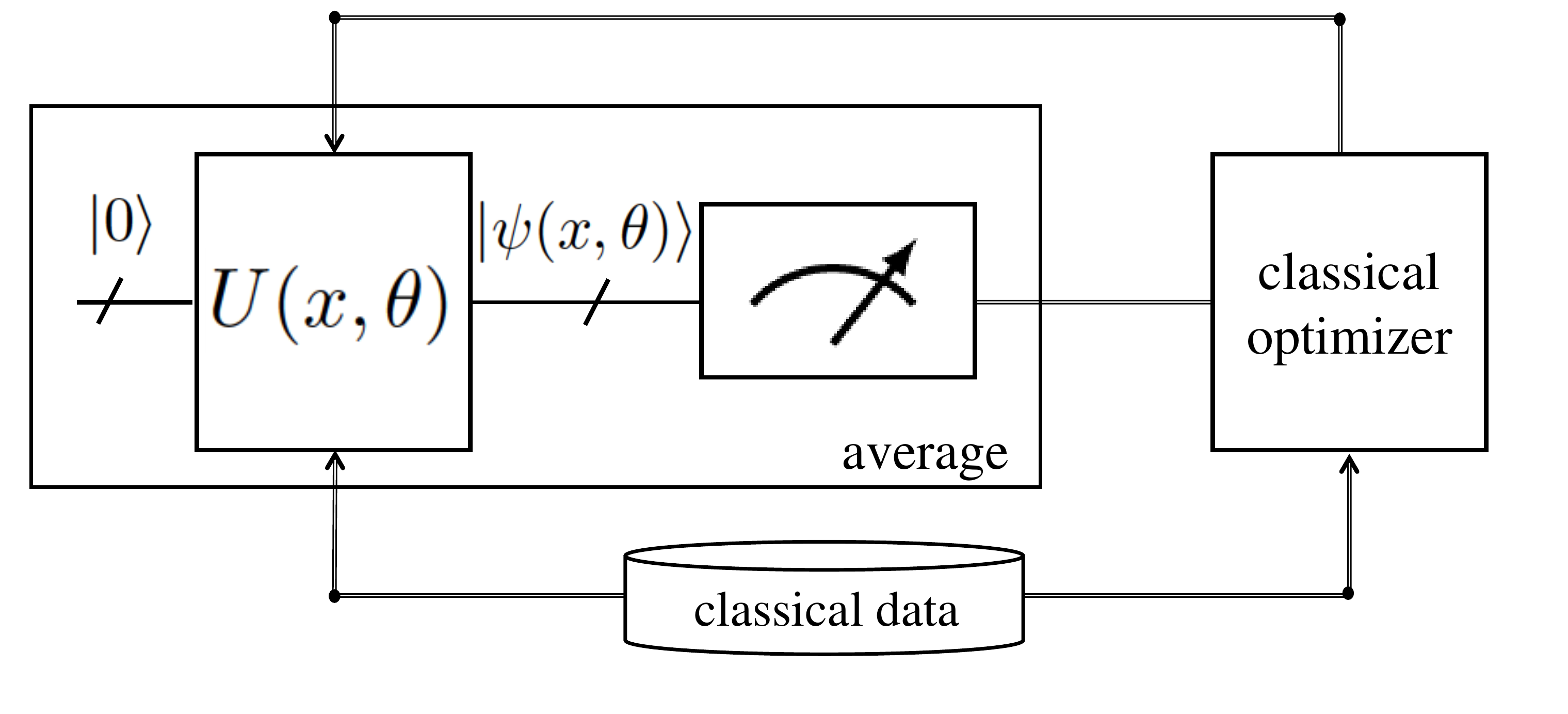}
	\par\end{center}\vspace{-0.4cm}\caption{An illustration of the ``CQ'' setting of quantum machine learning, in which data are classical and processing is quantum.\label{fig:ch6:cq}}
\end{figure}

Currently, the most common quantum machine learning setting is characterized by \textbf{classical data and quantum processing (CQ)}. Taking the case of supervised learning as an example,  as illustrated in Fig. \ref{fig:ch6:cq}, 
a quantum model is typically defined by a PQC that implements a unitary transformation $U(x,\theta)$. The unitary $U(x,\theta)$  depends on both classical input $x$ and 
 model parameter vector $\theta$. 
The input to the PQC is a set of qubits in the
ground state $|0\rangle$, and the output state produced by the PQC is given by  \begin{equation}\label{eq:ch6:embedding}|\psi(x,\theta)\rangle=U(x,\theta)|0\rangle.
\end{equation} The output state (\ref{eq:ch6:embedding}) is also known as quantum embedding of the classical data $x$, as it encodes the classical input $x$ into a quantum state. The measurement output, possibly after averaging, encodes the model's prediction of the target variable $y$.  
The classical optimizer minimizes some cost function over the model parameter vector $\theta$, which measures how well the output of the measurement of the quantum state (\ref{eq:ch6:embedding}) matches the target $y$. The cost function is expressed in terms of the expectation of an observable for the quantum embedding 
$|\psi(x,\theta)\rangle$. Estimating the expectation of the given cost-defining observable requires running the PQC multiple times in order to evaluating an empirical average.

\begin{figure}
	\begin{center}
		\includegraphics[clip,scale=0.35]{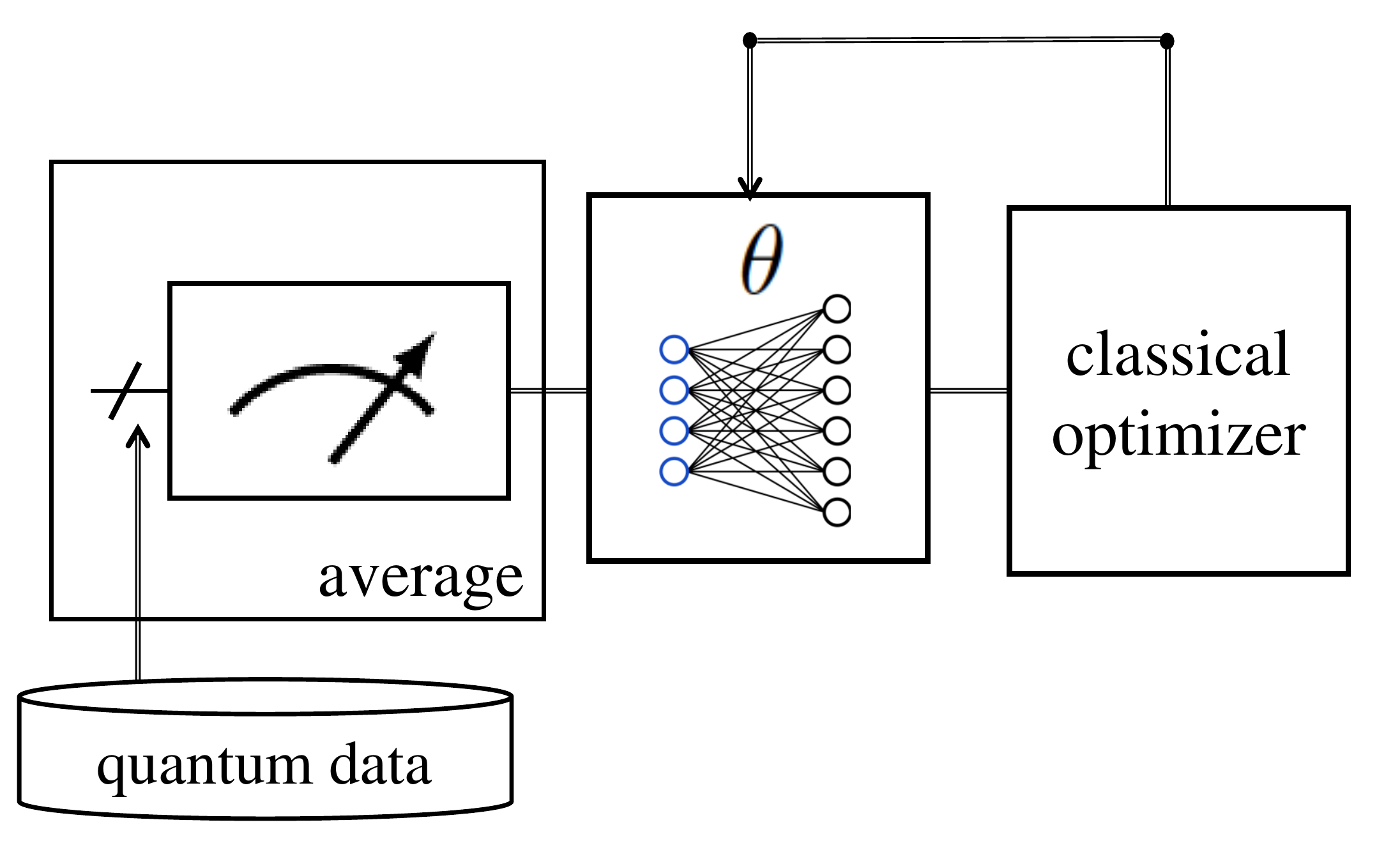}
		\par\end{center}\vspace{-0.4cm}\caption{An illustration of the ``QC'' setting of quantum machine learning, in which data are quantum and processing is classical.\label{fig:ch6:qc}}
\end{figure}

As illustrated in Fig. \ref{fig:ch6:qc}, with \textbf{quantum data and classical processing (QC)}, the learner has access to quantum data.  Quantum data consist of a collection of quantum systems, whose state is described by a
density matrix. Quantum data may be produced by a quantum sensor and stored in a quantum memory -- with both quantum sensing and quantum storage being two active areas of research.  In the QC case, quantum data are first measured, and then the classical measurement outputs are processed by a classical machine learning model. QC machine learning may be applied to quantum tomography, which is the problem of inferring properties of a quantum state based on measurement results. 

\begin{figure}
\begin{center}
	\includegraphics[clip,scale=0.35]{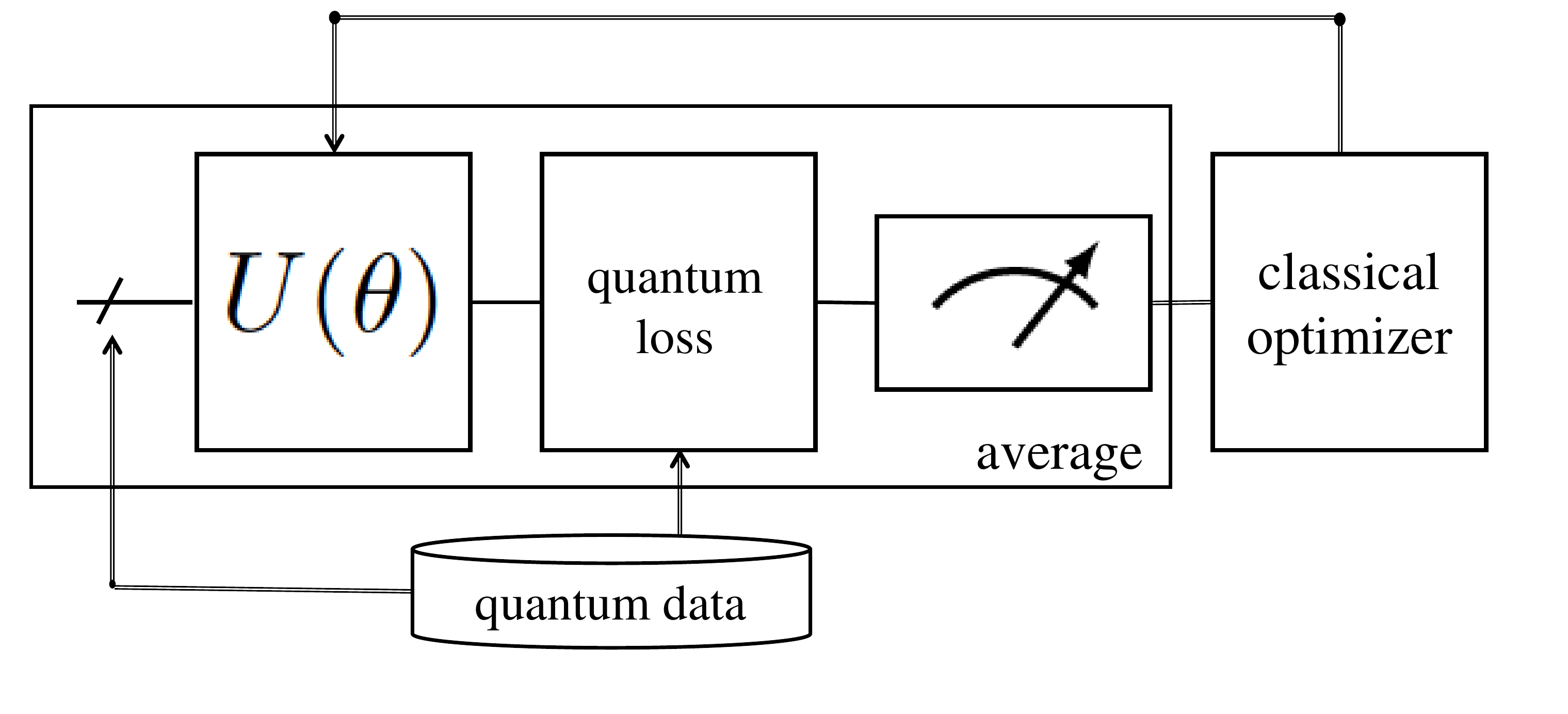}
	\par\end{center}\vspace{-0.4cm}\caption{An illustration of the ``QQ'' setting of quantum machine learning, in which data are quantum and processing is quantum.\label{fig:ch6:qq}}
\end{figure}

In the \textbf{quantum data and quantum processing (QQ)} case, as depicted in Fig. \ref{fig:ch6:qq}, quantum data are processed by a PQC. The cost function is evaluated within the quantum computer by comparing, using a suitable quantum circuit (see, e.g., Sec. \ref{sec:ch3:spiking}), the output of the PQC with  target quantum data. An instance of QQ machine learning is given by  \textbf{quantum generative adversarial networks}, in which the quantum
loss is computed via a detector that attempts to distinguish real
quantum data from quantum states generated by the PQC. Another example is provided by \textbf{quantum variational autoencoders}, which aim at compressing a quantum state into a smaller set of qubits. We refer to the recommended resources (Sec. \ref{sec:ch6:rr}) for pointers to the literature.

In this chapter, we will focus exclusively on the CQ case, which is better studied and potentially more suitable for near-term engineering applications. That said, the discussion in the next section regarding PQCs is also relevant for the QQ solutions. 

\section{Ansatz and Parametrized Quantum Circuits}\label{sec:ch6:PQC}

As discussed in the previous section, at the core of quantum machine learning methods are PQCs, which are the topic of this section.

A PQC implements a unitary transformation $U(\theta)$ whose operation on $n$ qubits depends
on a tunable (classical) parameter vector $\theta$ consisting of $D$ real numbers. As  seen in the previous section, in the CQ case, the unitary may also depend on input data $x$, and we
will discuss this situation in the context of supervised learning in Sec. \ref{sec:ch6:sl}.
The unitary $U(\theta)$ is typically specified as a sequence of one-
or two-qubit gates, with each gate possibly dependent on the parameter
$\theta.$


The choice of the PQC architecture as a sequence of specific
gates is akin to the choice of the \textbf{model class} in classical machine
learning, e.g., of a neural network architecture.
In quantum machine learning, we refer to the architecture of the PQC as the \textbf{ansatz}.  The term comes from the German word used for ``approach'' or ``attempt''. As for the model class in machine learning, one should choose the
ansatz, if possible, based on domain knowledge. 
For instance, in quantum chemistry problems,  there are known algorithmic structures that
are known to reflect well the nature of certain phenomena.

In the rest of this section, we review typical ansatzes by proceeding in order of complexity, and hence also of expressivity, of the resulting PQC. Specifically, we first review parametrized single-qubit gates and the mean-field ansatz; then we describe the more complex hardware-efficient ansatz; and finally we discuss parametrized multi-qubit gates.

\subsection{Parametrized Single-Qubit Gates}

The most common type of parametrized quantum gates is given by single-qubit \textbf{Pauli rotation matrices}. A Pauli $P$-rotation matrix $R_P(\theta)$, with $P\in\{X,Y,Z\}$, is a unitary matrix whose generator  is given by $\theta/2\cdot P$, i.e., \begin{equation}\label{eq:ch6:Pauli}
	R_{P}(\theta)=\exp\left(-i\frac{\theta}{2} P\right)=\cos\left(\frac{\theta}{2}\right)I-i\sin\left(\frac{\theta}{2}\right)P,
	\end{equation} where  $\theta$ is a real parameter (see Sec. \ref{sec:ch1:gen} and  Appendix A). The parameter $\theta$ is also known as the \textbf{rotation angle} for the gate $R_P(\theta)$. The three Pauli rotations are summarized in Table \ref{table:ch6:Pauli}. 

By (\ref{eq:ch6:Pauli}), we have the equalities
\begin{equation}\label{eq:ch6:Rx}
R_{X}(-\pi)=iX,
\end{equation}
\begin{equation}
R_{Y}(-\pi)=iY,
\end{equation}
and 
\begin{equation}\label{eq:ch6:Rz}
R_{Z}(-\pi)=iZ.
\end{equation}
Therefore, apart from a global phase, the Pauli rotations recover as special
cases, i.e., for a specific choice of the rotation angle $\theta$, the Pauli gates $X$, $Y$, and $Z$.

\begin{table}

\begin{center}
	{\small{}}%
	\begin{tabular}{cc}
		\toprule 
		{\small{}Pauli rotation} & {\small{}definition}\tabularnewline
		\midrule
		\midrule 
		{\small{}$R_{X}(\theta)$} & {\small{}$\exp\left(-i\frac{\theta}{2}X\right)=\left[\begin{array}{cc}
				\cos\left(\frac{\theta}{2}\right) & -i\sin\left(\frac{\theta}{2}\right)\\
				-i\sin\left(\frac{\theta}{2}\right) & \cos\left(\frac{\theta}{2}\right)
			\end{array}\right]$}\tabularnewline
		\midrule 
		{\small{}$R_{Y}(\theta)$} & {\small{}$\exp\left(-i\frac{\theta}{2}Y\right)=\left[\begin{array}{cc}
				\cos\left(\frac{\theta}{2}\right) & -\sin\left(\frac{\theta}{2}\right)\\
				\sin\left(\frac{\theta}{2}\right) & \cos\left(\frac{\theta}{2}\right)
			\end{array}\right]$}\tabularnewline
		\midrule 
		{\small{}$R_{Z}(\theta)$} & {\small{}$\exp\left(-i\frac{\theta}{2}Z\right)=\left[\begin{array}{cc}
				\cos\left(\frac{\theta}{2}\right)-i\sin\left(\frac{\theta}{2}\right) & 0\\
				0 & \cos\left(\frac{\theta}{2}\right)+i\sin\left(\frac{\theta}{2}\right)
			\end{array}\right]$}\tabularnewline
		\bottomrule
	\end{tabular}{\small\par}
	\par\end{center}

\vspace{-0.4cm}\caption{Single-quit Pauli rotations are the most commonly used parametrized gates. \label{table:ch6:Pauli}}
\end{table}

A \textbf{general parametrized single-qubit gate} can be expressed, apart from a global
phase, as the following cascade of two types of Pauli rotations 
\begin{equation}\label{eq:ch6:gensq}
	R(\theta^{1},\theta^{2},\theta^{3})=R_{P}(\theta^{1})R_{P'}(\theta^{2})R_{P}(\theta^{3}),
\end{equation}where $P'\neq P$ with $P,P' \in \{X,Y,Z\}$. A typical choice is $P=Z$ and $P'=Y$. The general single-qubit gate (\ref{eq:ch6:gensq}) is specified by the three parameters, or rotation angles, $\theta^{1}$, $\theta^{2}$, and $\theta^{3}$.

Single-qubit gates can be applied in parallel to the qubits of a quantum
system. To indicate the qubit $k\in \{0,1,...,n-1\}$ to which a single-qubit gate is applied, it is common to add subscript $k$ to the notation for the gate. 
As an example, the notation $Z_{k}$ describes a unitary that applies the Pauli $Z$ gate only on the $k$-th qubit. This transformation amounts to a Pauli string containing all 
identity matrices $I$, of dimension $2 \times 2$, except for the position $k$, where
we have the Pauli operator $Z$, i.e., 
\begin{equation}\label{eq:ch6:Zk}
	Z_{k}=\underbrace{I\otimes\cdots\otimes I}_{k-1}\otimes Z\otimes\underbrace{I\otimes\cdots\otimes I}_{n-k}.
\end{equation}
Similarly, we can write 
\begin{equation}
	R_{k}(\theta_{k}^{1},\theta_{k}^{2},\theta_{k}^{3})=\underbrace{I\otimes\cdots\otimes I}_{k-1}\otimes R(\theta_{k}^{1},\theta_{k}^{2},\theta_{k}^{3})\otimes\underbrace{I\otimes\cdots\otimes I}_{n-k}
\end{equation}
for a general single-qubit gate with parameters $(\theta_{k}^{1},\theta_{k}^{2},\theta_{k}^{3})$
applied to the $k$th qubit.

\subsection{Mean-Field Ansatz}

\begin{figure}
\begin{center}
		\begin{quantikz} 
			\lstick[wires=4]{$\ket{0}$}\qw &\gate{R(\theta_{0}^{1},\theta_{0}^{2},\theta_{0}^{3})}& \qw\\
			\qw &\gate{R(\theta_{1}^{1},\theta_{1}^{2},\theta_{1}^{3})}& \qw \\	
			\qw &\gate{R(\theta_{2}^{1},\theta_{2}^{2},\theta_{2}^{3})}& \qw\\
			\qw &\gate{R(\theta_{3}^{1},\theta_{3}^{2},\theta_{3}^{3})}& \qw
			
		\end{quantikz}
\end{center}\vspace{-0.4cm}\caption{The mean-field ansatz consists of general single-qubit gates applied in parallel to the $n$ qubits, which are initially in the ground state $|0\rangle$ ($n=4$ in the figure). \label{fig:ch6:mf}}
\end{figure}
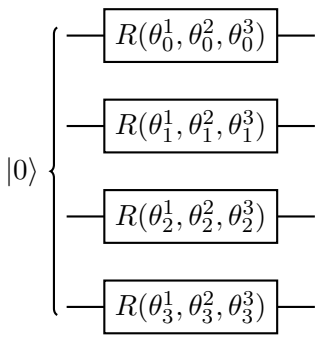

An ansatz $U(\theta)$ that uses only single-qubit gates is known
as a \textbf{mean-field ansatz}. As illustrated in Fig. \ref{fig:ch6:mf}, the mean-field ansatz defines unitaries of the form
\begin{align}
	U(\theta) & =R(\theta_{0}^{1},\theta_{0}^{2},\theta_{0}^{3})\otimes\cdots\otimes R(\theta_{n-1}^{1},\theta_{n-1}^{2},\theta_{n-1}^{3})\nonumber\\
	& =\prod_{k=0}^{n-1}R_{k}(\theta_{k}^{1},\theta_{k}^{2},\theta_{k}^{3}),
\end{align}
with $(\theta_{k}^{1},\theta_{k}^{2},\theta_{k}^{3})$ for $k\in\{0,1,...,n-1\}$
being scalar parameters. Overall, the parameter vector $\theta=[\theta_{0}^{1},\theta_{0}^{2},\theta_{0}^{3},...,\theta_{n-1}^{1},\theta_{n-1}^{2},\theta_{n-1}^{3}]^{T}$ contains $D=3n$ parameters.

Given the input state $|0\rangle$, a PQC following the mean-field ansatz outputs the separable state 
\begin{align}
	U(\theta) & |0\rangle=R(\theta_{0}^{1},\theta_{0}^{2},\theta_{0}^{3})|0\rangle\otimes\cdots\otimes R(\theta_{n-1}^{1},\theta_{n-1}^{2},\theta_{n-1}^{3})|0\rangle,
\end{align}where we have written as $|0\rangle$ both the multi-qubit ground state and the corresponding single-qubit states with some abuse of notation. 

While simple, the mean-field ansatz may be sufficient for some applications. This may be the case, for instance, if the desired output of a computation is a vector in the computational basis. In fact, a PQC following the mean-field ansatz can produce all possible vectors in the computational basis. This result can be obtained by setting
$\theta_{k}^{1}=0$ and $\theta_{k}^{3}=0$ for all qubits $k\in\{0,1,...,n-1\}$, and by  choosing parameters $\theta_{k}^{2}$ as either $\theta_{k}^{2}=0$ or $\theta_{k}^{2}=\pi$ 
to encode the two computational-basis vectors. 

More generally, however, the solution of many practical problems can benefit from ansatzes that can produce entanglement among the qubits. This can be done by introducing entangling, multi-qubit, gates in the PQC, hence moving beyond the mean-field ansatz.

\subsection{Hardware-Efficient Ansatz}

Moving one level up along the complexity axis for ansatzes, the \textbf{hardware-efficient ansatz} includes both parametrized single-qubit gates
and a \textbf{fixed entangling unitary} $U_{ent}$, which does not depend on the parameter vector $\theta$.  As detailed below, the entangling unitary $U_{ent}$ typically consist of CNOT or CZ
gates (see Sec. \ref{sec:ch2:controlled}). 
 A fixed entangling unitary is adopted for simplicity of implementation,
since two-qubit gates are generally hard to implement on quantum computers.
This motivation justifies the name ``hardware-efficient'' for this
ansatz.

\begin{figure}
\begin{center}
	\includegraphics[clip,scale=0.32]{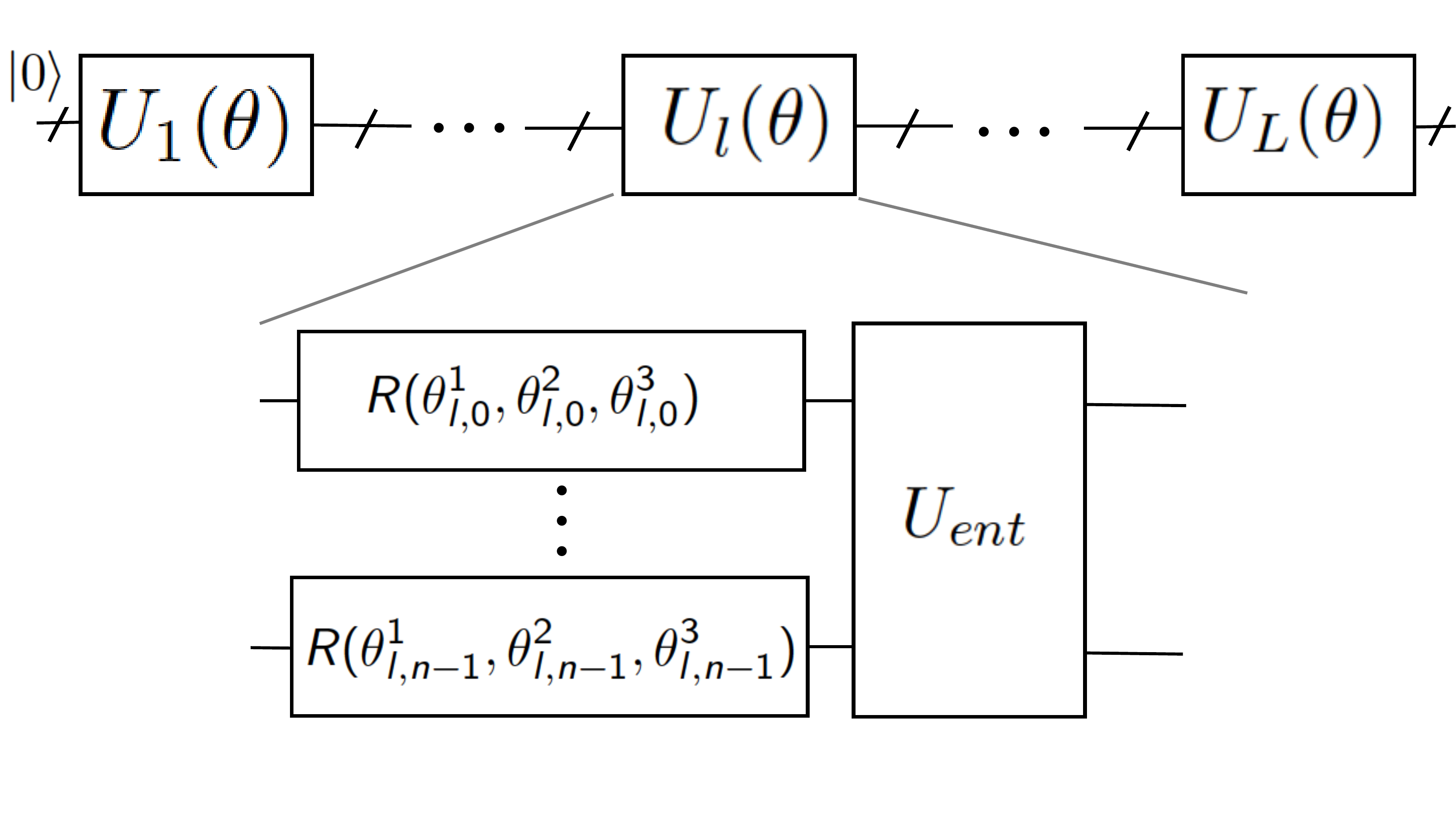}
	\par\end{center}\vspace{-0.4cm}\caption{The hardware-efficient ansatz consists of $L$ layers, each encompassing general single-qubit gates applied in parallel to the $n$ qubits and a fixed entangling circuit $U_{ent}$. The qubits are initially in the ground state $|0\rangle$. \label{fig:ch6:he}}
\end{figure}

Specifically, as illustrated in Fig. \ref{fig:ch6:he}, the hardware-efficient ansatz prescribes PQCs that implement a cascade of $L$ layers
of unitaries as in\begin{equation}\label{eq:ch6:hwa1}
	U(\theta)=U_{L}(\theta)\cdot U_{L-1}(\theta)\cdots U_{1}(\theta),
\end{equation}
where each unitary matrix $U_{l}(\theta)$ at the $l$th layer can
be expressed as
\begin{equation}\label{eq:ch6:hwa2}
	U_{l}(\theta)=U_{ent}\left(R(\theta_{l,0}^{1},\theta_{l,0}^{2},\theta_{l,0}^{3})\otimes\cdots\otimes R(\theta_{l,n-1}^{1},\theta_{l,n-1}^{2},\theta_{l,n-1}^{3})\right),
\end{equation}
with $(\theta_{l,k}^{1},\theta_{l,k}^{2},\theta_{l,k}^{3})$ for $k\in\{0,1,...,n-1\}$. The parameter vector $\theta$ hence includes  $D=3nL$ parameters, with $3n$ parameters describing each layer.

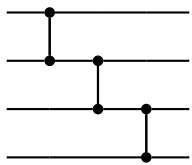
\begin{figure}
\begin{center}
		\begin{quantikz} 
			\qw & \ctrl{1} & \qw & \qw & \qw & \\
			\qw &\ctrl{-1}& \ctrl{1} & \qw & \qw &\\	
			\qw &\qw& \ctrl{-1} & \ctrl{1} &\qw \\
			\qw &\qw& \qw & \ctrl{-1} & \qw 
			
		\end{quantikz}
\end{center}\vspace{-0.4cm} \caption{A linear entangling circuit $U_{ent}$ for $n=4$ qubits implemented via CZ gates.\label{fig:ch6:linent}}
\end{figure}

\begin{figure}
\begin{center}
		\begin{quantikz} 
			\qw & \ctrl{1} & \qw & \qw & \ctrl{2} & \qw & \ctrl{3} & \qw\\
			\qw &\ctrl{-1}& \ctrl{1} & \qw & \qw & \ctrl{2} & \qw & \qw\\	
			\qw &\qw& \ctrl{-1} & \ctrl{1} &\ctrl{-2} & \qw & \qw & \qw\\
			\qw &\qw& \qw & \ctrl{-1} & \qw & \ctrl{-2} & \ctrl{-3} & \qw
			
		\end{quantikz}
\end{center}\vspace{-0.4cm} \caption{A full entangling circuit $U_{ent}$ for $n=4$ qubits implemented via CZ gates.\label{fig:ch6:fullent}}
\end{figure}
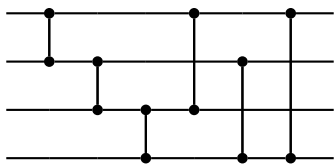

By (\ref{eq:ch6:hwa1})-(\ref{eq:ch6:hwa2}),  a PQC following the hardware-aware ansatz applies, at each layer $l$, separate single-qubit gates 
to all the $n$ qubits, as in the mean-field ansatz, followed by a 
fixed entangling unitary $U_{ent}$.  The entangling unitary $U_{ent}$ is typically implemented using linear, circular, 
or full entangling circuits: \begin{itemize}\item 
\textbf{Linear entangling circuits} implement two-qubit gates between
successive qubits. The example in Fig. \ref{fig:ch6:linent} uses CZ gates, and it can be expressed as
\begin{equation}
	U_{ent}=\prod_{k=0}^{n-2}C^{Z}_{k,k+1},
\end{equation} where $C^{Z}_{k,k+1}$ is the CZ gate between qubits $k$ and $k+1$.
\item \textbf{Circular entangling circuits} add to the architecture of linear entangling circuit an additional two-quit gate between the first and the last qubit.
\item \textbf{Full entangling circuits} implement  two-qubit gates between
all pairs of qubits, as illustrated in Fig. \ref{fig:ch6:fullent} with CZ gates. \end{itemize}
In this regard, we observe that CZ gates have the advantage over CNOT gates that they can be applied in any order since they commute.

The hardware-efficient ansatz is generic, and it is often deemed to play a similar role
to fully connected classical neural networks in the design of quantum
machine learning.
That said, it is important to stress that a hardware-efficient
ansatz does not have analogous properties to fully connected classical
neural networks when it comes to dependence on model parameters. 
Notably, in fully connected classical neural networks, one has significant
freedom in optimizing the connectivity between neurons by designing
the individual synaptic weights are biases.
In contrast, in the hardware-efficient ansatz, one can control only
the rotations applied to each individual qubit, and dependencies among
qubits are dictated by fixed entangling circuits.

\subsection{Parametrized Two-Qubit Gates}

At a higher level of complexity with respect to the hardware-efficient ansatz are PQCs that include also parametrized two-qubit gates. In this subsection, we describe two common types of parametrized two-qubit gates, namely parametrized controlled gates and two-qubit Pauli rotations.

\noindent \textbf{Parametrized two-qubit controlled gates}: Using the notation introduced in Sec. \ref{sec:ch2:controlled}, parametrized two-qubit controlled gates  are of the form
$C_{jk}^{U(\theta)}$, where $j$ is the index of controlling qubit, 
 $k$ is the index of the controlled qubit, and $U(\theta)$ is a parametrized single-qubit gate. 
Considering for simplicity of notation qubit $0$ as the controlling qubit and qubit $1$ as the controlled qubit, we can write (cf. (\ref{eq:ch2:cu}))
\begin{equation}
	C_{01}^{U(\theta)}=|0\rangle\langle0|\otimes I+|1\rangle\langle1|\otimes U(\theta).
\end{equation} Therefore, on the branch of the computation in which the controlling qubit has state $|0\rangle$, no operation is applied to controlled qubit; while, on the branch in which the controlling qubit has state $|1\rangle$, the parametrized single-qubit gate $U(\theta)$ is applied.

Instances of parametrized two-qubit gates include the \textbf{parametrized CNOT gate}
\begin{equation}
	\textrm{CNOT}_{jk}(\theta)=C_{jk}^{R_{X}(\theta)},
\end{equation}
where the controlled single-qubit gate is a Pauli $X$-rotation $R_{X}(\theta)$; and
the \textbf{parametrized CZ gate}
\begin{equation}
	\textrm{CZ}_{jk}(\theta)=C_{jk}^{R_{Z}(\theta)},
\end{equation}
where the controlled single-qubit gate is a Pauli $Z$-rotation $R_{Z}(\theta)$. Note that, by (\ref{eq:ch6:Rx})-(\ref{eq:ch6:Rz}),  the parametrized CNOT and CZ gates recover the standard CNOT and CZ gates by setting $\theta=-\pi$, apart from a phase term.

\noindent \textbf{Two-qubit Pauli rotations}: For a system with two qubits, a two-qubit Pauli $PQ$-rotation is defined as the unitary matrix
\begin{equation}\label{eq:ch6:tqPR}
	R_{PQ}(\theta)=\exp\left(-i\frac{\theta}{2}(P\otimes Q)\right),
\end{equation}
where $P$ and $Q$ are selected as one of the four Pauli operators
$\{I,X,Y,Z\}$.
A common example is the \textbf{Pauli $ZZ$-rotation gate} defined
by the unitary
\begin{equation}
	R_{ZZ}(\theta)=\exp\left(-i\frac{\theta}{2}Z\otimes Z\right).
\end{equation}

\begin{figure}
\begin{center}
		\begin{quantikz} 
			\qw & \ctrl{1} & \qw & \ctrl{1} & \qw & \\
			\qw &\gate{X}& \gate{R_Z(\theta)} & \gate{X}& \qw 
			
		\end{quantikz}
\end{center}\vspace{-0.4cm} \caption{An implementation of the Pauli $ZZ$-rotation gate \label{fig:ch6:Rzz}}
\end{figure}
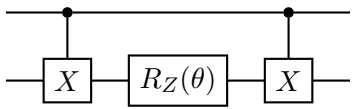

Despite the presence of the product $P\otimes Q$ in (\ref{eq:ch6:tqPR}), two-qubit
Pauli rotations are generally not local operators. In this regard, note that we have the equality  $R_{ZZ}(-\pi)=i(Z\otimes Z)$, so
that with the special choice of $\theta=-\pi$ the gate $R_{ZZ}(-\pi)$
consists of two local single-qubit operations, while this is not the case in general. For any value of $\theta$, the Pauli $ZZ$-rotation can be implemented as the
cascade of CNOT, Pauli $Z$-rotation, and CNOT gate depicted in Fig. \ref{fig:ch6:Rzz}.

\section{Cost Functions for Quantum Machine Learning}\label{sec:ch6:estgrad}

At the core of quantum machine learning methods (with quantum processing) is the minimization, using a classical optimizer, of a cost function $L(\theta)$ over the vector of PQC parameters $\theta$. As we will see in this chapter, for the CQ case, the cost function  $L(\theta)$ is typically a function of the expected values of observables evaluated under the state produced by the PQC, as well as (possibly) of classical data. In this section, we first formalize this notion of cost function, and describe some implications for optimization. Then,  we discuss the problem of estimating the gradient of the cost function $L(\theta)$. This problem is central to many implementations of quantum machine learning techniques in which the cost function $L(\theta)$ is minimized via gradient descent.

\subsection{Cost Functions via the Expected Values of Observables}

Let us denote as $|\psi(\theta)\rangle$ the state of the $n$ qubits upon application of the PQC defined by unitary matrix $U(\theta)$. Conventionally, the input of the PQC is given by the ground state $|0\rangle$, so that the output state can be written as \begin{equation} |\psi(\theta)\rangle=U(\theta)|0\rangle.
\end{equation} More generally, the operation of the PQC may also depend on input data (see Fig. \ref{fig:ch6:cq}), but for the purpose of the discussion in this section we can ignore this aspect. The cost function $L(\theta)$ to be minimized via a classical optimizer may depend on a classical data set $\mathcal{D}$, as well as on the expected values of a set of $N_O$ observables $O_1,O_2,...,O_{N_O}$, i.e., \begin{equation}\label{eq:ch6:lossfunge}
	L(\theta)=f(\mathcal{D},\{\langle O_j \rangle_{|\psi(\theta)\rangle}\}_{j=1}^{N_O}),
\end{equation} where $f(\cdot)$ is some function. 

To proceed, we recall a few basic facts about observables from Sec. \ref{sec:ch5:obs}. An observable is defined by a Hermitian matrix $O$. Measuring the observable amounts to implementing a measurement in the eigenbasis $\{|v_x\rangle\}_{x=0}^{2^n-1}$ of matrix $O$. Furthermore, each possible measurement outcome, indexed by an $n$-cbit string $x\in \{0,1\}^n$, produces a real-valued observation given by the corresponding eigenvalue $o_x$. Let us denote as $p(x|\theta)$ the probability of measuring output $x$ and hence obtaining the observation $o_x$. By Born's rule, this probability can be expressed as \begin{equation}\label{eq:ch6:probtheta}
	p(x|\theta)=|\langle v_x | \psi(\theta) \rangle|^2,
	\end{equation}where we recall that $\langle v_x|\psi(\theta)\rangle$ is the complex amplitude of the state $|\psi(\theta)\rangle$ associated with the  basis vector $|v_x\rangle$.  Importantly, the probability $p(x|\theta)$ in (\ref{eq:ch6:probtheta}) depends on the parameter vector $\theta$, since it describes measurements of the quantum state $|\psi(\theta)\rangle$. The outcome $x$ of a measurement is denoted as $x\sim p(x|\theta)$ to indicate that sample $x$ is distributed according to the distribution $p(x|\theta)$.

The expected value $\langle O \rangle_{|\psi(\theta)\rangle}$ of observable $O$ evaluated for the quantum state $|\psi(\theta)\rangle$ is defined as the average \begin{equation}
\langle O \rangle_{|\psi(\theta)\rangle} = \langle \psi(\theta)| O |\psi(\theta)\rangle = \sum_{x=0}^{2^n-1} p(x|\theta) o_x.
\end{equation}  To emphasize that this expectation is the average of random variable $o_x$ with distribution $p(x|\theta)$, we may also use the notation \begin{equation}\label{eq:ch6:notavg}
\langle O \rangle_{|\psi(\theta)\rangle} = \langle o_x \rangle_{p(x|\theta)},
\end{equation}where the subscript on the right-hand side indicates that the average is computed with respect to distribution  $p(x|\theta)$.

To elaborate further on cost functions of the form (\ref{eq:ch6:lossfunge}), in the rest of this section we will consider the basic situation in which the cost function equals the expectation of a single observable $O$ as $L(\theta)=\langle O \rangle_{|\psi(\theta)\rangle}$.  The problem of interest is the minimization \begin{equation}\label{eq:ch6:obsdiff}
\min_{\theta}\left\{L(\theta)=\langle O\rangle_{|\psi(\theta)\rangle}=\langle o_x \rangle_{p(x|\theta)}\right\}.
\end{equation}

\subsection{Gradient-Based Optimization}

Problems of the form (\ref{eq:ch6:obsdiff}) amount to the minimization of an expectation over the averaging distribution. We will refer to this type of problems as \textbf{stochastic optimization}.  Stochastic optimization is common in classical machine learning, where it underlies techniques such as variational inference, Bayesian learning, and generative modelling. In some of these methods, the distribution $p(x|\theta)$ under optimization is available explicitly and can be differentiated. In others, the distribution $p(x|\theta)$ is only available \emph{implicitly}: The model can produce samples $x \sim p(x|\theta)$, but one does not have direct access to the distribution $p(x|\theta)$ itself. 

Quantum machine learning models fall in the latter category. A PQC produces random samples $x$ distributed according to $p(x|\theta)$, by Born's rule as in (\ref{eq:ch6:probtheta}), but the distribution $p(x|\theta)$ is not explicitly computable. 

For both explicit and implicit models, solutions of stochastic problems  (\ref{eq:ch6:probtheta}) are often addressed via \textbf{stochastic gradient descent}. Accordingly,  at the current iterate $\theta$ for the model parameter vector, the classical optimizer obtains an \emph{estimate} $\hat{\nabla}L(\theta)$
of the gradient \begin{equation}\label{eq:ch6:grad}
	\nabla L(\theta)=\left[\frac{\partial L(\theta)}{\partial\theta_{1}}\cdots\frac{\partial L(\theta)}{\partial\theta_{D}}\right]^{T}
\end{equation}of the cost function $L(\theta)$. This estimate is obtained based on a number $N_S$ of samples $x\sim p(x|\theta')$ for some values $\theta'$ of the model parameter vector. Therefore, the estimate is subject to \textbf{shot noise}, that is, to the randomness of the sample generation process. 

Following the direction of the stochastic gradient estimate $\hat{\nabla}L(\theta)$, the classical optimizer  updates the parameter vector as \begin{equation}\theta\leftarrow\theta-\gamma\hat{\nabla}L(\theta) \end{equation} 
for some learning rate $\gamma>0$. Adaptive learning rates can also be readily implemented. The procedure is repeated for a number of iterations.



In classical machine learning, for explicit models, standard methods to estimate the gradient of the cost function $L(\theta)$ include the REINFORCE algorithm; while certain implicit models can be addressed by using the so-called reparametrization trick. We refer the reader to the recommended resources for details. All of these methods make use of backpropagation in the typical case where classical models are implemented via neural networks.  As we will discuss, none of these approaches are  directly applicable to quantum machine learning.


\subsection{Numerical Differentiation}
To proceed, let us first pursue the natural idea of estimating the gradient $\nabla L(\theta)$ via numerical differentiation.
Accordingly, the $d$-th partial derivative in the gradient (\ref{eq:ch6:grad}), for $d\in\{1,...,D\}$,  is first approximated as 
\begin{equation}\label{eq:ch6:approxgrad}
	\frac{\partial L(\theta)}{\partial\theta_{d}}\simeq\frac{L(\theta+\epsilon e_{d})-L(\theta-\epsilon e_{d})}{2\epsilon},
\end{equation}
where $\epsilon>0$ is a small number and $e_{d}$ is a one-hot $D$-dimensional
vector with a 1 in the $d$-th position. It can be readily checked that the error of the approximation (\ref{eq:ch6:approxgrad}) is of the order $\mathcal{O}(\epsilon^2)$, and hence it can be controlled by selecting a sufficiently small $\epsilon$. In words, the approximation (\ref{eq:ch6:approxgrad}) is obtained by \textbf{perturbing} each parameter $\theta_{d}$ by a small number $\epsilon$ in both positive and negative directions in order to evaluate
the corresponding partial derivative. 

\begin{figure}
	\begin{center}
		\includegraphics[clip,scale=0.32]{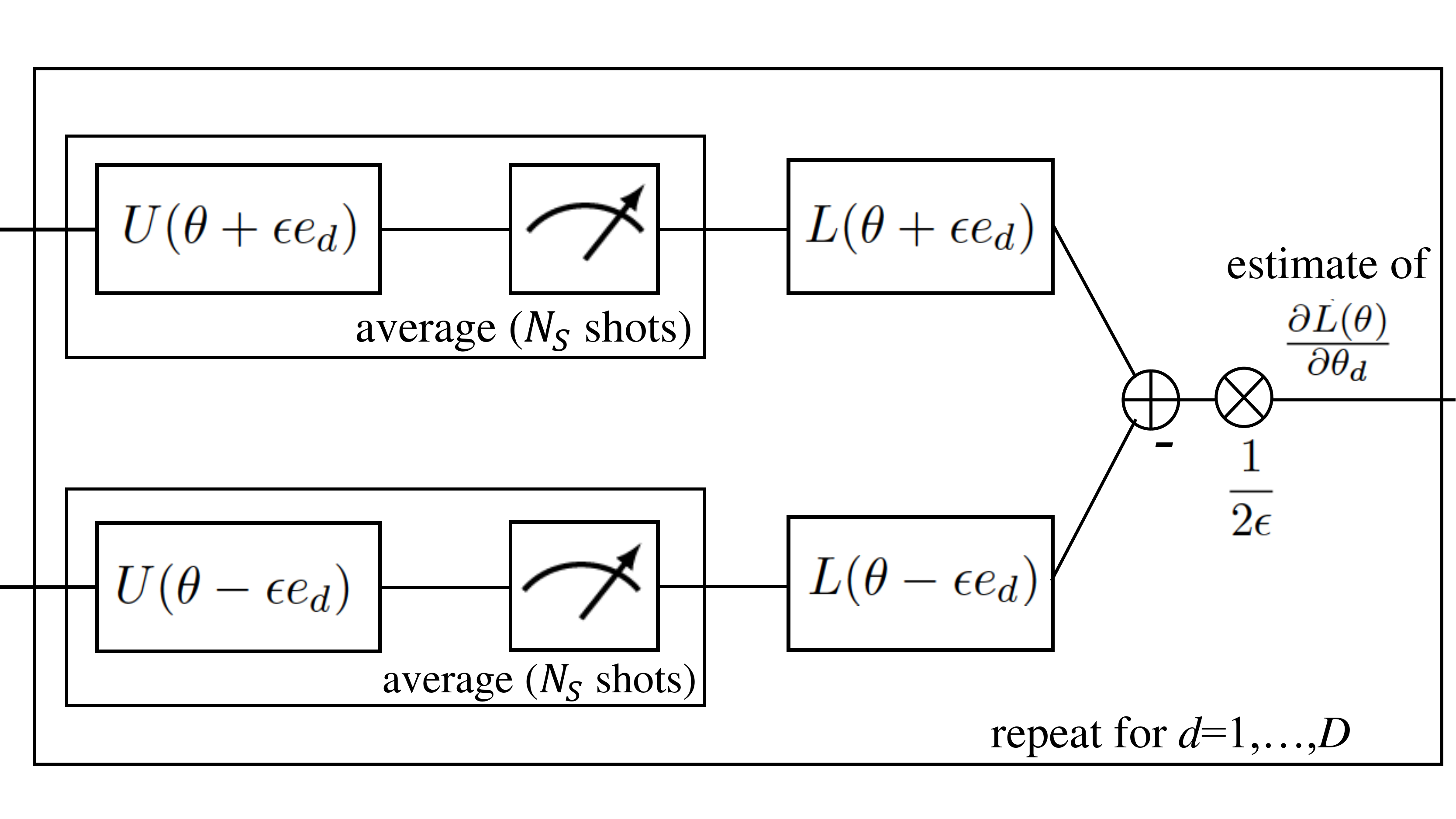}
		\par\end{center}\vspace{-0.4cm} \caption{Numerical differentiation method to estimate the gradient of the expectation of an observable (\ref{eq:ch6:obsdiff}): For each $d$-th entry $\partial L(\theta)/ \partial \theta_d$ of the gradient, with $d=1,...,D$, one needs to carry out $2N_S$ separate measurements of the observable. $N_S$ measurements, or shots, are used to estimate the expected value $L(\theta+\epsilon e_d)$, and $N_S$ measurements are used to estimate the expected value $L(\theta-\epsilon e_d)$. \label{fig:ch6:numdiff} }
\end{figure}

The second step is to estimate both terms in the approximation (\ref{eq:ch6:approxgrad}) via empirical averages obtained from multiple measurements of the quantum state $|\psi(\theta)\rangle$ produced by the PQC. As illustrated in Fig. \ref{fig:ch6:numdiff},  for each $d$-th parameter $\theta_d$, with $d\in\{1,...,D\}$, we run the quantum circuit
$2N_{S}$ times, where $N_{S}$ is the number of runs, or shots, used to estimate
the expected value of the observable. In particular, $N_S$ measurements are used to estimate the expected value \begin{equation} L(\theta+\epsilon e_d)=\langle O \rangle_{|\psi(\theta+\epsilon e_d)\rangle=U(\theta+\epsilon e_d)|0\rangle},
\end{equation} and $N_S$ measurements are used to estimate the expected value\begin{equation} L(\theta-\epsilon e_d)=\langle O \rangle_{|\psi(\theta-\epsilon e_d)\rangle=U(\theta-\epsilon e_d)|0\rangle}.
\end{equation} It follows that a total number of $2DN_S$ measurements are needed to estimate the gradient vector at any given model parameter vector $\theta$.

\subsection{Parameter Shift Rule}

In classical machine learning, with neural networks used as models, the gradient of the loss function can be computed at once via \textbf{backpropagation}. Backpropagation consists of two runs through the model, once in the forward direction and once backwards. Backpropagation is hence significantly more efficient than numerical differentiation, since the latter requires a number of runs through the model that is proportional to $D$. No quantum counterpart of backpropagation exists, since in quantum machine learning one does not have access to the internal workings of the model, i.e., the PQC, which is only accessible via measurements. (That is, unless one relies on a simulation of the quantum computer on a classical computer, which is only feasible for small values of $n$.) Hence, the computation of the gradient is an inherently more complex, and less scalable, operation than in classical machine learning. 

That said, when only, possibly multi-qubit, Pauli rotations are used as parametric
gates in the PQC, a related approach, known as the \textbf{parameter shift rule}, can be proved to provide the exact gradient as the number of shots $N_S$ goes to infinity. As we elaborate in this subsection, the parameter shift rule does \emph{not} solve the scalability problem highlighted above, as it still requires a number of runs through the model proportional to $D$. The only advantage as compared to numerical differentiation is that the approximate equality in (\ref{eq:ch6:approxgrad}) can be turned into an equality without requiring the numerically problematic step of reducing the value of $\epsilon$. 

To elaborate, assume that each parameter $\theta_d$ contributes to the PQC through
a gate of the form
\begin{equation}\label{eq:ch6:UPauli}
	U_{d}(\theta)=\exp\left(-i\frac{\theta_d}{2}P_{d}\right),
\end{equation}
where the generator $P_{d}$ is a Pauli string. Note that the gate (\ref{eq:ch6:UPauli}) may apply to an arbitrary number of qubits, and that it encompasses as special cases the single- and two-qubit Pauli rotation gates discussed in Sec. \ref{sec:ch6:PQC}. Therefore, mean-field ansatz, hardware-efficient ansatz, as well as generalizations thereof involving parametrized two-qubit Pauli rotations  satisfy the assumption at hand.

Under this assumption, it can be proved that the partial derivative can be
exactly computed as (see Appendix B)
\begin{equation}\label{eq:ch6:psrule}
	\frac{\partial L(\theta)}{\partial\theta_{d}}=\frac{L\left(\theta+\frac{\pi}{2}e_{d}\right)-L\left(\theta-\frac{\pi}{2}e_{d}\right)}{2}.
\end{equation}
Therefore, the use of the ``numerical differentiation-like'' expression in (\ref{eq:ch6:psrule}) does not entail any approximation, unlike (\ref{eq:ch6:approxgrad}). 

\begin{figure}
	\begin{center}
		\includegraphics[clip,scale=0.32]{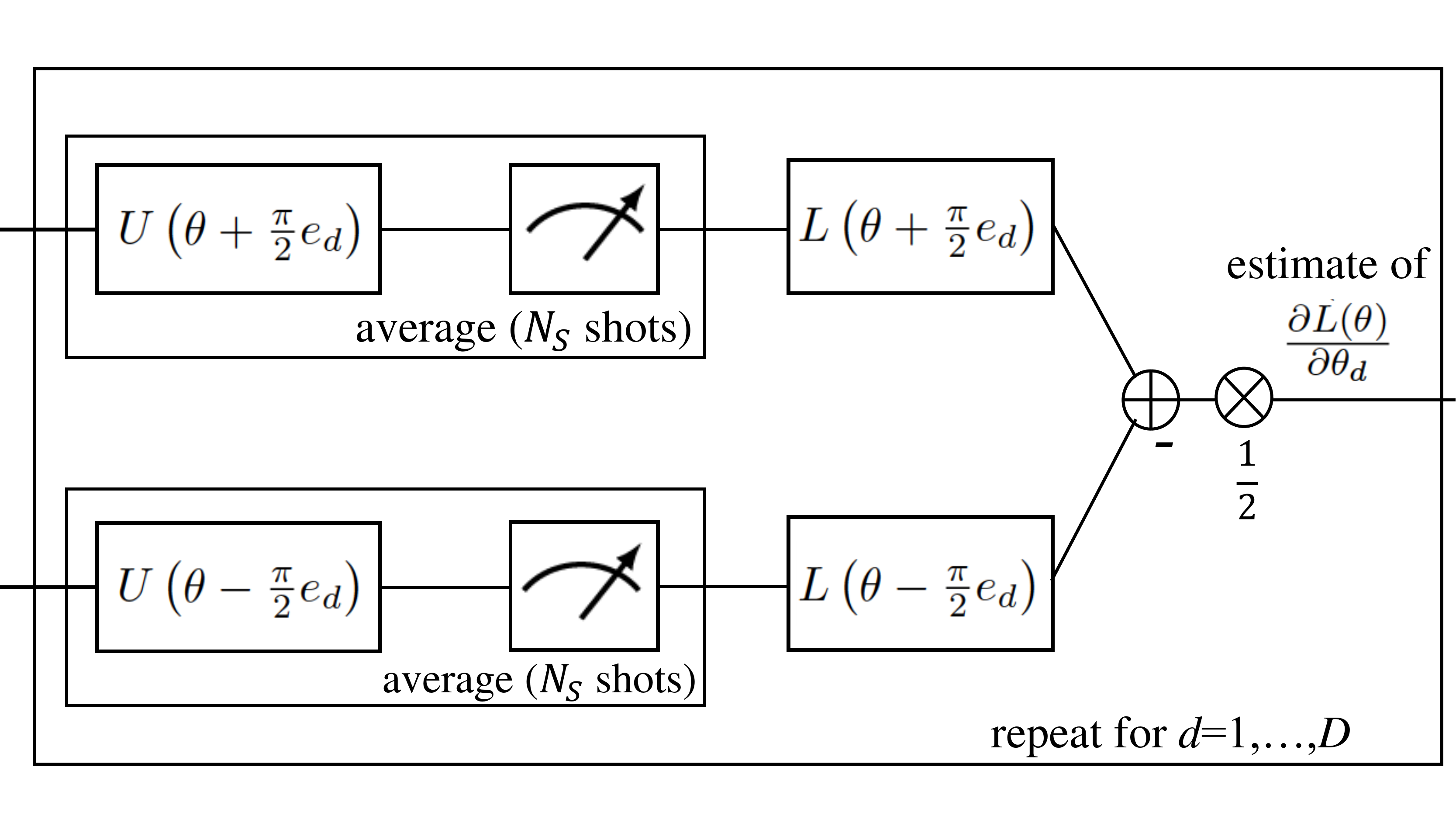}
		\par\end{center}\vspace{-0.4cm} \caption{Parameter shift rule to estimate the gradient of the expectation of an observable (\ref{eq:ch6:obsdiff}): For each $d$-th entry $\partial L(\theta)/ \partial \theta_d$ of the gradient, with $d\in\{1,...,D\}$, one needs to carry out $2N_S$ separate measurements of the observable. $N_S$ measurements, or shots, are used to estimate the expected value $L(\theta+\pi/2 e_d)$, and $N_S$ measurements are used to estimate the expected value $L(\theta-\pi/2 e_d)$. \label{fig:ch6:psrule} }
\end{figure}

As anticipated, and as detailed in Fig. \ref{fig:ch6:psrule}, the parameter shift rule still entails the implementation of the same number $2DN_S$ of measurements. Furthermore, the procedure is equivalent to that depicted in Fig. \ref{fig:ch6:numdiff} for numerical differentiation, with the only caveat that the shift takes the specific value $\pi/2$ and that the normalization constant is given by $1/2$.

\subsection{Limitations of Gradient Descent}
An important practical problem related to the implementation of the outlined gradient descent method 
is that the landscape of the loss function $L(\theta)$ may not be well behaved as the number of qubits $n$ grows large.  Intuitively, this is due to the mismatch between the size of the parameter vector $\theta$ and the exponentially large dimension, $2^n$, of the Hilbert space.  A particularly important issue concerns the scaling of the partial derivatives of the cost function $L(\theta)$. In particular, it has been shown that, when the ansatz is sufficiently rich and the model parameters are randomly initialized, the partial derivatives tend exponentially
to zero in probability as $n$ grows.  
This is known as the \textbf{barren plateaus problem}. 

Possible solutions include selecting optimized initializations; implementing ansatzes that are well matched to the problem under study, e.g., by leveraging symmetries in the underlying distribution or mapping of interest; and exploring alternatives to gradient descent.

Based on the drawbacks just discussed, alternatives to gradient descent may also be considered that  encompass more advanced 
forms of local optimization, e.g., based on second-order information, as well as global optimization strategies. 
Global optimization may leverage \textbf{surrogate objective
	functions}, which extrapolate the value of the cost $L(\theta)$ across
different values of $\theta$. One such method is provided by Bayesian optimization, which updates the surrogate sequentially based on the outcomes of previous
measurements.

\section{Variational Quantum Eigensolver}

In this section, we describe a first application of the methodology illustrated in Fig. \ref{fig:ch6:qmlintro}. In it, the parameter vector $\theta$ of a PQC is designed via a classical optimizer that is fed measurements of the PQC's output state. Specifically, we introduce the  variational quantum eigensolver
(VQE) as a tool to address a class of binary combinatorial
optimization problems via the approach depicted in Fig. \ref{fig:ch6:qmlintro}. Broader applications of VQEs will be also mentioned later in this section. 

While VQEs are not data-driven, and hence may not qualify as ``machine learning'', they serve as a useful stepping stone to describe quantum machine
learning applications. In fact, VQEs optimize the model parameter $\theta$ by minimizing the expectation of an observable under the quantum state produced by the PQC. As anticipated in the previous section, this approach is common to many quantum machine learning methods.

\subsection{Quadratic Unconstrained Binary Optimization}

A \textbf{quadratic unconstrained binary optimization
	(QUBO)} problem can be formulated as the combinatorial optimization
\begin{align}\label{eq:ch6:qubo}
	\min_{\tilde{x}\in\{-1,1\}^{n}}\left\{ f(\tilde{x})=\sum_{k=0}^{n-1}\sum_{j=0}^{k-1}a_{k,j}\tilde{x}_{k}\tilde{x}_{j}+\sum_{k=0}^{n-1}b_{k}\tilde{x}_{k}\right\}
\end{align}
with coefficients $\{a_{k,j},b_{k}\}.$ QUBO has a wide range of applications from finance and economics to
machine learning. 
The exact solution of a QUBO generally requires a search over the exponentially
large space, in $n$, of the binary optimization vector $\tilde{x}$. We will assume for simplicity that the problem has a single optimum solution 
$\tilde{x}^{*}\in\{-1,1\}^{n}.$

Problem (\ref{eq:ch6:qubo}) can be equivalently formulated as an optimization over a binary vector $x\in\{0,1\}^{n}$. To this end, we can create a one-to-one mapping between each binary string  $x\in\{0,1\}^{n}$ and the corresponding signed binary
string $\tilde{x}\in\{-1,1\}^{n}$ by setting  $\tilde{x}_{k}=-1$
for $x_{k}=1$ and $\tilde{x}_{k}=1$ for $x_{k}=0$ for $k\in\{0,1,...,n-1\}$. The resulting  mapping between the two vectors is hence given by the entry-wise equality
\begin{equation}
	\tilde{x}_{k}=1-2x_{k}
\end{equation}
for $k\in\{0,1,...,n-1\}.$
We will write this assignment for short as $\tilde{x}=1-2x$, in which
the operations are applied element-wise.

\subsection{QUBO Objective as a Quantum Observable}

In order to address the QUBO problem (\ref{eq:ch6:qubo}) with a PQC, we start by expressing
the objective function $f(\cdot)$ in terms of a quantum observable
$F$ on $n$ qubits.
As we will see in the next subsections, the main idea underlying VQEs is to design the PQC to output $n$ qubits in a state
that (approximately) minimizes the expectation of the observable $F$.

The objective function $f(\cdot)$ can be associated with the following
observable defined on $n$ qubits
\begin{equation}\label{eq:ch6:obsF}
	F=\sum_{k=0}^{n-1}\sum_{j=0}^{k-1}a_{k,j}Z_{k}Z_{j}+\sum_{k=0}^{n-1}b_{k}Z_{k},
\end{equation}
which is also known as the \textbf{Ising Hamiltonian}. We recall from the previous section that the matrix $Z_k$ is defined as in (\ref{eq:ch6:Zk}). 
The observable $F$ is the sum of $n(n-1)/2+n$ observables, each given
by a Pauli string consisting of Pauli operators $\{I,Z\}$. Each individual observable in (\ref{eq:ch6:obsF}) corresponds to the term in (\ref{eq:ch6:qubo}) obtained by replacing the Pauli string $Z_k$ with the binary variable $\tilde{x}_k$.

The association between objective function $f(\cdot)$ in (\ref{eq:ch6:qubo}) and the observable
$F$ in (\ref{eq:ch6:obsF}) has the following properties:\begin{itemize}
\item \textbf{Eigenvectors as solutions}: The observable $F$ has eigenvectors given by the vectors $\{|x\rangle\}_{x=0}^{2^{n}-1}$
of the computational basis. Each eigenvector $|x\rangle$ can be thus associated with a possible solution $\tilde{x}=1-2x$ of the QUBO problem.
\item \textbf{Eigenvalues as objective function values}: Each eigenvector $|x\rangle$ has corresponding eigenvalue $f(\tilde{x})$,
with $\tilde{x}=1-2x$, i.e., \begin{equation}\label{eq:ch6:eigVQE}
	F|x\rangle=f(\tilde{x})|x\rangle.
\end{equation} 
\end{itemize} Accordingly, the value $f(\tilde{x})$ of the objective function at any candidate solution $\tilde{x}\in\{-1,1\}^{n}$ can be computed by obtaining the eigenvalue of the observable $F$ corresponding to the eigenvector $|x\rangle$ with $\tilde{x}=1-2x.$

To prove the key equality (\ref{eq:ch6:eigVQE}), note that the eigendecomposition of the
Pauli $Z$ operator implies the equalities
\begin{equation}
	Z_{k}Z_{j}|x\rangle=\tilde{x}_{k}\tilde{x}_{j}|x\rangle\textrm{ and }Z_{k}|x\rangle=\tilde{x}_{k}|x\rangle,
\end{equation}where $\tilde{x}_k$ is the $k$th element of vector $\tilde{x}=1-2x.$ 
It follows that we have the equalities \begin{align}
	F|x\rangle & =\sum_{k=0}^{n-1}\sum_{j=0}^{k-1}a_{k,j}\left(Z_{k}Z_{j}|x\rangle\right)+\sum_{k=0}^{n-1}b_{k}Z_{k}|x\rangle \nonumber\\
	& =\sum_{k=0}^{n-1}\sum_{j=0}^{k-1}a_{k,j}\tilde{x}_{k}\tilde{x}_{j}|x\rangle+\sum_{k=0}^{n-1}b_{k}\tilde{x}_{k}|x\rangle \nonumber \\
	& =f(\tilde{x})|x\rangle.
\end{align}

\subsection{QUBO as the Minimization of the Expected Value of an Observable}\label{sec:ch6:mineig}

Given the equality (\ref{eq:ch6:eigVQE}), the minimum eigenvalue of the observable $F$ corresponds to the minimum value $f(\tilde{x}^{*})$ of the QUBO problem. It is a well-known result from linear algebra that the minimum eigenvalue of a Hermitian matrix can be obtained by solving the problem\begin{equation}\label{eq:ch6:QUBOeig}
	\min_{|\psi\rangle}\langle\psi|F|\psi\rangle 
\end{equation}  over all possible normalized $2^n$ vectors $|\psi\rangle$. Therefore, the QUBO problem can be equivalently solved by addressing the \textbf{minimum-eigenvalue problem} (\ref{eq:ch6:QUBOeig}) in the sense that we have the equalities \begin{equation}\label{eq:ch6:QUBOeig1}
\min_{|\psi\rangle}\langle\psi|F|\psi\rangle=\min_{\tilde{x}\in\{-1,1\}^{n}}f(\tilde{x})=f(\tilde{x}^{*}).
\end{equation}Moreover, the optimal solution $|\psi^{*}\rangle$ of problem (\ref{eq:ch6:QUBOeig}) provides, by (\ref{eq:ch6:eigVQE}), the solution of the QUBO problem, since we have the equality $|\psi^{*}\rangle=|x^*\rangle$ with $\tilde{x}^{*}=1-2x^{*}$.

The expression $\langle\psi|F|\psi\rangle$ equals the expected value $\langle F\rangle_{|\psi\rangle}$ of the observable $F$, and hence  the minimum-eigenvalue problem (\ref{eq:ch6:QUBOeig}) amounts to searching for the quantum state $|\psi\rangle$ that minimizes the expectation $\langle F\rangle_{|\psi\rangle}$, i.e., 
\begin{align}\label{eq:ch6:QUBOexp}
	\min_{|\psi\rangle} \langle F\rangle_{|\psi\rangle}. 
\end{align} 


\subsection{Variational Quantum Eigensolver}\label{sec:ch6:vqe}

\begin{figure}
	\begin{center}
		\includegraphics[clip,scale=0.32]{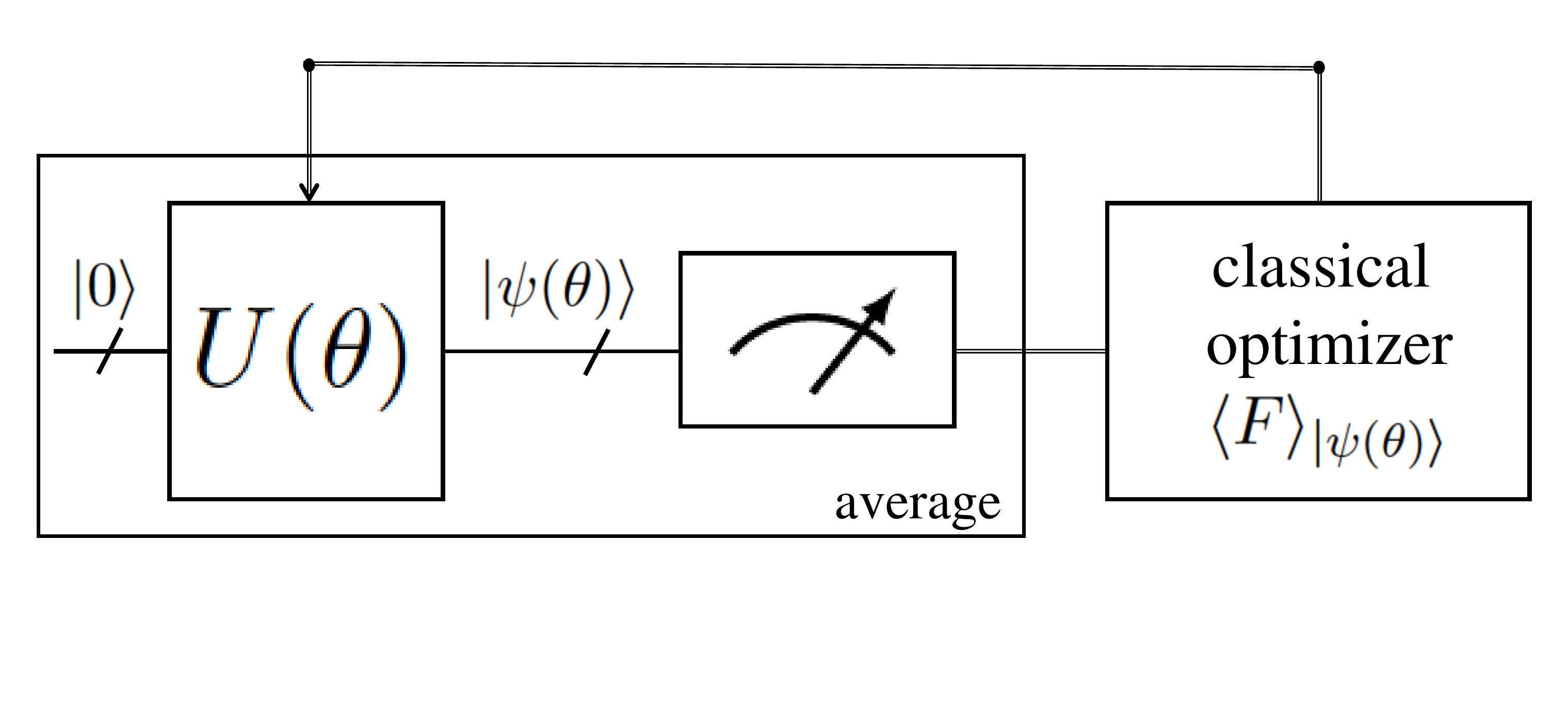}
		\par\end{center}\vspace{-0.4cm}\caption{Illustration of the operation of a VQE. The classical optimizer aims at addressing the stochastic optimization problem of minimizing the expected value $\langle F \rangle_{\psi(\theta)\rangle} = \langle \psi(\theta)|F| \psi(\theta)\rangle$ of the observable $F$.\label{fig:ch6:vqe}}
\end{figure}

The key idea underlying VQEs is to replace the optimization over state
$|\psi\rangle$ in (\ref{eq:ch6:QUBOexp}) with an optimization over the parameters $\theta$
of the PQC. 
Specifically, as illustrated in Fig. \ref{fig:ch6:vqe}, the VQE design methodology starts off by assuming a given ansatz for the PQC implementing a parametrized unitary $U(\theta).$ The PQC produces the parametrized quantum state
\begin{equation}
	|\psi(\theta)\rangle=U(\theta)|0\rangle.
\end{equation}
For any vector $\theta$, the expected value of the observable $F$ is
given by $\langle F\rangle_{|\psi(\theta)\rangle}=\langle\psi(\theta)|F|\psi(\theta)\rangle$. The goal of the classical optimizer in Fig. \ref{fig:ch6:vqe} is to use estimates of the expected value $\langle F\rangle_{|\psi(\theta)\rangle}$ to address the problem of minimizing
$\langle F\rangle_{|\psi(\theta)\rangle}$ over vector $\theta$, i.e.,
\begin{equation}\label{eq:ch6:vqeopt}
	\min_{\theta}\left\{ L(\theta)=\langle F\rangle_{|\psi(\theta)\rangle}=\langle\psi(\theta)|F|\psi(\theta)\rangle=\langle0|U(\theta)^{\dagger}FU(\theta)|0\rangle\right\} .
\end{equation}

The solution of the VQE problem (\ref{eq:ch6:vqeopt}) generally provides an upper bound on the minimum
value of the cost function of the QUBO problem, i.e., 
\begin{equation}
	\min_{\theta}L(\theta)\geq\min_{|\psi\rangle}\langle\psi|F|\psi\rangle=f(\tilde{x}^{*}),
\end{equation}
and the tightness of the bound depends on the flexibility of the ansatz underlying the PQC $U(\theta)$. In fact, if there exists a model parameter $\theta^{*}$ such that the PQC produces the optimal state, i.e., such that we have 
$|\psi(\theta^{*})\rangle=|x^{*}\rangle$, the solution of the VQE
problem coincides with the optimal solution of the QUBO problem. 
As we discussed in Sec. \ref{sec:ch6:PQC}, this is the case even with a simple mean-field ansatz
involving only Pauli $Y$-rotations, since the latter can produce any vector in the computational basis. 

That said, in practice, exact optimization of problem (\ref{eq:ch6:vqeopt}) is not possible, and hence
one is not guaranteed to recover the optimal solution of the QUBO problem. Rather, one typically implements stochastic gradient descent, or variants thereof, as discussed in the previous section.

Having found a model parameter $\theta$ that approximately solves problem (\ref{eq:ch6:vqeopt}), how can one extract a solution for the original QUBO problem?
If the goal is merely to compute the minimum value $f(\tilde{x}^{*})$ of the objective
function, an estimate of the optimal objective $f(\tilde{x}^{*})$
can be obtained by estimating the expectation $\langle F \rangle_{|\psi(\theta^{*})\rangle}$
for the optimized parameters $\theta^{*}.$ Approximating the solution $\tilde{x}^{*}$ is instead possible by making
a single standard measurement of the state $|\psi(\theta^{*})\rangle$. In fact, by Born's rule, such a measurement returns a vector $|x\rangle$ in the computational basis, whose corresponding signed vector $\tilde{x}=1-2x$ can be taken to be an approximation of the optimal solution $\tilde{x}^{*}$. 


\subsection{Addressing the VQE Optimization Problem}

The VQE problem (\ref{eq:ch6:vqeopt}) is a stochastic optimization problem in the form (\ref{eq:ch6:obsdiff}) with observable given by $F$. Therefore, classical optimization can be done via gradient descent by following the perturbation-based gradient descent approach detailed in the previous section. As mentioned therein, such methods require estimates of the 
expected value $\langle F\rangle_{|\psi(\theta)\rangle}$ for given candidate solutions $\theta$. 

By the decomposition of the observable $F$ into Pauli strings, we can write this expectation as
	\begin{equation}\label{eq:ch6:mF}
		\langle F\rangle_{|\psi(\theta)\rangle}=\sum_{k=0}^{n-1}\sum_{j=0}^{k-1}a_{k,j}\langle Z_{k}Z_{j}\rangle_{|\psi(\theta)\rangle}+\sum_{k=0}^{n-1}b_{k}\langle Z_{k}\rangle_{|\psi(\theta)\rangle}.
	\end{equation}Following Sec. \ref{sec:ch5:projmeas}, all the  $n(n-1)/2+n$ expectations in (\ref{eq:ch6:mF}) can be approximated via empirical averages obtained
from multiple measurements in the computational basis of the $n$
qubits in state $|\psi(\theta)\rangle$. Note, in fact, that the $n(n-1)/2+n$ observables in (\ref{eq:ch6:mF}) are compatible (see Sec. \ref{sec:ch5:compatible}).

\subsection{Other Applications of VQEs}

In this section, we have presented VQE as a means to address the QUBO problem. More generally, following the discussion in Sec. \ref{sec:ch6:mineig}, VQE can be adopted as a tool to tackle the problem of estimating
the minimum eigenvalue of an observable $F$. While the observable arising from the solution of a QUBO is given in terms of Pauli $Z$ operators, one may be more broadly interested in observables that can
be written as arbitrary linear combinations of Pauli strings. 
Therefore, the observable $F$ may include also the $X$ and $Y$
Pauli operators, as well as terms including more than two Pauli operators.
 When implementing VQE for this more general class of observables, it should be kept in mind that it is only possible to measure simultaneously, on the same qubit, observables derived
from the same Pauli operator (see Sec. \ref{sec:ch5:compatible}). Therefore, a general observable of this form would 
requires three separable, i.e., qubit by qubit, measurements in the
$X$, $Y$, and $Z$ bases.

\section{Unsupervised Learning for Generative Models}

As a first machine learning application of PQCs, this section studies the unsupervised
learning problem of training probabilistic models generating
random binary strings $x\in\{0,1\}^{n}$. We proceed by first describing the operation of such generative models at a high level, and by then detailing an implementation using PQCs. 

\subsection{Generative Models}\label{sec:ch6:genmod}

Given a training set $\text{\ensuremath{\mathcal{D}}=}\{x_{1},...,x_{|\mathcal{D}|}\}$ of $|\mathcal{D}|$ $n$-cbit strings ($|\mathcal{D}|$ is the cardinality of set $\mathcal{D}$), we would like to train a probabilistic model that is able to generate binary strings that are distributed in a manner that ``resembles'' the underlying, unknown, distribution of the data. 

As an example, one may think of $x$ as a black-and-white image, flattened into an $n$ dimensional vector, with black and white pixels identified by entries of the vector equal to $1$ and $0$, respectively. The goal of unsupervised learning in this context is that of training a model that can produce black-and-white images that are ``similar'' to those in the training set $\mathcal{D}$.

In the rest of this subsection, we review the two main phases of operation of a parametrized generative model. At run time, when the model is used with a fixed parameter vector $\theta$, the goal is to sample binary strings $x$ from a desired distribution; while the training phase aims at optimizing the model parameter vector $\theta$. 

\noindent \textbf{Sample generation}: Probabilistic generative models produce randomized $n$-cbit strings  from some parametric distribution $p(x|\theta)$, where $\theta$ is the vector of model parameters. Once trained, that is, once vector $\theta$ is optimized, the model is used to  generate  one, or more, random strings $x$ -- e.g., images -- drawn from distribution $p(x|\theta)$. 

\noindent \textbf{Training}: Training relies on the definition of a loss function $\ell(x|\theta)$, which describes how suitable the generative model defined by distribution $p(x|\theta)$ is as a mechanism to generate the specific string $x$. Loss functions for probabilistic models are also known as \textbf{scoring functions}. A scoring function should be a non-increasing function of $p(x|\theta)$: A model $p(x|\theta)$ that assigns a low probability to a data point $x$ should be assigned a larger loss than a model $p(x|\theta)$ that assigns a high probability to a data point $x$. 

The most common scoring function in classical machine learning is the \textbf{log-loss} \begin{equation}
	\ell(x|\theta)=-\log p(x|\theta),
\end{equation}also known as \textbf{cross-entropy loss} or \textbf{information theoretic suprise}, which is indeed a decreasing function of $p(x|\theta)$. An issue with the log-loss is that it is not well defined when the probability $p(x|\theta)$ is zero. To avoid this problem, one can modify the log-loss as \begin{equation}\label{eq:ch6:llpractical}\ell(x|\theta)=-\log \max(\epsilon,p(x|\theta))
\end{equation} for some small $\epsilon > 0$. Alternatively, a popular option for quantum machine learning is the \textbf{linear loss} \begin{equation}\label{eq:ch6:llpractical1}\ell(x|\theta)=1-p(x|\theta).
\end{equation}

Once a loss function is determined, the \textbf{training loss} is defined as \begin{equation}\label{eq:ch6:trloss}
	L_{\mathcal{D}}(\theta)=\frac{1}{|\mathcal{D}|}\sum_{x\in\mathcal{D}}\ell(x|\theta)
\end{equation} as the empirical average of the loss accrued on the examples in the training set $\mathcal{D}$. The goal of training is typically defined as the minimization of the training loss (\ref{eq:ch6:trloss}) -- a problem known as \textbf{empirical risk minimization}\begin{equation}\label{eq:ch6:erm}
\min_{\theta} L_{\mathcal{D}}(\theta).
\end{equation}

The training loss (\ref{eq:ch6:trloss}) can be interpreted as a ``distance'' measure between  the empirical  distribution of the training data and the distribution $p(x|\theta)$ of the samples produced by the model. This viewpoint will be further elaborated on in the next subsection. 

\subsection{Quantum Generative Models: Born Machines}\label{sec:ch6:genquant}

A PQC implementing a unitary $U(\theta)$ on $n$ qubits can be readily used as a generative model of randomized $n$-cbit strings. In this subsection, we describe this application of PQCs by discussing both sample generation and training phases.

\begin{figure}
	\begin{center}
		\includegraphics[clip,scale=0.32]{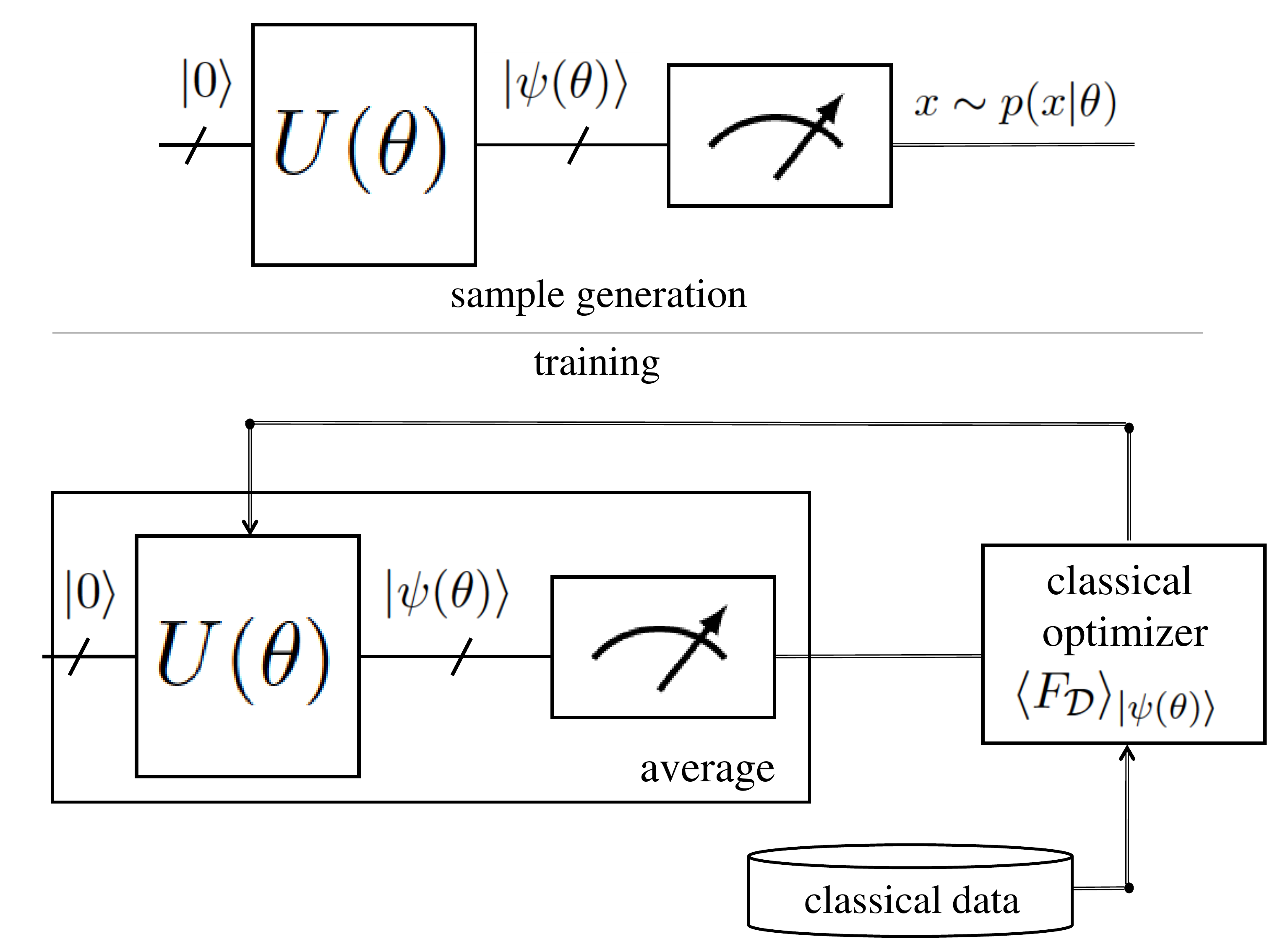}
		\par\end{center}\vspace{-0.4cm} \caption{(top) For data generation, a Born machine implements a single run of the PQC $U(\theta)$, producing as output of the measurement a random cbit string $x\sim p(x|\theta)$. (bottom) For training, a classical optimizer can address the problem of minimizing the expected value of a training data set-dependent observable $F_{\mathcal{D}}$, or the minimization of a distance metric between the empirical data distribution and the distribution $p(x|\theta)$ (not shown). \label{fig:ch6:Born}}
\end{figure}

\noindent \textbf{Sample generation}: As illustrated in the top part of Fig. \ref{fig:ch6:Born}, in order to generate a random $n$-cbit string, the PQC is run \emph{once} and  a standard measurement is made on the $n$ qubits. The measurement produces an $n$-cbit string $x \sim p(x|\theta)$, where the distribution \begin{equation}\label{eq:ch6:probgen}
	p(x|\theta)=|\langle x|\psi(\theta)\rangle|^{2},
\end{equation} is obtained from (\ref{eq:ch6:probtheta}) by considering the computational basis $|v_x\rangle=|x\rangle$ with $x\in\{0,1\}^{n}$.  As a result, generative models based on PQCs are also known as \textbf{Born machines}.

It is worth reiterating that a sample $x \sim p(x|\theta)$ is produced with a \emph{single} run of the PQC, hence not requiring multiple runs to average out the shot noise associated with quantum measurements. In this sense,  shot noise is actually \emph{useful} for generative models, as it is leveraged to produce random
samples.
This is a common feature of probabilistic models, as we will also
see in the next section for supervised learning.

Before turning to the training phase, we briefly describe a useful perspective on scoring functions dependent on the probability  (\ref{eq:ch6:probgen}). The quantity  $|\langle x|\psi(\theta)\rangle|^{2}$ in (\ref{eq:ch6:probgen}) can be interpreted 
as the \textbf{fidelity} between the two states $|x\rangle$
and $|\psi(\theta)\rangle$ (see Sec. \ref{sec:ch3:spiking}). 
The fidelity is an important metric in quantum theory that measures the degree to which one state can be ``confused''
for the other state based on measurement outputs. In this regard,  a loss function defined as a non-increasing function of the fidelity $|\langle x|\psi(\theta)\rangle|^{2}$ can be thought of as measuring how well -- or, rather, how poorly -- the state $|\psi(\theta)\rangle$
represents the input data $x$ encoded in state $|x\rangle$.


\noindent \textbf{Training via the ERM problem}: We now address the ERM problem (\ref{eq:ch6:erm}) for the linear loss
\begin{equation}\label{eq:ch6:ll}
	\ell(x|\theta)=1- p(x|\theta)=1-|\langle x|\psi(\theta)\rangle|^{2}.
\end{equation} Following the discussion in Sec. \ref{sec:ch6:estgrad},  a Born machine is an \textbf{implicit} probability model: It can generate samples $x\sim p(x|\theta)$, but it does not provide an explicit way to compute the probability $p(x|\theta)$. Therefore, the loss (\ref{eq:ch6:ll}) cannot be explicitly computed and differentiated. We show next how the ERM problem (\ref{eq:ch6:erm}) can be formulated as a stochastic problem in the form (\ref{eq:ch6:obsdiff}). 

To this end, we first write the probability $p(x|\theta)$ as  \begin{equation}\label{eq:ch6:logloss0}	p(x|\theta)=\langle\psi(\theta)|F_{x}|\psi(\theta)\rangle=\langle F_{x}\rangle_{|\psi(\theta)\rangle},
\end{equation} i.e., as the expected value of the observable \begin{equation} F_{x}=|x\rangle\langle x|.
\end{equation}  It follows that the training loss can be written as \begin{align}L_{\mathcal{D}}(\theta) & =\frac{1}{|\mathcal{D}|}\sum_{x\in\mathcal{D}}(1-\langle\psi(\theta)|F_{x}|\psi(\theta)\rangle)\nonumber\\
& =\left\langle \psi(\theta)\left|I-\frac{1}{|\mathcal{D}|}\sum_{x\in\mathcal{D}}F_{x}\right|\psi(\theta)\right\rangle \nonumber \\
& =\langle\psi(\theta)|F_{\mathcal{D}}|\psi(\theta)\rangle=\langle F_{\mathcal{D}}\rangle_{|\psi(\theta)\rangle}.
\end{align}Therefore, the training loss can be expressed as the expected value of the training data set-dependent observable \begin{equation} F_{\mathcal{D}}=I-\frac{1}{|\mathcal{D}|}\sum_{x\in\mathcal{D}}F_{x}.
\end{equation} We hence conclude that the ERM problem (\ref{eq:ch6:erm}) is indeed in the form  (\ref{eq:ch6:obsdiff}), and thus it can be addressed using the methods described in Sec. \ref{sec:ch6:estgrad} as illustrated in the bottom part of Fig. \ref{fig:ch6:Born}.

The observable $F_{\mathcal{D}}$ can be equivalently written as \begin{equation} F_{\mathcal{D}}=\sum_{x=0}^{2^{n}-1}(1-p_{\mathcal{D}}(x))|x\rangle\langle x|,
	\end{equation} where \begin{align}\label{eq:ch6:emp1}
	p_{\mathcal{D}}(x)&=\frac{\textrm{number of data points equal to $x$ in $\mathcal{D}$}}{|\mathcal{D}|}
\end{align} is the \emph{empirical distribution} of the training data. Therefore, the observable $F_{\mathcal{D}}$ is measured by making a standard measurement of the output state $|\psi(\theta)\rangle$, and by associating the observed value $1-p_{\mathcal{D}}(x)$ to each possible output $x\in \{0,1\}^n$. 

From this observation, we can conclude that, by minimizing the expected value of the observable $F_\mathcal{D}$ as in the ERM problem, we seek to produce a state $|\psi(\theta)\rangle$ that is aligned with the states $|x\rangle$ that have the maximal empirical distribution $p_{\mathcal{D}}(x)$. We can think of this objective as the minimization of a specific distance measure between the empirical distribution $p_{\mathcal{D}}(x)$ and the model distribution $p(x|\theta)$.

\noindent \textbf{Training via the minimization of a distance metric}: Given its focus on capturing the peaks of the empirical distribution $p_{\mathcal{D}}(x)$, the approach described above may fail to capture complex multi-modal distributions. Note that similar problems  arise also when considering  loss functions other than the linear loss studied above. We now review an alternative approach that aims at addressing such limitations.

This class of methods optimizes alternative distance measures between the distribution $p(x|\theta)$ of the samples produced by the model and the empirical distribution $p_{\mathcal{D}}(x)$ in (\ref{eq:ch6:emp1}). Importantly, one should select a distance measure that can be estimated based solely on samples from both distributions, without requiring knowledge of either distribution. Such methods are known as \textbf{two-sample estimators}.

 A key example of such measures is given by \textbf{integral probability metrics (IPMs)}. An IPM between two distributions $p(x)$ and $q(x)$ is defined by the  maximization \begin{equation}\label{eq:chap12:IPM}
	\textrm{IPM}(p(x),q(x))=\max_{T(\cdot)\in\mathcal{T}}\{\langle T(x) \rangle_{p(x)}-\langle T(x)\rangle_{q(x)}\},
\end{equation}where different IPMs are obtained by choosing distinct optimization domains $\mathcal{T}$. In (\ref{eq:chap12:IPM}), we have used the notation introduced in (\ref{eq:ch6:notavg}) for the expectation of random variables, i.e., $\langle T(x) \rangle_{p(x)}= \sum_x p(x) T(x)$. Intuitively, for a fixed \textbf{test function} $T(\cdot)$, the metric (\ref{eq:chap12:IPM}) measures how distinguishable the averages of random variables $T(x)$ are when evaluated under the two distributions $p(x)$ and $q(x)$. The maximization in (\ref{eq:chap12:IPM}) selects the test function that yields maximally distinct averages within the class of functions $\mathcal{T}$.   

A  notable example of IPMs is given by the \textbf{Wasserstein distance},
for which the optimization domain $\mathcal{T}$ is the set of all functions that are 1-Lipschitz. A function $T(\cdot)$ is 1-Lipschitz if it satisfies the inequality
$||T(x)-T(x')||\leq||x-x'||$ for all $x,x'$ in its domain. For differentiable functions, this condition is equivalent
to the inequality $||\nabla T(x)||\leq1$ on the gradient. Other important examples are given by the \textbf{maximum mean discrepancy} and the \textbf{Stein discrepancies}. 

When the optimization in (\ref{eq:chap12:IPM}) can be computed explicitly, as for the maximum mean discrepancy, a two-sample estimate can be directly obtained by averaging over the two sets of samples. More generally, the test function $T(\cdot)$ can be  parametrized, e.g., using a classical neural network, and the parameters of the test function optimized to address problem (\ref{eq:chap12:IPM}) using standard methods from classical machine learning. 

Going back to the original problem of training a generative quantum model, once one has chosen a particular IPM through the selection of the domain  $\mathcal{T}$, the learning process alternates between the following two steps: \begin{enumerate}
	\item Given the current model parameter $\theta$, we generate samples $x\sim p(x|\theta)$ from measurements of the PQC. We then use such samples, along with data set $\mathcal{D}$, to estimate the distance metric	$\textrm{IPM}(p(x|\theta),p_{\mathcal{D}}(x))$. This step requires optimizing over the test function $T(\cdot)$, which may in turn entail the use of classical optimization tools.
	\item  Given the test function $T(\cdot)$, the distance metric	$\textrm{IPM}(p(x|\theta),p_{\mathcal{D}}(x))$ is optimized with respect to the PQC's parameter vector  $\theta$. Since, for a fixed function  $T(\cdot)$,  the right-hand side of (\ref{eq:chap12:IPM}) is in the form of the stochastic optimization problem (\ref{eq:ch6:obsdiff}), this can be done using the tools described in Sec. \ref{sec:ch6:estgrad}.
\end{enumerate}


%
\section{Supervised Learning}\label{sec:ch6:sl}

In this section, we study supervised learning tasks. In such problems, the training data set  $\text{\ensuremath{\mathcal{D}}=}\{(x_{1},y_{1}),...,(x_{|\mathcal{D}|},y_{|\mathcal{D}|})\}$ contains $|\mathcal{D}|$ pairs $(x,y)$ consisting of classical
input $x$, in the form of a real-valued vector,  and  classical target variable $y$.
The target variable $y$ is discrete for classification problems, and a real-valued for  regression problems.  The goal is to optimize a predictive model mapping input $x$ to output $y$ in such a way that the optimized model  reflects well the underlying, unknown, data distribution. 

In this section, we start by discussing the most common way to encode the input $x$ into the quantum state produced by the PQC. Then we cover separately probabilistic and deterministic quantum models defined from the output of a PQC.

\subsection{Angle Encoding}\label{sec:ch6:angleenc}
A common and efficient way to encode the input $x$ into a quantum state is via \textbf{angle encoding}. As illustrated in the top part of Fig. \ref{fig:ch6:psup},  angle encoding defines a unitary transformation $U(x,\theta)$ that depends on the input vector $x$ in a manner similar to the model parameters $\theta.$ The output state of the PQC $U(x,\theta)$, i.e., \begin{equation}
	|\psi(x,\theta)\rangle=U(x,\theta)|0\rangle,
\end{equation} is a function of the input $x$, and is known as the \textbf{quantum
	embedding} of the classical input $x$. The quantum embedding $|\psi(x,\theta)\rangle$ ``lifts'' the classical input $x$ from the space of $M$-dimensional real vectors to a quantum state in the $2^n$-dimensional Hilbert space.
Note that the embedding can be controlled via the model parameter
$\theta.$

Angle encoding adopts a PQC $U(x,\theta)$ that alternates between unitaries dependent on input 
$x$ and unitaries dependent on the model parameter vector $\theta$ as 
\begin{equation}
	U(\theta)=W_{L}(\theta)V_{L}(x)\cdot W_{L-1}(\theta)V_{L-1}(x)\cdots W_{1}(\theta)V_{1}(x).
\end{equation} Note that, in this architecture, the data $x$ is potentially entered multiple times, i.e., at each layer $l$.   This process is also known as \textbf{data re-uploading}. Alternatively, one could also enter data $x$ progressively, with a different fragment of vector $x$ encoded at each layer. 

The unitaries dependent on $x$ and $\theta$ -- i.e., $V_{l}(x)$
and $W_{l}(\theta)$ for $l\in\{1,...,L\}$ --  consist of a cascade of single-qubit rotations and, possibly parametrized, entangling gates as discussed in Sec. \ref{sec:ch6:PQC}. The name ``angle encoding'' reflects the standard choice of parameterized single- or two-qubit Pauli rotations for the unitaries  $V_{l}(x)$, so that the entries of vector $x$ play the role of angles defining such rotations. 
 
The two types of unitaries,  $V_{l}(x)$
and $W_{l}(\theta)$, may follow different ansatzes. The rationale for this differentiation is that the ansatz  selected for the model parameters-dependent unitaries $W_{l}(\theta)$ should facilitate training, particularly in light of the practical limitations of gradient-based optimization discussed in Sec. \ref{sec:ch6:estgrad}. In contrast, the choice of the ansatz for the input-dependent unitaries $V_{l}(x)$ is typically dictated by the expressivity of the resulting PQC as a mapping between input and output. For instance,
in some architectures, one uses single- and two-qubit rotations to define the architecture of 
the unitaries $V_{l}(x)$, while fixed entangling gates are used to design 
the unitaries $W_{l}(\theta)$ as in the hardware-efficient ansatz. It is also possible to consider unitaries that directly depend on both $x$ and $\theta$ as for a rotation gate of the form $\exp(-i(\theta x/2) P)$, where $P$ is a Pauli operator.

In order to encode an input $x=[x_0,...,x_{M-1}]$ containing $M$ real numbers, angle encoding typically uses at least $M$ qubits.  For example, one could use a simple mean-field ansatz for the unitary $V_{l}(x)$, with \begin{equation}\label{eq:ch6:anglex}
	V_{l}(x)=R_P(x_0)\otimes \cdots \otimes R_P(x_{M-1}),
	\end{equation} where $R_P(\cdot)$ is some Pauli rotation. Accordingly, a separate rotation -- each determined by an entry of the vector $x$ -- is applied to a distinct qubit. It is also common to follow local transformations of the form (\ref{eq:ch6:anglex}) with other local operations  \begin{equation}\label{eq:ch6:anglex1}
	R_P(g(x_0))\otimes \cdots \otimes R_P(g(x_{M-1})),
\end{equation} where $g(\cdot)$ is some non-linear function, i.e., $g(x)=x^2$. As mentioned, the ansatz for $V_{l}(x)$ can be made more complex, and expressive, by adding also two-qubit gates.

We refer to Sec. \ref{sec:ch6:beyondangle} for a discussion of alternative encoding methods that differ from angle encoding in terms of number of required qubits and information processing models.

\subsection{Probabilistic Models}

 In this subsection and in the next, we consider probabilistic models that relate input $x$ and output $y$ through a stochastic mapping. In a manner that generalizes generative models, the stochastic mapping between $x$ and $y$ is implicitly described by a conditional distribution $p(y|x,\theta)$. This is in the sense that the model produces a randomized output $y\sim p(y|x,\theta)$ when fed with input $x$. We begin in this subsection by describing the operation of such probabilistic models at a high level, and details related to the implementation via PQC are provided in the next subsection.

We specifically focus on probabilistic models for \textbf{classification}, in which the
target variable is a binary string $y\in\{0,1\}^{m}$ with $m\leq n$. This accounts for settings in which the input $x$ belongs to one of  
$2^{m}$ classes, which are indexed by the binary string $y$. In practice, the value of $m$ is typically much smaller than $n$. For instance, for binary classification, one would have $m=1$. 


The operation of the system distinguishes between inference and training phases. In the inference phase, the model parameters $\theta$ are kept fixed, and the model is used to classify a new, test, input $x$ via the randomized mapping $y\sim p(y|x,\theta)$. In contrast, training aims at optimizing the model parameter vector $\theta$.

\noindent \textbf{Inference}: Inference in probabilistic models amounts to the random generation of a label $y\sim p(y|x,\theta)$ given the input $x$ to be classified. A single sample is sufficient to make a point prediction, and multiple samples can be used if one is also interested in quantifying uncertainty. A classifier can be deemed to be more uncertain if different runs of the algorithm tend to yield distinct predictions $y$.

\noindent  \textbf{Training}: Training of probabilistic models relies on the definition of a loss function $\ell(x,y|\theta)$ that accounts for the accuracy of the model $p(y|x,\theta)$ when evaluated on an input-output pair $(x,y)$. As for generative models, the loss function is defined as a non-increasing scoring function of $p(y|x,\theta)$: A model $p(y|x,\theta)$ that assigns a low probability to the label $y$ of a data point $(x,y)$ is assigned a larger loss than a model $p(y|x,\theta)$ that assigns a high probability to $y$. The most common scoring function in classical machine learning is the \textbf{log-loss} \begin{equation}
	\ell(x,y|\theta)=-\log p(y|x,\theta),
\end{equation}which is also known as \textbf{cross-entropy loss}. In quantum machine learning, one often adopts the \textbf{linear loss} \begin{equation}\label{eq:ch6:linearsl}\ell(x,y|\theta)=1-p(y|x,\theta),
\end{equation} which measures the probability of error for the model when the true label is $y$ given input $x$.

 The goal of training is typically defined in terms of the \textbf{empirical risk minimization} problem (\ref{eq:ch6:erm}) with  \textbf{training loss}  \begin{equation}\label{eq:ch6:trloss1}
	L_{\mathcal{D}}(\theta)=\frac{1}{|\mathcal{D}|}\sum_{(x,y)\in\mathcal{D}}\ell(x,y|\theta).
\end{equation}
\subsection{Probabilistic Quantum Models}

\begin{figure}
\begin{center}
	\includegraphics[clip,scale=0.32]{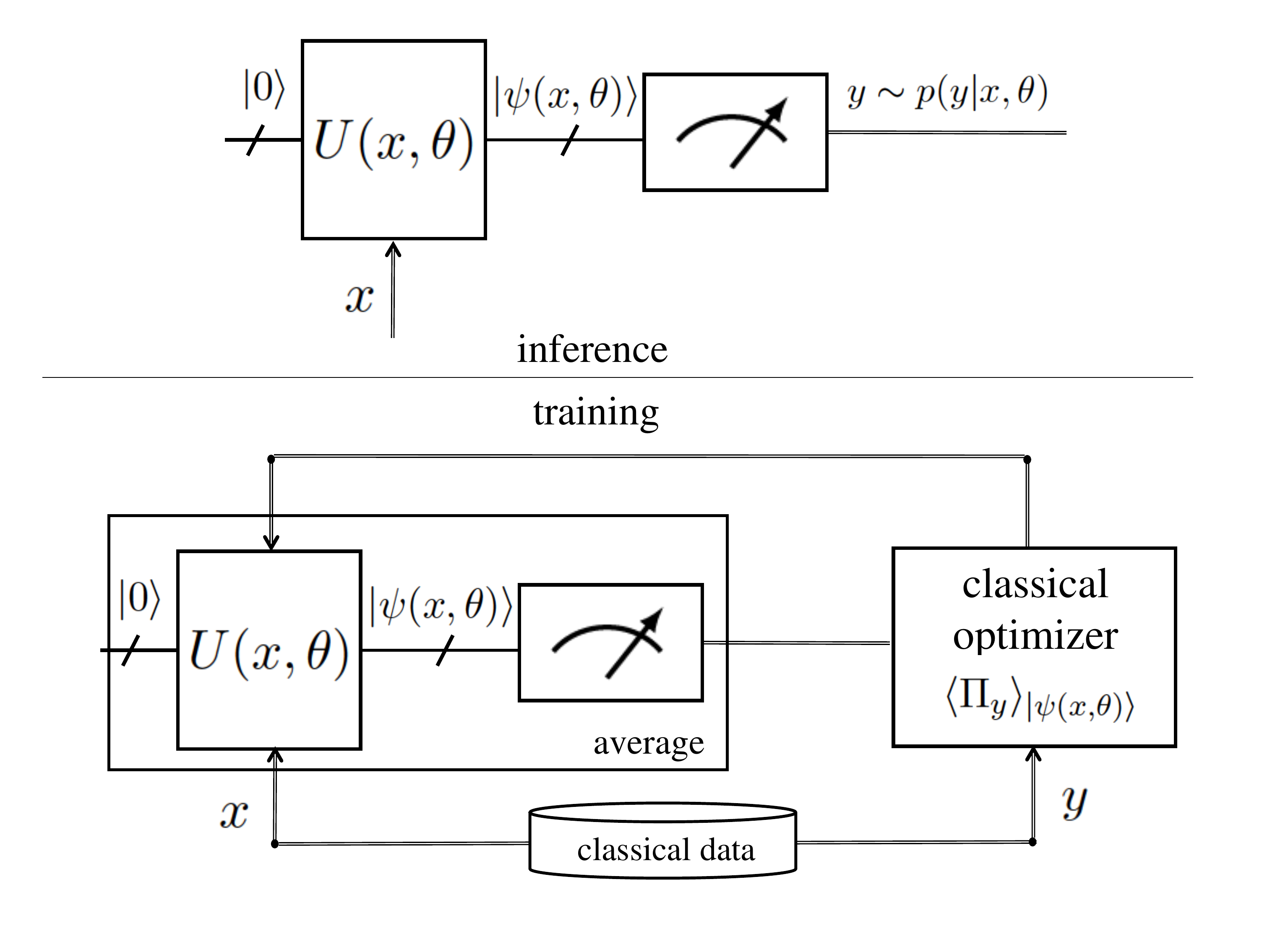}
	\par\end{center}\vspace{-0.4cm} \caption{(top) For inference, a probabilistic quantum model implements a single run of the PQC $U(x,\theta)$, which produces a discrete label $y$ via a projective measurement. (b) For training, a classical optimizer applies  doubly stochastic gradient descent based on estimates the gradients of the expected values $\langle \psi(x,\theta)| I-\Pi_{y} | \psi(x,\theta) \rangle$ for all examples $(x,y)$ in a mini-batch.   \label{fig:ch6:psup}}
\end{figure}

As illustrated in the top part of Fig. \ref{fig:ch6:psup}, a probabilistic model for classification can be implemented via a PQC in a manner similar to a Born machine. In this subsection, we first discuss how to carry out inference via probabilistic quantum models and then we describe the problem of training such models.

\noindent \textbf{Inference}: Consider first inference, and hence assume that the model parameter vector $\theta$ is fixed. Given any input $x$ to be classified, the PQC $U(x,\theta)$ leaves the $n$ qubits in the quantum embedding state $|\psi(x,\theta)\rangle$. To perform classification, we need  to extract a binary string $y$ of $m$ cbits -- the label assigned by the model to input $x$ -- from this state. This can be done by implementing a projective measurement with $2^m$ possible outputs (see Sec. \ref{sec:ch5:projmeas}). For instance,  for binary classification ($m=1$), one can apply a parity measurement, whereby label $y=0$ is assigned to the subspace of states with even parity, and label $y=1$ to the odd-parity subspace. We refer to Sec. \ref{sec:ch5:paritymeas} for a description of parity measurements the case $n=2$.

To formalize the outlined approach, each possible output $y\in\{0,1\}^{m}$ is assigned to  a projection matrix $\Pi_{y}$, defining a projective measurement $\{\Pi_{y}\}_{y=0}^{2^{m}-1}.$ As illustrated in the top part of Fig. \ref{fig:ch6:psup}, a randomized label $y$ is produced by the circuit at inference time by making a single measurement of the quantum embedding $|\psi(x,\theta)\rangle$.  This way, the model leverages shot noise for sampling, and a single run of the algorithm is sufficient for inference. Note that one could also adopt a more general POVM in lieu of a projective measurement.

By the  Born rule  (\ref{eq:ch5:Bornpure}), the outlined quantum probabilistic models implements a stochastic 
mapping between $x$ and $y$ that is defined by the conditional probability
\begin{equation}\label{eq:ch6:probsl}
	p(y|x,\theta)=\langle\psi(x,\theta)|\Pi_{y}|\psi(x,\theta)\rangle=\langle\Pi_{y}\rangle_{|\psi(x,\theta)\rangle}.
\end{equation} As in the case of generative models, by (\ref{eq:ch6:probsl}), the probability $p(y|x,\theta)$ equals the expectation of a data-dependent observable.  Specifically, here the observable depends on the label $y$, and is given by the projector $\Pi_y$. Recall that projection matrices are Hermitian, and hence they are also  observables.

\noindent \textbf{Training}: Adopting the linear loss (\ref{eq:ch6:linearsl}), i.e.,
\begin{equation}
	\ell(x,y|\theta)=1- p(y|x,\theta)=1-\langle\Pi_{y}\rangle_{|\psi(x,\theta)\rangle},
\end{equation}
the training loss can be written as \begin{equation}
	L_{\mathcal{D}}(\theta)=\frac{1}{|\mathcal{D}|} \sum_{(x,y)\in \mathcal{D}} \langle \psi(x,\theta)| I-\Pi_{y} | \psi(x,\theta) \rangle,
	\end{equation}which is the average of the expected values of observables $I-\Pi_{y}$ under the embedding $| \psi(x,\theta) \rangle$ across all data points $(x,y) \in \mathcal{D}$. It follows that the ERM problem is again in the form of a stochastic optimization, with the caveat that the cost function involves multiple observables as per the general form   (\ref{eq:ch6:lossfunge}).

In practice the ERM problem is addressed via \textbf{doubly stochastic gradient descent}. Accordingly, one draws a random mini-batch of examples from the data set, and estimates the gradients of the corresponding expected value $\langle \psi(x,\theta)| I-\Pi_{y} | \psi(x,\theta)\rangle$ for each example $(x,y)$ in the mini-batch (see the bottom part of Fig. \ref{fig:ch6:psup}). The ``double'' randomness arises from the processes of sampling both data and measurement outcomes.



%
\subsection{Deterministic Models}

In this subsection, we focus on deterministic models for supervised learning. We consider a general formulation of such models, leaving the details of a quantum implementation to the next subsection. 

A deterministic model implements a parametrized functions $h(x|\theta)$ of the input $x$. The function $h(x|\theta)$ generally returns a continuous-valued vector. Deterministic models can be used for both regression and classification.

\noindent \textbf{Inference}: In the case of \textbf{regression}, the target variable $y$ is generally a real-valued vector. Therefore, at inference time, given an input $x$, the model can directly use the output  of function $h(x|\theta)$ as the estimate of target variable $y$. 

In the case of \textbf{classification}, the target variable is the discrete-valued index of the class. To produce an estimate of $y$, the output of function $h(x|\theta)$ is  passed through an activation function to produce either a hard decision or a probability distribution over the possible outcome values. As an example, for binary classification,  the sign of the scalar-valued function $h(x|\theta)$ can be used to obtain a hard decision; or we can  ``squash'' the output 
$h(x|\theta)$ using, e.g., a sigmoid activation function $\sigma(\cdot)$, to produce a
probability
\begin{equation}\label{eq:ch6:logreg}
	\Pr[y=1|x,\theta]=\sigma(h(x|\theta))=\frac{1}{1+\exp(-h(x|\theta))}
\end{equation} as in logistic regression. Similarly, for multi-class classification, the  real-valued vector output by function $h(x|\theta)$ can be fed to a softmax activation function. This produces  a vector of probabilities on the class indices.

\noindent \textbf{Training}: For training, the standard approach is to tackle the ERM problem (\ref{eq:ch6:erm}) with training loss (\ref{eq:ch6:trloss1}).  For example, for  regression with a scalar target variable $y$, one can adopt the squared loss
\begin{equation}\label{eq:ch6:quadratic}
	\ell(x,y|\theta)=(y-h(x|\theta))^{2};
\end{equation}
while, for binary classification with activation function (\ref{eq:ch6:logreg}), one can choose the cross-entropy loss
\begin{equation}
	\ell(x,y|\theta)=-y\log\sigma(h(x|\theta))-(1-y)\log\sigma(-h(x|\theta)).
\end{equation}

\subsection{Deterministic Quantum Models} 

In this subsection, we discuss how PQCs can be used to implement deterministic models for supervised learning by covering both inference and training phases.

\begin{figure}
	\begin{center}
		\includegraphics[clip,scale=0.32]{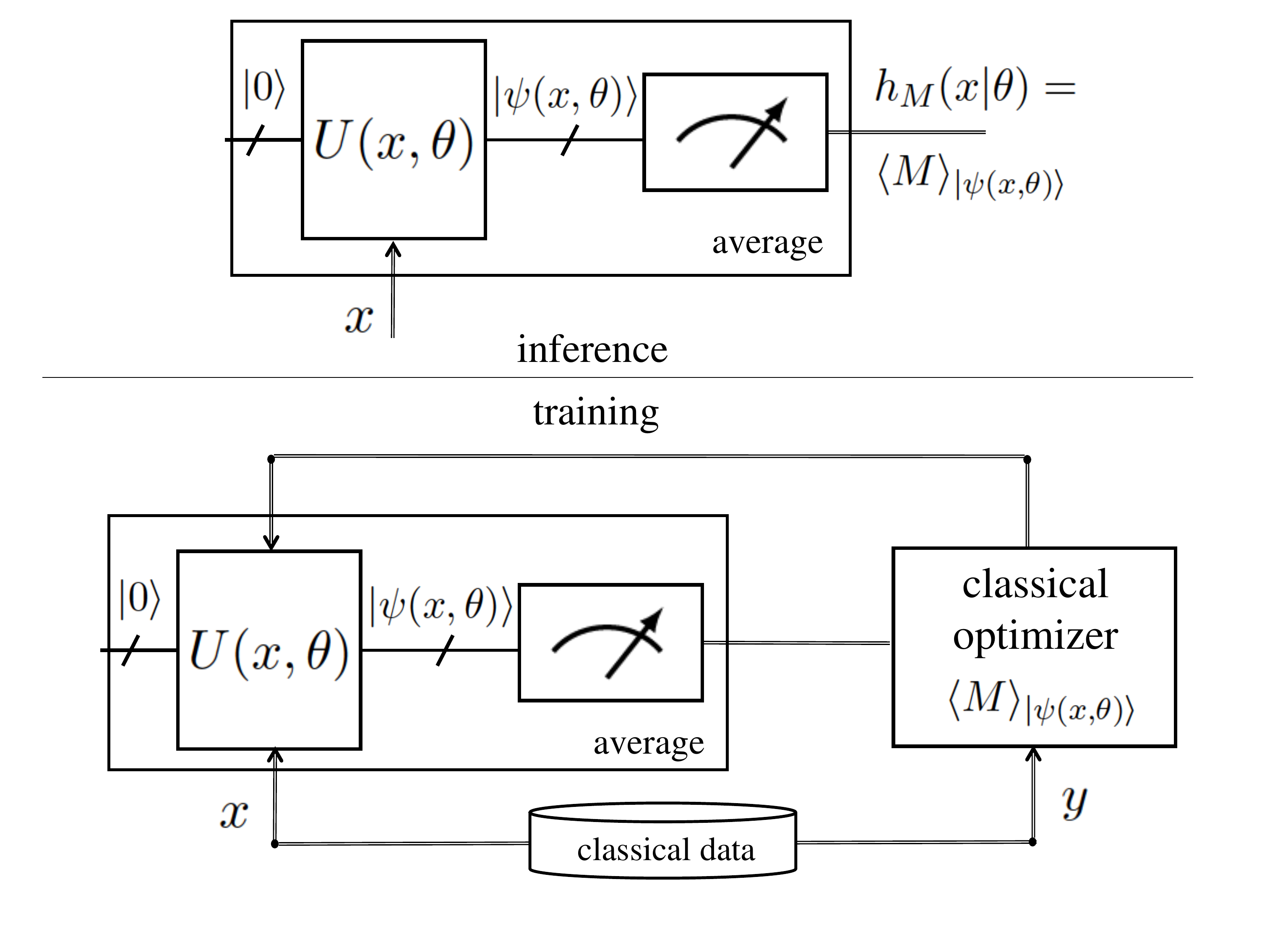}
		\par\end{center}\vspace{-0.4cm} \caption{(top) For inference, a deterministic quantum model implements multiple runs of the PQC $U(x,\theta)$ to estimate the expected value of an observable $M$ (or multiple observables). (bottom) For training, a classical optimizer can apply gradient descent based on estimates of the expectations of observable $M$.  \label{fig:ch6:dsup}}
\end{figure}

\noindent \textbf{Inference}: As illustrated in the top part of Fig. \ref{fig:ch6:dsup}, a PQC can be used to implement  a deterministic function $h_M(x|\theta)$ by measuring the expected value of an observable $M$ under the quantum embedding $|\psi(x,\theta)\rangle$. Accordingly, given an input $x$, the PQC is run multiple times in order to estimate the expectation 
\begin{equation}\label{eq:ch6:expd}
	h_{M}(x|\theta)=\langle\psi(x,\theta)|M|\psi(x,\theta)\rangle=\langle M\rangle_{|\psi(x,\theta)\rangle},
\end{equation}where we have made explicit the dependence of the function $h_{M}(x|\theta)$ on the observable $M$. The function $h_{M}(x|\theta)$ can then be used as described in the previous subsection to carry out regression and classification decisions. Note that vector-valued functions 	$h_M(x|\theta)$ can be similarly implemented by measuring multiple observables, one for each entry of the output vector. 

Importantly, unlike  probabilistic models, deterministic models require multiple runs of the PQC in order to mitigate the effect of shot noise in estimating the expectation (\ref{eq:ch6:expd}) even during inference.

Function $h_{M}(x|\theta)$ in (\ref{eq:ch6:expd}) can be rewritten as \begin{equation}\label{eq:ch6:expd1}
	h_{M}(x|\theta)=\textrm{tr}(M\rho(x,\theta)),
\end{equation} where we have defined the density matrix $\rho(x,\theta)=|\psi(x,\theta)\rangle \langle \psi(x,\theta)|$. This expression makes it clear that the function $h_{M}(x|\theta)$ is linear in the embedding density state $\rho(x,\theta)$. Therefore, a  deterministic quantum model effectively implements a linear
discriminative function in the space of quantum density  embeddings, whose 
weights are defined by the choice of the observable $M$.
As such, deterministic quantum models are akin to classical \textbf{kernel}
methods in that they operate over a large feature space -- the Hilbert
space of dimension $2^{n}$ -- via linear operations.

\noindent \textbf{Training}: Having defined the model output function as in (\ref{eq:ch6:expd}), the loss functions can be directly introduced as detailed in the previous subsection. To elaborate, let us consider the problem of scalar regression with the quadratic loss function (\ref{eq:ch6:quadratic}). In this case, we have \begin{align}\label{eq:ch6:ql}
	\ell(x,y|\theta)&=(y-h_M(x|\theta))^{2}\nonumber\\
	& = (y-\langle M\rangle_{|\psi(x,\theta)\rangle})^{2},
\end{align}which is a function of the expected value $\langle M\rangle_{|\psi(x,\theta)\rangle}$ of observable $M$, as well as of the target variable $y$. This is illustrated in the bottom part of Fig. \ref{fig:ch6:dsup}.

Given this dependence on the expectation of an observable, the ERM problem can be addressed by using the optimization methods discussed in  Sec. \ref{sec:ch6:estgrad}. For example, for the quadratic loss (\ref{eq:ch6:ql}), the gradient of the loss is given as \begin{equation}
	\nabla \ell(x,y|\theta) = - 2(y-\langle M\rangle_{|\psi(x,\theta)\rangle})\nabla \langle M\rangle_{|\psi(x,\theta)\rangle}, 
\end{equation} where the gradient of the observable can be estimated using the parameter shift rule.

\section{Beyond Generic Ansatzes}

In Sec. \ref{sec:ch6:PQC}, we have discussed generic ansatzes that may be broadly applied to a variety of problems. In some settings, one can choose specialized ansatzes
that follow a more principled approach that is tailored to the given problem. In this section, we review two  ways to pursue this approach: \begin{itemize}
\item mimic the operation of classical neural networks, while maintaining the superposition of the quantum states being processed; and
\item relax and parametrize a  quantum algorithm designed using a traditional, formal methodology.
\end{itemize}

\subsection{Mimicking Classical Neural Networks}

Consider first the class of ansatzes that aims at mimicking the way in
which neurons operate in a classical neural network.
A classical neuron produces an output   as a function $g\left(\sum_{k}w_{k}x_{k}\right)$ of a linear combination
$\sum_{k}w_{k}x_{k}$, with
weights $\{w_{k}\}$, of the scalar inputs $\{x_{k}\}$ to the neuron. Activation functions $g(\cdot)$ include the sigmoid function 
$\sigma(a)=(1+\exp(-a))^{-1}$ and the rectified linear unit (ReLU) $\max(0,\cdot)$.

A quantum neuron mimicking the operation of a classical deterministic
neuron would produce an output encoding the quantity  $g\left(\sum_{k}w_{k}x_{k}\right)$
when the inputs are in the computational basis, i.e., classical, while
also being able to operate on superposition states.
Implementing such a neuron is not straightforward due to
the linearity of quantum evolutions, and it typically requires the
introduction of mechanisms involving repeated measurements and measurement-controlled operations. 

As an alternative, a  classical \emph{probabilistic} neuron produces randomized outputs whose probabilities
depend on non-linear functions of the linear combination $\sum_{k}w_{k}x_{k}$. 
An example is given by a binary neuron that outputs
$1$ with probability $\sigma\left(\sum_{k}w_{k}x_{k}\right)$, and $0$ otherwise. 
Classical neurons of this type can be implemented by using measurements as explained in Sec. \ref{sec:ch3:spiking}. Note that such measurements destroy superposition and entanglement.

In both cases, multiple neurons can be combined in different architectures, which may involve multiple layers and/or recurrent processing.

\subsection{Relaxing Handcrafted Quantum Algorithms}
An example of the second type of approaches, which relax and parametrize
a handcrafted quantum algorithm, is given by the Quantum Approximate
Optimization Algorithm (QAOA). Specifically, QAOA provides a problem-dependent ansatz for the VQE as applied to an Ising
Hamiltonian (see Sec. \ref{sec:ch6:vqe}).
QAOA uses knowledge of the loss function observable $F$ in (\ref{eq:ch6:obsF}) 
to define the ansatz of the quantum circuit $U(\theta)$ producing
the state $|\psi(\theta)\rangle$.
The ansatz assumed by QAOA consists of a first layer of Hadamard gates, producing the 
state $|+\rangle \otimes |+\rangle \cdots |+\rangle,$ followed by multiple layers. 
Each layer alternates between the unitary $U(F,\gamma)=\exp(-i\gamma F)$
dependent on the Ising Hamiltonian $F$ and the unitary $U(B,\beta)=\exp(-i\beta B),$ where
$B=\sum_{k=0}^{n-1}X_{k}$ with $X_{k}$ being the Pauli $X$ operator applied
to the $k$-th qubit. Parameters $\gamma$ and $\beta$ can be optimized. 
The circuit provides an asymptotic approximation of the so-called quantum adiabatic
algorithm as the number of layers goes to infinity.

\section{Beyond Angle Encoding}\label{sec:ch6:beyondangle}

In this chapter we have so far adopted angle encoding as a mechanism to convert classical information,  in the form of a real vector $x$, into a quantum state. As we have introduced in Sec. \ref{sec:ch6:angleenc}, angle encoding is an \textbf{analog} strategy in which the $M$ entries of vector $x$ are directly
mapped into the continuous-valued angles of at least $M$ parametrized quantum gates. In this section, we briefly review two alternative ways to encode classical information in a quantum state, and we also give an example of the type of inference methods that may be enabled by such methods.

\subsection{Amplitude Encoding}

Like angle encoding, amplitude encoding is an \textbf{analog} strategy, which is, however, exponentially more efficient in terms of the number of required qubits. Amplitude encoding maps each entry $x_{k}$ of vector $x$ into the $k$-th amplitude of a quantum
state. Accordingly, the encoded state, or quantum embedding, for data vector $x$ is 
\begin{equation}\label{eq:ch6:aa}
	|\psi(x)\rangle=\sum_{k=0}^{M-1}f(x_{k})|k\rangle,
\end{equation}
where $f(\cdot)$ is some function that ensures the normalization condition $\sum_{k=0}^{M-1}|f(x_{k})|^{2}=1$. This equality may be guaranteed, e.g., by setting $f(x_k) = 1/(2^{n/2})\exp(i x_k)$; or by normalizing vector $x$ and then setting $f(x)=\bar{x}$, where $\bar{x}=x/||x||_2$.

Amplitude encoding requires a number of qubits, $n$, equal to the logarithm
of the size of the input, i.e., $n=\log_2(M)$ (assuming that $\log_2(M)$ is an integer). This provides an exponential improvement in terms of efficiency
of the input representation with respect to angle encoding. The computational complexity of amplitude encoding -- that is, the number of quantum operations necessary to produce state (\ref{eq:ch6:aa}) -- is, however, generally linear in the size $M$ of the input 
 as for angle encoding.

\subsection{Basis Encoding}

Unlike angle and amplitude encoding, basis encoding is a \textbf{digital} strategy, and
it enables the encoding of an \emph{entire data set} $\mathcal{D}$
of data points $x$ into a single quantum state. As we detail next, the number of qubits dictates the resolution of the representation of the input $x$.

Basis encoding first converts each data point $x$ in the binary domain, producing a binary
string $x_{b}\in\{0,1\}^{n}$. Note that the number of bits $n$ determines the precision level of the binary representation. Then, the entire data set $\mathcal{D}$ is mapped to the quantum state
\begin{equation}\label{eq:ch6:npqs}
	|\psi(\mathcal{D})\rangle=\frac{1}{Z}\sum_{x\in\mathcal{D}}|x_{b}\rangle,
\end{equation}
where $Z$ is a suitable constant to ensure a unitary norm. State (\ref{eq:ch6:npqs}) encodes a superposition of all entries of the data set $\mathcal{D}$.

\subsection{Non-Parametric Quantum Machine Learning}

How can quantum states such as (\ref{eq:ch6:npqs}) be used in machine learning applications? In this section, we outline an approach for \textbf{density estimation} that is akin to \textbf{non-parametric} methods in classical machine learning. This is in the sense that the approach makes decisions by using the entire data set, in the form of state (\ref{eq:ch6:npqs}), without requiring a training stage. 

Given a data set $\text{\ensuremath{\mathcal{D}}=}\{x_{1},...,x_{|\mathcal{D}|}\}$ of $|\mathcal{D}|$ samples, density estimation aims at estimating the probability distribution $p(x)$ underlying the generation of the data. Consider a new point $x^*$ at which we would like to estimate the distribution $p(x^*)$. Intuitively, the probability  $p(x^*)$ should be larger the more ``similar'' vector $x^*$ is to the data points in the training set $\mathcal{D}$. Therefore, one can conceive the use of the inner product 
\begin{equation}
\hat{p}(x^{*})	\propto |\langle\psi(x^{*})|\psi(\mathcal{D})\rangle|^{2}
\end{equation}
as a non-parametric (unnormalized) estimate of probability distribution $p(x^{*})$. The estimate $\hat{p}(x^{*})$ can be evaluated as described in Sec. \ref{sec:ch3:spiking}.


%
%
%
%
%
%

\section{Conclusions}
This chapter has provided an introduction to quantum machine learning, focusing on the most common setting in which data are classical and processing is quantum. For this ``CQ'' case, a parametrized quantum circuit (PQC) is optimized via a classical processor by minimizing functions of the expected values of some observables, as well as of (classical) data. Gradient-based optimization for PQCs is significantly less efficient than its counterpart  for standard machine learning models. In fact, estimating the gradient entails a complexity that scales linearly with the number of model parameters. Furthermore, it requires multiple runs through the model in order to mitigate shot noise due to the inherent randomness of quantum measurements.

The inefficiency in estimating gradients is  one of the many differences between the fields of quantum and classical machine learning. While classical machine learning is typically applied to unstructured, large data sets with the goal of maximizing accuracy, ``CQ'' quantum machine learning is often applied to small, well-structured, data sets, and \textbf{quantum advantages} are often measured in terms of efficiency. Efficiency is evaluated in terms of the number of physical resources needed to implement a model: number of model parameters, number of bits/ qubits, training time, inference  time, and so on. While it is possible to prove such advantages for very specific problems, the current practice appears to be that of making limited comparisons via heuristic arguments and experiments, or by sidestepping the question of ``quantum advantages'' altogether. 

Rather than focusing on establishing formal advantages, research in the field of quantum machine learning may focus on establishing effective building blocks for quantum machine learning models; understanding the relative merits of deterministic and probabilistic models; evaluating the impact of different data encoding techniques; developing effective software platforms; and investigating at a theoretical level the problem of generalization for given ansatzes.

\section{Recommended Resources}\label{sec:ch6:rr}

Quantum machine learning is an emerging, very active, field of research. A useful starting point is the book \cite{schuld21}, which contains many pointers to papers published before 2021. Monte Carlo estimates of gradients in classical machine learning are reviewed in \cite{mohamed2020monte,simeone22}. The barren plateaus problem is detailed in \cite{mcclean2018barren}. An example of QC machine learning is provided by \cite{youssry2019efficient}. Quantum generative adversarial networks are presented in \cite{lloyd2018quantum}; while quantum variational autoencoders for quantum compression are studied in \cite{romero2017quantum}. Hybrid classical-quantum machine learning models are studied in \cite{benedetti2018quantum}. For more recent work, the reader is referred to the online repository arXiv (https://arxiv.org/), in which most up-to-date pre-prints can be found.

\section{Problems}
\begin{enumerate}
\item Write the matrix representation (in the computational basis) of
the three unitaries $Z\otimes Z$, $CZ$, and $R_{ZZ}(\theta)=\exp\left(-i\frac{\theta}{2}Z\otimes Z\right)$.
Which one(s) amount to local operations and why? Obtain eigenvalues and eigenvectors.
Which one(s) are also observables?
\item  Show that the Pauli $ZZ$-rotation can be implemented using two
CNOT gates and a Pauli $Z$-rotation.
\item  Prove that using only Pauli $Y$-rotations allows the mean-field
ansatz to produce all possible vectors in the computational basis.
\item  Describe the quantum adiabatic theorem and the quantum adiabatic
algorithm (see recommended resources).
\item  Derive the parameter shift rule (see Appendix B).
\item  Demonstrate some practical examples of applications of quantum
machine learning methods. (You can browse arXiv or other repositories.)
\item  Describe an example of a QQ machine learning problem.
\end{enumerate}

\section*{Appendix A: On Unitaries with Pauli Strings as Generators}
%
%
In this appendix, we derive some useful properties of parametrized unitaries whose generators are given by Pauli strings. Examples include single-qubit and two-qubit Pauli rotation gates. Such unitaries can be written as 
\begin{equation}
	U(\theta)=\exp\left(-i\frac{\theta}{2}P\right),
\end{equation}
where  $P$ is a Pauli string. 

A Pauli string $P$ has two 
distinct eigenvalues $+1$ and $-1$. Therefore, it can be expressed as  $P=\Pi_{0}-\Pi_{1},$ where $\Pi_{0}=\sum_{x:\textrm{ \ensuremath{\lambda_{x}=1}}}v_{x}$
is the projection matrix into the subspace spanned by the eigenvectors $|v_x\rangle$ of matrix $P$ that are  
associated with eigenvalues $\lambda_x=1$, and $\Pi_{1}=\sum_{x:\textrm{ \ensuremath{\lambda_{x}=-1}}}v_{x}$
is the projection matrix into the subspace spanned by the eigenvectors
$|v_x\rangle$ associated with eigenvalues $\lambda_x=-1$. 
Note that we have the equality $\Pi_{0}+\Pi_{1}=I$, since the set of all eigenvectors
span the entire Hilbert space.

Given the above, and using the definition (\ref{eq:ch1:matfun}) of a function of a normal matrix, we obtain the equality
\begin{align}
	U(\theta) & =\sum_{x=0}^{2^{n}-1}\exp\left(-i\frac{\theta}{2}\lambda_{x}\right)v_{x}\nonumber\\
	& =\Pi_{0}\exp\left(-i\frac{\theta}{2}\right)+\Pi_{1}\exp\left(i\frac{\theta}{2}\right)\nonumber\\
	& =\Pi_{0}\left(\cos\left(\frac{\theta}{2}\right)-i\sin\left(\frac{\theta}{2}\right)\right)+\Pi_{1}\left(\cos\left(\frac{\theta}{2}\right)+i\sin\left(\frac{\theta}{2}\right)\right)\nonumber\\
	& =\cos\left(\frac{\theta}{2}\right)\left(\Pi_{0}+\Pi_{1}\right)-i\sin\left(\frac{\theta}{2}\right)\left(\Pi_{0}-\Pi_{1}\right)\nonumber\\
	& =\cos\left(\frac{\theta}{2}\right)I-i\sin\left(\frac{\theta}{2}\right)P.
\end{align}
Note that if we set $\theta=\pm\pi/2$, we get the equality
\begin{equation}\label{eq:ch6:pi2}
	U\left(\pm\frac{\pi}{2}\right)=\frac{1}{\sqrt{2}}\left(I\mp iP\right).
\end{equation}

\section*{Appendix B: The Parameter Shift Rule}

Consider one of the scalar parameters of the parameter vector $\theta=[\theta_{1},...,\theta_{D}]^{T}$, say $\theta_d$ for some $d\in \{1,...,D\}$.
We are interested in computing the partial derivative $\partial\langle O\rangle_{|\psi(\theta)\rangle}/\partial\theta_{d}$
for the expectation 
\begin{equation}\label{eq:ch6:app1}
	\langle O\rangle_{|\psi(\theta)\rangle}=\langle\psi(\theta)|O|\psi(\theta)\rangle=\langle0|U^{\dagger}(\theta)OU(\theta)|0\rangle
\end{equation}
of some observable $O$ under the output state $U(\theta)|0\rangle$
produced by a PQC. Note that, in some cases, the unitary $U(\theta)$ may also be a function of the  input data $x$, and this can be easily accommodated in the analysis
by considering unitaries of the form $U(x,\theta)$. 
We also observe that loss functions of the form (\ref{eq:ch6:app1}) are also relevant
for QQ problems, in which the observable $O$ may depend on the target
quantum state.

As discussed in this chapter, in many PQCs of interest, parameter $\theta_{d}$ enters
as part of the overall unitary in a single gate of the form 
\begin{equation}\label{eq:ch6:Udapp}
	U_{d}(\theta_{d})=\exp\left(-i\frac{\theta_{d}}{2}G_{d}\right),
\end{equation}
where $G_{d}$ is the generator Hermitian matrix.
Examples of gates (\ref{eq:ch6:Udapp}) include the Pauli rotation matrices used in the hardware-efficient
ansatz. If the parameter $\theta_d$ entered multiple gates of the form (\ref{eq:ch6:Udapp}), the corresponding partial derivatives to be calculated in this appendix would need to be summed.

According to the assumption stated in the previous paragraph, we can write the overall unitary transformation as
\begin{equation}\label{eq:ch6:overall}
	U(\theta)=V(\theta_{-d})U_{d}(\theta_{d})W(\theta_{-d})
\end{equation}
for some unitaries $V(\theta_{-d})$ and $W(\theta_{-d})$ representing the gates applied before
and after the $d$-th gate $U_d(\theta_{d})$. 
Unitaries $V(\theta_{-d})$ and $W(\theta_{-d})$ generally depend
on the parameters in vector $\theta$ other than the parameter $\theta_{d}.$
This dependence is captured by the notation \textrm{$\theta_{-d}$. }

Using (\ref{eq:ch6:overall}), we write the expected observable as 
\begin{equation}
	\langle O\rangle_{|\psi(\theta)\rangle}=\langle\phi_{d}|U_{d}^{\dagger}(\theta_{d})R_{d}U_{d}(\theta_{d})|\phi_{d}\rangle,
\end{equation}
where $|\phi_{d}\rangle=W(\theta_{-d})|0\rangle$ and $R_{d}=V(\theta_{-d})^{\dagger}OV(\theta_{-d})$.
Note that matrix $R_{d}$ is Hermitian.
This way, we have incorporated gates following and preceding the $d$-th
gate in the quantum state $|\phi_{d}\rangle$ and in the observable
$R_{d}$, respectively.
The subscript in $|\phi_{d}\rangle$ and $R_{d}$ is used to emphasize
that both quantities depend on the index $d$.

The partial derivative is computed as 
	\begin{align}\label{eq:ch6:pd}
		\frac{\partial\langle O\rangle_{|\psi(\theta)\rangle}}{\partial\theta_{d}}= & \frac{\partial\langle\phi_{d}|U_{d}^{\dagger}(\theta_{d})R_{d}U_{d}(\theta_{d})|\phi_{d}\rangle}{\partial\theta_{d}}\nonumber \\
		= & \langle\phi_{d}|U_{d}^{\dagger}(\theta_{d})R_{d}\nabla_{\theta_{d}}U_{d}(\theta_{d})|\phi_{d}\rangle \nonumber\\ & +\langle\phi_{d}|\nabla_{\theta_{d}}U_{d}^{\dagger}(\theta_{d})R_{d}U_{d}(\theta_{d})|\phi_{d}\rangle,
	\end{align}
where $\nabla_{\theta_{d}}U_{d}(\theta_{d})$ represents the matrix
of derivatives of the unitary $U_{d}(\theta_{d}),$ and a similar definition
applies to $\nabla_{\theta_{d}}U_{d}^{\dagger}(\theta_{d}).$
We have 
\begin{align}
	\nabla_{\theta_{d}}U_{d}(\theta_{d}) & =\nabla_{\theta_{d}}\exp\left(-i\frac{\theta_{d}}{2}G_{d}\right)\nonumber \\
	& =-\frac{1}{2}iG_{d}\exp\left(-i\frac{\theta_{d}}{2}G_{d}\right)=-\frac{1}{2}iG_{d}U_{d}(\theta_{d})
\end{align}
and 
\begin{equation}
	\nabla_{\theta_{d}}U_{d}^{\dagger}(\theta_{d})=\frac{i}{2}G_{d}U_{d}^{\dagger}(\theta_{d}).
\end{equation}

We conclude that the following equalities hold
\begin{align}\label{eq:ch6:r1}
	U_{d}(\theta_{d})+2\nabla_{\theta_{d}}U_{d}(\theta_{d}) & =U_{d}(\theta_{d})-iG_{d}U_{d}(\theta_{d})\nonumber\\
	& =(I-iG_{d})U_{d}(\theta_{d})\nonumber\\
	& =\sqrt{2}U_{d}(\pi/2)U_{d}(\theta_{d})\nonumber\\
	& =\sqrt{2}U_{d}(\theta_{d}+\pi/2),
\end{align}where we have used (\ref{eq:ch6:pi2}) for the third equality. 
Similarly, we have 
	\begin{equation}\label{eq:ch6:r2}
		U_{d}(\theta_{d})-2\nabla_{\theta_{d}}U_{d}(\theta_{d})=\sqrt{2}U_{d}(-\pi/2)U_{d}(\theta_{d})=\sqrt{2}U_{d}(\theta_{d}-\pi/2).
	\end{equation}

Finally, we write the partial derivative (\ref{eq:ch6:pd}) as 
	\begin{align}
		\frac{\partial\langle O\rangle_{|\psi(\theta)\rangle}}{\partial\theta_{d}}= & \frac{1}{4}\langle\phi_{d}|(U_{d}(\theta_{d})+2\nabla_{\theta_{d}}U_{d}(\theta_{d}))^{\dagger}R_{d}(U_{d}(\theta_{d})+2\nabla_{\theta_{d}}U_{d}(\theta_{d}))|\phi_{d}\rangle \nonumber\\
		& -\frac{1}{4}\langle\phi_{d}|(U_{d}(\theta_{d})-2\nabla_{\theta_{d}}U_{d}(\theta_{d}))^{\dagger}R_{d}(U_{d}(\theta_{d})-2\nabla_{\theta_{d}}U_{d}(\theta_{d}))|\phi_{d}\rangle,\nonumber\\
		= & \frac{1}{2}\langle\phi_{d}|U_{d}(\theta_{d}+\pi/2){}^{\dagger}R_{d}U_{d}(\theta_{d}+\pi/2)|\phi_{d}\rangle\nonumber\\
		& -\frac{1}{2}\langle\phi_{d}|U_{d}(\theta_{d}-\pi/2){}^{\dagger}R_{d}U_{d}(\theta_{d}-\pi/2)|\phi_{d}\rangle, 
	\end{align} where the first equality can be checked by direct comparison with (\ref{eq:ch6:pd}), and the second equality follows from (\ref{eq:ch6:r1})-(\ref{eq:ch6:r2}). This concludes the derivation.

\begin{acknowledgements}
I would like to thank Prof. Lajos Hanzo for the inspiration, as well as Prof. Bipin Rajendran, Dr. Hari Chittoor, Dr. Sharu Jose, and Dr. Ivana Nikoloska, who have been ideal companions during my ongoing journey of discovery of the field of quantum machine learning. My gratitude goes also to the other members of my research team at King's  who have provided useful feedback, comments, and encouragement: Kfir Cohen,  Dr. Sangwoo Park, Clement Ruah, and Matteo Zecchin.

\end{acknowledgements}

\backmatter  

\printbibliography

@book{mermin2007quantum,
	title={Quantum computer science: an introduction},
	author={Mermin, N. David},
	year={2007},
	publisher={Cambridge {U}niversity {P}ress}
}

@book{lipton2021introduction,
	title={Introduction to quantum algorithms via linear algebra},
	author={Lipton, Richard J and Regan, Kenneth W},
	year={2021},
	publisher={MIT {P}ress}
}

@article{wood2011tensor,
	title={Tensor networks and graphical calculus for open quantum systems},
	author={Wood, Christopher J and Biamonte, Jacob D and Cory, David G},
	journal={arXiv preprint arXiv:1111.6950},
	year={2011}
}

@book{susskind2014quantum,
	title={Quantum mechanics: the theoretical minimum},
	author={Susskind, Leonard and Friedman, Art},
	year={2014},
	publisher={Basic {B}ooks}
}

@book{penrose2005road,
	title={The road to reality: A complete guide to the laws of the universe},
	author={Penrose, Roger},
	year={2005},
	publisher={Random {H}ouse}
}

@book{wilde2013quantum,
	title={Quantum information theory},
	author={Wilde, Mark M},
	year={2013},
	publisher={Cambridge {U}niversity {P}ress}
}

@book{nielsen10,
	title={Quantum computation and quantum information},
	author={Nielsen, Michael A. and Chuang, Isaac L.},
	year={2010},
	publisher={Cambridge {U}niversity {P}ress}
}

@book{watrous2018theory,
	title={The theory of quantum information},
	author={Watrous, John},
	year={2018},
	publisher={Cambridge {U}niversity {P}ress}
}

@book{kaye2006introduction,
	title={An introduction to quantum computing},
	author={Kaye, Phillip and Laflamme, Raymond and Mosca, Michele},
	year={2006},
	publisher={Oxford {U}university {P}ress}
}

@book{bergou2021quantum,
	title={Quantum information processing},
	author={Bergou, J{\'a}nos A and Hillery, Mark and Saffman, Mark},
	year={2021},
	publisher={Springer}
}

@book{hidary2019quantum,
	title={Quantum computing: an applied approach},
	author={Hidary, Jack D and Hidary, Jack D},
	volume={1},
	year={2019},
	publisher={Springer}
}

@book{rohde2021quantum,
	title={The quantum Internet: The second quantum revolution},
	author={Rohde, Peter P},
	year={2021},
	publisher={Cambridge {U}niversity {P}ress}
}

@book{schuld21,
	title={Machine learning with quantum computers},
	author={Schuld, Maria and Petruccione, Francesco},
	year={2021},
	publisher={Springer}
}

@article{mcclean2018barren,
	title={Barren plateaus in quantum neural network training landscapes},
	author={McClean, Jarrod R and Boixo, Sergio and Smelyanskiy, Vadim N and Babbush, Ryan and Neven, Hartmut},
	journal={Nature {C}ommunications},
	volume={9},
	number={1},
	pages={1--6},
	year={2018},
	publisher={Nature {P}ublishing {G}roup}
}

@article{romero2017quantum,
	title={Quantum autoencoders for efficient compression of quantum data},
	author={Romero, Jonathan and Olson, Jonathan P and Aspuru-Guzik, Alan},
	journal={Quantum {S}cience and {T}echnology},
	volume={2},
	number={4},
	year={2017},
	publisher={{IOP} Publishing}
}

@article{lloyd2018quantum,
	title={Quantum generative adversarial learning},
	author={Lloyd, Seth and Weedbrook, Christian},
	journal={Physical {R}eview {L}etters},
	volume={121},
	number={4},
	year={2018},
	publisher={{APS}}
}

@article{youssry2019efficient,
	title={Efficient online quantum state estimation using a matrix-exponentiated gradient method},
	author={Youssry, Akram and Ferrie, Christopher and Tomamichel, Marco},
	journal={New {J}ournal of {P}hysics},
	volume={21},
	number={3},
	year={2019},
	publisher={{IOP} Publishing}
}

@book{simeone22,
	title={Machine learning for engineers},
	author={Simeone, Osvaldo},
	year={2022},
	publisher={Cambridge {U}niversity {P}ress}
}

@article{mohamed2020monte,
	title={Monte {C}arlo Gradient Estimation in Machine Learning},
	author={Mohamed, Shakir and Rosca, Mihaela and Figurnov, Michael and Mnih, Andriy},
	journal={Journal {M}achine {L}earning Research},
	volume={21},
	number={132},
	pages={1--62},
	year={2020}
}

@article{benedetti2018quantum,
	title={Quantum-assisted {H}elmholtz machines: A quantum--classical deep learning framework for industrial datasets in near-term devices},
	author={Benedetti, Marcello and Realpe-G{\'o}mez, John and Perdomo-Ortiz, Alejandro},
	journal={Quantum {S}cience and {T}echnology},
	volume={3},
	number={3},
	pages={034007},
	year={2018},
	publisher={{IOP} Publishing}
}

\end{document}